\documentclass[twoside,11pt]{article}

\usepackage{jmlr2e}

\usepackage{stackengine}
\usepackage{filecontents}

\setstackgap{S}{8pt}

\usepackage{amsmath,amssymb,comment,tikz,pgf,bm}
\usepackage{amsopn, amsfonts, leftidx}
\usetikzlibrary{shapes}

\usepackage{algorithm}
\usepackage{algpseudocode}
\usepackage{graphicx}
\usepackage{booktabs}
\usepackage{listings}
\usepackage[mode=buildnew]{standalone}

\usepackage{xr}

\newtheorem{thm}{Theorem}[section]
\newtheorem{lem}{Lemma}[section]

\DeclareMathOperator{\pa}{pa}

\DeclareMathOperator{\pre}{pre}

\DeclareMathOperator{\nb}{nb}
\DeclareMathOperator{\post}{post}

\DeclareMathOperator{\E}{\mathbb{E}}

\newcommand{\G}{\mathop{{\cal G}}}

\newcommand{\cond}{\,|\,}
\newcommand{\COND}{\,\bigg|\,}
\newcommand{\Cond}{\,\Big|\,}
\newcommand{\con}{;}
\newcommand{\IF}{\texttt{IF}}
\DeclareMathOperator*{\argmin}{arg\,min}
\newcommand{\EXP}{\mathbb{E}}

\newcommand{\N}{\mathcal{N}}
\newcommand{\bC}{\vec{C}}
\newcommand{\bc}{\vec{c}}
\newcommand{ \ETA}{^{(\eta)}}

\newcommand{\bO}{\vec{O}}
\usepackage{enumitem} 

\def\ci{\perp\!\!\!\perp}

\newcommand{\ilya}[1]{{\color{red!70!blue} #1}}
\usepackage[normalem]{ulem}
\newcommand{\stkout}[1]{\ifmmode\text{\sout{\ensuremath{#1}}}\else\sout{#1}\fi}

\usepackage{lastpage}
\jmlrheading{1}{2020}{1-\pageref{LastPage}}{3/27}{??/??}{shpitser20}{Ilya Shpitser, Chan Park,  Ryan Andrews, and Eric Tchetgen Tchetgen}

\ShortHeadings{Symmetric Treatment Decomposition}{Shpitser, Park, Andrews, and Tchetgen Tchetgen}
\firstpageno{1}

\begin{document}

\title{
Symmetric Treatment Decomposition Of Spillover Effects}


\author{\name Ilya Shpitser \email ilyas@cs.jhu.edu\\
\addr Department of Computer Science\\
Johns Hopkins University\\
Baltimore, MD 21218, USA
\AND 
\name Chan Park \email chanpk@wharton.upenn.edu\\
\addr Department of Statistics and Data Science\\
The Wharton School, University of Pennsylvania\\
265 South 37th Street, Philadelphia, PA 19104, USA 
\AND
\name Ryan M.\ Andrews \email 
ryana@bu.edu\\
\addr Department of Epidemiology\\
Boston University\\
715 Albany Street, Boston, MA 02118, USA
\AND
\name Eric J.\ Tchetgen Tchetgen \email ett@wharton.upenn.edu\\
\addr Department of Statistics and Data Science\\
The Wharton School, University of Pennsylvania\\
265 South 37th Street, Philadelphia, PA 19104, USA  
}

\editor{Peter Spirtes}

\maketitle

\begin{abstract}
Classical causal inference assumes treatments meant for a given unit do not
have an effect on other units.  This assumption is violated in \emph{interference problems},
where new types of spillover causal effects arise, and causal inference becomes
much more difficult.  In addition, interference introduces a unique complication where
variables may transmit treatment influences to each other, which is a relationship that
has some features of a causal one, but is symmetric.

In this paper, we develop a new approach to decomposing the spillover effect into unit-specific components that extends the DAG based treatment decomposition approach to mediation of Robins and Richardson to causal models that admit stable symmetric relationships among variables in a network.  We discuss two interpretations of such models:
a network 
structural model interpretation, and an interpretation based on equilibrium of structural equation models discussed in \citep{lauritzen02chain}.  We show that both interpretations yield identical identification theory, and
give conditions for components of the spillover effect to be identified.

We discuss statistical inference for identified components of the spillover effect, including a maximum likelihood estimator, and a doubly robust estimator for the special case of two interacting outcomes. We verify consistency and robustness of our estimators via a simulation study,
and illustrate our method by assessing the causal effect of education attainment on depressive symptoms using the data on households from the Wisconsin Longitudinal Study.
\end{abstract}

\begin{keywords}
chain graphs; graphical models; interference; mediation analysis; semi-parametric inference
\end{keywords}

\section{Introduction}

A standard assumption in causal inference is absence of unit interference,
which asserts that giving treatment to a particular unit only affects the response
of that unit.  While a sensible assumption in many statistical applications, there are
settings where this assumption is not reasonable.  A classic example from infectious
disease epidemiology is herd immunity: vaccinating a subset of a population may grant
immunity to the unvaccinated members of the population.


The presence of interference introduces a number of conceptual difficulties.
First, unlike classical causal inference, variables associated with experimental units can no longer
be viewed as independent realizations
of some underlying distribution. Second, new types of causal effects called \emph{spillover effects}
arise, which quantify the degree to which treatments for one unit affect the outcome of another unit.
Like total causal effects from classical causal inference, it may be of scientific interest to decompose
spillover effects into direct and indirect components, and more generally into components that arise due to
unit interactions in a network.

In the context of infectious disease epidemiology, the direct and indirect components of the spillover effect are called the \emph{infectiousness effect}, and the \emph{contagion effect}, respectively \citep{vanderweele12components}.
In the context of data-driven online marketing, decomposing the effect of
an advertisement on purchasing or voting behavior of a set of people forming a social network into
a set of unit-specific components may also be of substantive interest.  In particular, the magnitude of these
unit-specific effects can help quantify which sorts of people drive the overall response to an
advertisement in a network.


Prior work has used ideas from the mediation analysis literature to obtain decompositions of
spillover effects \citep{vanderweele12components}.  Such an approach is not appropriate in
interference settings where unit outcomes do not form a natural causal ordering.
We propose an alternative approach to
the decomposition of spillover effects in
interference problems that does not require such an ordering.

This approach is based on causal models that impose Markov restrictions represented by chain graphs \citep{lauritzen96graphical}.  We discuss two interpretations of such models: sampling of a structural equation model with feedback leading to an equilibrium \citep{lauritzen02chain}, or 
a network structural model imposed on counterfactual distributions derived from a standard causal model defined on blocks of variables and represented by a directed acyclic graph (DAG).  Using these models, we define a symmetric generalization of interventionist mediation analysis \citep{robins10alternative,rrs20volume_mediation_arxiv}.


\subsection{A Motivating Example And Outline Of Contributions}
\label{subsec:example}

We begin with an example described in \citep{vanderweele12components}, and motivated by a study described in \citep{trollfors98immunization}.
In this hypothetical example, one-year-old children at a day care center are randomized to receive a vaccine (denoted by $A=a$) or placebo (denoted by $A=a'$) against a particular pathogen serotype prevalent in children attending day care.  A number of questions may be of interest in such a study.  A question of primary interest may be the causal effect of vaccination on pathogen colonization status in the child (denoted by $Y_1$).  A secondary question would be a similar causal effect: that of vaccination on the pathogen colonization status in the care provider (e.g. mother) of the child, denoted by $Y_2$.  Note that since the child and care provider live in the same household, the potential for disease spread implies the outcomes $(Y_1,Y_2)$ should be modelled as dependent random variables.  In other words, variable pairs pertaining to children and care providers should not be viewed as independent realizations of an underlying distribution, but as forming a dependent \emph{dyad} data structure \citep{kenny06dyadic}.

In addition to variables explicitly mentioned, that is the treatment $A$, and the outcomes of the child/care provider dyad $(Y_1,Y_2)$, the study may also record a set of relevant baseline covariates
$\vec{C}$, for both the child and the care provider.  These covariates may be used to assign the vaccine or placebo treatment (corresponding to values of $A$) via a distribution $p(A \mid \vec{C})$
corresponding to a known design rule, or an assignment probability that must be learned from data.

Causal effects are often conceptualized via potential outcome random variables \citep{neyman23app,rubin76inference}.  For example, the potential outcome $Y_1(a)$ denotes colonization status in the child had, possibly contrary to fact, the child been vaccinated.  Causal effects are generally defined using potential outcomes as contrasts on the mean scale.  For instance, the \emph{average causal effect (ACE)} of vaccination on colonization status of the child would be defined as: $\E[Y_1(a)] - \E[Y_1(a')]$, while the similar \emph{average spillover effect} of vaccination of the child on colonization status of the care provider would be defined as $\E[Y_2(a)] - \E[Y_2(a')]$.

Decomposition of an established spillover effect into components is of interest in cases where these components can be isolated and have a substantive interpretation.
In our example, the presence of an indirect component of the spillover effect, known as the contagion effect, indicates that vaccinating units directly lessens the chance of infection of those units, and thus the chance of those units passing the infection on.  Similarly, the presence of a direct component of the spillover effect, known as the infectiousness effect, indicates that vaccinating units may modify the chance of infection in some other way, perhaps by suppressing more virulent strains from propagating.


There are two complementary views of causal relationships underlying variables in the example we outlined, which influence how the spillover effect and its 
components are defined, identified and estimated.
The distinction between the two views concerns the causal relationships of $A$ and $\vec{C}$ and outcomes of the child and the caregiver: $Y_1$ and $Y_2$.  The modeling choice made in \citep{vanderweele12components} proceeds by assuming that a child's caregiver is only likely to get infected with the pathogen through their child, who in turn would have obtained the infection from daycare.  This assumption, which is sensible if the pathogen is a childhood disease such as pertussis, imposes a natural causal ordering where variables $A$ and $\vec{C}$ cause both $Y_1$ and $Y_2$, and $Y_1$ causes $Y_2$.  A popular representation of causal models with variables that follow a known ordering is via directed acyclic graphs (DAGs).  Such a graph for our model is shown in Fig.~\ref{fig:med} (a), with vertices representing random variables in the problem, and directed edges between vertices meaning, in the sense to be made precise below, ``direct causation.''
In this view, spillover effects can be defined as standard causal effects, and direct and indirect components of the spillover effects can be defined using tools of mediation analysis, as described in \citep{vanderweele12components}, and below.

However, this approach is less sensible for pathogens that could be caught by either the child or the caregiver (such as COVID-19), since there is no unambiguous causal order on the outcomes $Y_1,Y_2$ in such cases.  A popular approach for causal models of this sort has been developed in the partial interference literature \citep{hudgens08toward,tchetgen12on}.
In the partial interference view, outcome variables $(Y_1,Y_2)$ (and their corresponding counterfactuals $(Y_1(a),Y_2(a))$) are defined \emph{jointly} as a block, with no clear causal ordering on variables within the block.

In this paper, we show that the spillover effect, and its 
components can be formally represented as potential outcomes in
causal models that do not require a total causal ordering on variables, and allow jointly defined counterfactuals of the above sort.
In particular, we consider causal models allowing
variable relationships that are symmetric, and stable (meaning that they remain invariant under interventions).
We discuss two alternative versions of such models: a feedback process generated by structural equations \citep{lauritzen02chain}, and a
network structural model that places restrictions on potential outcomes represented by a directed acyclic graph (DAG) defined on blocks of variables.  
Both models imply Markov restrictions on counterfactual distributions that correspond to graphical models that allow both asymmetric (directed) and symmetric (undirected) relationships between variables \cite{lauritzen96graphical}.
An example of a graph associated with such models, called a \emph{chain graph}, is shown in Fig.~\ref{fig:med} (d).
See \citep{lauritzen96graphical} and \citep{cox93linear} for additional discussion of graphical models with symmetric relationships between variables.

We further show how these effects
may be identified via a key 
assumption that generalizes assumptions made in interventionist mediation analysis \citep{robins10alternative,rrs20volume_mediation_arxiv}.
In the dyadic context, we call the resulting identifying functional for direct and indirect components of the spillover effect \emph{the symmetric mediation formula},
due to the fact that it can be viewed as an appropriate generalization of the mediation formula in 
DAG models \citep{pearl11cmf}.
In general network contexts, unit-specific effects we define may be viewed as natural analogues of general edge specific interventions arising in mediation analysis in
DAGs \citep{shpitser15hierarchy}.

In addition, we demonstrate that identifying assumptions in our models impose restrictions
on the observed data law, which leads to falsifiability (but not testability) of our models,
a 
feature not present in the classical mediation setting.  Finally, we consider
estimation of functionals identifying components of spillover effects 
as an inference problem in statistical chain graph models.
We derive maximum likelihood estimators that are straightforward to implement, as well as 
a semi-parametric doubly robust estimator for the symmetric mediation formula.

Finally, we apply our derived estimators to both a real-world example and simulated data. We use data taken from the Wisconsin Longitudinal Study, a longitudinal cohort of Wisconsin high school graduates and their spouses, to decompose the effect of educational attainment on depressive symptoms, taking into account covariates and likely interference between husband and wife pairs. To illustrate the behavior of our doubly robust estimator, we designed a simulation study for spillover effect components in both randomized treatment and non-randomized treatment settings.

\begin{figure}
\begin{center}
\begin{tikzpicture}[>=stealth, node distance=0.9cm]
\tikzstyle{format} = [draw, very thick, circle, minimum size=5.0mm,
inner sep=0pt]
\tikzstyle{square} = [draw, very thick, rectangle, minimum size=5mm]

\begin{scope}
\path[very thick, ->]
node[format] (C) {$\vec{C}$}
node[format, below of=C] (A) {$A$}
node[below of=A] (d) {}
node[format, below of=d] (M) {$Y_1$}
node[format, right of=M] (Y) {$Y_2$}

(A) edge[blue] (M)
(A) edge[blue] (Y)
(M) edge[blue,->] (Y)

(C) edge[blue] (A)
(C) edge[blue, bend right=30] (M)
(C) edge[blue, bend left=20] (Y)

node[below of=M,yshift=0.4cm,xshift=0.45cm]
(l) {$(a)$}
;
\end{scope}
\begin{scope}[xshift=2.0cm]
\path[very thick, ->]
node[format] (C) {$\vec{C}$}
node[format, below of=C] (A) {$A$}
node[format, below of=A] (A1) {$\tilde{A}^1$}
node[format, right of=A1] (A2) {$\tilde{A}^2$}
node[format, below of=A1] (M) {$Y_1$}
node[below of=A2] (dummy) {}
node[format, right of=dummy] (Y) {$Y_2$}

(A) edge[red] (A1)
(A) edge[red] (A2)
(A1) edge[blue] (M)
(A2) edge[blue] (Y)
(M) edge[blue,->] (Y)

(C) edge[blue] (A)
(C) edge[blue, bend right=30] (M)
(C) edge[blue, bend left=20] (Y)

(A1) edge[blue] (Y)
(A2) edge[blue] (M)

node[below of=M,yshift=0.4cm,xshift=0.45cm]
(l) {$(b)$}
;
\end{scope}
\begin{scope}[xshift=4.6cm]
\path[very thick, ->]
node[format] (C) {$\vec{C}$}
node[format, below of=C] (A) {$A$}
node[format, below of=A] (A1) {$\tilde{A}^1$}
node[format, right of=A1] (A2) {$\tilde{A}^2$}
node[format, below of=A1] (M) {$Y_1$}
node[below of=A2] (dummy) {}
node[format, right of=dummy] (Y) {$Y_2$}

(A) edge[red] (A1)
(A) edge[red] (A2)
(A1) edge[blue] (M)
(A2) edge[blue] (Y)
(M) edge[blue,->] (Y)

(C) edge[blue] (A)
(C) edge[blue, bend right=30] (M)
(C) edge[blue, bend left=20] (Y)

node[below of=M,yshift=0.4cm,xshift=0.45cm]
(l) {$(c)$}
;
\end{scope}
\begin{scope}[xshift=7.15cm]
\path[very thick, ->]
node[format] (C) {$\vec{C}$}
node[format, below of=C] (A) {$A$}
node[below of=A] (d) {}
node[format, below of=d] (M) {$Y_1$}
node[format, right of=M] (Y) {$Y_2$}

(A) edge[blue] (M)
(A) edge[blue] (Y)
(M) edge[-] (Y)

(C) edge[blue] (A)
(C) edge[blue, bend right=30] (M)
(C) edge[blue, bend left=20] (Y)

node[below of=M,yshift=0.4cm,xshift=0.45cm]
(l) {$(d)$}
;
\end{scope}
\begin{scope}[xshift=9.15cm]
\path[very thick, ->]
node[format] (C) {$\vec{C}$}
node[format, below of=C] (A) {$A$}
node[format, below of=A] (A1) {$\tilde{A}^1$}
node[format, right of=A1] (A2) {$\tilde{A}^2$}
node[format, below of=A1] (M) {$Y_1$}
node[below of=A2] (dummy) {}
node[format, right of=dummy] (Y) {$Y_2$}

(A) edge[red] (A1)
(A) edge[red] (A2)
(A1) edge[blue] (M)
(A2) edge[blue] (Y)
(M) edge[-] (Y)

(C) edge[blue] (A)
(C) edge[blue, bend right=30] (M)
(C) edge[blue, bend left=20] (Y)

(A1) edge[blue] (Y)
(A2) edge[blue] (M)

node[below of=M,yshift=0.4cm,xshift=0.45cm]
(l) {$(e)$}
;
\end{scope}
\begin{scope}[xshift=11.75cm]
\path[very thick, ->]
node[format] (C) {$\vec{C}$}
node[format, below of=C] (A) {$A$}
node[format, below of=A] (A1) {$\tilde{A}^1$}
node[format, right of=A1] (A2) {$\tilde{A}^2$}
node[format, below of=A1] (M) {$Y_1$}
node[below of=A2] (dummy) {}
node[format, right of=dummy] (Y) {$Y_2$}

(A) edge[red] (A1)
(A) edge[red] (A2)
(A1) edge[blue] (M)
(A2) edge[blue] (Y)
(M) edge[-] (Y)

(C) edge[blue] (A)
(C) edge[blue, bend right=30] (M)
(C) edge[blue, bend left=20] (Y)

node[below of=M,yshift=0.4cm,xshift=0.45cm]
(l) {$(f)$}
;
\end{scope}
\end{tikzpicture}
\end{center}
\caption{
(a) A directed acyclic graph (DAG) representing a common causal model for mediation analysis.
(b) An elaboration of the DAG in (a) that allows interventions on components of the treatment $A$.  Red edges are deterministic.
(c) A causal submodel of (b) considered in \citep{robins10alternative}
which represents a counterfactual used to define direct and indirect effects as a response to two edge-specific interventions.
(d) A causal model representing a partial interference causal inference setting with a dyad outcome.
(e) An elaboration of the graph in (d) that allows interventions on components of the treatment $A$.  Red edges are deterministic.
(f) A causal submodel of (e) which represents an interference
counterfactual as a response to two dyad outcome specific interventions.
}
\label{fig:med}
\end{figure}

\section{Notation and Preliminaries}
\label{sec:notation}

Here we describe the necessary preliminaries: causal models, mediation analysis, and extensions of causal models that permit reasoning about interference.



\subsection{Classical Causal Inference}
\label{subseq:classical}


Causal inference aims to use realizations of the observed data distribution $p(\vec{V})$
to make inferences about parameters defined using potential outcome random variables.
In our running example, a potential outcome $Y_1(a)$ denotes what would happen to the outcome $Y_1$
(the child's colonization status) had the treatment $A$ been set, possibly contrary to fact, to $a$ (vaccination).


The difficulty with 
causal parameters such as the average causal effect (ACE) is that they are
a function of responses that occur \emph{contrary to fact}.  The fundamental problem of causal inference is that
we only observe the response actually assigned.  A link between counterfactual
contrasts such as the ACE, and observed data is typically made by means of
the \emph{consistency assumption} stating that observed $Y$ and counterfactual $Y(a)$ are equal if it is the case that $A=a$,
and additional assumptions forming a \emph{causal model}, often conceptualized by means
of directed acyclic graphs (DAGs), where vertices represent variables of interest, and directed edges represent direct causal relationships.
%

The DAG representing the causal model for the vaccination study example described in Section~\ref{subsec:example} is shown in Fig.~\ref{fig:med} (a).
Formally, such a model corresponds to a set of independence assumptions on potential outcome random variables.
It is common to assume, explicitly or implicitly, the structural causal model (SCM) also known as the non-parametric
structural equation model with independent errors (NPSEM-IE) of \citet{pearl09causality}.
This model associates a set of variables and a set of vertices ${\vec{V}} = \{ V_1, \ldots, V_k \}$
in a DAG, and for each variable $V_i \in {\vec{V}}$ assumes a noise variable
$\epsilon_{V_i}$, and an arbitrary, invariant causal mechanism
$f_{V_i} : {\mathfrak X}_{\pa_{\G}(V_i) \cup \{ \epsilon_{V_i} \}} \mapsto {\mathfrak X}_{V_i}$
mapping values of parents of $V_i$ in the graph ($\pa_{\G}(V_i)$)
and $\epsilon_{V_i}$ to values of $V_i$.\footnote{Here ${\mathfrak X}_{\vec{S}}$ denotes the state space of the set of variables $\vec{S}$.}
It is assumed $f_{V_i}$ determines the value of $V_i$ regardless of how
the values of $\pa_{\G}(V_i)$ were assigned.  Moreover, it is assumed
the noise variables are mutually independent:
$p(\epsilon_{V_1}, \ldots \epsilon_{V_k}) = \prod_{i=1}^k p(\epsilon_i)$.
The arbitrary nature of $f_{V_i}$ justifies the word ``non-parametric,'' and
this property justifies the phrase ``independent errors'' in the name of the model.
Interventions
are represented by replacing certain mechanisms by constant values.


The four 
variable example in Fig. \ref{fig:med} (a) is represented by four
functions $f_{\vec{C}} : {\mathfrak X}_{\epsilon_{\vec{C}}} \mapsto {\mathfrak X}_{\vec{C}}$, 
$f_A : {\mathfrak X}_{\vec{C}, \epsilon_A} \mapsto {\mathfrak X}_A$,
$f_{{Y_1}} : {\mathfrak X}_{\{ \vec{C}, A, \epsilon_{{Y_1}} \}} \mapsto {\mathfrak X}_{{Y_1}}$, and
$f_{{Y_2}} : {\mathfrak X}_{\{ \vec{C}, A, {Y_1, \epsilon_{Y_2}} \}} \mapsto {\mathfrak X}_{{Y_2}}$.
The intervention that sets $A$ to $a$ is conceptualized by replacing the structural equation
$f_A$ by a constant function $\tilde{f}_A$ that outputs the value $a$, regardless of the input
values of $\vec{C}$ and $\epsilon_A$.



An alternative definition of the NPSEM-IE model for a DAG $\G$ with a
vertex set ${\vec{V}}$, given in \citep{thomas13swig}, uses one step ahead
counterfactuals of the form $V(\vec{a}_V)$, for any
$\vec{a}_V \in {\mathfrak X}_{\pa_{\G}(V)}$, to define all other variable\ilya{s},
factual or counterfactual, using \emph{recursive substitution}.  Specifically,
for any $\vec{A} \subseteq {\vec{V}}$, and any $\vec{a} \in {\mathfrak X}_{\vec{A}}$,
we have for every $V \in {\vec{V}}$
{\small
\begin{align}
V(\vec{a}) \equiv V(\vec{a}_{\pa_{\G}(V)},
\{ W(\vec{a}) : W \in \pa_{\G}(V) \setminus \vec{A} \})
\label{eqn:rec-sub}
\end{align}
}In this definition, and subsequently, we will use a notation convention where for a subset $\vec{S}$
of variables in $\vec{A}$, $\vec{a}_{\vec{S}}$ denotes a subset of values $\vec{a}$ of
$\vec{A}$ pertaining to $\vec{S}$.

Given an arbitrary $V(\vec{a})$, recursive substitution (\ref{eqn:rec-sub}) implies $V(\vec{a})$ is
only a function of values $\vec{a}^*$ in $\vec{a}$ corresponding to elements in $\vec{A}$ with a
directed path to $V$ not through other elements in $\vec{A}$.  Such constraints are sometimes
termed \emph{exclusion restrictions}.  As an example, $Y_1(a,y_2)$ in the DAG $\G$ in
Fig~\ref{fig:med} (a) is only a function of $a$ and not of $y_2$.

Other restrictions defining the NPSEM-IE model
are entailed by a set of assumptions on one-step-ahead counterfactuals, as a kind of
causal version of the local Markov property.
These assumptions state that
{\small
\begin{align}
\text{``sets of variables in the set}
\left\{
\{ V(\vec{a}_V) \mid \vec{a}_V \in {\mathfrak X}_{\pa_{\G}(V)}
\} \middle| V \in {\vec{V}}
\right\}\text{are mutually independent.''}
\label{eqn:npsem-ie}
\end{align}
}This assumption is equivalent to the independent errors assumption above.
In Fig.~\ref{fig:med} (a), this assumption states that the following sets of ``cross-world'' variables
are mutually independent
{\small
\begin{align}
\vec{C}
\ci
\{ A(\vec{c}) \!:\! \vec{c} \in {\mathfrak X}_{\vec{C}} \}
\ci
\{ Y_1(a,\vec{c}) \!:\! a \in {\mathfrak X}_A, \vec{c} \in {\mathfrak X}_{\vec{C}} \}
\ci
\{ Y_2(y_1, a, \vec{c}) \!:\! y_1 \in {\mathfrak X}_{Y_1}, a \in {\mathfrak X}_A, \vec{c} \in {\mathfrak X}_{\vec{C}} \}.
\label{eqn:indep-ex}
\end{align}
}A weaker model known as the \emph{finest fully randomized causally interpretable structured tree graph (FFRCISTG)} model for a DAG $\G$, described in \citep{robins86new}, entails a weaker set of assumptions than (\ref{eqn:npsem-ie})
{\small
\begin{align}
\text{``variables in the set}
\left\{
V(\vec{v}_{\vec{V}})
\middle| V \in \vec{V}
\right\}\text{are mutually independent for any }\vec{v} \in {\mathfrak X}_{\vec{V}}\text{.''}
\label{eqn:ffrcistg}
\end{align}
}In Fig.~\ref{fig:med} (a), this assumption states that the following sets of variables
are mutually independent for any
$\vec{v} \in {\mathfrak X}_{\vec{C} \cup \{ A, Y_1, Y_2 \}}$
{\small
\begin{align}
\{ \vec{C}, A(\vec{v}_{\vec{C}}), Y_1(\vec{v}_{\vec{C} \cup \{ A \}}), Y_2(\vec{v}_{\vec{C} \cup \{ A, Y_1 \}}) \}.
\label{eqn:indep-ex-2}
\end{align}
}Note that (\ref{eqn:ffrcistg}) is a subset of assumptions in (\ref{eqn:npsem-ie}), meaning that the NPSEM-IE is a submodel of the FFRCISTG model.

It has been shown in \citep{thomas13swig} that (\ref{eqn:rec-sub})
and (\ref{eqn:ffrcistg}) entail that the observed data law obeys
the standard Markov factorization with respect to the DAG $\G$
{\small
\begin{align}
p({\vec{V}}) =
\prod_{V \in {\vec{V}}} p(V(\pa_{\G}(V))) =
\prod_{V \in {\vec{V}}} p(V \mid \pa_{\G}(V)),
\label{eqn:dag_fact}
\end{align}
}and every interventional distribution of the form $p(\{ W(\vec{a}) : W \in \vec{V} \setminus \vec{A} \})$ is identified by a truncated
Markov factorization of $\G$ known as the \emph{g-formula}
{\small
\begin{align}
p(\{ W(\vec{a}) : W \in \vec{V} \setminus \vec{A} \}) =
\prod_{V \in \vec{V} \setminus \vec{A}} p(V \mid \pa_{\G}(V)) \vert_{\{ W = \vec{a}_W : W \in \pa_{\G}(V) \cap \vec{A} \}}.
\label{eqn:g}
\end{align}
}A simple consequence of (\ref{eqn:g}) is that $\E[Y_1(a) - Y_2(a')]$ is identified in the causal model corresponding to Fig.~\ref{fig:med} (a)
via the \emph{adjustment functional} $\E[\E[Y_1 \mid a, \vec{C}] - \E[Y_1 \mid a', \vec{C}]]$.

In the presence of hidden variables, not every interventional distribution is identified, and identification theory for
identified interventional distributions becomes considerably more complicated.  A graphical characterization with
corresponding identification algorithms has been given in \citep{tian02on,shpitser06id,huang06do,shpitser06idc}.
A reformulation of these identification algorithms that synthesize graphical and potential outcomes based approaches to
causal inference are given in \citep{malinsky19po,rrs20volume_id_arxiv}.

\subsection{Mediation Analysis Via Treatment Decomposition}
\label{subsec:med-dag}

Given the overall effect, as quantified by the ACE,
we may wish to decompose it into 
a direct effect and
an indirect effect (mediated by a third variable on a causal pathway from treatment to
outcome), or more generally into effects associated with bundles of causal pathways
connecting the treatment and the outcome.  Defining such a decomposition and recovering
it from observed data is the goal of mediation analysis.



Here we describe an interventionist formulation of mediation analysis
outlined in \citep{robins10alternative}, where direct, indirect, and path-specific effects
are conceptualized as counterfactual responses to interventions on \emph{treatment
components}.  This formulation
uses 
ordinary intervention operations, and assumptions on counterfactuals
defined by such operations.  An alternative approach, based on nested counterfactuals, is
outlined in \citep{robins92effects}, while identification strategies for direct and indirect
effects defined in \citep{robins92effects} based on often difficult to justify ``cross-world''
independence assumptions is described in \citep{pearl01direct} and \citep{shpitser13cogsci}.


Consider as an example a hypothetical study of the effect of smoking on health, where
smoking ($A$) affects a health outcome $Y_2$ either directly via smoke inhalation or
indirectly via nicotine content, mediated by cardiovascular disease $Y_1$.  Note that
while both components of the treatment are present in smokers and absent in non-smokers,
we can
imagine intervening on these components separately, by means of smokeless cigarettes
or nicotine patches; see discussion in Section 5 of \citet{robins10alternative}.


We can represent these treatment components explicitly in an expanded causal diagram obtained
from Fig.~\ref{fig:med} (a), shown in Fig.~\ref{fig:med} (b), where components of the treatment are
``copies'' $\tilde{A}^1, \tilde{A}^2$ of $A$ that in ordinary circumstances (represented by data elements obtained
from the study) have the same value as $A$, but whose values can in principle be set separately.
The larger causal model can be viewed as a 
FFRCISTG model with a deterministic relationship between $\tilde{A}^1, \tilde{A}^2$ and $A$.
While variables $\tilde{A}^1$ and $\tilde{A}^2$ share the state space with $A$ by construction,
to avoid confusion we will denote values of the former by 
$\tilde{a}^1$ and $\tilde{a}^2$, and the
latter by 
$a$.

Without further assumptions, the existence of components $\tilde{A}^1, \tilde{A}^2$ of $A$,
both of which influences $M$ and $Y$ preclude identification of any causal contrast which sets these components
to distinct values.  This is due to the fact that the observed data exhibits a \emph{positivity violation}, where
the values of $\tilde{A}^1$ and $\tilde{A}^2$ always coincide.  Thus, no information is available in the observed
data on situations where these values no longer coincide.
However, Fig.~\ref{fig:med} (c) which represents a refinement of the causal model in Fig.~\ref{fig:med} (b), posits, in addition to
treatment components, additional restrictions described below, corresponding to missing edges between $\tilde{A}^1$
and $Y_2$ and $\tilde{A}^2$ and $Y_1$.

Though it might appear that Fig.~\ref{fig:med} (c) is a simple recoding of Fig. \ref{fig:med} (a),
this is not the case, and the models entail different assumptions.
In particular, NPSEM-IE assumptions in (\ref{eqn:npsem-ie}) applied to Fig.~\ref{fig:med} (a) imply an
untestable 
assumption ($Y_2(a,y_1) \ci Y_1(a') \mid \vec{C})$. 
On the other hand, 
assumptions 
implied by Fig. \ref{fig:med} (c) contain the following exclusion restrictions:
{\small
\begin{align}
\label{eqn:dag_y_const}
p(Y_2(\tilde{a}^1, \tilde{a}^2) = y_2 \mid Y_1(\tilde{a}^1, \tilde{a}^2) = y_1,\vec{C}=\vec{c}) \text{ is only a function of } y_2, y_1, \tilde{a}^2,\vec{c}\\
\label{eqn:dag_m_const}
p(Y_1(\tilde{a}^1, \tilde{a}^2) = y_1 | \vec{C}=\vec{c}) \text{ is only a function of } y_1, \tilde{a}^1,\vec{c},
\end{align}
}for any values $\tilde{a}^1, \tilde{a}^2 \in {\mathfrak X}_A$.
These assumptions are \emph{testable in principle} by an experiment that intervenes on components $\tilde{A}^1, \tilde{A}^2$ of $A$ in the model in Fig.~\ref{fig:med} (c).
Note that assumptions (\ref{eqn:dag_y_const}) and (\ref{eqn:dag_m_const}) follow from both the NPSEM-IE and FFRCISTG interpretations of Fig.~\ref{fig:med} (c), though the latter weaker interpretation always suffices to obtain them.  Note also that the assumption $(Y_2(\tilde{a}, y_1) \ci Y_1(\tilde{a}') \mid \vec{C})$ will also hold under either the NPSEM-IE or the FFRCISTG interpretation of Fig.~\ref{fig:med} (c), whereas this assumption will only hold under the NPSEM-IE interpretation of Fig.~\ref{fig:med} (a).  
As noted above, this assumption is untestable under the NPSEM-IE corresponding to Fig.~\ref{fig:med} (a), but becomes testable under either the FFRCISTG or the NPSEM-IE interpretation of Fig.~\ref{fig:med} (c) since it is possible, in principle, to set $\tilde{A}^1,\tilde{A}^2$ to different values, even if in the observed data the values of these variables always coincide.
See Section 3.6 in \citep{rrs20volume_mediation_arxiv} for additional discussion.

Assumptions (\ref{eqn:dag_y_const}) and (\ref{eqn:dag_m_const}) imply that
equation (\ref{eqn:dag_fact}) for the observed data law of the FFRCISTG
model of Fig. \ref{fig:med} (c) is
{\small
\begin{align}
p(Y_1, Y_2, \tilde{A}^1, \tilde{A}^2, A,\vec{C}) = p(Y_2 \mid Y_1, \tilde{A}^2,\vec{C}) p(Y_1 \mid \tilde{A}^1, \vec{C})
p(\tilde{A}^1 \mid A)
p(\tilde{A}^2 \mid A)
p(A \mid \vec{C}) p(\vec{C}),
\label{eqn:dag_fact_2}
\end{align}
}where factors $p(\tilde{A}^1 \mid A)$ and $p(\tilde{A}^2 \mid A)$ are deterministic.



The key idea behind the treatment decomposition approach to mediation is to consider a contrast between the response to a treatment value, for example $Y_2(a)$, and a response to a hypothetical experiment where one treatment component is set to an active value, while another is set to a baseline value, yielding a counterfactual such as $Y_2(\tilde{a}^1, \tilde{a}^2{'})$.
In the model corresponding to Fig.~\ref{fig:med} (c),
the intuition is that setting the treatment component $\tilde{A}^2$ 
to baseline 
``turns off'' the direct causal pathway from $A$ to $Y_2$ 
and leaves active the indirect causal pathway from $A$ to
$Y_2$ mediated by $Y_1$. 
Given this intuition, we can define an direct effect contrast as
{\small
\begin{align}
\mathbb{E}[Y_2(\tilde{a}^1, \tilde{a}^2)] - \mathbb{E}[Y_2(\tilde{a}^1, \tilde{a}^2{'})]
\label{eqn:dir}
\end{align}
}(subtracting off the counterfactual where the direct path is ``turned off'' from the counterfactual where it is active),
and an indirect effect contrast as
{\small
\begin{align}
\mathbb{E}[Y_2(\tilde{a}^1, \tilde{a}^2{'})] - \mathbb{E}[Y_2(\tilde{a}^1{'}, \tilde{a}^2{'})]
\label{eqn:indir}
\end{align}
}(subtracting off the counterfactual where all paths are ``turned off'' from one where only the indirect path mediated by $Y_{1}$ is active).
The ACE decomposes into a sum of these contrasts, by a simple telescoping argument:
{\small
\begin{align*}
\mathbb{E}[Y_2(a)] - \mathbb{E}[Y_2(a')]
&=
\mathbb{E}[Y_2(\tilde{a}^1, \tilde{a}^2)] - \mathbb{E}[Y_2(\tilde{a}^1{'}, \tilde{a}^2{'})]\\
&=
\left( \mathbb{E}[Y_2(\tilde{a}^1, \tilde{a}^2{'})] - \mathbb{E}[Y_2(\tilde{a}^1{'}, \tilde{a}^2{'})] \right) +
\left( \mathbb{E}[Y_2(\tilde{a}^1, \tilde{a}^2)] - \mathbb{E}[Y_2(\tilde{a}^1, \tilde{a}^2{'})] \right).
\end{align*}
}By definition, $a$ and $a'$ are equivalent to $(\tilde{a}^1, \tilde{a}^2)$ and $(\tilde{a}^1{'}, \tilde{a}^2{'})$.

Given this set of assumptions,
it is straightforward to verify the following identifying functionals for the direct and indirect effects:
{\small
\begin{align}
\notag
\hspace{0.4cm} \E[Y_2(\tilde{a}^1, \tilde{a}^2)] - \E[Y_2(\tilde{a}^1, \tilde{a}^2{'})]
&=
\sum_{y_1,\vec{c}} \E[Y_2 | Y_1 = y_1, \tilde{a}^2,\vec{C}=\vec{c}] p(Y_1 = y_1 | \tilde{a}^1,\vec{C}=\vec{c})p(\vec{C}=\vec{c})
\\
\label{eqn:g-med-1}
& \hspace{0.4cm} -
\E[Y_2 | Y_1 = y_1, \tilde{a}^2{'},\vec{C}=\vec{c}] p(Y_1 = y_1 | \tilde{a}^1,\vec{C}=\vec{c})p(\vec{C}=\vec{c}),\\
\notag
\hspace{0.4cm} \E[Y_2(\tilde{a}^1, \tilde{a}^2{'})] - \E[Y_2(\tilde{a}^1{'}, \tilde{a}^2{'})]
&=
\sum_{y_1,\vec{c}} \E[Y_2 | Y_1 = y_1, \tilde{a}^2{'},\vec{C}=\vec{c}] p(Y_1 = y_1 | \tilde{a}^1,\vec{C}=\vec{c})p(\vec{C}=\vec{c})
\\
\label{eqn:g-med-2}
& \hspace{0.4cm} -
\E[Y_2 | Y_1 = y_1, \tilde{a}^2{'},\vec{C}=\vec{c}] p(Y_1 = y_1 | \tilde{a}^1,\vec{C}=\vec{c})p(\vec{C}=\vec{c}),
\end{align}
}where $a = \tilde{a}^1 = \tilde{a}^2$ and $a' = \tilde{a}^1{'} = \tilde{a}^2{'}$.
The resulting functionals are known as the \emph{mediation formula} \citep{pearl11cmf}, and
may be viewed as a modified version of the g-formula (\ref{eqn:g}), where different factors are evaluated at different
values of the treatment $A$.  Recall that the g-formula itself is a modified factorization of a DAG.

\subsection{Partial Interference And Spillover Effects}

We now describe extensions of causal models to interference problems, meant to represent studies where experimental units do not yield independent identically distributed data, but instead yield data where units can be grouped into blocks.  In such problems, units across blocks are assumed to be independent, while units within blocks are assumed to be potentially dependent.
Assume we are analyzing data from a randomized controlled trial with 
$B$ blocks
with $N$ labelled units each.


We are interested in effects of treatments applied to a certain
subset of units on outcomes for another
subset of units.
Obtaining spillover effects of child vaccinations on mothers is an example of such a setting.
In addition, we may be interested in obtaining summaries of effects averaged over all units
in a network, which are 
called \emph{network average effects}.

We will denote realizations associated with a unit $n$ in block $b$ as $\vec{v}_n^b$.
We 
view values $\vec{v}_n^b$ ($b = 1, \ldots, B$) for every unit $n$
as i.i.d. realizations of a single set of random variables $\vec{V}_n$
representing aspects of the 
$n$th unit in the network.
Denote all variables for all units in the network as $\vec{V} \equiv (\vec{V}_1, \ldots, \vec{V}_N)$,
with $\vec{A} \subseteq \vec{V}$ denoting treatments, and $\vec{Y} \subseteq \vec{V}$ denoting outcomes by convention,
with $\vec{A}_n, {Y}_n$ denoting treatments and outcome for unit $n$, respectively.




For any value assignment $\vec{a}$, denote
denote $Y_n(\vec{a})$ to be the random variable potential response of unit $n$ to variables $\vec{A} = \{ \vec{A}_1, \ldots, \vec{A}_N \}$
being set, possibly contrary to fact, to values $\vec{a}$.  This notation emphasizes that the response of unit $n$ may depend on
treatment values not only of unit $n$, but other units in the network.
For the moment, we allow treatments within a single block to affect units within that block in an arbitrary way.

Following
\citet{halloran95causal} and \citet{tchetgen12on}, we define the \emph{
main effect}\footnote{This effect is sometimes called the \emph{direct effect} in the interference literature.  Here we eschew this term to avoid confusion with direct effects in mediation analysis.}
(on the mean difference scale) of treatments $\vec{A}_n$ on ${Y}_n$ as
\[
\text{ME}_n(\vec{a}_{-n},\vec{1},\vec{0}) \equiv \mathbb{E}\{Y_n(\vec{a}_{-n},{a}_n = \vec{1})\} - \mathbb{E}\{Y_n(\vec{a}_{-n}, {a}_n = \vec{0})\},
\]
where $\vec{1}$ and $\vec{0}$ are sets of active and baseline treatment
values, and $\vec{a}_{-n}$ is all values in a treatment value assignment $\vec{a}$ other than $\vec{a_n}$, the value
assignment for unit $n$.
Similarly, we define the
\emph{
spillover effect}
(on the mean difference scale) of treatments \emph{other than} $\vec{A}_n$ on ${Y}_n$ as
\[
\text{SE}_n(\vec{a}_{-n},\vec{a}'_{-n},{0}) \equiv \mathbb{E}\{Y_n(\vec{a}_{-n},{a}_n = {0})\} -
\mathbb{E}\{Y_n(\vec{a}'_{-n},{a}_n = {0})\}.
\]

Given a fixed set of active treatments $\vec{a} = \vec{1}$ 
and a fixed set of baseline treatments $\vec{a}' = \vec{0}$ 
the network average versions of the main effect and the spillover effect are defined in the natural way as
{\small
\begin{align*}
\text{NAME}(\vec{a}, \vec{a}') \equiv \frac{1}{N} \sum_{n=1}^N \text{ME}_n(\vec{a}_{-n}, \vec{a}_n, \vec{a}'_n); \hspace{1.0cm}
\text{NASE}(\vec{a}, \vec{a}') \equiv \frac{1}{N} \sum_{n=1}^N \text{SE}_n(\vec{a}_{-n},\vec{a}'_{-n},{a}'_n).
\end{align*}
}

\section{
The Symmetric Spillover Effect Decomposition}
\label{sec:cg}

Using standard mediation analysis to model direct and indirect components of the spillover effect
runs into difficulties in settings where a sensible causal ordering on variables for different units may not exist.
For example, an endemic disease may infect the child's caretaker first, or the child first.
Such situations thus cannot be represented with DAG models, as such models assume a valid causal ordering.
This difficulty is perhaps resolvable if we are able to collect very detailed information on
the temporal order in which variables influence each other (perhaps representing fine grained temporal
information on infection transmission in our example).
Problems where such information is available can be well-modeled by a DAG ``unrolled'' in time.
See \citep{martinussen21estimation} for one approach to mediation
analysis in such cases.

This 
approach is also sometimes taken in the analysis of longitudinal data with
interference \citep{ogburn14interference}. 
However, in practice such detailed temporal information is rarely available, and
instead information on outcomes is collected in such a way that detailed information on
transmission dynamics is lost.
This means we cannot use standard causal models representable by DAGs,
such as the NPSEM-IE in most settings we are interested in, where
we wish to define the decomposition of \emph{any} spillover effect
within a block of units in a coherent way, such that any outcome may act either
as 
a mediator or an outcome.

We propose a new modeling approach to such situations by extending the treatment decomposition approach to mediation analysis \citep{robins10alternative,rrs20volume_mediation_arxiv} 
using causal models that yield interventional and observational distributions
Markov with respect to chain graphs \citep{lauritzen96graphical}.
Such distributions may be obtained as equilibria obtained from sampling
certain types of structural equation (equivalently counterfactual) models,
as described in \citep{lauritzen02chain}.
Alternatively, 
Markov restrictions may be imposed directly on distributions derived from causal models associated with directed acyclic graphs (DAGs), where individual vertices in DAGs represent vectors of variables.  We refer to both types of models as \emph{causal chain graph models,} since they both
induce distributions that obey Markov properties associated with chain graphs \citep{lauritzen96graphical}, though their substantive interpretations are different.



%



\subsection{Representation of Interference Problems With Chain Graph Models}
\label{subsec:single-block-cg}

For purposes of illustration, we will consider a 
partial interference setting where we wish to decompose spillover effects in a set of $N$ labelled units forming a network, where each unit $n$ contains a set of
baseline covariates $\vec{C}_n$, a treatment variable $A_n$, and an outcome $Y_n$.  This network is represented by an undirected graph ${\cal N}$, with
with vertices corresponding to units, and edges corresponding to ``network ties,'' such as friendship or family links.
In a graph ${\cal N}$ with undirected edges, the set of vertices sharing an undirected edge with a vertex $Y_i$ will be denoted by $\nb_{\cal N}(Y_i)$ (neighbors of $Y_i$).
A simple example of such a network on four units is shown in Fig.~\ref{fig:auto-g} (a), where
$\nb_{\cal N}(2) = \{ 1, 3 \}$.

To model interference, we allow baseline covariates $\vec{C}_n$ and treatments $A_n$ of any unit $n$ to affect variables of other units in the network.   A common approach 
is to allow any unit's variables to influence any other units' variables, and obtain identification by generalizing common identifying assumptions in causal inference to network settings.
An example of such assumptions is the network version of conditional ignorability, which states that $(\vec{Y}(\vec{a}) {\ci} \vec{A} \mid \vec{C})$ for every set of values $\vec{a}$, along with network versions of positivity: $p(\vec{A} | \vec{C}) > 0$, and consistency: $\vec{Y}(\vec{A}) = \vec{Y}$ \citep{tchetgen20auto}.
Such assumptions may be represented by a causal DAG in Fig.~\ref{fig:auto-g} (b), where baseline factors, treatments, and outcomes of every unit are collected into three vector-valued variables: $\vec{C}$, $\vec{A}$, and $\vec{Y}$, respectively.

While above assumptions suffice for identification and estimation if multiple i.i.d. realizations from the network are available, in high dimensional settings, or in settings with low sample sizes (with the extreme case of a \emph{single} sample being the subject of \emph{full interference} problems \citep{tchetgen20auto}) additional 
structural assumptions are needed for inference.

For many types of network data, a reasonable 
assumption would be that variables of any unit $n$ are only influenced by variables of their neighbors in ${\cal N}$.
If these additional causal relationships among variables of neighboring units obey the causal order of variables, where $\vec{C}_n$ precede $A_n$, and both precede $Y_n$, the resulting model corresponds to a causal DAG on labelled units, shown in Fig.~\ref{fig:auto-g} (c).


However, a disadvantage of such a model is that it imposes independences among outcomes that are not 
realistic in applications.  In particular, in the graph in Fig.~\ref{fig:auto-g} (c), outcomes of all units (even units that are neighbors in the network) are independent conditional on treatments and covariates.
Such restrictions are difficult to justify in many settings
where substantial sources of network homophily are suspected to exist \citep{shalizi2011homophily}.
Chain graph (CG) models provide an approach for relaxing unrealistic restrictions among variables, and may be viewed as a tradeoff between two kinds of causal models for interference problems:
realistic but difficult to work with models that treat a network as inducing unrestricted blocks of variables as shown in Fig.~\ref{fig:auto-g} (b) \citep{hudgens08toward}, and sparse and tractable, but substantively less realistic DAG models as shown in Fig.~\ref{fig:auto-g} (c) \citep{ogburn14interference}.




Chain graphs (CGs) that represent models we will discuss are mixed graphs with directed edges ($\to$), representing causal relationships, and undirected edges ($-$), representing symmetric relationships induced by the network.  Other types of mixed graphs and their associated models, suitable for representing unobserved confounding between units or within variables of a unit 
are described in \citep{richardson02ancestral,evans13marginal,richardson23nested}.

We use the network in Fig.~\ref{fig:auto-g} (a) to illustrate two alternative interpretations of causal CG models, with the corresponding CG shown in Fig.~\ref{fig:auto-g} (d).
To simplify the presentation, we discuss differences between DAG models and CG models only with respect to the block $\vec{Y}$.
Note that each element $\vec{C}_n \in \vec{C}$ has no direct causes (parents), whereas the direct causes (parents) of every element $A_n \in \vec{A}_n$ are $\vec{C}_{\nb_{\cal N}(n) \cup \{ n \}}$.
Unlike the DAG model shown in Fig.~\ref{fig:auto-g} (c), any outcome $Y_n \in \vec{Y}$ has both parents $\pa_{\cal G}(Y_n)$ equal to $\vec{C}_{\nb_{\cal N}(n) \cup \{ n \}} \cup \vec{A}_{\nb_{\cal N}(n) \cup \{ n \}}$, and neighbors $\nb_{\cal G}(Y_n)$ equal to $\vec{Y}_{\nb_{\cal N}(n)}$.

Just as in a DAG model, missing edges among elements in $\vec{Y}$, and missing directed edges into $\vec{Y}$ in the CG in Fig.~\ref{fig:auto-g} (d) are associated with restrictions on the counterfactual distribution $p(\vec{Y}(\vec{a},\vec{c})) = p(\{ Y_n(\vec{a},\vec{c}) : Y_n \in \vec{Y} \})$.  One approach of obtaining these restrictions is via a chain graph generalization of DAG structural equation models described in \citep{lauritzen02chain}.

In these models, every variable $Y_n$ is assumed to have an independent noise variable $\epsilon_{Y_n}$ invariant causal mechanism $f_{Y_n} : {\mathfrak X}_{\pa_{\cal G}(Y_n) \cup \nb_{\cal G}(Y_n) \cup  \{ \epsilon_{V_i} \}} \mapsto {\mathfrak X}_{Y_n}$.  In other words, $f_{Y_n}$ maps values of parents and neighbors of $Y_n$ in ${\cal G}$, as well as $\epsilon_{Y_n}$ to values of $Y_n$.  Importantly, just as in DAG models, $f_{Y_n}$ acts as an invariant mechanism and does not change regardless of whether inputs were observed to hold, or were set by intervention.

Given a set of values $\vec{c},\vec{a}$, the counterfactual distribution $p(\vec{Y}(\vec{a},\vec{c}))$ representing responses of $\vec{Y}$ had $\vec{A},\vec{C}$ been set to $\vec{a},\vec{c}$, is obtained as an equilibrium distribution of a stochastic process defined using the set of structural equations $\{ f_{Y_n} : Y_n \in \vec{Y} \}$, and noise variables $\{ \epsilon_{Y_n} : Y_n \in \vec{Y} \}$.  A number of such processes are discussed in \citep{lauritzen02chain}, the simplest being a Gibbs sampler which uses conditional distributions
$p(Y_n \mid \pa_{\cal G}(Y_n) \cup \nb_{\cal G}(Y_n))$, for every $Y_n \in \vec{Y}$ evaluated at appropriate subsets of values in $\vec{c},\vec{a}$.
Each such conditional may be obtained in a straightforward way from $f_{Y_n}$ and $\epsilon_{Y_n}$.

It follows as a special case of results in \citep{lauritzen02chain} that if all $\vec{c},\vec{a}$ were intervened on, the
distribution 
$p(\vec{Y}(\vec{a},\vec{c}))$ that is obtained as the equilibrium distribution of the Gibbs sampler obeys the following Markov restrictions
{\small
\begin{align}
\label{eqn:markov-y}
p(y_n(\vec{c},\vec{a})  \mid \{ y_k(\vec{c},\vec{a})  : k \neq n \})
&=
p(y_n(\vec{c},\vec{a}) \mid \{ y_k(\vec{c},\vec{a}) : Y_k \in \nb_{\cal G}(Y_n) \})\\
\notag
&= g(y_n, \{ y_k : Y_k \in \nb_{\cal G}(Y_n) \}, \{ a_k,\vec{c}_k : A_k,\vec{C}_k \in \pa_{\cal G}(Y_n) \}),
\end{align}
}for each $n$, where $y_n(\vec{c},\vec{a})$ is a shorthand for an event $Y_n(\vec{c},\vec{a})=y_n$.

In words, this states that in the 
distribution 
$p(\vec{Y}(\vec{a},\vec{c}))$
each counterfactual outcome $Y_n(\vec{c},\vec{a})$ is conditionally independent of any outcome that isn't a neighbor of $Y_n$ conditional on outcomes that are neighbors of $Y_n$.  In addition,
for each $Y_n$, the conditional distribution 
$p(Y_n(\vec{a},\vec{c}) \mid \{ Y_m(\vec{a},\vec{c}) : Y_m \in \vec{Y} \setminus \{ Y_n \} \})$ 
only depends on $\{ Y_m(\vec{a},\vec{c}) \mid Y_m \in \nb_{\cal G}(Y_n) \}$, and the subset of values of $\vec{c},\vec{a}$ that have a direct causal influence on $Y_n$, namely those in $\pa_{\cal G}(Y_n)$.

Furthermore, since results in \citep{lauritzen02chain} imply that $p(\vec{Y}(\vec{a},\vec{c})) = p(\vec{Y}(\vec{a}) \mid \vec{c})$, the
restrictions in (\ref{eqn:markov-y}) may be rephrased as:
{\small
\begin{align}
\label{eqn:markov-y-2}
p(y_n(\vec{a}) \mid \vec{c}, \{ y_k(\vec{a})  : k \neq n \})
&=
p(y_n(\vec{a}) \mid \vec{c},\{ y_k(\vec{a}) : Y_k \in \nb_{\cal G}(Y_n) \})\\
\notag
&= g(y_n, \{ y_k : Y_k \in \nb_{\cal G}(Y_n) \}, \{ a_k,\vec{c}_k : A_k,\vec{C}_k \in \pa_{\cal G}(Y_n) \}),
\end{align}
}for each $n$, where $y_n(\vec{a})$ is a shorthand for an event $Y_n(\vec{a})=y_n$.

Unlike causal DAG models, independences or potential dependences among outcomes in $\vec{Y}$ are symmetric in the sense that they cannot be associated with a total ordering on variables.
Furthermore, dependences among outcomes are induced by a stochastic process operating using causal mechanisms that remain invariant after interventions (in this case interventions on $\vec{A},\vec{C}$).
In this sense, these symmetric dependences are invariant in the same sense that causal relationships are invariant in a structural equation model of a DAG.


The advantage of the stochastic process interpretation for causal chain graphs described in \citep{lauritzen02chain} and outlined above is it provides a clear generalization of structural equation models for DAGs, while allowing for symmetric relationships between variables.  The disadvantage is that causal models with these semantics can only reasonably be applied in settings where underlying dynamics creating observational distributions and distributions corresponding to responses to interventions involve an appropriate stochastic process.

An alternative approach that eschews equilibrium semantics starts with an unrestricted 
distribution
over the counterfactual outcomes $\vec{Y}(\vec{a},\vec{c})$
and imposes restrictions (\ref{eqn:markov-y}) implied by the network 
as a \emph{
network structural model (NSM)}.  Common examples of structural models are \emph{marginal structural models (MSMs)} that impose restrictions on a marginal counterfactual distribution, and \emph{structural nested models (SNMs)} that impose restrictions on a counterfactual blip function \citep{robins99marginal}. 
In this approach, CG causal models arise from a standard DAG model where vertices correspond to blocks of units, with further structural assumptions within blocks imposed by the NSM recovering the CG Markov property.
Naturally, such an NSM ought to be substantively justified, just as all 
structural models ought to be.
Such a justification may involve appealing to the network of units inducing a screening off property, where neighbors of a unit in a network screen off variables of that unit from variables of other units in the network in a particular way.


Both the DAG model corresponding to Fig.~\ref{fig:auto-g} (c), and the CG models corresponding to Fig.~\ref{fig:auto-g} (d) are submodels of the model corresponding to Fig.~\ref{fig:auto-g} (b), where the network version of conditional ignorability holds. As a result, the counterfactual distribution $p(\vec{Y}(\vec{a}))$, which can be used to define causal effects on the network, is identified in both models by the adjustment functional:
$p(\vec{Y}(\vec{a})) = \sum_{\vec{c}} p(\vec{Y} | \vec{a}, \vec{c}) p(\vec{c})$.

If $p(\vec{Y}(\vec{a},\vec{c})) = p(\vec{Y}(\vec{a}) | \vec{c})$
is a positive distribution, restrictions in (\ref{eqn:markov-y}) imply that $p(\vec{Y}(\vec{a},\vec{c})) = p(\vec{Y}(\vec{a}) | \vec{c})$ further obeys the \emph{conditional Markov random field (CMRF)} factorization by the conditional version of the Hammersley-Clifford theorem.  A conditional extension of the proof of this theorem in \citep{lauritzen96graphical} is found in \citep{shpitser23lc}.

This factorization associates factors with cliques (pairwise connected subsets of vertices) in the undirected graph obtained from ${\cal G}$ by restricted it to vertices in $\vec{Y}$ and edges among them.
We denote the set of all such cliques by ${\cal C}_{\vec Y}$.  The CMRF factorization is:
{\small
\begin{align}
p(\vec{Y}(\vec{a},\vec{c})) = p(\vec{Y}(\vec{a}) = \vec{y} \mid \vec{c})
= p(\{ y_n : Y_n \in \vec{Y} \} \mid \vec{a}, \vec{c}) = \frac{1}{Z(\vec{c},\vec{a})} \prod_{\vec{S} \in {\cal C}_{\vec{Y}}} \phi_{\vec{S}}(\vec{s}, \vec{w}_{\vec{S}}),
\label{eqn:cmrf-g-term}
\end{align}
}where $\vec{w}_{\vec{S}}$ is defined to be the set of values $\{ \vec{c}_k,a_m : \vec{C}_k,A_m \in \bigcap_{S \in \vec{S}} \pa_{\cal G}(\vec{S}) \}$, and
values $\vec{s}$ for every clique factor $\phi_{\vec{S}}(\vec{s},\vec{w}_{\vec{S}})$ are consistent with $\{ y_n : Y_n \in \vec{Y} \}$.
Since $p(\vec{Y}(\vec{a},\vec{c})) = p(\vec{Y}(\vec{a}) \mid \vec{c})$ is identified as $p(\vec{Y} \mid \vec{a}, \vec{c})$, the CMFR factorization also applies to $p(\vec{Y} \mid \vec{a}, \vec{c})$, and
the adjustment functional may be further rewritten as:
{\small
\begin{align}
p(\vec{Y}(\vec{a})) = \sum_{\vec{c}} p(\vec{Y} \mid \vec{a},\vec{c}) p(\vec{c}) = \sum_{\vec{c}} \left( \frac{1}{Z(\vec{c},\vec{a})} \prod_{\vec{S} \in {\cal C}_{\vec{Y}}} \phi_{\vec{S}}(\vec{s}, \vec{w}_{\vec{S}}) \right) p(\vec{c}).
\label{eqn:g-cmrf}
\end{align}
}

Identification theory in CG models follows from restrictions of the form (\ref{eqn:markov-y}), and thus is not affected by model interpretation.  However, substantive interpretation of components of spillover effects differs depending on the CG model interpretation.  We discuss this issue further below.
We defer the discussion of general causal CG models, as well as general decompositions of the spillover effects in blocks of arbitrary size to the Appendix.

\begin{figure}[t]
\begin{tikzpicture}[>=stealth, node distance=1.0cm]
\tikzstyle{format} = [draw, very thick, circle, minimum size=6mm,
inner sep=0pt]

\begin{scope}[xshift=0cm]

\path[->, line width=.3mm]

node[format, yshift=-1.0cm] (v1) {$1$}
node[format, right of=v1] (v2) {$2$}
node[format, right of=v2] (v3) {$3$}
node[format, right of=v3] (v4) {$4$}

(v1) edge[-] (v2)
(v2) edge[-] (v3)
(v3) edge[-] (v4)

node[below of=v2, xshift=0.5cm] (l) {(a)}
;

\end{scope}

\begin{scope}[xshift=4.5cm]
\path[->, line width=.3mm]
node[format] (a) {$\vec{A}$}
node[format, above of=a] (c) {$\vec{C}$}
node[format, below of=a] (y) {$\vec{Y}$}

(a) edge[blue] (y)
(c) edge[blue] (a)
(c) edge[blue, bend left] (y)

node[below of=y, xshift=0cm] (l) {(b)}
;
\end{scope}

\begin{scope}[xshift=6.0cm]
\path[->, line width=.3mm]
node[format] (a1) {$\tilde{a}^1$}
node[format, above of=a1] (c1) {$\vec{C}_1$}
node[format, below of=a1] (y1) {$Y_1$}

node[format, right of=a1] (a2) {$A_2$}
node[format, above of=a2] (c2) {$\vec{C}_2$}
node[format, below of=a2] (y2) {$Y_2$}

node[format, right of=a2] (a3) {$A_3$}
node[format, above of=a3] (c3) {$\vec{C}_3$}
node[format, below of=a3] (y3) {$Y_3$}

node[format, right of=a3] (a4) {$A_4$}
node[format, above of=a4] (c4) {$\vec{C}_4$}
node[format, below of=a4] (y4) {$Y_4$}

(c1) edge[blue] (a1)
(c1) edge[blue, bend right=30] (y1)
(a1) edge[blue] (y1)

(c2) edge[blue] (a2)
(c2) edge[blue, bend right=40] (y2)
(a2) edge[blue] (y2)

(c3) edge[blue] (a3)
(c3) edge[blue, bend left=40] (y3)
(a3) edge[blue] (y3)

(c4) edge[blue] (a4)
(c4) edge[blue, bend left=30] (y4)
(a4) edge[blue] (y4)

(c1) edge[blue] (a2)
(c2) edge[blue] (a1)
(a1) edge[blue] (y2)
(a2) edge[blue] (y1)

(c2) edge[blue] (a3)
(c3) edge[blue] (a2)
(a2) edge[blue] (y3)
(a3) edge[blue] (y2)

(c3) edge[blue] (a4)
(c4) edge[blue] (a3)
(a3) edge[blue] (y4)
(a4) edge[blue] (y3)

(c2) edge[blue] (y1)

(c1) edge[blue] (y2)
(c3) edge[blue] (y2)

(c2) edge[blue] (y3)
(c4) edge[blue] (y3)

(c3) edge[blue] (y4)

node[below of=y2, xshift=0.5cm] (l) {(c)}

;
\end{scope}

\begin{scope}[xshift=10.5cm]
\path[->, line width=.3mm]
node[format] (a1) {$\tilde{a}^1$}
node[format, above of=a1] (c1) {$\vec{C}_1$}
node[format, below of=a1] (y1) {$Y_1$}

node[format, right of=a1] (a2) {$A_2$}
node[format, above of=a2] (c2) {$\vec{C}_2$}
node[format, below of=a2] (y2) {$Y_2$}

node[format, right of=a2] (a3) {$A_3$}
node[format, above of=a3] (c3) {$\vec{C}_3$}
node[format, below of=a3] (y3) {$Y_3$}

node[format, right of=a3] (a4) {$A_4$}
node[format, above of=a4] (c4) {$\vec{C}_4$}
node[format, below of=a4] (y4) {$Y_4$}

(c1) edge[blue] (a1)
(c1) edge[blue, bend right=30] (y1)
(a1) edge[blue] (y1)

(c2) edge[blue] (a2)
(c2) edge[blue, bend right=40] (y2)
(a2) edge[blue] (y2)

(c3) edge[blue] (a3)
(c3) edge[blue, bend left=40] (y3)
(a3) edge[blue] (y3)

(c4) edge[blue] (a4)
(c4) edge[blue, bend left=30] (y4)
(a4) edge[blue] (y4)

(c1) edge[blue] (a2)
(c2) edge[blue] (a1)
(a1) edge[blue] (y2)
(a2) edge[blue] (y1)

(c2) edge[blue] (a3)
(c3) edge[blue] (a2)
(a2) edge[blue] (y3)
(a3) edge[blue] (y2)

(c3) edge[blue] (a4)
(c4) edge[blue] (a3)
(a3) edge[blue] (y4)
(a4) edge[blue] (y3)

(y1) edge[-] (y2)
(y2) edge[-] (y3)
(y3) edge[-] (y4)

(c2) edge[blue] (y1)

(c1) edge[blue] (y2)
(c3) edge[blue] (y2)

(c2) edge[blue] (y3)
(c4) edge[blue] (y3)

(c3) edge[blue] (y4)

(a1) edge[-] (a2)
(a2) edge[-] (a3)
(a3) edge[-] (a4)

(c1) edge[-] (c2)
(c2) edge[-] (c3)
(c3) edge[-] (c4)

node[below of=y2, xshift=0.5cm] (l) {(d)}

;
\end{scope}

\end{tikzpicture}
\caption{
(a) An undirected graph representing friendship ties in a simple four unit network.
(b) A causal DAG model representing partial interference among units forming a network in (a), where baseline covariates, treatments, and outcomes of all units are treated as single variables.
(c) A causal model imposing additional network restrictions on the model in (b) that may be represented as a causal DAG.
(d) A causal model imposing additional network restrictions on the model in (b) that may be represented as a chain graph.
}
\label{fig:auto-g}
\end{figure}

\subsection{Interpretation of Missing Edges in Causal Chain Graph Models}
\label{subsec:missing-edges}

Consider the counterfactual outcomes $\vec{Y}(\vec{a},\vec{c})$ in the model shown in Fig.~\ref{fig:auto-g} (d).
Under the equilibrium interpretation of causal CGs given in \citep{lauritzen02chain}, absences of edges in this CG
are interpreted as \emph{individual level} restrictions.  This is because restrictions on counterfactual or observed distributions
due to absent edges are a logical consequence of restricted inputs to structural equations used to generate these distributions.
For example, the missing edge from $A_3$ to $Y_1$ implies that the outcome $Y_1(\vec{a},\vec{c})$ is not a function of values of $A_3$
for every individual in the data arising from this causal model.  This restriction arises from the fact the structural equation $f_{Y_1}$ only has inputs
$C_1,C_2,A_1,A_2,Y_2$ (along with similar restrictions on inputs for structural equations $f_{Y_2},f_{Y_3},f_{Y_4}$.
Such an interpretation of missing edges is analogous to and a generalization of exclusion restrictions in causal DAG models
\citep{thomas13swig}.

On the other hand, under the NSM interpretation of causal CGs outlined above, missing edges correspond to distribution level statements, since NSMs restrict counterfactual distributions, rather than structural equations.  As an example, a causal CG under an NSM interpretation in Fig.~\ref{fig:auto-g} (d) imposes (\ref{eqn:markov-y}) on the distribution of $\vec{Y}(\vec{a},\vec{c})$.
This is consistent with the structural equation for the block variable $\vec{Y}$ depending on all variables in $\vec{A} \cup \vec{C}$,
provided this dependence still yields appropriate distribution level restrictions in (\ref{eqn:markov-y}).  Thus, if we interpret edges as arising due to potential causal dependence induced by structural equations and their inputs, and absence of causal dependence due to absent inputs in structural equations, then the appropriate causal graph for the causal CG under the MSM interpretation is the ``DAG of blocks'' in Fig.~\ref{fig:auto-g} (b).  In this view, Fig.~\ref{fig:auto-g} (d) does not represent the structure of the causal model, but does represent the Markov structure of the NSM imposed in addition to the causal model in Fig.~\ref{fig:auto-g} (b).  In other words, just as is the case for causal DAG models, a single chain graph may correspond to very different causal models.

\subsection{Symmetric Treatment Decomposition on a Two Outcome Example}
\label{subsec:two-example}

Having described causal models associated with chain graphs for interference problems, 
we now generalize the treatment decomposition approach to mediation analysis advocated in \citep{robins10alternative}
and described above, to these models.  We illustrate our proposal by reconceptualizing the dyadic partial interference setting of a vaccine trial discussed in Section~\ref{subsec:example}, using an elaboration of the model in Fig.~\ref{fig:med} (d).


In our example, we have blocks with two outcomes, $Y_1$ and $Y_2$, a single treatment $A$
administered to $Y_1$, and a set of baseline factors $\vec{C}$.
As before, we split $A$ into two components $\tilde{A}^1$ and $\tilde{A}^2$ that always
assume the same values in the observed data,
but can in principle be intervened on separately.
The CG corresponding to this elaboration of Fig.~\ref{fig:med} (d) is shown in Fig.~\ref{fig:med} (e).
As was the case with DAGs, treatment decomposition without additional restrictions leads to a situation
where any causal contrast where components $\tilde{A}^1$ and $\tilde{A}^2$ are intervened on to different
values is not identified from the observed data, where the values of $\tilde{A}^1$ and $\tilde{A}^2$ must always
coincide.

However, a refinement of the causal model in Fig.~\ref{fig:med} (e) shown in Fig.~\ref{fig:med} (f) posits, in addition to
treatment components, a special case of restrictions in (\ref{eqn:markov-y}) corresponding to missing edges between
$\tilde{A}^1$ and $Y$ and $\tilde{A}^2$ and $M$:
{\small
\begin{align}
\label{eqn:cg_y_const}
p(Y_1(\tilde{a}^1, \tilde{a}^2,\vec{c}) = y_1 \mid Y_2(\tilde{a}^1, \tilde{a}^2,\vec{c})=y_2) \text{ is only a function of }y_1, \tilde{a}^1,y_2,\vec{c}
\\
\label{eqn:cg_m_const}
p(Y_2(\tilde{a}^1, \tilde{a}^2,\vec{c}) = y_2 \mid Y_1(\tilde{a}^1, \tilde{a}^2,\vec{c})=y_1) \text{ is only a function of }y_2, \tilde{a}^2, y_1,\vec{c}.
\end{align}
}These restrictions may
be viewed as symmetric versions of constraints
(\ref{eqn:dag_y_const}) and (\ref{eqn:dag_m_const}), corresponding to Fig.~\ref{fig:med} (c).

If $p(Y_1(\tilde{a}^1, \tilde{a}^2), Y_2(\tilde{a}^1, \tilde{a}^2) \mid \vec{c})$ (for $\tilde{a}^1 = \tilde{a}^2$) is positive, 
the above restrictions translate into the following special case of
(\ref{eqn:cmrf-g-term})
{\small
\begin{align}
\notag
p(Y_1(\tilde{a}^1, \tilde{a}^2), Y_2(\tilde{a}^1, \tilde{a}^2) \mid \vec{c}) &= p(Y_1(a), Y_2(a) \mid \vec{c}) = p(Y_1, Y_2 \mid a,\vec{c})\\
&= \frac{1}{Z(a,\vec{c})} \phi_{Y_1}(y_1, a,\vec{c}) \phi_{Y_2}(y_2, a,\vec{c}) \phi_{Y_1, Y_2}(y_1, y_2, \vec{c}),
\label{eqn:dyad-missing-or}
\end{align}
}where $a$ is consistent with $(\tilde{a}^1, \tilde{a}^2)$.

Missing edges from $\tilde{A}^1$ to $Y_2$ and from $\tilde{A}^2$ to $Y_1$ imply that the third term in (\ref{eqn:dyad-missing-or}) $\phi_{Y_1, Y_2}(y_1, y_2, \vec{c})$ is not a function of $a$.
Since $p(Y_1(\tilde{a}^1, \tilde{a}^2), Y_2(\tilde{a}^1, \tilde{a}^2) \mid \vec{c})$ is identified and equal to $p(Y_1, Y_2 \mid a,\vec{c})$, this represents a testable restriction on the observed data distribution.  We discuss the implications of this in Section \ref{sec:false}.

Given a single treatment $A$ meant for $Y_1$,
the spillover effect $\mathbb{E}[Y_2(a)] - \mathbb{E}[Y_2(a')]$
admits precisely the same decomposition into two components
shown in (\ref{eqn:indir}) and (\ref{eqn:dir}) as the spillover effects in Section \ref{subsec:med-dag}.
As discussed in Section \ref{subsec:med-dag}, these components had a natural mediation interpretation 
in a DAG model in Fig.~\ref{fig:med} (c), since the \emph{direct effect} (\ref{eqn:dir}) corresponds to a contrast where a treatment component $\tilde{A}_2$ of $A$ directly causing $Y_2$ is changed, whereas the \emph{indirect effect} corresponds to a contrast where a treatment component $\tilde{A}^1$
of $A$ indirectly causing $Y$ via a path $\tilde{A}^2 \to M \to Y$ is changed.

In a DAG model, the statement ``$\tilde{A}^2$ directly causes $Y_2$'' has a natural interpretation as ``values of $\tilde{A}^2$ serve as inputs for the structural equation producing values of $Y_2$,''
while the statement ``$\tilde{A}^1$ indirectly causes $Y_2$'' has a similarly natural interpretation as ``there is a sequence of variables
that starts at $\tilde{A}^1$, ends at $Y_2$, and each intermediate variable in the sequence is directly caused by a variable just before it in the chain.''
This sense of ``direct'' and ``indirect'' is reflected in the modified DAG factorization corresponding to the mediation formula (\ref{eqn:g-med-1}), (\ref{eqn:g-med-2}), where values of $\tilde{A}^2$ occur in the Markov factor for $Y_2$, while values of $\tilde{A}^1$ do not, and instead occur in the Markov factor for $Y_1$, while $Y_1$ occurs in the Markov factor for $Y_2$.


Notions of direct and indirect influence may be generalized to a CG model in Fig.~\ref{fig:med} (f) under the stochastic process semantics in \citep{lauritzen02chain}.
Specifically, the statement ``$\tilde{A}^2$ directly causes $Y_2$'' now means ``values of $\tilde{A}^2$ serve as inputs for the structural equation producing values of $Y_2$,''
and the statement ``$\tilde{A}^1$ indirectly causes $Y_2$'' now means ``there is a sequence of variables 
that starts at $\tilde{A}^1$, ends at $Y_2$, and each intermediate
variable in the sequence is directly caused by a variable just before it in the chain.''
Note, however, that in a CG model the existence of the above chain of variables establishing that $\tilde{A}^1$ indirectly causes $Y_2$ may not preclude some variables
later in the chain from causing variables earlier in the chain, something that cannot happen in a DAG model.  In addition, $\tilde{A}^2$ directly
causing $Y_2$ may involve a stochastic process involving the structural equation for $Y_2$ rather than just the structural equation for $Y_2$ itself.
In other words, direct causation is ``less direct'' in CG models compared to DAG models, and indirect causation in CG models does not necessarily impose
a unique causal ordering on variables, unlike in DAG models.

In addition to a different interpretation for direct and indirect effects compared to the DAG model in Fig.~\ref{fig:med} (c),
the identifying functionals 
for counterfactuals involved are different in the models in Fig.~\ref{fig:med} (f). For example, $p(Y_2(\tilde{a}^1, \tilde{a}^2{'})=y_2)$ is equal to:
{\small
\begin{align}
\notag
&
\sum_{y_1,\vec{c}} p(Y_2(\tilde{a}^1, \tilde{a}^2{'}) = y_2, Y_1(\tilde{a}^1, \tilde{a}^2{'}) = y_1 | \vec{C}=\vec{c}) p(\vec{C}=\vec{c})
\\
&=
\notag
\sum_{y_1,\vec{c}} p(Y_2 = y_2, Y_1 = y_1 \mid \tilde{a}^1, \tilde{a}^2{'},\vec{C}=\vec{c}) p(\vec{C}=\vec{c})
\\
&=
\sum_{y^1,\vec{c}}
\frac{
\phi_{Y_1,Y_2,\vec{C}}(y_1,y_2,\vec{c}) \phi_{Y_1,\tilde{a}^1,\vec{C}}(y_1,\tilde{a}^1,\vec{c}) \phi_{Y_2,A_2,\vec{C}}(y_2,\tilde{a}^2{'},\vec{c})
}{Z(\tilde{a}^1,\tilde{a}^2{'},\vec{c})} p(\vec{C}=\vec{c}),
\label{eqn:sym_med}
\end{align}
}
where $a = \tilde{a}^1 = \tilde{a}^2$ and $a' = \tilde{a}^1{'} = \tilde{a}^2{'}$.
We derive an identification result that yields (\ref{eqn:sym_med}) as a special case in the Appendix, where we describe how causal CG models apply in general network settings. 
Unlike similar results in DAG models, obtaining results such as (\ref{eqn:sym_med}) rely on assumptions of positivity for $p(Y_2(\tilde{a}^1, \tilde{a}^2{'})=y_2 | \vec{c})$ for any combination of values $\tilde{a}^1,\tilde{a}^2{'},\vec{c}$, even if such a combination does not occur in the data.

Just as the mediation formula in (\ref{eqn:g-med-1}) and (\ref{eqn:g-med-2}) may be viewed as a modified version of the g-formula (\ref{eqn:g}) with different DAG factorization terms evaluated at
different values of the treatment $A$, the functional (\ref{eqn:sym_med}) may be viewed as a modified version of the g-formula containing a CMRF factorization (\ref{eqn:g-cmrf}) with different CMRF factors evaluated at different values of the treatment $A$.  Unlike factors in the g-formula, which is a modified DAG factorization, factors in (\ref{eqn:sym_med}) are symmetric, and represents an effect arising in part due to
symmetric relationships among outcomes.  We thus call (\ref{eqn:sym_med}) the \emph{symmetric mediation formula}.

As was the case in DAG models, the notions of direct and indirect influence are reflected in the modified CG factorization corresponding to the symmetric mediation formula (\ref{eqn:sym_med}).  Specifically, that $\tilde{A}^2$ directly causes $Y_2$ is reflected in the term $\phi_{Y_2}$ being a function of values of $\tilde{A}^2$.  Similarly, that $\tilde{A}^1$ indirectly causes $Y_2$ is reflected in the term $\phi_{Y_2}$ not being a function of values of $\tilde{A}^1$, while
being a function of values of $Y_1$, with the term $\phi_{Y_1}$ being a function of values of $\tilde{A}^1$.
This is directly analogous to the factorization interpretation of direct and indirect influence in DAGs via the mediation formula, discussed above.


If we adopt the network structural model (NSM) interpretation for CG models, only the above interpretation of direct and indirect influence remains meaningful.  Since this interpretation is based on the factorization of a counterfactual distribution, it involves population level statements, rather than individual level statements in the structural equation based interpretation described earlier.

In particular, since the decomposition of the spillover effect under the NSM interpretation relies exclusively in distribution level assumptions, there is no interpretation of terms of the decomposition as direct or indirect effects that hold for all individuals.  In fact, under some definitions of ``effect,'' there terms are not effects at all.



\subsection{Assigning Substantive Meaning to Treatment Components}
\label{subsec:component-meaning}

Just as in the interventionist approach to mediation analysis based on treatment decomposition in causal models of a DAG, the approach we outline only yields meaningful causal quantities if it is possible to define components of the treatments that may, in principle, be manipulated in isolation.  Furthermore, identifiability may only be obtained if appropriate exclusion restrictions corresponding to missing edges in Fig.~\ref{fig:med} (f) hold.
We outline how treatment components may be used to define substantively important causal effects using the pertussis vaccination example outlined in Section~\ref{subsec:example}.

Recent studies using animal models have shed light on two primary types of pertussis vaccines: whole-cell vaccines (wP) and acellular vaccines (aP).  Both vaccines have demonstrated efficacy in protecting against infection by \emph{Bordetella pertussis (BP)} (the bacterium causing pertussis) by triggering T-cell activation in the lungs, which helps the body dramatically reduce the presence of pertussis bacteria in the lungs. However, while whole-cell vaccines effectively prevent the nasal carriage of pertussis bacteria by inducing T-cell activation in the nasal cavity, acellular vaccines fail to do so. The absence of T-cell activation induction in the nasal cavity with acellular vaccines results in prolonged nasal carriage of pertussis bacteria, potentially heightening the risk of bacterial transmission.

More specifically, prior work \citep{warfel14acellular,dubois21suppression}
has shown that pertussis infection induces T helper 17 (Th17) and T helper 1 (Th1) memory responses, as does wP vaccination (although to a lesser extent), while aP vaccination induces T helper 2 (Th2) and Th1 memory responses, but not Th17 memory responses.  Further, it has been shown that aP vaccination is protective against pertussis, but does not prevent transmission, while Th17 is key for controlling nasal colonization by BP.  These findings suggest that protection against symptomatic pertussis infection is related to protection from lung colonization, while protection against transmission is related to protection from nasal colonization, and moreover wP vaccination offers substantial protection from both types of colonization, while aP vaccination offers substantial protection only from lung colonization.

Consider a randomized trial on cohabiting pairs of individuals, where the treatment group is vaccinated with the wP vaccine, and the control group with a placebo treatment.  The above results suggest that we can consider the effect of the wP vaccine administered to unit $1$ in the study (denoted by $A_1$) as consisting of two components, the component affecting colonization of the lungs ($\tilde{A}_1^1$), and the component affecting the colonization of the nasal passages ($\tilde{A}_1^2$).
Since wP vaccination is protective against both nasal and lung colonization, in the randomized trial the treatment group receives $A_1 = 1$, and thus $\tilde{A}_1^1 = \tilde{A}_1^2 = 1$, while the control group receives $A_1 = 0$, and thus $\tilde{A}_1^1 = \tilde{A}_1^2 = 0$.  However, since aP vaccination protects against only lung colonization, and not against nasal colonization, we may be interested in assessing the \emph{contagion effect} of aP vaccination of unit $1$ on infection status of unit $2$, given the data on wP vaccination RCT.

Provided that the only relevant difference between aP and wP vaccinations pertain to activation or deactivation of treatment components of $A_1$, we may represent the contagion effect of aP vaccination using the treatment decomposition framework we described, specifically as:
\begin{align*}
\mathbb{E}[Y_2(\tilde{a}_1^1, \tilde{a}_1^2{'})]
-
\mathbb{E}[Y_2(\tilde{a}_1^1{'}, \tilde{a}_1^2{'})],
\end{align*}
which may be identified as outlined in the previous section, provided assumptions (\ref{eqn:cg_y_const}) and (\ref{eqn:cg_m_const}) hold, which we believe are reasonable assumptions given the nature of treatment components outlined.

Note that the substantive meaning of this effect is influenced by the semantics of the causal CG model we consider.  The equilibrium structural equation semantics yield an individual level effect, while the structural model semantics yield a population level effect.






\section{Model Falsifiability} 
\label{sec:false}

One advantage of the treatment decomposition approach to mediation we adopt here, compared
to classical mediation analysis based on nested counterfactuals and cross-world restrictions, is that
assumptions necessary for identification 
may in principle be tested by a randomized experiment on treatment components.
The same is true in our symmetric treatment decomposition model represented by causal chain graphs.
However, an additional 
property holds in causal CGs, but not in causal DAGs --
identifying assumptions for direct and indirect effects may be falsified using observed data.
As an example, consider the factorization of the observed law 
$p(y_1, y_2, a, \tilde{a}^1, \tilde{a}^2, \vec{c})$
corresponding to the model in Fig.~\ref{fig:med} (f).  Including the treatment component variables, this factorization is:
{\small
\begin{align}
\frac{
\phi_{Y_1,Y_2,\vec{C}}{(y_1,y_2,\vec{c})}
\phi_{Y_1,A,\vec{C}}{(y_1,\tilde{a}^1,\vec{c})}
\phi_{Y_2,A,\vec{C}}{(y_2,\tilde{a}^2,\vec{c})}
}{Z({\tilde{a}^1,\tilde{a}^2},\vec{c})}
{p(\tilde{a}^1 | a) p(\tilde{a}^2 | a)} p(a | \vec{c}) p(\vec{c}).
\label{eqn:restricted-0}
\end{align}
}Since values of treatment components $\tilde{A}^1,\tilde{A}^2$ always correspond to the values of $A$ in the observed data,
this factorization may be rewritten by dropping treatment components, yielding:
{\small
\begin{align}
p(y_1, y_2, a,\vec{c}) =
\frac{
\phi_{Y_1,Y_2,\vec{C}}{(y_1,y_2,\vec{c})}
\phi_{Y_1,A,\vec{C}}{(y_1,{a},\vec{c})}
\phi_{Y_2,A,\vec{C}}{(y_2,{a},\vec{c})}
}{Z({a},\vec{c})}
p(a | \vec{c}) p(\vec{c}).
\label{eqn:restricted}
\end{align}
}This factorization differs from the saturated observed data factorization, given by
{\small
\begin{align}
p(y_1, y_2, a,\vec{c}) =
\frac{
\phi_{Y_1,Y_2,\vec{C}}{(y_1,y_2,a,\vec{c})}
\phi_{Y_1,A,\vec{C}}{(y_1,{a},\vec{c})}
\phi_{Y_2,A,\vec{C}}{(y_2,{a},\vec{c})}
}{Z({a},\vec{c})}
p(a | \vec{c}) p(\vec{c}).
\label{eqn:saturated}
\end{align}
}In particular, $\phi_{Y_1,Y_2,\vec{C}}{(y_1,y_2,\vec{c})}$, the first term in the numerator of (\ref{eqn:restricted}), does not
depend on values of $A$, while $\phi_{Y_1,Y_2,A,\vec{C}}{(y_1,y_2,a,\vec{c})}$, the first term in the numerator of (\ref{eqn:saturated}), does.

This implies that we may falsify our model by checking whether the data supports the restriction imposed on the observed law
via a hypothesis testing procedure.
If the submodel corresponding to (\ref{eqn:restricted}) is not supported by the data, this implies that it is not possible to set up a randomized controlled trial, where the decomposition of the treatment $A$ into components is represented by a chain graph where 
the appropriate exclusion restrictions, corresponding by missing edges in Fig. \ref{fig:med} (f), hold.  We contrast this situation with what happens with mediation analysis in a DAG model.  In such a model, assumptions underlying identification of mediation functionals in DAG models do not
place any restrictions on the observed data law.  This implies that decomposability of the treatment $A$ into components that satisfy exclusion restrictions, represented by missing edges in Fig. \ref{fig:med} (c), must be verified entirely using background knowledge, or a subsequent randomized experiment on the treatment components.

Note that since the treatment component model represented by a CG is falsifiable while the treatment component model represented by a DAG is not, if the factorization (\ref{eqn:restricted}) can be ruled out by the data,
it is still possible to set up a randomized controlled trial, where the decomposition of the treatment $A$ into components is represented by a \emph{DAG} where 
the appropriate exclusion restrictions, corresponding by missing edges in Fig. \ref{fig:med} (c), hold.  However, while such a causal model is not (and cannot) be falsified using observed data, it may nevertheless not be appropriate for situations involving dyads, network data, or symmetry among units in a study that would render imposing an order on unit outcomes substantively inappropriate.


Another implication of the fact that in our setting contagion and infectiousness components of the spillover effect are only identified in causal models consistent with a strict submodel of the saturated observed data model is that if the observed data law does not lie in this submodel, the 
functionals 
that would have corresponded to identified
contagion and infectiousness effects, had the model been true, do \emph{not} add up to the spillover effect.  This is in contrast to classical mediation analysis settings where functionals given by the mediation formula corresponding to natural direct and indirect effects always add up to the functional corresponding to the average causal effect (this follows by a simple telescoping sum argument), even in cases where direct and indirect effects are \emph{not identifiable}, and thus not equal to those functionals.

Finally, we note that the existence of observable implications of assumptions (\ref{eqn:cg_y_const}) and (\ref{eqn:cg_m_const}) imply that the network structural model (NSM) interpretation of CG models differs in
a crucial respect from structural models imposed on counterfactual quantities in causal DAG models.  Standard structural models are employed as smoothing assumptions to make inference tractable in the presence of the curse of dimensionality.  However, identifiability of a counterfactual quantity on which a structural model is imposed does not rely on structural model assumptions. 
This is an important property as it allows potentially unrealistic smoothing assumptions of a structural model to be relaxed if more data becomes available without substantially changing prior steps of the analysis, including identification arguments.

By contrast, a causal model corresponding to DAG of blocks, such as that shown in Fig.~\ref{fig:auto-g} (b), along with the NSM imposed on it that yields a CG model allowing treatment decomposition, such as that shown in Fig.~\ref{fig:med} (f), together must impose a testable implication for identification to be recovered.
Thus, a CG model imposed as an NSM on a ``DAG of blocks'' for partial interference problems may be viewed either as a regular structural model, but one that only applies to ``DAG of blocks'' causal submodels where an appropriate interaction term in the observed data likelihood is absent,
or as a structural model that applies to a causal model implying the saturated model on the observed data distribution, but which partly restricts this model, with ramifications for identifiability.

\section{Statistical Inference For Symmetric Treatment Decompositions}
\label{sec:stat}

We now consider two approaches to statistical inference for the
symmetric mediation formula in the dyad (block of size $2$) setting,
one based on maximum likelihood estimation, and one on doubly robust
semi-parametric estimation.

Assume a dataset with $B$ two unit blocks (dyads), with labeled
outcomes $\vec{Y} = ({Y_1, Y_2})$, and baseline covariates
${\vec{C}} = ({C_1, C_2})$ for each of the two units.  For simplicity, we assume only
a single treatment $A$ is assigned.  Without loss of generality, assume $A$ is
assigned to unit $1$.  We also assume the causal CG model shown in Fig.~\ref{fig:med} (f).
We wish to estimate direct and indirect components of the spillover effect of $A$ on
$Y_{2}$ ``mediated'' by $Y_{1}$, in the sense described above.

\subsection{Maximum likelihood inference}
\label{subsec:mle}


We first describe estimation of the target parameter $\psi(\tilde{a}^1,\tilde{a}^2{'}) =\E[Y_{2}(\tilde{a}^1,\tilde{a}^2{'})]$, which is the expectation with respect to
the distribution identified 
by the functional in (\ref{eqn:sym_med}).

Without loss of generality, we denote $\phi_{Y_1,Y_2,{\vec{C}}}(y_1,y_2,\vec{c})$ by the conditional odds ratio function
$\gamma(y_1,y_2 | \vec{c}) = \frac{p(Y_1=y_1,Y_2=y_2 | \vec{c}) \cdot p(Y_1=0,Y_2=0 | \vec{c})}{ p(Y_1=y_1,Y_2=0 | \vec{c}) \cdot p(Y_1=0,Y_2=y_2 | \vec{c})}$, $\phi_{Y_2,A,{\vec{C}}}(y_2,\tilde{a}^{2}{'},\vec{c})$ by the conditional density $f(y_2 | \tilde{a}^{2}{'}, Y_1 = 0, {\vec{c}})$, and
$\phi_{Y_1,A,{\vec{C}}}(y_1,\tilde{a}^{1},\vec{c})$ by the conditional density $f(y_1 | \tilde{a}^{1}, Y_2 = 0, {\vec{c}})$ \citep{shpitser23lc}.

For any fixed function $h(y_{1},y_{2})$, define
$\beta_{h}(\tilde{a}^1, \tilde{a}^2{'},{\vec{c}})$ as 
{\small
\begin{equation}
\sum_{y_{1},y_{2}}h(y_{1},y_{2}) f(y_{1}\mid \tilde{a}^1,Y_{2}=0,{\vec{c}}) \gamma(y_{1},y_{2}\mid{\vec{c}}) f(y_{2}\mid \tilde{a}^2{'},Y_{1}=0,{\vec{c}}),
\label{eqn:beta}
\end{equation}
}where $\sum$ may be interpreted as integration for continuous variables.

We obtain the following representation: 
$\psi(\tilde{a}^1,\tilde{a}^2{'}) = \sum_{{\vec{c}}}\theta(\tilde{a}^1,\tilde{a}^2{'},{\vec{c}})p({\vec{c}})$, for
$\theta(\tilde{a}^1,\tilde{a}^2{'},{\vec{c}}) =\frac{\beta _{y_{2}}(\tilde{a}^1,\tilde{a}^2{'},{\vec{c}}) }{\beta _{1}(\tilde{a}^1,\tilde{a}^2{'},{\vec{c}})}$. 
Note that $\beta_{1}(\tilde{a}^1,\tilde{a}^2{'},{\vec{c}})$, for $h(y_1,y_2) = 1$, serves as the normalizing function $Z(\tilde{a}^{1}, \tilde{a}^{2}{'}, {\vec{c}})$
in (\ref{eqn:sym_med}).



This type of parameterization is described in more detail in \citep{chen07semiparametric,shpitser23lc}.
An advantage of this parameterization is that it decomposes the joint outcome distribution
into variation independent components, including conditional distributions, which are easy to specify using standard
regression models if the outcomes are binary.  
Another alternative for binary models is the standard log-linear parameterization.

Maximum likelihood estimation of $\psi(\tilde{a}^1,\tilde{a}^2{'}) $ requires the
correct specification of models for $f(y_{1}\mid \tilde{a}^1,Y_{2}=0,{\vec{c}})$,
$\gamma(y_{1},y_{2}\mid \vec{c})$, and $f(y_{2}\mid \tilde{a}^2{'},Y_{1}=0,\vec{c})$.
Given parametric models $f(y_{1}\mid \tilde{a}^1,Y_{2}=0,\vec{c};\omega _{1})$,
$\gamma( y_{1},y_{2}\mid \vec{c};\nu)$, and $f(y_{2}\mid \tilde{a}^2{'},Y_{1}=0,\vec{c};\omega_{2})$,
and a data matrix ${\cal D}$, the maximum likelihood estimator
$(\widehat{\omega }_{1}, \widehat{\omega }_{2},\widehat{\nu})$
of $(\omega _{1},\omega_{2},\nu)$
maximizes the following log-likelihood
$\log {\cal L}_{\vec{Y},A,{\vec{C}}}({\cal D}; {(} \omega_1, \omega_2, \nu {)})$
{\small
\begin{equation*}
\sum_{b=1}^B\log \frac{
f\left( y^{b}_{1}\mid a^{b},Y_{2}=0,{\vec{c}}^{b};\omega _{1}\right) \gamma \left(
y^{b}_{1},y^{b}_{2}\mid {\vec{c}}^{b};\nu \right) f\left(
y^{b}_{2}\mid a^{b},Y_{1}=0,{\vec{c}}^{b};\omega _{2}\right) }{\sum_{y_{1},y_{2}}f\left(
y_{1}\mid a^{b},Y_{2}=0,{\vec{c}}^{b};\omega _{1}\right) \gamma \left(
y_{1},y_{2}\mid {\vec{c}}^{b};\nu \right) f\left( y_{2}\mid a^{b},Y_{1}=0,{\vec{c}}_{i};\omega
_{2}\right) }.
\end{equation*}
}The corresponding score equation does not yield closed form solutions, even for binary data, but standard iterative algorithms may be used \citep{lauritzen96graphical}.
The corresponding maximum likelihood estimator of
$\psi(\tilde{a}^1,\tilde{a}^2{'})$ is given by 
{\small
\begin{equation}
\widehat{\psi}(\tilde{a}^1,\tilde{a}^2{'}) =B^{-1}\sum_{b=1}^B\theta \left(\tilde{a}^1,\tilde{a}^2{'},{\vec{c}}^{b};
\widehat{\omega }_{1},\widehat{\omega}_{2},\widehat{\nu}\right)
\label{eqn:mle}
\end{equation}
}where $\theta(\tilde{a}^1,\tilde{a}^2{'},{\vec{c}}; \widehat{\omega }_{1},
\widehat{\omega}_{2},\widehat{\nu})$ is equal to $\theta(\tilde{a}^1,\tilde{a}^2{'},{\vec{c}})$
evaluated at $\gamma(y_{1},y_{2}\mid {\vec{c}}; \widehat{\nu})$,
$f(y_{1}\mid \tilde{a}^1,Y_{2}=0,{\vec{c}}; \widehat{\omega}_{1})$,
and $f(y_{2}\mid \tilde{a}^2{'},Y_{1}=0,{\vec{c}}; \widehat{\omega}_{2})$.
Under standard regularity conditions, $B^{1/2}( \widehat{\psi}-\psi)$
is approximately normal for large $B$ with mean zero and variance given by
$\dot{\widehat{\psi}}\widehat{I}^{-1}{\dot{\widehat{\psi}}}\,^{T}$, where
{\small
\[
\dot{\widehat{\psi}}=B^{-1}\sum_{b=1}^B
\frac{
\partial \theta \left(\tilde{a}^1,\tilde{a}^2{'},{\vec{c}}^{b};\omega_1,\omega_2,\nu\right)
}{
\partial \left(\omega_1,\omega_2,\nu\right)^{{T}}
}
\bigg|_{(\omega_1,\omega_2,\nu)=(\widehat{\omega}_{1},
\widehat{\omega}_{2},\widehat{\nu})}
\ ,
\]
}and $\widehat{I}$ is the second derivative of
$\log {\cal L}_{\vec{Y},A,{\vec{C}}}({\cal D}; {(} \omega_1, \omega_2, \nu {)})$
with respect of ${(} \omega_{1},\omega _{2},\nu {)}$ evaluated at
${(} \widehat{\omega }_{1}, \widehat{\omega }_{2},\widehat{\nu }{)}$.
In other words, (\ref{eqn:mle}) is a maximum likelihood plug-in estimator for (\ref{eqn:sym_med}),
with the factorization in (\ref{eqn:mle}) given in the form described by \citet{chen07semiparametric}.


\subsection{
Semi-parametric inference via doubly robust estimation}
\label{subsec:doubly}

We develop a robust statistical inference method for the target parameter $\psi(\tilde{a}^1, \tilde{a}^2{'})$, by constructing an estimator using the theory of influence functions.
To do so, we first find a collection of influence functions for $\psi$ in the statistical model ${\cal M}_{\text{sym}}$ corresponding to the graph in Fig.~\ref{fig:med} (f), which is the set of all joint distributions admitting the factorization in (\ref{eqn:restricted-0}).

To simplify subsequent presentation, we will define the following functions. First, let $\gamma_{1A}(y_1, a \cond \vec{c} )$ and $\gamma_{2A}(y_2, a \cond \vec{c})$ be the odds ratio functions associated with $Y_1$ and $Y_2$ defined as follows:
{\small
\begin{align*}
&
\gamma_{1A} (y_1 , a \cond \vec{c})
=
\frac{ f(y_1 \cond a, Y_2=0, \vec{c}) }{ f(y_1 \cond \tilde{a}^1, Y_2=0, \vec{c}) }
\frac{ f(0 \cond \tilde{a}^1, Y_2=0, \vec{c}) }{ f(0 \cond a, Y_2=0, \vec{c}) }
\ , \\
&
\gamma_{2A} (y_2 , a \cond \vec{c})
=
\frac{ f(y_2 \cond a, Y_1=0, \vec{c}) }{ f(y_2 \cond \tilde{a}^2{'}, Y_1=0, \vec{c}) } 
\frac{ f(0 \cond \tilde{a}^2{'}, Y_1=0, \vec{c}) }{ f(0 \cond a, Y_1=0, \vec{c}) } 
\ .
\end{align*}
}Note that $\gamma_{1A}$ and $\gamma_{2A}$ use $A=\tilde{a}^1$ and $A=\tilde{a}^2{'}$ as the 
baseline values, respectively. 
Next, let $Q_1(y_1 \cond \vec{c})$ and $Q_2(y_2 \cond \vec{c})$ be the following functions:
{\small
\begin{align}	\label{eq-Qfunctions}
& 
Q_1(y_1 \cond \vec{c})
=
\sum_{y_2} 
\big\{ h(y_1,y_2) - \theta(\tilde{a}^1,\tilde{a}^2{'}, \vec{c}) \big\}
\gamma(y_1,y_2 \cond \vec{c} )
f_2(y_2 \cond \tilde{a}^2{'},Y_1=0,\vec{c})
\ ,
\\
&
Q_2(y_2 \cond \vec{c})
=
\sum_{y_1} 
\big\{ h(y_1,y_2) - \theta(\tilde{a}^1,\tilde{a}^2{'}, \vec{c}) \big\}
\gamma(y_1,y_2 \cond \vec{c} )
f_1(y_1 \cond \tilde{a}^1,Y_2=0,\vec{c})
\ .
\nonumber
\end{align}
}where $\sum$ may be interpreted as integration for continuous variables. Lastly, let $\delta_1(\vec{C})$ and $\delta_2(\vec{C})$ be
{\small
\begin{align*}
    &
    \delta_1(\vec{C})
    =
    \mathbb{E} \bigg\{ \frac{ \mathbb{I}(A=\tilde{a}^1) }{ \gamma_{2A}(Y_2, \tilde{a}^1 \cond \vec{C}) } \, \bigg| \, \vec{C} \bigg\}
    \ ,
    &&
    \delta_2(\vec{C})
    =
    \mathbb{E} \bigg\{ \frac{ \mathbb{I}(A=\tilde{a}^2{'}) }{ \gamma_{1A}(Y_1, \tilde{a}^2{'} \cond \vec{C}) } \, \bigg| \, \vec{C} \bigg\}
\end{align*}
}

The following theorem presents a class of influence functions for $\psi(\tilde{a}^1,\tilde{a}^2{'})$ in the model ${\cal M}_{\text{sym}}$. 
\begin{theorem}
\label{thm:IF}
Let $\mathcal{W}$ be the space of all square-integrable functions of $\vec{C}$.
Then, for any $w(\vec{C}) \in {\mathcal W}$, the following function $\IF_w$
is an influence function for $\psi(\tilde{a}^1,\tilde{a}^2{'})$ in the model ${\cal M}_{\text{sym}}$. 
{\small
\begin{align*}
& \IF_w(Y_1,Y_2,A,\vec{C})
\\
&
=
w(\vec{C}) 
\left[
\!\!
\begin{array}{l}			
\frac{ \mathbb{I}(A=\tilde{a}^1) }{ \gamma_{2A}(Y_2, A \cond \vec{C})  } \frac{h(Y_1,Y_2) - \theta(\tilde{a}^1,\tilde{a}^2{'}, \vec{C})}{  \delta_2(\vec{C})  }
+
\Big\{
\frac{ \mathbb{I}(A=\tilde{a}^2{'}) }{  \gamma_{1A} (Y_1, A \cond \vec{C})  } \frac{1}{ \delta_1(\vec{C}) }
-
\frac{ \mathbb{I}(A=\tilde{a}^1) }{\gamma_{2A} (Y_2, A \cond \vec{C}) } \frac{1}{ \delta_2(\vec{C}) }
\Big\}
\frac{Q_2(Y_2 \cond \vec{C})}{\gamma(Y_1,Y_2 \cond \vec{C})}
\end{array}
\!\!
\right]
\\
&
+
\big\{ 1 - w(\vec{C})  \big\}
\left[
\!\!
\begin{array}{l}			
\frac{  \mathbb{I}(A=\tilde{a}^2{'}) }{ \gamma_{1A} (Y_1, A \cond \vec{C}) } \frac{ h(Y_1,Y_2) - \theta(\tilde{a}^1,\tilde{a}^2{'}, \vec{C}) }{ \delta_1(\vec{C}) }
+
\Big\{
\frac{  \mathbb{I}(A=\tilde{a}^1) }{ \gamma_{2A} (Y_2, A \cond \vec{C}) } \frac{1}{ \delta_2(\vec{C}) }
-
\frac{  \mathbb{I}(A=\tilde{a}^2{'}) }{  \gamma_{1A} (Y_1, A \cond \vec{C}) } \frac{1}{ \delta_1(\vec{C}) }
\Big\}
\frac{Q_1(Y_1 \cond \vec{C})}{\gamma(Y_1,Y_2 \cond \vec{C})}
\end{array}
\!\!
\right] 
\\
&
+
\theta(\tilde{a}^1,\tilde{a}^2{'},\vec{C})
-
\psi(\tilde{a}^1,\tilde{a}^2{'}).
\end{align*}
}
\end{theorem}

Theorem \ref{thm:IF} provides a collection of influence functions for the estimand $\psi(\tilde{a}^1,\tilde{a}^2{'})$, indexed by the function $w(\vec{C})$.  For any $\IF_w$, the first group of terms, weighted by $w(\vec{C})$, can be viewed as a functional obtained by treating the outcome $Y_1$ of the first unit as the outcome in a mediation problem, and the outcome $Y_2$ of the second unit as the mediator. Likewise, the second group of terms, weighted by $1-w(\vec{C})$, can be viewed as a functional obtained by treating the outcome $Y_2$ of the second unit as the outcome in a mediation problem, and the outcome $Y_1$ of the first unit as the mediator. The first and third terms, involving the term $\{ h(Y_1,Y_2) - \theta(\tilde{a}^1,\tilde{a}^2{'},\vec{C}) \}$, resemble the augmentation term found in the augmented inverse probability-weighted estimator for the average treatment effect in the i.i.d. setting. The second and fourth terms, involving $Q_1$ and $Q_2$ functions, can be viewed as augmentation terms that are required due to the nature of the functional. 
Influence functions in this class exhibit a symmetric structure, due to the symmetric factorization (\ref{eqn:restricted-0}) of the model ${\cal M}_{\text{sym}}$. 

Any choice of $w(\vec{C}) \in {\cal W}$ yields a valid influence function in the class described above.
A simple choice of $w(\vec{C})$ sets it as a constant (e.g., $w(\vec{C})=0.5$).  The optimal choice, which we denote by $w_{\text{opt}}(\vec{C})$, 
minimizes the conditional variance of the influence function given $\vec{C}$. 
In other words, $w_{\text{opt}}(\vec{C}) = \argmin_{w(\vec{C}) \in \mathcal{W}} \text{Var}\{ \IF_w (Y_1,Y_2,A,\vec{C} ) \mid \vec{C} \}$. Moreover, it is important to note that
$\IF_w$ is not the entire collection of influence functions for $\psi(\tilde{a}^1,\tilde{a}^2{'})$, that is, some influence functions do not have a form of $\IF_w$. 


One can characterize the efficient influence function for $\psi(\tilde{a}^1,\tilde{a}^2{'})$ by improving upon influence functions $\IF_w$. Specifically, let the tangent space of model corresponding to (\ref{eqn:restricted-0})
be $\mathcal{T}$ and $\Pi( \cdot \cond \mathcal{T} )$ be a projection operator onto $\mathcal{T}$.
By modern semiparametric efficiency theory, the efficient influence function for $\psi(\tilde{a}^1,\tilde{a}^2{'})$ is characterized as $\IF^*(Y_1,Y_2,A,\vec{C}) =  \Pi \{ \IF_w(Y_1,Y_2,A,\vec{C}) \mid \mathcal{T} \}$ for any
$\IF_w$. Since the efficient influence function is unique, the projection of any influence function coincides with the efficient influence function, i.e., even though $w \neq w'$, we have $\IF^*(Y_1,Y_2,A,\vec{C}) =  \Pi \{ \IF_w(Y_1,Y_2,A,\vec{C}) \mid \mathcal{T} \} = \Pi \{ \IF_{w'} (Y_1,Y_2,A,\vec{C}) \mid \mathcal{T} \}$. Unfortunately, a closed-form representation of the efficient influence function may not be available in general, especially when the outcomes are continuous. For readers interested in technical details, Section \ref{subsec:eif} of the Appendix provides additional discussions on (i) the characterization of the efficient influence function and (ii) the closed-form representation of the efficient influence function under binary outcomes as a special case. We briefly note that these results effectively extend the semiparametric efficiency theory obtained by \cite{tchetgen12semi2} for the standard mediation formula under a causal DAG model to the symmetric mediation formula for a causal chain graph model.  

Using the influence function in Theorem \ref{thm:IF} as a basis, we can construct a robust estimator for $\psi(\tilde{a}^1,\tilde{a}^2{'}) $. We posit parametric models for the nuisance functions as $\gamma(y_{1},y_{2} \cond \vec{c} ;\nu) $, $f(y_{1} \cond a,Y_{2}=0, \vec{c};\omega _{1})$, $f(y_{2}\cond a,Y_{1}=0, \vec{c};\omega _{2})$, $\delta_1 (\vec{c} \con \kappa_1)$, and $\delta_2 (\vec{c} \con \kappa_2)$, respectively. We then define submodels $\mathcal{M}_{\gamma}$, $\mathcal{M}_{f_1}$, $\mathcal{M}_{f_2}$, $\mathcal{M}_{\delta_1}$, and $\mathcal{M}_{\delta_2}$ of $\mathcal{M}_{\text{sym}}$ as follows:
{\small
\begin{align*}
&
\mathcal{M}_{\gamma} = \big\{ f \in \mathcal{M}_{\text{sym}} \cond \gamma(y_{1},y_{2} \cond \vec{c} ;\nu) \text{ is correctly specified} \big\}
\\
&
\mathcal{M}_{f_1} = \big\{ f \in \mathcal{M}_{\text{sym}} \cond f(y_{1} \cond a, Y_{2}=0, \vec{c} ; \omega_1) \text{ is correctly specified} \big\}
\\
&
\mathcal{M}_{f_2} = \big\{ f \in \mathcal{M}_{\text{sym}} \cond f(y_{2} \cond a, Y_{1}=0, \vec{c} ; \omega_2) \text{ is correctly specified} \big\}
\\
&
\mathcal{M}_{\delta_1} = \big\{ f \in \mathcal{M}_{\text{sym}} \cond \delta_1(\vec{c} \con \kappa_1) \text{ is correctly specified} \big\}
\\
&
\mathcal{M}_{\delta_2} = \big\{ f \in \mathcal{M}_{\text{sym}} \cond \delta_2(\vec{c} \con \kappa_2) \text{ is correctly specified} \big\} \ .
\end{align*}
}We then define a model $\mathcal{M}^\dagger = \mathcal{M}_\gamma \cap \big\{ \mathcal{M}_{f_1} \cup \mathcal{M}_{f_1} \big\}$ where (i) the odds ratio $\gamma$ is correctly specified and either (ii.a) the conditional density $f(y_1 \cond a,Y_2=0,\vec{c})$ or (ii.b) the conditional density $f(y_2 \cond a,Y_1=0,\vec{c})$, but not necessarily both, is correctly specified. Likewise, we define a model $\mathcal{M}^* = \mathcal{M}_\gamma \cap \big\{ 
\{ \mathcal{M}_{f_1} \cap \mathcal{M}_{\delta_1}\} \cup \{ \mathcal{M}_{f_2} \cap \mathcal{M}_{\delta_2}\} \big\}$ where (i) the odds ratio $\gamma$ is correctly specified and either (ii.a) the nuisance functions of the first individual (i.e., $f(y_1 \cond a,Y_2=0,\vec{c})$ and $\delta_1$) or (ii.b) those of the second individual (i.e.,  $f(y_2 \cond a,Y_1=0,\vec{c})$ and $\delta_2$), but not necessarily both, are correctly specified. Note that $\mathcal{M}^*$ is a submodel of $\mathcal{M}^\dagger$. In addition, when $\vec{C}$ is empty or categorical, $\mathcal{M}_{\delta_1}$ and $\mathcal{M}_{\delta_2}$ are readily satisfied under $\mathcal{M}_{f_1}$ and $\mathcal{M}_{f_2}$, respectively, by specifying $\kappa_1$ and $\kappa_2$ as the stratum-specific parameters.

The proposed estimator remains consistent under model $\mathcal{M}^*$, so it offers the analyst two opportunities to obtain a consistent estimator for $\psi(\tilde{a}^{1},\tilde{a}^2{'})$. In order to exhibit such an estimator requires successfully completing the following tasks:
\begin{enumerate}
\item First, obtaining a consistent estimator of $\gamma(y_{1},y_{2}; \nu)$ under model $\mathcal{M}^\dagger$;
\item Second, obtaining a consistent estimator of $\psi(\tilde{a}^1,\tilde{a}^2{'})$ under model $\mathcal{M}^*$.
\end{enumerate}
\cite{tchetgen11double}
have previously characterized a large class
of doubly robust estimators that accomplish task 1, in the sense that any
estimator of $\gamma(y_{1},y_{2}; \nu)$ in their class (which
includes {
the} semiparametric locally efficient estimator) is guaranteed to
remain consistent and asymptotically normal under model $\mathcal{M}^\dagger$ denote the
conditional MLE that maximizes the conditional log likelihood $\sum_{b=1}^B\log
f( y^{b}_{1}\mid a^{b},y^{b}_{2},{\vec{c}}^{\,b};\omega _{1},\nu )$, where
{\small
\begin{align*}
f(y_1^{b} \cond a^b, y_2^{b}, {\vec{c}}^{\,b} \con \omega_1,\nu)
=
\frac{ 
\gamma(y_1^{b},y_2^{b} \cond {\vec{c}}^{\,b} \con \nu)
f(y_1^{b} \cond a^b, Y_2=0, {\vec{c}}^{\,b} \con \omega_1)	
}{ 
\sum_{y_1}
\gamma(y_1,y_2^{b} \cond {\vec{c}}^{\,b} \con \nu)
f(y_1 \cond a^b, Y_2=0, {\vec{c}}^{\,b} \con \omega_1)
}
\end{align*}
}Likewise, let $\widetilde{\omega }_{2}\left( \nu \right)$ denote the
corresponding conditional MLE of $\omega _{2}$.  \cite{tchetgen11double}
proved that the solution $\widehat{\nu }_{dr}$ to the following class
of estimating equations is doubly robust, i.e. consistent and asymptotically
normal under (i) and (ii): 
{\small
\begin{align} \label{eqn:dr_eta}
0
=
\sum_{b=1}^{B}
\frac{
g_{\gamma} ({\vec{c}}^{b})
}{
\gamma(y_1^{b}, y_2^{b} \cond {\vec{c}}^{b} \con \widehat{\nu}_{dr} )
}
\left[
\begin{array}{l}
\big[
y_1^{b} - \EXP \big\{ Y_1 \cond a^b, Y_2=0,{\vec{c}}^{\,b} \con \widetilde{\omega}_1(\widehat{\nu}_{dr}) \big\} 
\big]	
\\
\times
\big[
y_2^{b} - \EXP \big\{ Y_2 \cond a^b, Y_1=0, {\vec{c}}^{\,b} \con \widetilde{\omega}_2(\widehat{\nu}_{dr}) \big\} 
\big]	
\end{array}
\right] 
\end{align}
}where $g_{\gamma}$ is a user-specified function of dimension matching that of $\nu$.
\cite{tchetgen11double}
developed a more general class of doubly
robust estimators including locally semiparametric efficient estimators for
polytomous, count or continuous $Y,$ we refer the reader to the original
manuscript for more details.

Next, we turn to task $2$. We first consider the following estimating functions for $\kappa_1$ and $\kappa_2$ where
{\small
\begin{align*}
&
U_{1} ( \kappa_1 , \omega_1 )
=
g_1(\vec{C} )
\bigg\{
\frac{ \mathbb{I}(A = \tilde{a}^2{'}) }{  \gamma_{1A} \big( Y_1, \tilde{a}^2{'} \cond \vec{C} \con \omega_1 \big)   }
-
\delta_{1}(\vec{C} \con \kappa_1 )
\bigg\}
\\
&
U_{2} ( \kappa_2, \omega_2 )
=
g_2(\vec{C} )
\bigg\{
\frac{
\mathbb{I}(A = \tilde{a}^{1}) }{  \gamma_{2A} \big( Y_2, \tilde{a}^1 \cond \vec{C} \con \omega_2 \big)    } 
-
\delta_{2} (\vec{C} \con \kappa_2 )
\bigg\}
\ ,
\end{align*}
}where $g_{1}$ and $g_{2}$ are user-specified functions of dimension matching that of $\kappa_1$ and $\kappa_2$, respectively. Let $\widetilde{\kappa}_{1}$ and
$\widetilde{\kappa}_{2}$ be the solutions to the estimating equations $0 = \EXP \big\{ U_{1} (\kappa_1, \omega_1) \big\}$ and $0 = \EXP \big\{ U_{2} (\kappa_2, \omega_2) \big\}$ at $\omega_1 = \widetilde{\omega}_1(\widehat{\nu}_{dr})$ and $\omega_2 = \widetilde{\omega}_2(\widehat{\nu}_{dr})$, respectively. 

Lastly, we use the influence function in Theorem \ref{thm:IF} to obtain the estimating function for $\psi:= \psi(\tilde{a}^1,\tilde{a}^2{'})$:
{\small
\begin{align}
& U_{\psi}
( \psi, 
\kappa_{1},
\kappa_{2},	
\omega_{1}, \omega_{2}, \nu )
\label{eq-EE psi}
\\
&
=
\left[
\begin{array}{l}
w(\vec{C}) 
\frac{1}{ \delta_{2} (\vec{C} \con \kappa_2) }
\frac{ \mathbb{I}(A=\tilde{a}^1) }{   \gamma_{2A} (Y_2, \tilde{a}^1 \cond \vec{C} \con \omega_2 )  }
\{ h(Y_1,Y_2) - \theta(\tilde{a}^1,\tilde{a}^2{'}, \vec{C} \con \omega_1,\omega_2,\nu ) \}
\\
+
w(\vec{C}) 
\Big[
\frac{1}{ \delta_{1} (\vec{C} \con \kappa_1) }
\frac{ \mathbb{I}(A=\tilde{a}^2{'})   }{ \gamma_{1A} (Y_1, \tilde{a}^2{'} \cond \vec{C} \con \omega_1 ) }
-
\frac{1}{ \delta_{2} (\vec{C} \con \kappa_2) }
\frac{ \mathbb{I}(A=\tilde{a}^1) }{ \gamma_{2A} (Y_2, \tilde{a}^1 \cond \vec{C} \con \omega_2 )  }
\Big]
\frac{Q_2(Y_2 \cond \vec{C} \con \omega_1, \omega_2, \nu) }{\gamma(Y_1,Y_2 \cond \vec{C} \con \nu)}
\\ 
+
\big\{ 1 - w(\vec{C}) \big\}
\frac{1}{ \delta_{1} (\vec{C} \con \kappa_1) }
\frac{  \mathbb{I}(A=\tilde{a}^2{'}) }{  \gamma_{1A} (Y_1 , \tilde{a}^2{'} \cond \vec{C}  \con \omega_1 )  }
\{ h(Y_1,Y_2) - \theta(\tilde{a}^1,\tilde{a}^2{'}, \vec{C} \con \omega_1,\omega_2,\nu ) \}
\\
+
\big\{ 1 - w(\vec{C} )  \big\}
\Big[
\frac{1}{ \delta_{2} (\vec{C} \con \kappa_2) }
\frac{ \mathbb{I}(A=\tilde{a}^1)   }{ \gamma_{2A} (Y_2, \tilde{a}^1 \cond \vec{C} \con \omega_2 )  }
-
\frac{1}{ \delta_{1} (\vec{C} \con \kappa_1) }
\frac{ \mathbb{I}(A=\tilde{a}^2{'})  }{ \gamma_{1A}(Y_1, \tilde{a}^2{'} \cond \vec{C} \con \omega_1 )  }
\Big]
\frac{Q_1(Y_1 \cond \vec{C} \con \omega_1, \omega_2,  \nu) }{\gamma(Y_1,Y_2 \cond \vec{C} \con \nu)}
\\ 
+
 \theta(\tilde{a}^1,\tilde{a}^2{'}, \vec{C} \con \omega_1,\omega_2,\nu )
-
\psi
\end{array}
\right] 
\nonumber
\end{align}
}where $\theta(\tilde{a}^1,\tilde{a}^2{'},\vec{C} \con \omega_1,\omega_2,\nu)$, $Q_1(Y_1 \cond \vec{C} \con \omega_1, \omega_2, \nu)$ and $Q_2(Y_2 \cond \vec{C} \con \omega_1, \omega_2, \nu)$ are obtained from \eqref{eqn:beta} and \eqref{eq-Qfunctions} with parametrized functions. Let $\widehat{\psi}_{dr}$ be the solution to the estimating equation
$0= \EXP \big\{ U_{\psi}( \psi, \kappa_{1},\kappa_{2},\omega_{1}, \omega_{2}, \nu ) \big\}$ at
$(\widetilde{\kappa}_{1}, \widetilde{\kappa}_{2}, \widetilde{\omega}_1(\widehat{\nu}_{dr}),  \widetilde{\omega}_2 (\widehat{\nu}_{dr}), \widehat{\nu}_{dr})$.
In Section~\ref{subsec:eif} of the Appendix, we present the form of the estimator for the case when $(Y_1,Y_2) \cond A$ follows a bivariate normal distribution. Additionally,
we show that $\widehat{\psi}_{dr}$ is multiply robust in the sense that it is consistent for $\psi(\tilde{a}^1,\tilde{a}^2{'})$ under model $\mathcal{M}^*$. We remark that the robustness property does not depend on the choice of $w(\vec{C})$. Confidence intervals for these estimates can be obtained via the standard nonparametric bootstrap. 

\section{Simulation Studies}
\label{sec:sims}

We conducted simulation studies to illustrate the behavior of estimators proposed in Section~\ref{sec:stat} for the components of
the spillover effect.
All figures displaying our results are deferred to the supplement, in the interests of space. We considered the following data generating mechanism where $p(Y_1,Y_2,A,\vec{C})$ consisted of a vector $\vec{C}$ of $2$ baseline variables, a single treatment variable, and two dependent outcomes $Y_1,Y_2$.  In all cases, we assumed binary treatments,
continuous outcomes, and in the second case continuous baseline variables.

To ensure the constraint on the observed data law of the type shown in (\ref{eqn:restricted}) held, the data
generating mechanisms were selected from the conditional Gaussian mixed interaction model class, described in
\citep{hojsgaard12graphical}.  Given a vector of discrete variables $\vec{X}$ and continuous variables $\vec{Y}$,
a conditional Gaussian joint distribution $p(\vec{Y},\vec{X})$ is specified as
{\small
\begin{align}
\frac{p(\vec{x})}{(2 \pi)^{q/2} \text{det}(\Sigma)^{1/2}} \exp\left\{
- \frac{\{ \vec{y} - \mu(\vec{x}) \}^T \Sigma^{-1} (\vec{y} - \mu(\vec{x}) \}}{2} 
\right\} 
= \exp \left\{
g(\vec{x}) + h(\vec{x})^T \vec{y} - \frac{(\vec{y})^T K \vec{y}}{2} 
\right\},
\label{eqn:canonical}
\end{align}
}where $\mu(\vec{x})$ is a vector of mean parameters for $\vec{Y}$ that depend on $\vec{X}$, $\Sigma$ is the
covariance matrix for $\vec{Y}$ (that is assumed to not depend on $\vec{X}$), and $K = \Sigma^{-1}$,
$h(\vec{x}),g(\vec{x})$ are the canonical parameters for the exponential family representation of
this class of densities.


In the first case, we specified the model corresponding to a subgraph of Fig.~\ref{fig:med} (f) where $\vec{C}$, the set of
baseline covariates, is absent.  In other words, we ensured that the conditional independence constraints
$(Y_1 \ci \tilde{A}^{2} \mid Y_2, \tilde{A}^{1})$, and $(Y_2 \ci \tilde{A}^{1} \mid Y_1, \tilde{A}^{2})$, hold.
In the second case, we specified the model in such a way that the conditional independence constraints in Fig. \ref{fig:med} (f) itself,
namely $(Y_1 \ci \tilde{A}^{2} \mid Y_2, \tilde{A}^{1},\vec{C})$, and $(Y_2 \ci \tilde{A}^{1} \mid Y_1, \tilde{A}^{2},\vec{C})$, hold.

We accomplish this by considering the parameter vector
$(h_{y_1}(\tilde{a}^1,\tilde{a}^2), h_{y_2}(\tilde{a}^1,\tilde{a}^2))$ specified as the following mixed interaction model:
{\small
\begin{align*}
h^{y_1}(\tilde{a}^1,\tilde{a}^2) &= v + v_j^{\tilde{a}^1} + v_k^{\tilde{a}^2} + v_{jk}^{\tilde{a}^1\tilde{a}^2} = v + v_j^{\tilde{a}^1}\\
h^{y_2}(\tilde{a}^1,\tilde{a}^2) &= w + w_j^{\tilde{a}^1} + w_k^{\tilde{a}^2} + w_{jk}^{\tilde{a}^1\tilde{a}^2} = w + w_k^{\tilde{a}^2}.
\end{align*}
}for the first case, and the parameter vector $(h^{y_1}(\tilde{a}^1,\tilde{a}^2,c_1,c_2), h^{y_2}(\tilde{a}^1,\tilde{a}^2,c_1,c_2))$ specified as the following mixed interaction model:
{\small
\begin{align*}
h^{y_1}(\tilde{a}^1,\tilde{a}^2,c_1,c_2) &= \sum_{\vec{s} \subseteq \{ \tilde{a}^1,\tilde{a}^2,c_1,c_2 \}} v_{\vec{s}}^{\left( \prod_{s \in \vec{s}} s \right)} = v + v_j^{\tilde{a}^1} + v_l^{c_1} + v_m^{c_2}\\
h^{y_2}(\tilde{a}^1,\tilde{a}^2,c_1,c_2) &= \sum_{\vec{t} \subseteq \{ \tilde{a}^1,c_2,c_1,c_2 \}} w_{\vec{t}}^{\left( \prod_{t \in \vec{t}} t \right)} = w + w_k^{\tilde{a}^2} + v_l^{c_1} + v_m^{c_2}.
\end{align*}
}

In other words, we specify $h^{y_1}$ and $h^{y_2}$ via a set of interaction parameters, and set some of these parameters to zero in such a way that
the appropriate independence
constraints 
hold; see Section \ref{sec:sim-details} of the Appendix for details of the data generating mechanism.

Our simulation study considered sample sizes from $B \in \{1000, 2000, 5000, 10000\}$ with $500$ replicates at each sample size. We implemented the maximum likelihood estimator in equation (\ref{eqn:mle}) with
the following specifications for $\gamma_{12}$, $f_1 := f(y_1 \cond a, Y_2=0,\vec{c} ; \omega_1)$, and $f_2 := f(y_2 \cond a, Y_1=0,\vec{c} ; \omega_2)$:
\begin{itemize}
    \item[] {\bf (MLE-CC)} $\gamma_{12}$, $f_{1}$,  $f_{2}$ are correctly specified;
    \item[] {\bf (MLE-CM)} $\gamma_{12}$, $f_{1}$ are correctly specified, and $f_{2}$ is mis-specified.
\end{itemize}
In addition, we also implemented the influence function-based estimator following the approach described in Section \ref{subsec:doubly}. In particular, for the influence function-based estimator, we considered the following four model specifications for the nuisance functions:
\begin{itemize}
    \item[] {\bf (IF-CC)} $\gamma_{12}$, $f_{1}$, $\delta_1$, $f_{2}$, $\delta_2$ are correctly specified;
    \item[] {\bf (IF-CM)} $\gamma_{12}$, $f_{1}$, $\delta_1$ are correctly specified, and $f_{2}$, $\delta_2$ are mis-specified;
    \item[] {\bf (IF-MC)} $\gamma_{12}$, $f_{2}$, $\delta_2$ are correctly specified, and $f_{1}$, $\delta_1$ are mis-specified;
    \item[] {\bf (IF-MM)} $\gamma_{12}$ is correctly specified, and $f_{1}$, $\delta_1$, $f_{2}$, $\delta_2$ are mis-specified.
\end{itemize}
Of note, the first three specification scenarios are submodels of $\mathcal{M}^*$ whereas the last one is not. Therefore, the influence function-based estimator is expected to be consistent in the first three scenarios. We then estimated $\mathbb{E}[Y_2(\tilde{a}^1,\tilde{a}^2{'})]$ for $(\tilde{a}^1,\tilde{a}^2{'}) \in \{(0,0),(0,1),(1,0),(1,1)\}$, and obtained confidence intervals from nonparametric bootstrap with 400 replicates.

Figures 1-4 
in the appendix provide the visual summaries of the result. The MLE and influence function-based estimator behave as expected.
In particular, the MLE appears to be consistent in scenario (MLE-CC), but not in scenario (MLE-CM). Likewise,
the influence-function-based estimator remains consistent in scenarios (IF-CC), (IF-CM), and (IF-MC), whereas it is no longer consistent in scenario (IF-MM).
In terms of efficiency, when all nuisance components are correctly specified (i.e., (MLE-CC) and (IF-CC)), the influence function-based estimator performs quite competitively even compared to the MLE.
Empirical coverage rates based on bootstrap percentile confidence intervals appear to attain a nominal coverage if estimators are expected to be consistent.

\section{
Application: The Wisconsin Longitudinal Study}
\label{sec:dataexample}

We applied our derived maximum likelihood estimators to assess the spillover effect components of $A_1$, educational attainment of unit $1$ (``the ego''), on $Y_2$, depressive symptoms of unit $2$ (``the alter'') in the presence of interference among spousal dyads.
Our interpretation of components of $A_1$ is as follows.  The component $\tilde{A}_1^1$ pertaining to unit $1$ influences psychological coping strategies learned via education, as well as socioeconomic status of unit $1$, while the component $\tilde{A}_1^2$ pertaining to unit $2$ influences spousal friction due to difference in educational attainment, and financial dependence on the spouse.

Our data comes from the Wisconsin Longitudinal Study (WLS), which has followed a random sample of Wisconsin-area high school graduates from the class of 1957 for over 50 years. The WLS collected information on a wide range of socioeconomic and psychological factors, including occupation, physical and mental well-being, and health in later life. The WLS participants were interviewed roughly every 10-15 years between 1957 and 2011, with several interview questions pertaining to their spouses (if married). The WLS participants' spouses themselves were interviewed in 2004. For further details on the WLS, we refer the reader to \citep{herd2014cohort}. Our exposure of educational attainment $A$ was based on the 1975 WLS interview, where the participant was asked about the highest level of education completed after high school. We dichotomized responses based on a cutoff of a four-year college/university degree or higher ($A=1$) versus less than a four-year degree. Depressive symptoms for the WLS participant ($Y_1$) and his/her spouse ($Y_2$) was ascertained in 2003-2005, when a random 80\% sample of WLS participants and their spouses were asked the question ``Have you ever had a time in life lasting two weeks or more when nearly every day you felt sad, blue, depressed, or when you lost interest in most things like work, hobbies, or things you usually liked to do for fun?''  An affirmative response to this question and a negative response to a follow-up question about the depressive episode being due to alcohol, drugs, medications, or physical illness was considered evidence of depressive symptoms ($Y_1=1$ and/or $Y_2=1$); otherwise, it was assumed that depressive symptoms were absent ($Y_1=0$ and/or $Y_2=0$). In all models, we adjusted for sex, the highest educational attainment of the WLS participant's/spouses' ``head of household'' when he or she was 16 years old, and the Duncan  Socioeconomic Status Index score of the ``head of household.'' For the WLS participant, we additionally adjusted for his/her 1957 IQ score. After excluding observations for missing treatment, outcome, and/or covariate information, our analytic sample was $B=1,768$ dyads where 618 graduates had a four-year college/university degree.


As a first step, we tested whether the restriction on the observed data law given in \eqref{eqn:restricted} held for data. Specifically, we used a likelihood ratio test to validate the null hypothesis that the odds ratio for $Y_1$ and $Y_2$ given $\vec{C}$ was homogeneous across $A=1$ and $A=0$, and we found insufficient evidence to reject this null (test statistic = 4.80, df=7, p-value=0.32). Therefore, it is reasonable to assume that \eqref{eqn:restricted} holds. We then estimated the direct component of the spillover effect ${\psi}(1, 0)-{\psi}(0, 0)$ and the indirect component ${\psi}(1, 1)-(1, 0)$ via the MLE and influence function-based estimator. We obtained 95\% confidence intervals by bootstrapping with 500 replicates. The estimates of the direct component of the spillover effect were 0.0486 (95\% CI: 0.0084, 0.0930) using the MLE and 0.0561 (95\% CI: 0.0046,0.1025) using the influence function-based estimator. In addition, those of the indirect component of the spillover effect were -0.0024 (95\% CI: -0.0075, 0.0021) using the MLE and 0.0028 (95\% CI: -0.0068, 0.0210) using the influence function-based estimator. Based on these results, we conclude that the direct component of the spillover effect is statistically significant at the nominal $\alpha=0.05$ level, therefore accounting for most of the spillover effect. In Online Appendix 4, we provide SAS and R code to replicate the analysis.

\section{Conclusions}
\label{sec:conclusions}

In this paper, we proposed a new approach for decomposing the spillover effect in causal inference problems with partial 
interference among interacting units. 
We decomposed the spillover effect into direct, indirect and unit-specific components using an approach that considers outcomes to be on the same footing.  In particular, our approach yields a coherent
way for any one of the interacting outcomes to serve as the ``outcome'' for the spillover effect, with the other outcomes acting as ``mediators.''

To achieve this property, we use a generalization of causal models of the DAG \citep{pearl09causality} to chain graphs \citep{lauritzen02chain}, which allow both directed causal relationships between treatments and outcomes, and symmetric relationships between outcomes that arise in interference problems.  Given a causal chain graph model, we propose to view mediation analysis as ``splitting,'' or decomposition of treatments, as a generalization of the approach to mediation analysis
described in \citep{robins10alternative}.
We proposed two interpretations of causal chain graph models for treatment decomposition, models based on equilibrium semantics of stochastic processes described in \citep{lauritzen02chain}, and models with Markov assumptions imposed as a 
network structural model.

We show that under 
either interpretation, functionals corresponding to direct and indirect components of the spillover effects are identified via the \emph{symmetric mediation formula}, and that some of the assumptions that identification relies on can be falsified from observed data.  This falsifiability property is not present in mediation analysis in DAG models, and is implied by the symmetric structure of our proposed model.  We describe statistical inference for components of the spillover effect in our setting. We propose two estimators, one based on maximizing the log likelihood, and one which exhibits double robustness in a restricted version of our problem.





\clearpage

\appendix



\section{General Chain Graphs Under Equilibrium and Structural Model Semantics}
\label{sec:gen-cg}

A chain graph (CG) ${\cal G}$ is a mixed graph with directed ($\to$) and undirected ($-$) edges such that no partially directed cycles exist.
Much like a DAG, a CG may be used to define a statistical model via a factorization or a global Markov property.
Given a CG ${\cal G}$, a \emph{block} is a maximal undirected connected set of vertices.  We will denote the set of blocks in ${\cal G}$
by ${\cal B}({\cal G})$.  By definition, ${\cal B}({\cal G})$ partitions the set $\vec{V}$ in ${\cal G}$.

Given an undirected graph ${\cal G}$, denote by ${\cal C}({\cal G})$ the set of maximal cliques of vertices.  Note that unlike ${\cal B}({\cal G})$,
${\cal C}({\cal G})$ does not, in general, partition the set $\vec{V}$ in ${\cal G}$ because maximal cliques can intersect.


Given a CG ${\cal G}$ with a vertex set $\vec{V}$, a distribution $p(\vec{V})$ is said to lie in a statistical model of ${\cal G}$ if it can be written as:
\begin{align}
p(\vec{V} = \vec{v}) &= \prod_{\vec{B} \in {\cal B}({\cal G})} p(\vec{v}_{\vec{B}} \mid \vec{v}_{\pa_{\cal G}(\vec{B})})
= \left(
\frac{1}{Z(\vec{v}_{\pa_{\cal G}(\vec{B})})} \prod_{ \vec{C} \in {\cal C}({\cal G}_{\vec{B}}) } \phi_{\vec{C}}(\vec{v}_{\vec{C}},\vec{v}_{\pa^*_{\cal G}(\vec{C})})
\right),
\label{eqn:cg-f}
\end{align}
where ${\cal G}_{\vec{B}}$ is the graph containing only vertices in $\vec{B}$ and edges in ${\cal G}$ among elements of $\vec{B}$,
and $\pa^*_{\cal G}(\vec{C})$ is defined as $\cap_{V \in \vec{C}} \pa_{\cal G}(V)$.

A statistical CG model may be viewed as a ``DAG model defined on blocks.''  Indeed, the factorization in (\ref{eqn:cg-f}) may be viewed as a DAG factorization, where
each factor $p(\vec{v}_{\vec{B}} \mid \vec{v}_{\pa_{\cal G}(\vec{B})})$ may be further factorized according to a conditional Markov random field (CMRF) associated with the graph
${\cal G}_{\vec{B}}$.
This CMRF encodes independences in this factor induced by missing edges in ${\cal G}$ among elements in $\vec{B}$, as well as missing edges from $\pa_{\cal G}(\vec{B})$ to $\vec{B}$.
More details on such factorizations may be found in \citep{shpitser23lc}.

We briefly review 
causal CG models with equilibrium semantics found in \citep{lauritzen02chain}.
A causal CG model associated with a CG ${\cal G}$ with vertices in $\vec{V}$ associated with each $V \in \vec{V}$ an exogenous noise variable $\epsilon_V$, as well as a structural equation
$f_V : \mathfrak{X}_{\pa_{\cal G}(V) \cup \nb_{\cal G}(V) \cup \{ \epsilon_V \} } \mapsto \mathfrak{X}_V$.

A sample from the observed data distribution $p(\vec{V})$ induced by this model may be obtained as follows.  Fix a topological ordering $\prec$ on blocks in ${\cal B}({\cal G})$.  In other words,
if $\vec{B}_i \prec \vec{B}_j$, then there is no partially directed path from any element of $\vec{B}_j$ to any element in $\vec{B}_i$.  A sampling procedure may be defined on any block
$\vec{B} \in {\cal B}({\cal G})$ if a sample $\vec{v}_{\prec \vec{B}}$ on values of every variable in all blocks $\prec$-smaller than $\vec{B}$ has already been obtained.  This procedure uses a sampler
which obtains a sample of ${\vec{B}}$ given values of $\vec{v}_{\pa_{\cal G}(\vec{B})} \subseteq \vec{v}_{\prec \vec{B}}$.  A number of such samplers are described in \citep{lauritzen02chain}, with the
simplest being a Gibbs sampler with the Gibbs factors $p(B \mid \vec{v}_{\pa_{\cal G}(B)}, \nb_{\cal G}(B))$ obtained from $f_B,\epsilon_B$ for every $B \in \vec{B}$.

This structural equation based definition of a causal model of a CG has an advantage in that it allows a representation of interventions in a way that clearly generalizes structural equation based causal models associated with DAGs.  In particular, interventions that set variables $\vec{A} \subseteq \vec{V}$ to $\vec{a}$ are represented by replacing each $f_A$ by the value $\vec{a}_A$ for each $A \in \vec{A}$.
$p(\{ V(\vec{a}) : V \in \vec{V} \setminus \vec{A} \}) = p(\vec{V} \setminus \vec{A} \mid \text{do}(\vec{a}))$ may be obtained by simply rerunning the above procedure with the 
new (and modified) set of structural equations.

Note that because conditional and marginal independences in the observed or interventional distributions obtained from such a model arise due to missing inputs in the structural equation, these constraints are individual level in this model. 

The following generalization of the g-formula identification result for DAG models was derived in \citep{lauritzen02chain}:

\begin{lem}
Fix a CG ${\cal G}$ with a vertex set $\vec{V}$, and a causal model under the structural equation equilibrium semantics associated with ${\cal G}$.
For any $\vec{A} \subseteq \vec{V}$,
{\small
\begin{align}
p(\{ V(\vec{a}) = {\vec{v}^*_V} : V \in {\vec{V}} \setminus \vec{A} \}) = \prod_{{\vec{B}} \in {\cal B}({\cal G})} 
p(\vec{B} \setminus \vec{A} = \vec{v}^*_{\vec{B} \setminus \vec{A}} \mid \vec{a}_{\vec{A} \cap \vec{B}}, \vec{v}^*_{\pa_{\cal G}(\vec{B}) \setminus \vec{A}}, \vec{a}_{\pa_{\cal G}(\vec{B}) \cap \vec{A}} ).
\label{eqn:g-cg}
\end{align}
for any assignment $\vec{v}^*$ to $\vec{V} \setminus \vec{A}$, provided each term in (\ref{eqn:g-cg}) has support.
}
\label{lem:g-cg}
\end{lem}

An alternative specification of a causal model associated with a CG ${\cal G}$ may be obtained as follows.  Fix the completion $\overline{\cal G}$ of ${\cal G}$ to be any edge supergraph CG of ${\cal G}$.
We can define a causal model of a DAG on $\overline{\cal G}$ by defining a DAG ${\cal D}$ with vertices corresponding to elements in ${\cal B}(\overline{\cal G})$, and a directed edge from
$\vec{B}_i$ to $\vec{B}_j$ whenever such an edge exists from an element of $\vec{B}_i$ to an element of $\vec{B}_j$.  Such a causal model may be defined using standard semantics, including structural equation semantics, which treat elements of ${\cal B}(\overline{\cal G})$ as variables.

Causal models associated with DAGs of blocks are commonly used, explicitly or implicitly, in the literature on partial interference problems in causal inference \citep{halloran95causal,tchetgen12on}.
Given a causal DAG model a structural model associated with a CG ${\cal G}$, which is an edge subgraph of $\overline{\cal G}$ by construction, may be obtained by imposing an appropriate CMRF factorization
with respect to ${\cal G}[\vec{B}]$ on sets of distributions $\{ \vec{B}(\vec{a}) : \vec{v} \in {\mathfrak X}_{\pa_{\cal G}(\vec{B})} \}$:
\begin{align}
p(\vec{B}(\vec{a})) = \frac{1}{Z(\vec{a})} \prod_{ \vec{C} \in {\cal C}({\cal G}_{\vec{B}}) }
\phi_{\vec{C}}(\vec{v}_{\vec{C}},\vec{a}_{\pa^*_{\cal G}(\vec{C})}),
\label{eqn:cg-c-f}
\end{align}
where, as before, ${\cal G}_{\vec{B}}$ is the graph containing only vertices in $\vec{B}$ and edges in ${\cal G}$ among elements of $\vec{B}$,
and $\pa^*_{\cal G}(\vec{C})$ is defined as $\cap_{V \in \vec{C}} \pa_{\cal G}(V)$.

Because conditional and marginal independences in the causal model are logical consequences of (\ref{eqn:cg-c-f}), these constraints are population level in this model.

Despite the fact that the CG model under the structural model semantics is far weaker than the CG model under the structural equation model equilibrium semantics, some of the identification theory still holds.
In particular, we have the following weaker analogue Lemma \ref{lem:g-cg}.
\begin{lem}
Fix a CG ${\cal G}$ with a vertex set $\vec{V}$, and a causal model under the structural model semantics associated with ${\cal G}$.
For any $\vec{A} \subseteq \vec{V}$ such that there exists a set of blocks $\vec{B}_1, \ldots, \vec{B}_k$ such that $\vec{A} = \bigcup_{i} \vec{B}_i$,
{\small
\begin{align*}
p(\{ V(\vec{a}) = {\vec{v}^*_V} : V \in {\vec{V}} \setminus \vec{A} \}) &= \prod_{{\vec{B}} \in {\cal B}({\cal G}); \vec{A} \cap \vec{B} = \emptyset}
p(\vec{B} =  \vec{v}^*_{\vec{B}} \mid \vec{v}^*_{\pa_{\cal G}(\vec{B}) \setminus \vec{A}}, \vec{a}_{\pa_{\cal G}(\vec{B}) \cap \vec{A}} )\\
&= \prod_{{\vec{B}} \in {\cal B}({\cal G}); \vec{A} \cap \vec{B} = \emptyset} \frac{1}{Z(\vec{v}^*_{\pa_{\cal G}(\vec{B}) \setminus \vec{A}}, \vec{a}_{\pa_{\cal G}(\vec{B}) \cap \vec{A}})}
\prod_{ \vec{C} \in {\cal C}({\cal G}_{\vec{B}}) } \phi_{\vec{C}}(\vec{v}_{\vec{C}},\vec{w}_{\pa^*_{\cal G}(\vec{C})}),
\end{align*}
for any assignment $\vec{v}^*$ to $\vec{V} \setminus \vec{A}$, provided each term in (\ref{eqn:g-cg}) has support.
Here values $\vec{w}_{\pa^*_{\cal G}(\vec{C})}$ of $\pa^*_{\cal G}(\vec{C})$ in each term are consistent with $\vec{a}$ and $\vec{v}^*$.
}
\label{lem:g-cg-2}
\end{lem}
\begin{proof}
This follows from the standard g-formula results for identification in fully observed DAGs, and further factorization of each g-formula term due to (\ref{eqn:cg-c-f}).
\end{proof}

\begin{figure}
\begin{center}
\begin{tikzpicture}[>=stealth, node distance=0.9cm]
\tikzstyle{format} = [draw, very thick, circle, minimum size=5.0mm,
inner sep=0pt]
\tikzstyle{square} = [draw, very thick, rectangle, minimum size=5mm]

\begin{scope}[xshift=6.8cm]
\path[very thick, ->]
node[format] (C1) {$\vec{C}_1$}
node[format, right of=C1] (C2) {$\vec{C}_2$}
node[format, right of=C2] (C3) {$\vec{C}_3$}
node[format, right of=C3] (C4) {$\vec{C}_3$}
node[format, yshift=-0.2cm, below of=C1, xshift=0.0cm] (A1) {$A_1$}
node[format, yshift=-0.7cm, below of=A1, xshift=-0.0cm] (Y1) {$Y_1$}
node[format, yshift=-0.2cm, below of=C2, xshift=0.0cm] (A2) {$A_2$}
node[format, yshift=-0.7cm, below of=A2, xshift=-0.0cm] (Y2) {$Y_2$}
node[format, yshift=-0.2cm, below of=C3, xshift=0.0cm] (A3) {$A_3$}
node[format, yshift=-0.7cm, below of=A3, xshift=-0.0cm] (Y3) {$Y_3$}
node[format, yshift=-0.2cm, below of=C4, xshift=0.0cm] (A4) {$A_4$}
node[format, yshift=-0.7cm, below of=A4, xshift=-0.0cm] (Y4) {$Y_4$}

(C1) edge[blue] (A2)
(C1) edge[blue] (A2)
(C1) edge[blue] (A3)

(C2) edge[blue] (A1)
(C2) edge[blue] (A2)
(C2) edge[blue] (A3)

(C3) edge[blue] (A1)
(C3) edge[blue] (A2)
(C3) edge[blue] (A4)

(C4) edge[blue] (A1)
(C4) edge[blue] (A2)
(C4) edge[blue] (A3)

(A1) edge[blue] (Y1)
(A2) edge[blue] (Y2)
(A3) edge[blue] (Y3)
(A4) edge[blue] (Y4)

(Y1) edge[-] (Y2)
(Y2) edge[-] (Y3)
(Y3) edge[-] (Y4)
(Y1) edge[-, bend right=30] (Y3)
(Y2) edge[-, bend right=30] (Y4)
(Y1) edge[-, bend right=40] (Y4)

(A1) edge[-] (A2)
(A2) edge[-] (A3)
(A3) edge[-] (A4)
(A1) edge[-, bend right=30] (A3)
(A2) edge[-, bend right=30] (A4)
(A1) edge[-, bend left=20] (A4)

(C1) edge[-] (C2)
(C2) edge[-] (C3)
(C3) edge[-] (C4)
(C1) edge[-, bend left=30] (C3)
(C2) edge[-, bend left=30] (C4)
(C1) edge[-, bend left=40] (C4)

(C1) edge[blue] (A1)
(C2) edge[blue] (A2)
(C3) edge[blue] (A3)
(C4) edge[blue] (A4)

(C1) edge[blue, bend right=20] (Y1)
(C2) edge[blue, bend right=20] (Y2)
(C3) edge[blue, bend left=20] (Y3)
(C4) edge[blue, bend left=20] (Y4)

(A1) edge[blue] (Y2)
(A1) edge[blue] (Y3)
(A1) edge[blue] (Y4)

(A2) edge[blue] (Y1)
(A2) edge[blue] (Y3)
(A2) edge[blue] (Y4)

(A3) edge[blue] (Y1)
(A3) edge[blue] (Y2)
(A3) edge[blue] (Y4)

(A4) edge[blue] (Y1)
(A4) edge[blue] (Y2)
(A4) edge[blue] (Y3)

node[below of=Y2,yshift=-0.2cm,xshift=0.45cm] (l) {$(a)$}
;
\end{scope}

\begin{scope}[xshift=11.9cm]
\path[very thick, ->]
node[format] (C) {$\vec{C}$}
node[format, below of=C, xshift=0.0cm] (A) {$A_1$}
node[format, below of=A, xshift=-0.45cm] (Y2) {$Y_2$}
node[format, left of=Y2] (Y1) {$Y_1$}
node[format, right of=Y2] (Y3) {$Y_3$}
node[format, right of=Y3] (Y4) {$Y_4$}

(A) edge[blue] (Y1)
(A) edge[blue] (Y2)
(A) edge[blue] (Y3)
(A) edge[blue] (Y4)

(Y1) edge[-] (Y2)
(Y2) edge[-] (Y3)
(Y3) edge[-] (Y4)

(Y1) edge[-, bend right=30] (Y3)
(Y2) edge[-, bend right=30] (Y4)
(Y1) edge[-, bend right=40] (Y4)

(C) edge[blue] (A)
(C) edge[blue, bend right=10] (Y1)
(C) edge[blue, bend right=20] (Y2)
(C) edge[blue, bend left=20] (Y3)
(C) edge[blue, bend left=10] (Y4)

node[below of=Y2,yshift=-1.1cm,xshift=0.45cm] (l) {$(b)$}
;
\end{scope}

\begin{scope}[xshift=15.7cm]
\path[very thick, ->]
node[format] (C) {$\vec{C}$}
node[format, below of=C, yshift=-0.0cm, xshift=0.0cm] (A) {$A_1$}
node[format, below of=A, yshift=0.0cm, xshift=-0.45cm] (A2) {$\tilde{A}_1^2$}
node[format, left of=A2] (A1) {$\tilde{A}_1^1$}
node[format, right of=A2] (A3) {$\tilde{A}_1^3$}
node[format, right of=A3] (A4) {$\tilde{A}_1^4$}
node[format, below of=A2] (Y2) {$Y_2$}
node[format, below of=A1] (Y1) {$Y_1$}
node[format, below of=A3] (Y3) {$Y_3$}
node[format, below of=A4] (Y4) {$Y_4$}
node[below of=A2] (dummy) {}

(A) edge[red] (A1)
(A) edge[red] (A2)
(A) edge[red] (A3)
(A) edge[red] (A4)

(A1) edge[blue] (Y1)
(A2) edge[blue] (Y2)
(A3) edge[blue] (Y3)
(A4) edge[blue] (Y4)

(Y1) edge[-] (Y2)
(Y2) edge[-] (Y3)
(Y3) edge[-] (Y4)

(Y1) edge[-, bend right=30] (Y3)
(Y2) edge[-, bend right=30] (Y4)
(Y1) edge[-, bend right=40] (Y4)

(C) edge[blue] (A)
(C) edge[blue, bend right=60] (Y1)
(C) edge[blue, bend right=40] (Y2)
(C) edge[blue, bend left=40] (Y3)
(C) edge[blue, bend left=60] (Y4)

node[below of=dummy,yshift=-0.2cm,xshift=0.45cm] (l) {$(c)$}
;
\end{scope}
\begin{scope}[xshift=19.7cm]
\path[very thick, ->]
node[format] (C) {$\vec{C}$}
node[format, below of=C, yshift=-0.0cm, xshift=0.0cm] (A) {$A_1$}
node[format, below of=A, yshift=0.0cm, xshift=-0.0cm] (A23) {\scriptsize $\tilde{A}_1^{23}$}
node[format, left of=A23, xshift=-0.45cm] (A1) {$\tilde{A}_1^1$}
node[format, right of=A23, xshift=0.45cm] (A4) {$\tilde{A}_1^4$}
node[format, below of=A23, xshift=-0.45cm] (Y2) {$Y_2$}
node[format, below of=A1] (Y1) {$Y^1$}
node[format, below of=A23, xshift=0.45cm] (Y3) {$Y_3$}
node[format, below of=A4] (Y4) {$Y_4$}
node[below of=A23] (dummy) {}

(A) edge[red] (A1)
(A) edge[red] (A23)
(A) edge[red] (A4)

(A1) edge[blue] (Y1)
(A23) edge[blue] (Y2)
(A23) edge[blue] (Y3)
(A4) edge[blue] (Y4)

(Y1) edge[-] (Y2)
(Y2) edge[-] (Y3)
(Y3) edge[-] (Y4)

(Y1) edge[-, bend right=30] (Y3)
(Y2) edge[-, bend right=30] (Y4)
(Y1) edge[-, bend right=40] (Y4)

(C) edge[blue] (A)
(C) edge[blue, bend right=60] (Y1)
(C) edge[blue, bend right=40] (Y2)
(C) edge[blue, bend left=40] (Y3)
(C) edge[blue, bend left=60] (Y4)

node[below of=dummy,yshift=-0.2cm,xshift=0.0cm] (l) {$(d)$}
;
\end{scope}
\end{tikzpicture}
\end{center}
\caption{
{
(a) An example of a causal model with partial interference and blocks of size $4$, where covariates $C_n$, treatments $A_n$, and outcomes $Y_n$ for $n = 1, \ldots 4$ are mutually associated, and each treatment $A_n$ potentially directly influences all outcomes.
(b) A simplified version of a model in (a) showing a single treatment $A_1$ for unit $1$, and with all baseline covariates joined into a single vertex $\vec{C}$.
(c) A CG model representing treatment decomposition assumptions necessary for the identification of unit-specific components of the network average spillover effect of $A_1$ on the outcomes in the model in (b).
(d) A relaxation of the model in (c) which does not permit identification of a unit-specific treatment decomposition of
the network average spillover effect of $A_1$ on the outcomes, but does permit identification of a coarser
decomposition where effects of $A_1$ on outcomes for units $2$ and $3$ are bundled together.
}
}
\label{fig:med2}
\end{figure}

\section{Treatment Decomposition In General Networks}
\label{subsec:gen-net}

%


We now consider how network average causal effects in the presence of interference may be decomposed in a general setting with partial interference.  Specifically, we consider the setting we introduced earlier, with $B$ blocks of $N$ units each.
We assume the size and structure of $B$ blocks are identical, with these variables then serving as independent realizations of underlying block level random variables $\vec{C}_n,A_n$, and $Y_n$ for $n \in \{ 1, \ldots, N \}$.  
As before, we assume network versions of conditional ignorability, positivity, and consistency.


These assumptions correspond to a CG model with 
three blocks of size $N$ corresponding to a block of covariates $\vec{C} = \{ \vec{C}_1, \ldots, \vec{C}_N \}$, a block of treatments $\vec{A} = \{ A_1, \ldots, A_N \}$, and a block of outcomes $\vec{Y} = \{ Y_1, \ldots, Y_N \}$.
The structure of undirected edges within each block is arbitrary, and represent dependence and independence among outcomes in the block.  In addition, we assume a causal ordering where covariates $\vec{C}$ precede treatments $\vec{A}$ and both $\vec{C}$ and $\vec{A}$ precede outcomes $\vec{Y}$.  Any variable in a block causally prior to another block may potentially causally influence any variable in that block.
An example of such a model for a block of size $4$ is shown in Fig.~\ref{fig:med2} (a).
In this model all covariates, treatments and outcomes are mutually associated, which is represented by vertices $C_n$, $A_n$, $Y_n$ (for $n = 1, \ldots 4$) forming cliques.
To avoid edge clutter, we omit directed edges from covariates $\vec{C}_i$ for any unit $i$ to any outcome other than the outcome $Y_i$ of unit $i$, while displaying all other causal relationships from variables in causally prior blocks to variables in causally subsequent blocks.
In particular,  each unit's treatment potentially influences outcomes of all units, allowing for the possibility of a spillover effect of any unit's treatment on any other unit's outcome.

In this model,
the main and spillover effects, and their network average versions, are identified by the standard argument for the conditionally ignorable model.  For instance, the 
$\text{SE}_n(\vec{a}_{-n},\vec{a}'_{-n},{0})$ for a particular unit $n$, where $\vec{a} \equiv 1$, and $\vec{a}' \equiv 0$ is identified as follows:
\begin{align*}
\mathbb{E}[Y_{n}(\vec{a}_{-n},a'_n)] - \mathbb{E}[Y_{n}(\vec{a}'_{-n},a'_{n})] =
\sum_{\vec{c}}
\left\{ \mathbb{E}[Y_{n} \mid \vec{a}_{-n}, a{_n}', \vec{c}] - \mathbb{E}[Y_{n} \mid \vec{a}', \vec{c}] \right\} p(\vec{c}).
\end{align*}

We are interested in decomposing the spillover effect, or possibly its network average version
into 
components
representing the 
causal influence of unit $i$'s treatment $A_i$ on the outcome $Y_j$ of another
unit $j$ within a block.

To simplify notation, we consider the spillover effect of unit $1$'s treatment $A_{1}$ 
on outcomes of other units $n \neq 1$, and suppress mention of treatments for all units other than $1$ (as all such treatments are either set to the baseline value, or are assigned the same value as the treatment of interest in the spillover effect).
In addition, we will use the vertex $\vec{C}$ to denote covariates of all units $\vec{C}_1,\vec{C}_2,\vec{C}_3,\vec{C}_4$.
The resulting simplified model is shown in Fig.~\ref{fig:med2} (b).


To obtain this decomposition, we generalize the assumption for the dyadic model described in Section~\ref{subsec:two-example}. 
Specifically, we assume the treatment $A_1$ may be decomposed into a set of components
$\tilde{A}_1^{n}$, where each component 
is connected via a directed edge to the outcome $Y_n$ of unit $n$ directly, and to no other outcomes.

As was the case in the dyadic example in the previous section, this assumption is 
represented by a chain graph where a treatment $A_1$ decomposes into a set of additional vertices $\tilde{A}_1^n$, one for each outcome $Y_n \in \vec{Y}$. 
In this chain graph, the vector of covariates $\vec{C}$ directly influences $A_1$ and each $Y_n \in \vec{Y}$, while each $\tilde{A}_1^n$ directly influences $Y_n$ (only), and is in turn influenced by $A_1$.  An example of such an \emph{extended} CG for the four variable example in Fig.~\ref{fig:med2} (b) is shown in Fig.~\ref{fig:med2} (c).  

Let $\tilde{A}_1 \equiv \{ \tilde{A}_1^n : n \in 1, \ldots, N \}$, and fix any $\{ \tilde{a}_1^n : n \in 1, \ldots N \} \equiv \tilde{a}_1 \in {\mathfrak X}_{\tilde{A}_1}$.
The analogues of assumptions in (\ref{eqn:markov-y}) encoded by the extended CG ${\cal G}^{\dag}$ in Fig.~\ref{fig:med2} (c) correspond to the following
\begin{align}
\label{eqn:general-model}
\notag
p(Y_n(\{ \tilde{a}_1^1, \ldots, \tilde{a}_1^N \}, \vec{c}) \mid \{ Y_m(\{ \tilde{a}_1^1, \ldots, \tilde{a}_1^N \}, \vec{c}) : Y_m \in \vec{Y} \setminus \{ Y_n \} \})\\
\text{ is only a function of }\nb_{\G^{\dag}}(Y_n) \cup \{ Y_n, \tilde{a}_1^n \} \cup \vec{c},
\end{align}
for all $n \in \{ 1, \ldots, N \}$. 
Note that these assumptions correspond to (\ref{eqn:cg_y_const}) and (\ref{eqn:cg_m_const}) in the special case of dyadic blocks.
In the model corresponding to Fig.~\ref{fig:med2} (c), the assumption in the above list corresponding to $Y_{1}$ is
\begin{align*}
p\{ Y_1(\tilde{a}_1^1,\tilde{a}_1^2,\tilde{a}_1^3,\tilde{a}_1^4, \vec{c}) \mid Y_2(\tilde{a}_1^1,\tilde{a}_1^2,\tilde{a}_1^3,\tilde{a}_1^4, \vec{c}),
Y_3(\tilde{a}_1^1,\tilde{a}_1^2,\tilde{a}_1^3,\tilde{a}_1^4, \vec{c}), Y_4(\tilde{a}_1^1,\tilde{a}_1^2,\tilde{a}_1^3,\tilde{a}_1^4, \vec{c}) \}\\
\text{ is only a function of $Y_1, Y_2, Y_3, Y_4, \tilde{a}_1^1, \vec{c}$},
\end{align*}
for any values $\tilde{a}_1^1,\tilde{a}_1^2,\tilde{a}_1^3,\tilde{a}_1^4, \vec{c}$.

We now show how to obtain a decomposition of the spillover effect
$\mathbb{E}[Y_{n}(\tilde{a}_1)] - \mathbb{E}[Y_{n}(\tilde{a}'_1)]$ of $A_1$ on $Y_n$ ($n \neq 1$)
into unit-specific components, and obtain identification via (\ref{eqn:general-model}).
We fix an ordering $\prec$ on units in a block, where for each unit $n$, we denote the set of units preceding $n$ according to the ordering as $\pre_{\prec}(n)$, and the set of units $n$ precedes according to
$\prec$ 
as $\post_{\prec}(n)$.
As before, let $\tilde{a}_1^{n} = 1$, and $(\tilde{a}_1^{n})' = 0$ for any $n$.  We consider the following decomposition of
$\mathbb{E}[Y_{n}(\tilde{a}_1)] - \mathbb{E}[Y_{n}(\tilde{a}'_1)]$ 
(recall that 
$\{ \tilde{A}_1^1, \ldots, \tilde{A}_1^N \}$ are components of a single unit treatment $A_{1}$) 
\begin{align*}
& \mathbb{E}[Y_{n}(\{ \tilde{a}_1^{l} \mid l \in \pre_{\prec}(n) \}, \tilde{a}_1^{n}, \{ (\tilde{a}_1^{m})' \mid m \in \post_{\prec}(n) \}] - \\
& \mathbb{E}[Y_{n}(\{ \tilde{a}_1^{l} \mid l \in \pre_{\prec}(n) \}, (\tilde{a}_1^n)', \{ (\tilde{a}_1^m)' \mid m \in \post_{\prec}(n) \}].
\end{align*}

In our four unit example, 
consider the decomposition of the spillover effect 
of $A_1$ on $Y_2$, which is equal to
$\mathbb{E}[Y_2({a}_1, {a}_2{'}, a_3, a_4)] - \mathbb{E}[Y_2({a}_1{'}, {a}_2{'}, a_3', a_4')]$ 
in the model corresponding to Fig.~\ref{fig:med2} (a), and equal to
$\mathbb{E}[Y_2(a_1)] - \mathbb{E}[Y_2(a_1{'})]$ in the simplified model in Fig.~\ref{fig:med2} (b), where treatments $A_2,A_3,A_4$ are suppressed from the notation.


Rewriting the spillover effect $\mathbb{E}[Y_2({a}_1)] - \mathbb{E}[Y_2({a}_1{'})]$ in terms of treatment components in the model corresponding to the extended CG in Fig.~\ref{fig:med2} (c),
we obtain
\begin{align*}
\mathbb{E}[Y_2({a}_1)] - \mathbb{E}[Y_2({a}_1{'})] &\equiv
\mathbb{E}[Y_2(\tilde{a}_1^1, \tilde{a}_1^2, \tilde{a}_1^3, \tilde{a}_1^4)] - \mathbb{E}[Y_2((\tilde{a}_1^1)', (\tilde{a}_1^2)', (\tilde{a}_1^3)', (\tilde{a}_1^4)')].
\end{align*}
Under an ordering $4 \prec 3 \prec 2 \prec 1$,
the effect 
$\mathbb{E}[Y_2(\tilde{a}_1^1, \tilde{a}_1^2, \tilde{a}_1^3, \tilde{a}_1^4)] - \mathbb{E}[Y_2((\tilde{a}_1^1)', (\tilde{a}_1^2)', (\tilde{a}_1^3)', (\tilde{a}_1^4)')]$
decomposes as follows:
{\small
\begin{align}
\notag
\underbrace{\mathbb{E}[Y_2({\tilde{a}_1^1,\tilde{a}_1^2,\tilde{a}_1^3,\tilde{a}_1^4})] - \mathbb{E}[Y_2((\tilde{a}_1^1)',\tilde{a}_1^2,\tilde{a}_1^3,\tilde{a}_1^4)]}_{\text{indirect effect through $Y_1$}}
+ \underbrace{\mathbb{E}[Y_2((\tilde{a}_1^1)',\tilde{a}_1^2,\tilde{a}_1^3,\tilde{a}_1^4)] - \mathbb{E}[Y_2((\tilde{a}_1^1)',(\tilde{a}_1^2)',\tilde{a}_1^3,\tilde{a}_1^4)]}_{\text{direct effect}}\\
\notag
+ \underbrace{\mathbb{E}[Y_2((\tilde{a}_1^1)',(\tilde{a}_1^2)',\tilde{a}_1^3,\tilde{a}_1^4)] - \mathbb{E}[Y_2((\tilde{a}_1^1)',(\tilde{a}_1^2)',(\tilde{a}_1^3)',\tilde{a}_4)]}_{\text{indirect effect through $Y_3$}}\\
+ \underbrace{\mathbb{E}[Y_2((\tilde{a}_1^1)',(\tilde{a}_1^2)',(\tilde{a}_1^3)',\tilde{a}_1^4)] - \mathbb{E}[Y_2((\tilde{a}_1^1)',(\tilde{a}_1^2)',(\tilde{a}_1^3)',(\tilde{a}_1^4)')]}_{\text{indirect effect through $Y_4$}}.
\label{eqn:4-way}
\end{align}
}
The notions of direct vs indirect influence discussed in Section~\ref{subsec:two-example} directly carries over to treatment components associated with multiple units' outcomes in a network.
That is, under the CG interpretation in \citep{lauritzen02chain}, a treatment component $\tilde{A}_1^n$ directly influences an outcome $Y_m$ if it appears
in its structural equation, and does not otherwise.  Similarly, a treatment component $\tilde{A}_1^n$ indirectly influences $Y_m$
if there is a chain of variables of size greater than 2 that starts at $\tilde{A}_1^n$, ends at $Y_m$, and each intermediate variable in the chain is directly caused by a variable just before it in the chain.
In this sense, the decomposition of the spillover effect of $A_1$ on $Y_2$ in (\ref{eqn:4-way}) may be viewed as a combination of four effects, 
the indirect effect through $Y_1$, the direct effect, the indirect effect through $Y_3$, and the indirect effect through $Y_4$.

Thus, the above decomposition serve as the symmetric generalization of decompositions of the total effect of $A$ on $Y$ along a set of mediators $M_1, \ldots, M_N$ in standard mediation analysis. 
However, in this decomposition, the outcome $Y$ is not distinguished from mediators $M_1, \ldots, M_N$. Instead, each outcome in the block serves as either an outcome or a mediator, depending on which component of which spillover effect is under consideration.

The arbitrary choice of ordering that yields the above decomposition parallels the choice of ordering for standard decompositions common in mediation analysis in DAG models.  In particular, for a single outcome $Y$ and mediator $M$ in a DAG model, the choice of ordering may yield different decompositions of the average causal effect into either the pure indirect effect, and the total direct effect, or alternatively the total indirect effect, and the pure indirect effect \citep{robins92effects}.

The ordering we chose here simply governs the order in which treatment components change from $0$ to $1$ in our decomposition.  This ordering does not correspond to a causal ordering on our model, does not entail a directed acyclic graph, and is still consistent with treating all block outcomes symmetrically, with some acting as mediators and some as outcomes, depending on which treatment component we consider.

We now show that all components of the above decomposition are identified given that assumptions (\ref{eqn:general-model}) hold.

Before doing so, we first review a number of relevant results for (conditional) Markov random fields (CMRFs).  Conditional Markov random fields are sets of distributions $p(\vec{V} \mid \vec{W})$ associated with conditional undirected graphs (CUGs) ${\cal G}(\vec{V},\vec{W})$ which contain undirected and directed edges, such that every directed edge is out of an element in $\vec{W}$ and into an element in $\vec{V}$, and all undirected edges are among elements in $\vec{V}$.
Given a CUG ${\cal G}(\vec{V},\vec{W})$, let ${\cal G}_{\vec{V}}$ be the (undirected) induced subgraph containing vertices in $\vec{V}$ and all and only edges in ${\cal G}(\vec{V},\vec{W})$ among elements in $\vec{V}$.
Let ${\cal C}({\cal G}_{\vec{V}})$ be the set of all cliques in ${\cal G}_{\vec{V}}$.

A conditional distribution $p(\vec{V} \mid \vec{W})$ is said to satisfy the pairwise Markov property with respect to a CUG ${\cal G}(\vec{V},\vec{W})$
if  for every $V \in \vec{V}$, $Z \in \vec{V} \cup \vec{W}$ such that $Z$ is non-adjacent to $V$ in $\G(\vec{V},\vec{W})$,
$p(V | (\vec{V} \cup \vec{W}) \setminus \{ V \})$ is only a function of values of $(\vec{V} \cup \vec{W}) \setminus \{ Z \}$.

The following theorem was derived in \citep{shpitser23lc}, based on an earlier result in \citep{lauritzen96graphical}.

\begin{thm}{\bf (Hammersly-Clifford for conditional MRFs)}
Assume a positive $p(\vec{v} | \vec{w})$ obeys the pairwise Markov property for a CUG $\G(\vec{V},\vec{W})$.
Then $p(\vec{V} | \vec{W})$ Markov factorizes with respect to $\G(\vec{V},\vec{W})$.
That is, 
\begin{align}
p(\vec{V} = \vec{v} | \vec{W} = \vec{w})
=
\prod_{\vec{C} \subseteq \bar{\cal C}(\G_{\vec{V}})} \phi_{\vec{C}}(\vec{v}_{\vec{C}}, \vec{w}_{\vec{C}^*}),
\label{eqn:cmrf-f}
\end{align}
where for every $\vec{C}$, $\vec{C}^* = \bigcap_{C \in \vec{C}} \pa_{\G}(C)$, and $\phi_{\vec{C}}(\vec{v}_{\vec{C}}, \vec{w})$, for any $\vec{C} \subseteq \vec{V}$ that corresponds to a clique in ${\cal G}_{\vec{V}}$, is defined as:
\begin{align}
\phi_{\vec{C}}(\vec{v}_{\vec{C}}, \vec{w}) &\equiv
\exp \left\{ \sum_{\vec{B} \subseteq \vec{C}} (-1)^{|\vec{C} \setminus \vec{B}|} H_{\vec{B}}(\vec{v}_{\vec{B}}, \vec{w}) \right\},
\label{eqn:phi}
\end{align}
and for any subset $\vec{C} \subseteq \vec{V}$, define
$H_{\vec{C}}(\vec{v}_{\vec{C}}, \vec{w}) \equiv \log p(\vec{v}_{\vec{C}}, \vec{v}^*_{\vec{V} \setminus \vec{C}} | \vec{w})$.
\label{thm:h-c-cmrf}
\end{thm}
Note that terms $\phi_{\vec{C}}(\vec{v}_{\vec{C}}, \vec{w})$ in (\ref{eqn:cmrf-f}) are only functions of $\vec{C}^* = \bigcap_{C \in \vec{C}} \pa_{\G}(C)$, and thus are written as $\phi_{\vec{C}}(\vec{v}_{\vec{C}}, \vec{w}_{\vec{C}^*})$.


\begin{theorem}
Fix an arbitrary value assignment $\tilde{a}_1$ to $\tilde{A}_1 \equiv \{ \tilde{A}_1^1, \ldots \tilde{A}_1^N \}$.  Assume every distribution $p(\vec{Y}(\tilde{a}_1) \mid \vec{c})$ 
is positive.

Then for every $Y_m$ ($m = 1, \ldots, N$), under the network versions of consistency, positivity, conditional ignorability, as well as (\ref{eqn:general-model}),
\begin{align}
p(y_m(\tilde{a}_1) 
) &= \sum_{\vec{c},\vec{y}_{-m}}
\frac{
\left(
\prod_{n=1}^N \phi_{Y_n}(y_n, \tilde{a}_1^n, \vec{c})
\right)
\prod_{\vec{S} \in {\cal C}_{\vec{Y}}} \phi_S(\vec{y}_{\vec{S}}, \vec{c})
}{
Z(\tilde{a}_1, \vec{c})
}
p(\vec{c}),
\label{eqn:network-sym-med}
\end{align}
where $\vec{y}_{\vec{S}}$ is the subset of $\vec{y}$ pertaining to the clique $\vec{S}$, and every term in (\ref{eqn:network-sym-med}) is a function only of $p(\vec{Y} \mid \tilde{a}^1, \vec{C}) p(\vec{C})$.
\label{thm:general-id}
\end{theorem}
Note that $\tilde{a}_1$ may potentially assign conflicting values to different components of $\tilde{a}^1$ in $\tilde{A}_1$.  As a result, positivity holds for $p(\tilde{a}^1 | \vec{C})$ but not for
$p(\{ \tilde{A}_1^1, \ldots, \tilde{A}_1^N \} | \vec{C})$.  Thus, this theorem does not follow without assuming (\ref{eqn:general-model}) in addition to standard assumptions.
\begin{proof}

By assumption $p(\vec{Y}(\tilde{a}_1)) = \sum_{\vec{c}} p(\vec{Y}(\tilde{a}_1) \mid \vec{c}) p(\vec{c}) = \sum_{\vec{c}} p(\vec{Y} \mid \tilde{a}_1, \vec{c}) p(\vec{c})$.

The extended CG associated with our model induces a CUG ${\cal G}(\vec{Y}, \tilde{A}_1 \cup \{ \vec{C} \} )$, where elements in $\vec{Y}$ are pairwise connected by undirected edges, the vertex $\vec{C}$ is a parent of every element in $\vec{Y}$, and each $\tilde{a}_1^n \in \tilde{A}_1$ is a parent of $Y_n \in \vec{Y}$.

Note that (\ref{eqn:general-model}) corresponds to the pairwise Markov property for this CUG, hence the distribution $p(\vec{Y}(\tilde{a}_1) \mid \vec{c})$ is in the CMRF model for ${\cal G}(\vec{Y}, \tilde{A}_1 \cup \{ \vec{C} \} )$.  By assumption and Theorem \ref{thm:h-c-cmrf}, $p(\vec{Y}(\tilde{a}_1) \mid \vec{c}) = p(\vec{Y} \mid \tilde{a}_1), \vec{c})$ factorizes with respec to ${\cal G}(\vec{Y}, \tilde{A}_1 \cup \{ \vec{C} \} )$, and thus can be written as
\begin{align*}
\frac{
\left(
\prod_{n=1}^N \phi_{Y_n}(y_n, \tilde{a}_1^n, \vec{c})
\right)
\prod_{\vec{S} \in {\cal C}_{\vec{Y}}} \phi_S(\vec{y}_{\vec{S}}, \vec{c})
}{
Z(\tilde{a}_1, \vec{c})
}
\end{align*}
The fact that each term in the above expression is only a function of $p(\vec{Y} \mid \tilde{a}_1{'},\vec{C})$ where $\tilde{a}_1{'}$ assign equal values to every element of
$\tilde{A}_1$ follows by the clique structure of ${\cal G}(\vec{Y}, \tilde{A}_1 \cup \{ \vec{C} \} )$, and in particular from the fact that every $a_1^n \in \tilde{A}_1$ only has a single child in $\vec{Y}$.

This concludes the proof.
\end{proof}

To illustrate this result, the term $p(y_2(\tilde{a}_1^1,\tilde{a}_1^2,\tilde{a}_1^3,\tilde{a}_1^4)) - p(y_2((\tilde{a}_1^1)',\tilde{a}_1^2,\tilde{a}_1^3,\tilde{a}_1^4))$ in the decomposition above is identified as
{\small
\begin{align*}
\sum_{y_1,y_3,y_4,\vec{c}}
\!\!\!
\frac{
\phi_{
Y_1}({\tilde{a}_1^1},y_1{,\vec{c}})
\phi_{
Y_2}({\tilde{a}_1^2},y_2{,\vec{c}})
\phi_{
Y_3}({\tilde{a}_1^3},y_3{,\vec{c}})
\phi_{
Y_4}({\tilde{a}_1^4},y_4{,\vec{c}})
\!\!\!\!
\prod\limits_{\vec{S} \subseteq \vec{Y}; |\vec{S}|>1}
\phi_{\vec{S}}(\{ y_i : Y_i \in \vec{S} \}, \vec{c})
}{
Z(\tilde{a}_1^1,\tilde{a}_1^2,\tilde{a}_1^3,\tilde{a}_1^4,\vec{c})
} p(\vec{c})\\
-
\sum_{y_1,y_3,y_4,\vec{c}}
\!\!\!
\frac{
\phi_{
Y_1}({(\tilde{a}_1^1)'},y_1{,\vec{c}})
\phi_{
Y_2}({\tilde{a}_1^2},y_2{,\vec{c}})
\phi_{
Y_3}({\tilde{a}_1^3},y_3{,\vec{c}})
\phi_{
Y_4}({\tilde{a}_1^4},y_4{,\vec{c}})
\!\!\!\!
\prod\limits_{\vec{S} \subseteq \vec{Y}; |\vec{S}|>1}
\phi_{\vec{S}}(\{ y_i : Y_i \in \vec{S} \}, \vec{c})
}{
Z((\tilde{a}_1^1)',\tilde{a}_1^2,\tilde{a}_1^3,\tilde{a}_1^4,\vec{c})
} p(\vec{c}),
\end{align*}
}
where $\vec{Y} = \{ Y_1, Y_2, Y_3, Y_4 \}$.

As discussed in Section \ref{subsec:two-example}, under the weaker structural model semantics of CGs, the notion of direct versus indirect influence of a treatment component may be recovered by considering the form of the modified factorization above.

Theorem~\ref{thm:general-id} implies a maximum likelihood plug-in estimation strategy that generalizes results in Section~\ref{subsec:mle}.  Such a strategy could based on a recent general likelihood for chain graphs described in \citep{shpitser23lc}.

\section{Extensions to Settings with Full Interference}
\label{sec:full}

Our identification result for treatment decompositions in general networks described in the previous section may be extended, in certain cases, to \emph{full interference settings}, where only a single realization of $n$ mutually dependent units is available.  While causal identification result in such settings remain unchanged, as they rely on the knowledge of the observed data distribution, achieving a usable estimator requires additional assumptions allowing statistical inference to be made from a single sample.

The auto-g-computation algorithm, described in \citep{tchetgen20auto}, for causal models of interference obeying the network version of conditional ignorability $(\vec{Y}(\vec{a}) \ci \vec{A} \mid \vec{C})$, where $\vec{Y}$, $\vec{A}$, $\vec{C}$, are vectors of outcomes, treatments and covariates for units in the network.  The method works by imposing additional Markov restrictions encoded by a chain graph associated with a chain graph on the observed data distribution from this model, an example of such a chain graph is shown in Fig.~\ref{fig:med2} (a).

In particular, given a network of $n$ units, if the chain graph ${\cal G}$ is sufficiently sparse, such that the Gibbs factor conditional distributions of the form $p(Y_i \mid \nb(Y_i), \pa(Y_i))$ depend only on variables of neighbors of unit $i$, and each unit has \emph{few neighbors}, statistical inference becomes possible by exploiting independence restrictions in the model via coding or pseudo-likelihood estimators \citep{besag75pseudo,tchetgen20auto}.

In addition, variable numbers of neighbors for units in the network may be modelled by imposing addition assumptions on models $p(Y_i \mid \nb(Y_i), \pa(Y_i))$, in particular that 
neighbors of each unit are exchangeable, and that the number of neighbors itself is a parameter.  Examples of coherent distributions that obey such assumptions are described in \citep{tchetgen20auto}.

Such assumptions also allow a natural modeling strategy for partial interference settings with heterogeneous block sizes by simply recasting them as full interference problems.

\section{Coarser Decompositions Of The Spillover Effect}
\label{subsec:coarser}
Consider the general network setting discussed in Section \ref{subsec:gen-net}, where, as before, we are interested in the spillover effect on unit $2$, and we suppress from the notation, without loss of generality, treatments of all units except unit $1$.

The factorization of the observed data distribution $p(y_1, y_2, y_3, y_4, \tilde{a}_1^1, \tilde{a}_1^2, \tilde{a}_1^3, \tilde{a}_1^4, {a}_1, \vec{c})$ 
consistent with the extended
CG containing treatment components 
in Fig.~\ref{fig:med2} (c) is: 
{\small
\begin{align}
\label{eqn:cg_fact}
\frac{
\phi_{
Y_1}({\tilde{a}_1^1},y_1{,\vec{c}})
\phi_{
Y_2}({\tilde{a}_1^2},y_2{,\vec{c}})
\phi_{
Y_3}({\tilde{a}_1^3},y_3{,\vec{c}})
\phi_{
Y_4}({\tilde{a}_1^4},y_4{,\vec{c}})
\!\!
\prod\limits_{Y_i \in \vec{Y}} \phi_{Y_i}(y_i, \tilde{a}_1^i, \vec{c})
\!\!\!\!
\prod\limits_{\vec{S} \subseteq \vec{Y}; |\vec{S}|>1}
\phi_{\vec{S}}(\{ y_i : Y_i \in \vec{S} \}, \vec{c})
}{
Z(\tilde{a}_1^1,\tilde{a}_1^2,\tilde{a}_1^3,\tilde{a}_1^4,\vec{c})
}
\\
\notag
p(\tilde{a}_1^1 | {a}_1)
p(\tilde{a}_1^2 | {a}_1)
p(\tilde{a}_1^3 | {a}_1)
p(\tilde{a}_1^4 | {a}_1)
p(a | \vec{c})
p(\vec{c}),
\end{align}
}
where $Z(\tilde{a}_1^1,\tilde{a}_1^2,\tilde{a}_1^3,\tilde{a}_1^4,\vec{c})$ is a normalizing constant, and, as before, factors $p(\tilde{a}_1^i \mid {a}_1)$
are deterministic. 

Dropping treatment components from the factorization allows us to rewrite it as:
{\small
\begin{align*}
\frac{
\phi_{
Y_1}(\tilde{a}^1,y_1{,\vec{c}})
\phi_{
Y_2}(\tilde{a}^1,y_2{,\vec{c}})
\phi_{
Y_3}(\tilde{a}^1,y_3{,\vec{c}})
\phi_{
Y_4}(\tilde{a}^1,y_4{,\vec{c}})
\!\!
\prod\limits_{Y_i \in \vec{Y}} \phi_{Y_i}(y_i, \tilde{a}^1, \vec{c})
\!\!\!\!
\prod\limits_{\vec{S} \subseteq \vec{Y}; |\vec{S}|>1}
\phi_{\vec{S}}(\{ y_i : Y_i \in \vec{S} \}, \vec{c})
}{
Z(\tilde{a}^1,\vec{c})
}
\\
\notag
p(\tilde{a}^1 | \vec{c})
p(\vec{c}).
\end{align*}
}
By contrast with above, the factorization of the saturated distribution would have every term $\phi_{\vec{S}}$ in the last product be a function of $A$.
Put another way, the above
factorization implies that interactions containing both elements of $Y$ and $\tilde{A}_1$ are of size at most two.  In the dyadic outcome case where both $Y$ and $\tilde{a}^1$ are binary variables, this constraint resulted in the loss of a single degree of freedom in the conditional log-linear model corresponding to the conditional factor of the CG model.  In a general network of size $N$, this results in many more restrictions on the observed law.  These restrictions may not be believable a priori, and some or even many may be ruled out by hypothesis tests.

To address this, we introduce a weaker treatment decomposition in the model described in Section \ref{subsec:gen-net},
where treatment components are not associated with specific outcomes, but with bundles of outcomes.  These weaker decompositions rely on correspondingly weaker restrictions on the observed data law, where arbitrary interaction terms between bundled outcomes and treatments are allowed.  Specifically, we partition $\vec{Y}$ into disjoint subsets $\vec{Y}_1, \ldots, \vec{Y}_K$, and decompose $A_1$ into components
$\tilde{A}_1^1, \ldots, \tilde{A}_1^K$, which now corresponds to these \emph{sets} of outcomes.  As before, let $\tilde{A}_1 \equiv \{ \tilde{A}_1^1, \ldots, \tilde{A}_1^K \}$, and fix any value $\tilde{a}_1$ in ${\mathfrak X}_{\tilde{A}_1}$.

We assume the following generalization of (\ref{eqn:general-model}):
\begin{align}
\label{eqn:general-model-2}
\notag
p(\vec{Y}_k(\{ \tilde{a}_1^1, \ldots, \tilde{a}_1^K \}, \vec{c}) \mid \{ \vec{Y}_m(\{ \tilde{a}_1^1, \ldots, \tilde{a}_1^K \}, \vec{c}) : \vec{Y}_m \in \vec{Y} \setminus \{ \vec{Y}_k \} \})\\
\text{ is only a function of }\nb_{\G^{\dag}}(\vec{Y}_k) \cup \{ \vec{Y}_k, \tilde{a}_1^k \} \cup \vec{c},
\end{align}
for all $k \in \{ 1, \ldots, K \}$. 

Graphically, this assumption states that the treatment $A_1$ may be decomposed into a set of components $\tilde{A}_1^k  (k = 1 \ldots K)$, where each copy only influences outcomes in the set $\vec{Y}_k$ directly, and other outcomes indirectly.  This assumption is encapsulated by a chain graph where the treatment $A_i$ (for unit $i$) decomposes into a set of additional vertices $\tilde{A}^k_i$, one for each outcome set $\vec{Y}_k$ above.  In this chain graph, the vector of covariates $\vec{C}$ directly influences $A_i$ and each $Y_k$, while each $\tilde{A}_i^k$ directly influences $\vec{Y}_k$ (only), and is in turn influenced by $A_i$.  An example of such a CG for the four variable example in Fig.~\ref{fig:med2} (b), where the sets are $\vec{Y}_1 = \{ Y_1 \}, \vec{Y}_{23} = \{ Y_2, Y_3 \}, \vec{Y}_4 = \{ Y_4 \}$ is shown in Fig.~\ref{fig:med2} (d).

Given this weaker decomposition, we obtain the following version of Theorem~\ref{thm:general-id}.
\begin{theorem}
Fix an arbitrary value assignment $\tilde{a}_1$ to $\tilde{A}_1 \equiv \{ \tilde{A}_1^1, \ldots \tilde{A}_1^K \}$.
Assume every distribution $p(\vec{Y}(\tilde{a}_1) \mid \vec{c})$ is positive.
Then for every $Y_m$, under the network versions of consistency, positivity, conditional ignorability,
as well as (\ref{eqn:general-model-2}),

\begin{align}
p(\ilya{y_m}(\tilde{a}_1)) &= \sum_{\vec{c},\vec{y}_{\ilya{-m}}}
\frac{
\left(
\prod_{k=1}^K \phi_{\vec{Y}_k,\tilde{A}^k_1}(\vec{y}_k, \tilde{a}_1^k{, \vec{c}})
\right)
\prod_{\vec{S} \in 
{\cal C}_{\vec{Y}} } \phi_{\vec{S}}(\vec{y}_{\vec{S}}{,\vec{c}})
}{
Z(\tilde{a}_1,\vec{c})
}
p(\vec{c}),
\label{eqn:network-sym-med-2}
\end{align}
where $\vec{y}_C$ is the subset of $\vec{y}$ pertaining to the clique $\vec{S}$.
\label{thm:general-id-2}
\end{theorem}
The proof is a direct analogue of the proof of Theorem~\ref{thm:general-id}.

The coarser decomposition obtained from this weaker model is obtained by a straightforward generalization of the outcome-specific decomposition.  Rather than fixing an ordering on units in a block, we fix an ordering on sets of outcomes $\vec{Y}_k$, $k \in 1, \ldots, K$.
For each set $\vec{Y}_k$, we denote the set of units preceding this set according to the ordering as $\pre_{\prec}(k)$, and the set of outcome sets $\vec{Y}_k$ precedes according to $\prec$ 
as $\post_{\prec}(k)$.
As before, let $\tilde{a} = 1$, and $\tilde{a}' = 0$.  We consider the following decomposition of
$\mathbb{E}[Y_i(\tilde{a})] - \mathbb{E}[Y_i(\tilde{a}')]$ (recall that $\tilde{A}_1$ are components of a single unit treatment $A_1$, with others treatments suppressed from the notation):
\begin{align*}
\sum_{{k}=1}^{{K}} &
\mathbb{E}[Y_i(\{ \tilde{a}_1^{l} \mid l \in \pre_{\prec}(k) \}, \tilde{a}_1^{k}, \{ (\tilde{a}_1^m){'} \mid {m} \in \post_{\prec}(k) \}] - \\
& \mathbb{E}[Y_i(\{ \tilde{a}_1^{l} \mid l \in \pre_{\prec}(k) \}, (\tilde{a}_1^k){'}, \{ (\tilde{a}_1^m){'} \mid {m} \in \post_{\prec}(k) \}].
\end{align*}

In the four unit example shown in Fig.~\ref{fig:med2} (c),
the effect $\mathbb{E}[Y_2(\tilde{a}^1_1, \tilde{a}^{23}_1, \tilde{a}^4_1)] - \mathbb{E}[Y_2((\tilde{a}^1_1)', (\tilde{a}^{23}_1)', (\tilde{a}^4_1)')]$ decomposes as follows:
\begin{align*}
&\underbrace{\mathbb{E}[Y_2({\tilde{a}^1_1,\tilde{a}^{23}_1,\tilde{a}^4_1})] - \mathbb{E}[Y_2({(\tilde{a}^1_1)',\tilde{a}^{23}_1,\tilde{a}^4_1})]}_{\text{indirect effect through $Y_1$}}
+ \underbrace{\mathbb{E}[Y_2({(\tilde{a}^1_1)',\tilde{a}^{23}_1,\tilde{a}^4_1} )] - \mathbb{E}[Y_2({(\tilde{a}^1_1)',(\tilde{a}^{23}_1)',\tilde{a}^4_1})]}_{\text{effect through $Y_2$ and $Y_3$}}\\
+ &
\underbrace{\mathbb{E}[Y_2({(\tilde{a}^1_1)',(\tilde{a}^{23}_1)',\tilde{a}^4_1})] - \mathbb{E}[Y_2({(\tilde{a}^1_1)',(\tilde{a}^{23}_1)',(\tilde{a}^4_1)'})]}_{\text{indirect effect through $Y_4$}}
\end{align*}
Note that given the coarser decomposition of the effect of $A_1$ on $Y_2$ which bundles $Y_2$ and $Y_3$ together, the second component of the above decomposition cannot be interpreted as either a direct or an indirect effect (mediated by $Y_3$), but is instead a combination of the two.  An analogue situation may occur in mediation problems in causal models associated with DAGs, where neither a direct effect, nor an indirect effect may be individually identifiable, but a ``bundle effect'' including them both may be.


To illustrate Theorem \ref{thm:general-id-2}, the term $\mathbb{E}[Y_2({\tilde{a}^1_1,\tilde{a}^{23}_1,\tilde{a}^4_1})] -
\mathbb{E}[Y_2({(\tilde{a}^1_1)',\tilde{a}^{23}_1,\tilde{a}^4_1})]$ in the above decomposition{, evaluated at $y_2$}
is identified as
\begin{align*}
&
\sum_{y_1,y_3,y_4{,\vec{c}}}
\frac{
\phi_{Y_1}(\tilde{a}^1_1,y_1) \phi_{Y_2, Y_3}(\tilde{a}^{23}_1,y_2,y_3) \phi_{Y_4}({\tilde{a}^4_1},y_4) \phi_{Y_1,Y_2,Y_3,Y_4}({y_1},y_2,y_3,y_{{4}})
}{
Z(\tilde{a}^1_1,\tilde{a}^{23}_1,\tilde{a}^4_1,\vec{c})
} p(\vec{c})\\
&-
\sum_{y_1,y_3,y_4{,\vec{c}}}
\frac{
\phi_{Y_1}({(\tilde{a}^1_1)'},y_1) \phi_{Y_2, Y_3}({\tilde{a}^{23}_1},y_2,y_3) \phi_{Y_4}({\tilde{a}^4_1},y_4) \phi_{Y_1,Y_2,Y_3,Y_4}({y_1},y_2,y_3,y_{{4}})
}{
Z((\tilde{a}^1_1)',\tilde{a}^{23}_1,\tilde{a}^4_1,\vec{c})
} p(\vec{c}).
\end{align*}




\section{Details on Results in Section \ref{subsec:doubly}}
\label{sec:appendix:doubly}

In this section, we provide technical details on the result in Section \ref{subsec:doubly}.

\subsection{Proof of Theorem \ref{thm:IF}}

We first prove that the influence function $\IF$ in Theorem \ref{thm:IF} is a valid influence function for $\psi(\tilde{a}^1,(\tilde{a}^2)')$ in the symmetric model $\mathcal{M}_{\text{sym}}$. The density in $\mathcal{M}_{\text{sym}}$ can be parametrized as follows:
\begin{align*}
& f(Y_1=y_1,Y_2=y_2,A=a,\vec{C}=\vec{c})
\\
&
=
\frac{ 
\gamma_{12} ( y_1,y_2 \cond \vec{c} )
\gamma_{1A} ( y_1,a \cond \vec{c} )
\gamma_{2A} ( y_2,a \cond \vec{c} ) 
f_1(y_1 \cond  \vec{c})
f_2(y_2 \cond  \vec{c})
f_A(a \cond  \vec{c})
}{\mathcal{N}(\vec{c})}
f_C (\vec{c})
 \ ,
\end{align*}
where 
\begin{align*}
& \gamma_{12} (y_1,y_2 \cond \bC)
=
\frac{ f(Y_1=y_1,Y_2=y_2,A=a,\vec{C}=\vec{c}) }
{ f(Y_1=y_1,Y_2=0,A=a,\vec{C}=\vec{c}) }
\frac{ f(Y_1=0,Y_2=0,A=a,\vec{C}=\vec{c}) } 
{ f(Y_1=0,Y_2=y_2,A=a,\vec{C}=\vec{c}) }
\\
& \gamma_{1A} (y_1,a \cond \bC)
=
\frac{ f(Y_1=y_1,Y_2=0,A=a,\vec{C}=\vec{c}) }
{ f(Y_1=y_1,Y_2=0,A=\tilde{a}^1,\vec{C}=\vec{c}) }
\frac{ f(Y_1=0,Y_2=0,A=\tilde{a}^1,\vec{C}=\vec{c}) }
{ f(Y_1=0,Y_2=0,A=a,\vec{C}=\vec{c}) }
\\
& \gamma_{2A} (y_2,a \cond \bC)
=
\frac{ f(Y_1=0,Y_2=y_2,A=a,\vec{C}=\vec{c}) }
{ f(Y_1=0,Y_2=y_2,A=\tilde{a}^2{'},\vec{C}=\vec{c}) }
\frac{ f(Y_1=0,Y_2=0,A=\tilde{a}^2{'},\vec{C}=\vec{c}) }
{ f(Y_1=0,Y_2=0,A=a,\vec{C}=\vec{c}) }
\\
&
f_1(y_1 \cond \vec{c}) = f_1(y_1  \cond \tilde{a}^1 , Y_2=0,\vec{c})
\\
&
f_2(y_2 \cond \vec{c}) = 
f_2(y_2 \cond \tilde{a}^2{'}, Y_1=0,\vec{c}) \\
&
f_A(a \cond \vec{c}) = 
f_A(a \cond Y_1=0,Y_2=0,\vec{c})
\\
&
\mathcal{N}(\vec{c}) = 
\int \gamma_{12} ( y_1,y_2 \cond \vec{c} )
\gamma_{1A} ( y_1,a \cond \vec{c} )
\gamma_{2A} ( y_2,a \cond \vec{c} ) 
f_1(y_1 \cond \vec{c})
f_2(y_2 \cond \vec{c})
f_A(a \cond \vec{c})
\, d(y_1,y_2,a)
\ .
\end{align*}
Therefore, any density in $\mathcal{M}_{\text{sym}}$ is parametrized by 7 nuisance functions, three odds ratios, three baseline densities, and the density of $\bC$.

The target estimand $\psi := \psi(\tilde{a}^1,\tilde{a}^2{'})$ is defined as $\psi = \EXP \big\{ \theta(\bC) \big\}$ where
\begin{align} \label{eq-theta}
& 
\theta(\bc)
=
\frac{\theta_N(\bc)}{\theta_D(\bc)}
\ ,
\\
&	 
\theta_N(\bc)
=
\int
h(y_1,y_2)
\gamma_{12} ( y_1,y_2 \cond \vec{c} ) 
f_1(y_1 \cond \vec{c})
f_2(y_2 \cond \vec{c}) 
\, d(y_1,y_2)
\ ,
\nonumber
\\
&
\theta_D(\bc)
=
\int
\gamma_{12}( y_1,y_2 \cond \vec{c} ) 
f_1(y_1 \cond \vec{c})
f_2(y_2 \cond \vec{c}) 
\, d(y_1,y_2)
\ .
\nonumber
\end{align}
Also, recall that $Q_1$ and $Q_2$ are defined as
\begin{align}	\label{eq-supp-Qfunctions}
& 
Q_1(y_1 \cond \vec{c})
=
\int 
\big\{ h(y_1,y_2) - \theta(  \vec{c}) \big\}
\gamma(y_1,y_2 \cond \vec{c} )
f_2(y_2 \cond \tilde{a}^2{'},Y_1=0,\vec{c})
d y_2
\ ,
\\
&
Q_2(y_2 \cond \vec{c})
=
\int
\big\{ h(y_1,y_2) - \theta( \vec{c}) \big\}
\gamma(y_1,y_2 \cond \vec{c} )
f_1(y_1 \cond \tilde{a}^1,Y_2=0,\vec{c})
d y_1
\ .
\nonumber
\end{align}

Before we prove the result, we consider a few useful results. First, for any function $\mathcal{G}$, we have
\begin{align*}
&
\EXP \big\{ \mathbb{I}(A=\tilde{a}^2{'}) \mathcal{G}(Y_1,Y_2) / \gamma_{1A} (Y_1, \tilde{a}^2{'} \cond \bc) \cond  \bc \big\} 
\\
&
=
\frac{1}{\N (\bc)}
\int 
\mathcal{G}(y_1,y_2) 
\gamma_{12} (y_1,y_2 \cond \bc)
f_1(y_1 \cond \bc)
f_2(y_2 \cond \bc)
f_A(\tilde{a}^2{'} \cond \bc)
\, d(y_1,y_2)
\ ,
\\
&
\EXP \big\{ \mathbb{I}(A=\tilde{a}^1) \mathcal{G}(Y_1,Y_2) / \gamma_{2A} (Y_2, \tilde{a}^1 \cond \bc) \cond  \bc \big\} 
\\
&
=
\frac{1}{\N (\bc)}
\int
\mathcal{G}(y_1,y_2) 
\gamma_{12}(y_1,y_2 \cond \bc)
f_1(y_1 \cond \bc)
f_2(y_2 \cond \bc)
f_A(\tilde{a}^1 \cond \bc)
\, d(y_1,y_2)
\ .
\end{align*} 
Therefore we obtain  
\begin{align}
\label{eq-alpha}
&
\EXP \bigg\{ 
\frac{\mathbb{I}(A=\tilde{a}^2{'})}{ \gamma_{1A}(Y_1, \tilde{a}^2{'}) } 
\COND \bc \bigg\}
=
\frac{\theta_D (\bc) f_A(\tilde{a}^2{'} \cond \bc)}{\N (\bc)}
=: \delta_1(\bc)
\ ,
\nonumber
\\
&
\EXP  \bigg\{	\frac{\mathbb{I}(A=\tilde{a}^1)}{ \gamma_{2A}(Y_2, \tilde{a}^1) } 
\COND \bc \bigg\}
=
\frac{\theta_D (\bc) f_A(\tilde{a}^1 \cond \bc)}{\N (\bc)}
=: \delta_2(\bc)
\ ,
\end{align}
and
\begin{align}		
&
\frac{ 
\EXP \big\{ 
\frac{ \mathbb{I}(A=\tilde{a}^2{'}) h(Y_1,Y_2) }{ \gamma_{1A}(Y_1, \tilde{a}^2{'} \cond \vec{c})  }
\cond \bc \big\}
}{ 
\EXP \big\{ \frac{ \mathbb{I}(A=\tilde{a}^2{'}) }{ \gamma_{1A}(Y_1, \tilde{a}^2{'} \cond \vec{c})  }
\cond \bc \big\}
}
=
\frac{ 
\EXP \big\{ \frac{
\mathbb{I}(A=\tilde{a}^1) h(Y_1,Y_2) }{ \gamma_{2A}(Y_1, \tilde{a}^1 \cond \vec{c}) }
\cond \bc \big\}
}{ 
\EXP \big\{ \frac{ \mathbb{I}(A=\tilde{a}^1) }{ \gamma_{2A}(Y_1, \tilde{a}^1 \cond \vec{c}) }
\cond \bc \big\}
}
=
\theta(\bc)  \ .
\label{eq-IPW}
\end{align}
Next, we will establish 
\begin{align}		\label{eq-Q-center}
&
\EXP \bigg\{
\frac{\mathbb{I}(A=\tilde{a}^2{'}) Q_1 (Y_1 \cond \bc)}{ \gamma_{12} (Y_1,Y_2 \cond \bc) \gamma_{1A} (Y_1,\tilde{a}^2{'} \cond \bc)}
\COND \bc
\bigg\}
=
0
,
&&
\EXP \bigg\{
\frac{ \mathbb{I}(A=\tilde{a}^1) Q_1 (Y_2 \cond \bc)}{\gamma_{12} (Y_1,Y_2 \cond \bc) 
\gamma_{2A} (Y_2,\tilde{a}^1 \cond \bc) }
\COND  \bC
\bigg\}
=
0
,
\nonumber
\\
&
\EXP \bigg\{
\frac{\mathbb{I}(A=\tilde{a}^2{'}) Q_2 (Y_1 \cond \bc)}{ \gamma_{12} (Y_1,Y_2 \cond \bc) \gamma_{1A} (Y_1,\tilde{a}^2{'} \cond \bc)}
\COND \bc
\bigg\}
=
0
,
&&
\EXP \bigg\{
\frac{ \mathbb{I}(A=\tilde{a}^1) Q_2 (Y_2 \cond \bc)}{\gamma_{12} (Y_1,Y_2 \cond \bc) 
\gamma_{2A} (Y_2,\tilde{a}^1 \cond \bc) }
\COND  \bC
\bigg\}
=
0
\end{align}
We only provide algebraic details of the first result because the others can be shown in a similar manner. 
\begin{align*}	
& 
\EXP \bigg\{
\frac{ \mathbb{I}(A=\tilde{a}^2{'}) Q_1 (Y_1 \cond \bc) }{ \gamma_{12} (Y_1,Y_2 \cond \bc) \gamma_{1A}(Y_1,\tilde{a}^2{'} \cond \bc) }
\COND \bc
\bigg\}
\\
&
\propto
\int
Q_1(y_1 \cond \bc)
f_1 (y_1 \cond \bc)
\underbrace{
\gamma_{2A} (y_2 , \tilde{a}^2{'} \cond \bc)
}_{=1}
f_2 (y_2 \cond \bc)
f_A(1 \cond \bc)
\, d(y_1,y_2)
\\
&
=
f_A(1 \cond \bc)
\int
Q_1(y_1 \cond \bc)
f_1 (y_1 \cond \bc)		
\, d y_1
\\
&
=
f_A(1 \cond \bc)
\int
\bigg[ \int \big\{ h(y_1,y_2) - \theta(\bc) \big\}
\gamma_{12}(y_1,y_2) 		
f_2 (y_2 \cond \bc)
\, dy_2
\bigg]
f_1 (y_1 \cond \bc)		
\, d y_1
\\
&
=
f_A(1 \cond \bc)
\bigg[
\begin{array}{l}
\int
h(y_1,y_2)
\gamma_{12}(y_1,y_2) 		
f_1 (y_1 \cond \bc)		
f_2 (y_2 \cond \bc)
\, d (y_1,y_2)
\\
-
\theta(\bc)
\int
\gamma_{12}(y_1,y_2) 		
f_1 (y_1 \cond \bc)		
f_2 (y_2 \cond \bc)
\, d (y_1,y_2)
\end{array} 
\bigg] 
\\
&
=
f_A(1 \cond \bc)
\big\{ \theta_N(\bc) - \theta(\bc) \theta_D(\bc) \big\}
\\
&
=
0
\ .
\end{align*}  

We consider a parametric submodel of $\mathcal{M}_{\text{sym}}$, parametrized by $\eta$, which is
\begin{align*}
&
f(Y_1=y_1,Y_2=y_2,A=a,\vec{C}=\vec{c} \con \eta)
\\
&
=
\frac{ 
\gamma_{12} ( y_1,y_2 \cond \vec{c} \con \eta )
\gamma_{1A} ( y_1,a \cond \vec{c} \con \eta )
\gamma_{2A} ( y_2,a \cond \vec{c} \con \eta ) 
f_1(y_1 \cond  \vec{c} \con \eta)
f_2(y_2 \cond  \vec{c} \con \eta)
f_A(a \cond  \vec{c} \con \eta)
}{\mathcal{N}(\vec{c} \con \eta)}
f_C (\vec{c} \con \eta)
\ .
\end{align*}
Without loss of generality, we assume that the true nuisance functions are recovered at $\eta^*$. Let $\EXP\ETA$ be the expectation operator with respect to $f(\cdot \con \eta)$. The target parameter evaluated at $\eta$ is $\psi(\eta) = \EXP\ETA \{ \theta(\bC \con \eta) \}$ where
\begin{align*}
& 
\theta(\bc \con \eta)
=
\frac{\theta_N(\bc \con \eta)}{\theta_D(\bc \con \eta)}
 \ ,
\\
&	 
\theta_N(\bc \con \eta)
=
\int
h(y_1,y_2)
\gamma_{12} ( y_1,y_2 \cond \vec{c}  \con \eta) 
f_1(y_1 \cond  \vec{c} \con \eta)
f_2(y_2 \cond  \vec{c} \con \eta) 
\, d (y_1,y_2)
 \ ,
\nonumber
\\
&
\theta_D(\bc \con \eta)
=
\int
\gamma_{12} ( y_1,y_2 \cond \vec{c}  \con \eta) 
f_1(y_1 \cond  \vec{c} \con \eta)
f_2(y_2 \cond  \vec{c} \con \eta) 
\, d (y_1,y_2)
 \ .
\nonumber
\end{align*}
The partial derivative of the target estimand is
\begin{align}
\frac{\partial \psi(\eta)}{\partial \eta}
&
=
\int 
\Big\{
\nabla_{\eta} \theta( \bc \con \eta) f_C(\bc \con \eta)
+
\theta( \bc \con \eta) s_C(\bc \con \eta) f_C(\bc \con \eta)
\Big\}
\, d \bc
\nonumber
\\
&
=
\int 
\theta( \bc \con \eta)
\Bigg\{
\frac{ 
\nabla_{\eta} \theta_N ( \bc \con \eta)
}{ 
\theta_N (\bc \con \eta)
}
-
\frac{ 
\nabla_{\eta} \theta_D ( \bc \con \eta)
}{ 
\theta_D ( \bc \con \eta)
}
+ s_C(\bc \con \eta)
\Bigg\}
f_C(\bc \con \eta)
\, d \bc
\nonumber
\\
&
=
\EXP \ETA
\Bigg[
\theta( \bC \con \eta)
\bigg\{
\frac{ 
\nabla_{\eta} \theta_N ( \bC \con \eta)
}{ 
\theta_N ( \bC \con \eta)
}
-
\frac{ 
\nabla_{\eta} \theta_D ( \bC \con \eta)
}{ 
\theta_D ( \bC \con \eta)
}
+ s_C(\bC \con \eta)
\bigg\}
\Bigg]
\ ,
\label{eq-path der 1}
\end{align}
where 
\begin{align*}
&
\nabla_{\eta} \theta_N(\bc \con \eta)
\\
&
=
\int h(y_1,y_2)
\left\{
s_{12}(y_1,y_2 \cond \bc \con \eta) 
+ 
s_{1} (y_1 \cond \bc \con \eta) 
+
s_{2} (y_2 \cond \bc \con \eta) 
\right\}
\gamma_{12} (y_1,y_2 \cond \bc \con \eta)
f_1(y_1 \cond \bc \con \eta)
f_2 (y_2 \cond \bc \con \eta ) \, d (y_1,y_2)
\ ,
\\
&
\nabla_{\eta} \theta_D(\bc \con \eta)
\\
&
=
\int 
\left\{
s_{12}(y_1,y_2 \cond \bc \con \eta) 
+ 
s_{1} (y_1 \cond \bc \con \eta) 
+
s_{2} (y_2 \cond \bc \con \eta) 
\right\}
\gamma_{12} (y_1,y_2 \cond \bc \con \eta) 
f_1(y_1 \cond \bc \con \eta)
f_2 (y_2 \cond \bc \con \eta ) \, d (y_1,y_2) 
\ .
\end{align*}
Here, $s_{12} = \nabla_\eta \gamma_{12}(\cdot) / \gamma_{12}(\cdot)$, $s_1 = \nabla_\eta f_1(\cdot) / f_1(\cdot)$, $s_2 = \nabla_\eta f_2(\cdot) / f_2(\cdot)$. Additionally, we define $s_{1A} =  \nabla_\eta \gamma_{1A}(\cdot) / \gamma_{1A}(\cdot)$, $s_{2A} = \nabla_\eta \gamma_{2A}(\cdot) / \gamma_{2A}(\cdot)$, $s_A = \nabla_\eta f_A(\cdot) / f_A(\cdot)$. Due to the boundary condition of the odds ratio functions (i.e., $\gamma_{12}(0,y_2 \cond \vec{c}) = \gamma_{12}(y_1,0 \cond \vec{c}) = \gamma_{1A}(y_1,\tilde{a}^1 \cond \vec{c}) = \gamma_{2A}(y_1,\tilde{a}^2{'} \cond \vec{c}) = 1$ for any $y_1,y_2$), we have
\begin{align}		\label{eq-boundary}
s_{12}(0,y_2 \cond \bc \con \eta)
=
s_{12}(y_1,0 \cond \bc \con \eta)
=
s_{1A}(y_1,\tilde{a}^1 \cond \bc \con \eta)
=
s_{2A}(y_2,\tilde{a}^2{'} \cond \bc \con \eta)
=
0
\ , \ \forall y_1,y_2 \ .
\end{align}

Therefore,
\begin{align*}
&
\frac{
\nabla_{\eta} \theta_N(\bc \con \eta)
}{
\theta_N(\bc \con \eta)
}
\\
&
=
\frac{ 
\EXP \ETA \big[
\mathbb{I}(A = \tilde{a}^2{'}) h(Y_1,Y_2) \big\{ 
s_{12}(Y_1,Y_2 \cond \bc \con \eta)
+
s_1 (Y_1 \cond \bc \con \eta)
+
s_2 (Y_2 \cond \bc \con \eta)
\big\} / \gamma_{1A} (Y_1, \tilde{a}^2{'} \cond \bc \con \eta)
\cond 
\bc
\big]
}{
\EXP\ETA \big[
\mathbb{I}(A = \tilde{a}^2{'}) h(Y_1,Y_2)  / \gamma_{1A} (Y_1, \tilde{a}^2{'} \cond \bc \con \eta)
\cond
\bc
\big]
}
 \ ,
\\
&
\frac{
\nabla_{\eta} \theta_D(\bc \con \eta)
}{
\theta_D(\bc \con \eta)
}
\\
&
=
\frac{ 
\EXP\ETA \big[
\mathbb{I}(A = \tilde{a}^2{'}) \big\{ 
s_{12}(Y_1,Y_2 \cond \bc \con \eta)
+
s_1 (Y_1 \cond \bc \con \eta)
+
s_2 (Y_2 \cond \bc \con \eta)
\big\} / \gamma_{1A} (Y_1, \tilde{a}^2{'} \cond \bc \con \eta)
\cond 
\bc
\big]
}{
\EXP\ETA \big[
\mathbb{I}(A = \tilde{a}^2{'}) / \gamma_{1A} (Y_1, \tilde{a}^2{'} \cond \bc \con \eta)
\cond
\bc 
\big]
}
 \ ,
\\
&
\theta( \bc \con \eta)
\frac{ 
\nabla_{\eta} \theta_N (\bc \con \eta)
}{ 
\theta_N (\bc \con \eta)
}
\\
&
=
\frac{ 
\EXP\ETA \big[
\mathbb{I}(A = \tilde{a}^2{'}) h(Y_1,Y_2) \big\{ 
s_{12}(Y_1,Y_2 \cond \bc \con \eta)
+
s_1 (Y_1 \cond \bc \con \eta)
+
s_2 (Y_2 \cond \bc \con \eta)
\big\} / \gamma_{1A} (Y_1, \tilde{a}^2{'} \cond \bc \con \eta)
\cond 
\bc 
\big]
}{
\EXP\ETA \big[
\mathbb{I}(A = \tilde{a}^2{'})  / \gamma_{1A} (Y_1, \tilde{a}^2{'} \cond \bc \con \eta)
\cond
\bc 
\big]
}
 \ ,
\\
&
\theta( \bc \con \eta)
\frac{ 
\nabla_{\eta} \theta_D (\bc \con \eta)
}{ 
\theta_D (\bc \con \eta)
} 
\\
&
=
\frac{ 
\EXP\ETA \big[
\theta(\bc \con \eta)
\tilde{I}(A = \tilde{a}^2{'}) \big\{ 
s_{12}(Y_1,Y_2 \cond \bc \con \eta)
+
s_1 (Y_1 \cond \bc \con \eta)
+
s_2 (Y_2 \cond \bc \con \eta)
\big\} / \gamma_{1A} (Y_1, \tilde{a}^2{'} \cond \bc \con \eta)
\cond 
\bc 
\big]
}{
\EXP\ETA \big[
\mathbb{I}(A = \tilde{a}^2{'})  / \gamma_{1A} (Y_1, \tilde{a}^2{'} \cond \bc \con \eta)
\cond
\bc 
\big]
}
 \ ,
\\
&
\theta( \bc \con \eta)
s_C(\bc \con \eta) 
\\
&
=
\frac{ 
\EXP\ETA \big[
\mathbb{I}(A = \tilde{a}^2{'}) h(Y_1,Y_2) s_C(\bc \con \eta) / \gamma_{1A} (Y_1, 1 \cond \bc \con \eta)
\cond 
\bc 
\big]
}{
\EXP\ETA \big[
\mathbb{I}(A = \tilde{a}^2{'})  / \gamma_{1A} (Y_1, \tilde{a}^2{'} \cond \bc \con \eta)
\cond
\bc 
\big]
}
 \ .
\end{align*}
Plugging these results in \eqref{eq-path der 1}, we obtain
\begin{align*}
&
\frac{\partial \psi(\eta)}{\partial \eta}
\\
&
=
\EXP \ETA 
\Bigg[
\frac{ 
\EXP\ETA \left[
\frac{   \mathbb{I}(A = \tilde{a}^2{'}) }{ \gamma_{1A} (Y_1, \tilde{a}^2{'} \cond \bC \con \eta) }
\left[ 
\begin{array}{l}
     \left\{ 
     \begin{array}{l}
      h(Y_1,Y_2) \\
      - \theta(\bC \con \eta)
     \end{array}
      \right\}
\left\{	
\begin{array}{l}			
s_{12}(Y_1,Y_2 \cond \bC \con \eta)
\\
+
s_1 (Y_1 \cond \bC \con \eta)
\\
+
s_2 (Y_2 \cond \bC \con \eta)
\end{array}
\right\}  
+
h(Y_1,Y_2) s_C(\bC \con \eta)  
\end{array} 
\right] 
\COND \bC \right]
}{ 
\EXP\ETA \big[
\mathbb{I}(A = \tilde{a}^2{'})  / \gamma_{1A} (Y_1, \tilde{a}^2{'} \cond \bC \con \eta)
\cond
\bC
\big]
}
\Bigg] \ .
\end{align*}

Next, we return to the influence function in Theorem, which we provide below for completeness:
\begin{align*}
&
\IF(\bO)
\\
& =
w(\bC)
\underbrace{
\Bigg[
\frac{ \big\{ h(Y_1,Y_2) - \theta(\bC) \big\} \mathbb{I}(A=\tilde{a}^1)  / \gamma_{2A} (Y_2,\tilde{a}^1 \cond \bC) }{  \EXP \big\{ \mathbb{I}(A=\tilde{a}^1)  / \gamma_{2A} (Y_2,\tilde{a}^1 \cond \bC) \cond  \bC \big\} }  
\Bigg]
}_{=\IF_1(\bO)}
\\
&
\hspace*{2cm}
+
w(\bC)
\underbrace{
\Bigg[
\frac{\mathbb{I}(A=\tilde{a}^2{'}) Q_2 (Y_2 \cond \bC) / \gamma_{1A} (Y_1,\tilde{a}^2{'} \cond \bC) }{\gamma_{12}(Y_1,Y_2 \cond \bC) 	\EXP \big\{ \mathbb{I}(A=\tilde{a}^2{'})/\gamma_{1A} (Y_1,\tilde{a}^2{'} \cond \bC) \cond \bC \big\}  } 
\Bigg]
}_{\IF_2(\bO)}
\\
&
\hspace*{2cm}
+
w(\bC)
\underbrace{
\Bigg[
-
\frac{ \mathbb{I}(A=\tilde{a}^1)Q_2 (Y_2 \cond \bC)  / \gamma_{2A} (Y_2,\tilde{a}^1 \cond \bC) }{\gamma_{12}(Y_1,Y_2 \cond \bC) 	\EXP \big\{ \mathbb{I}(A=\tilde{a}^1)/ \gamma_{2A} (Y_2,\tilde{a}^1 \cond \bC) \cond \bC \big\}  } 
\Bigg]
}_{\IF_3(\bO)}
\\
&
+
\Big\{ 1- w(\bC) \Big\}
\underbrace{
\Bigg[
\frac{ \big\{ h(Y_1,Y_2) - \theta(\bC) \big\} \mathbb{I}(A=\tilde{a}^2{'})/\gamma_{1A} (Y_1,\tilde{a}^2{'} \cond \bC) }{  \EXP \big\{ \mathbb{I}(A=\tilde{a}^2{'})/\gamma_{1A} (Y_1,\tilde{a}^2{'} \cond \bC) \cond  \bC \big\} }  
\Bigg]
}_{=\IF_1'(\bO)}
\\
&
\hspace*{2cm}
+
\Big\{ 1- w(\bC) \Big\}
\underbrace{
\Bigg[
-
\frac{\mathbb{I}(A=\tilde{a}^2{'}) Q_1 (Y_1 \cond \bC) / \gamma_{1A} (Y_1,\tilde{a}^2{'} \cond \bC) }{\gamma_{12}(Y_1,Y_2 \cond \bC) 	\EXP \big\{ \mathbb{I}(A=\tilde{a}^2{'})/\gamma_{1A} (Y_1,\tilde{a}^2{'} \cond \bC) \cond \bC \big\}  } 
\Bigg]
}_{\IF_2'(\bO)}
\\
&
\hspace*{2cm}
+
\Big\{ 1- w(\bC) \Big\}
\underbrace{
\Bigg[ 
\frac{ \mathbb{I}(A=\tilde{a}^1)Q_1 (Y_1 \cond \bC)  / \gamma_{2A} (Y_2,\tilde{a}^1 \cond \bC) }{\gamma_{12}(Y_1,Y_2 \cond \bC) 	\EXP \big\{ \mathbb{I}(A=\tilde{a}^1)/ \gamma_{2A} (Y_2,\tilde{a}^1 \cond \bC) \cond \bC \big\}  } 
\Bigg]
}_{\IF_3'(\bO)}
\\
&
\hspace*{2cm}
+
\underbrace{
\Big[
\theta (\bC) - \psi 
\Big]
}_{\IF_4(\bC)} \ .
\end{align*}
Below, we establish 
\begin{align}	\label{eq-path der}
\EXP \Big[ \IF(\bO) s(\bO \con \eta^*) \Big]
=
\frac{\partial \psi(\eta)}{\partial \eta} \bigg|_{\eta=\eta^*}
 \ ,
\end{align} 
i.e., $\psi(\eta)$ is pathwise differentiable with respect to the influence function $\IF(\bO)$.

First, we focus on $\EXP \big\{ \IF_1 \cdot s \big\}$. From \eqref{eq-IPW}, we establish that 
\begin{align*}
\EXP \big\{ \IF_1 (\bO) \cond \bC \big\} 
&
=
\frac{\EXP \big\{ \mathbb{I}(A = \tilde{a}^1)) h(Y_1,Y_2)   / \gamma_{2A}(Y_2,\tilde{a}^1 \cond \bC) \cond  \bC \big\}}{\EXP \big\{ \mathbb{I}(A = \tilde{a}^1)  / \gamma_{2A} (Y_2,\tilde{a}^1 \cond \bC) \cond  \bC \big\}}
-\theta (\bC)
=
0 
\ .
\end{align*}
Therefore, 
\begin{align}	\label{eq-IF1} 
&
\EXP \Big[
\IF_1 (\bO) s(\bO \con \eta^*)
\Big]
\\
\nonumber
&
=
\EXP \Bigg[
\IF_1 (\bO) 
\bigg\{ 
\begin{array}{l}		
s_{12}(Y_1,Y_2 \cond \bC \con \eta^*)
+
s_{1A}(Y_1,A \cond \bC \con \eta^*)
+
s_{2A}(Y_2,A \cond \bC \con \eta^*)
\\
+  
s_1(Y_1 \cond \bC\con \eta^*)
+ 
s_2(Y_2 \cond \bC\con \eta^*)
+ 
s_A(A \cond \bC \con \eta^*)
\end{array}
\bigg\}
\Bigg]
\\
\nonumber
& \hspace*{2cm}
+
\EXP \Big[
\IF_1 (\bO) 
\cdot  s_C(\bC\con \eta^*) 
\Big]
\\
\nonumber
&
=
\EXP \bigg[ 
\EXP \Big[ 
\IF_1 (\bO) 
\bigg\{ 
\begin{array}{l}		
s_{12}(Y_1,Y_2 \cond \bC \con \eta^*)
+
s_{1A}(Y_1,A \cond \bC \con \eta^*)
+
s_{2A}(Y_2,A \cond \bC \con \eta^*)
\\
+  
s_1(Y_1 \cond \bC\con \eta^*)
+ 
s_2(Y_2 \cond \bC\con \eta^*)
+ 
s_A(A \cond \bC \con \eta^*)
\end{array}
\bigg\}
\Cond  \bC
\Big]
\bigg] 
\\
\nonumber
&
\stackrel{\eqref{eq-boundary}}{=}
\EXP \Bigg[
\frac{ 
\EXP
\bigg[
\frac{\mathbb{I}(A = \tilde{a}^1) 
\big\{ h(Y_1,Y_2) -\theta (\bC) \big\}  }{\gamma_{2A}(Y_2,\tilde{a}^1 \cond \bC) } 
\bigg\{
\begin{array}{l}
s_{12}(Y_1,Y_2 \cond \bC \con \eta^*) 
+
s_{2A}(Y_2,\tilde{a}^1 \cond \bC \con \eta^*)
\\
+  
s_1(Y_1 \cond \bC\con \eta^*)
+ 
s_2(Y_2 \cond \bC\con \eta^*)
+ 
s_A(0 \cond \bC \con \eta^*)		
\end{array} 
\bigg\}
\, \bigg| \bC \bigg]
}{ \EXP \big\{ \mathbb{I}(A = \tilde{a}^1)  / \gamma_{2A} (Y_2,\tilde{a}^1 \cond \bC) \cond \bC \big\} }
\Bigg]		 
\\
\nonumber
&
=
\EXP  \Bigg[
\frac{ 
\EXP 
\bigg[
\frac{\mathbb{I}(A = \tilde{a}^1) 
\big\{ h(Y_1,Y_2) -\theta (\bC) \big\} }{  \gamma_{2A}(Y_2,\tilde{a}^1 \cond \bC)  }  
\bigg\{
\begin{array}{l}
s_{12}(Y_1,Y_2 \cond \bC \con \eta^*) 
+
s_{2A}(Y_2,\tilde{a}^1 \cond \bC \con \eta^*)
\\
+  
s_1(Y_1 \cond \bC\con \eta^*)
+ 
s_2(Y_2 \cond \bC\con \eta^*) 
\end{array} 
\bigg\}
\, \bigg| \bC \bigg]
}{ \EXP \big\{ (1-A)  / \gamma_{2A}(Y_2,0 \cond \bC) \cond \bC \big\} }
\Bigg]		 
\\
\nonumber
&
=
\EXP  \Bigg[
\frac{ 
\EXP 
\bigg[
\frac{
\mathbb{I}(A = \tilde{a}^2{'}) 
\big\{ h(Y_1,Y_2) -\theta (\bC) \big\}  }{\gamma_{1A} (Y_1,\tilde{a}^2{'} \cond \bC) }
\bigg\{
\begin{array}{l}
s_{12}(Y_1,Y_2 \cond \bC \con \eta^*) 
+
s_{2A}(Y_2,\tilde{a}^1 \cond \bC \con \eta^*)
\\
+  
s_1(Y_1 \cond \bC\con \eta^*)
+ 
s_2(Y_2 \cond \bC\con \eta^*) 
\end{array}
\bigg\} 
\, \bigg| \bC \bigg]
}{ \EXP \big\{ \mathbb{I}(A = \tilde{a}^1) / \gamma_{1A} (Y_1,\tilde{a}^2{'} \cond \bC) \cond \bC \big\} }
\Bigg]		 
\\
\nonumber
&
=
\EXP  \Bigg[
\frac{ 
\EXP 
\bigg[
\frac{
\mathbb{I}(A = \tilde{a}^2{'}) 
\big\{ h(Y_1,Y_2) -\theta (\bC) \big\}  }{\gamma_{1A} (Y_1,\tilde{a}^2{'} \cond \bC) }
\bigg\{
\begin{array}{l}
s_{12}(Y_1,Y_2 \cond \bC \con \eta^*) 
+
s_{2A}(Y_2,\tilde{a}^1 \cond \bC \con \eta^*)
\\
+  
s_1(Y_1 \cond \bC\con \eta^*)
+ 
s_2(Y_2 \cond \bC\con \eta^*) 
\end{array}
\bigg\} 
\, \bigg| \bC \bigg]
}{ \EXP \big\{ \mathbb{I}(A = \tilde{a}^1) / \gamma_{1A} (Y_1,\tilde{a}^2{'} \cond \bC) \cond \bC \big\} }
\Bigg]		 \ .
\end{align}
The identities are trivial from \eqref{eq-alpha} and \eqref{eq-IPW}. 

Second, we focus on $\EXP \big\{ \IF_2 \cdot s \big\}$. From \eqref{eq-Q-center}, we establish that 
\begin{align*}
& \EXP \big\{ \IF_2 (\bO) \cond  \bC \big\}
=
\frac{1}{\EXP \big\{ \mathbb{I}(A = \tilde{a}^2{'})/\gamma_{1A}(Y_1,\tilde{a}^2{'} \cond \bC) \cond \bC \big\}}
\EXP \bigg\{
\frac{\mathbb{I}(A = \tilde{a}^2{'}) Q_2 (Y_2 \cond \bC)}{ \gamma_{12} (Y_1,Y_2 \cond \bC) \gamma_{1A} (Y_1,\tilde{a}^2{'} \cond \bC)}
\COND \bC
\bigg\}
= 0 \ .
\end{align*}
Thus, we find 
\begin{align*}
&
\EXP \Big[
\IF_2 (\bO) s(\bO \con \eta^*)
\Big]
\\
&
=
\EXP \Bigg[
\IF_2 (\bO) 
\bigg\{ 
\begin{array}{l}		
s_{12}(Y_1,Y_2 \cond \bC \con \eta^*)
+
s_{1A}(Y_1,A \cond \bC \con \eta^*)
+
s_{2A}(Y_2,A \cond \bC \con \eta^*)
\\
+  
s_1(Y_1 \cond \bC\con \eta^*)
+ 
s_2(Y_2 \cond \bC\con \eta^*)
+ 
s_A(A \cond \bC \con \eta^*)
\end{array}
\bigg\}
\Bigg]
\\
& \hspace*{2cm}
+
\EXP \Big[
\IF_2 (\bO) 
\cdot  s_C(\bC\con \eta^*) 
\Big]
\\
&
=
\EXP \bigg[ 
\EXP \Big[ 
\IF_2 (\bO) 
\bigg\{ 
\begin{array}{l}		
s_{12}(Y_1,Y_2 \cond \bC \con \eta^*)
+
s_{1A}(Y_1,A \cond \bC \con \eta^*)
+
s_{2A}(Y_2,A \cond \bC \con \eta^*)
\\
+  
s_1(Y_1 \cond \bC\con \eta^*)
+ 
s_2(Y_2 \cond \bC\con \eta^*)
+ 
s_A(A \cond \bC \con \eta^*)
\end{array}
\bigg\}
\Cond  \bC
\Big]
\bigg] \ .
\end{align*} 
Each term is represented as follows:
\begin{itemize}[leftmargin=0cm]

\item $\IF_2  \times s_{12}$
\begin{align}
&
\EXP \Big\{ \IF_2  (\bO) s_{12} (Y_1,Y_2 \cond \bC \con \eta^*) \Cond  \bC \Big\}
\nonumber
\\
&
=
\bigg[
\frac{1}{\EXP \big\{ \mathbb{I}(A = \tilde{a}^2{'})/\gamma_{1A} (Y_1,\tilde{a}^2{'} \cond \bC) \cond \bC \big\}}
\bigg]
\Bigg[
\EXP 
\bigg\{
\frac{ \mathbb{I}(A = \tilde{a}^2{'}) Q_2 (Y_2 \cond \bC) s_{12} (Y_1,Y_2 \cond \bC \con \eta^*)}{\gamma_{12} (Y_1,Y_2 \cond \bC) \gamma_{1A} (Y_1 , \tilde{a}^2{'} \cond \bC) } 
\COND \bC
\bigg\}
\Bigg]
\nonumber
\\
&
\stackrel{\eqref{eq-alpha}}{=}
\frac{1}{\theta_D (\bC) f_A (\tilde{a}^2{'} \cond \bC)}
\int Q_2 (y_2 \cond \bC) 
s_{12} (y_1,y_2 \cond \bC \con \eta^*) 
f_1 (y_1 \cond \bC) 
f_2 (y_2 \cond \bC) 
f_A (\tilde{a}^2{'} \cond \bC)
\, d(y_1,y_2)
\nonumber
\\
&
=
\frac{1}{\theta_D (\bC) }
\int Q_2 (y_2 \cond \bC) 
s_{12} (y_1,y_2 \cond \bC \con \eta^*) 
f_1 (y_1 \cond \bC) 
f_2 (y_2 \cond \bC)  
\, d(y_1,y_2)\ .
\label{eq-IF2-term1}
\end{align}

\item $\IF_2 \times (s_{1A} + s_{1A} + s_A) $
\begin{align}
&
\EXP  \Big[ \IF_2(\bO) \big\{ s_{1A} (Y_1 , A \cond \bC \con \eta^*) + s_{1} (Y_1 \cond \bC \con \eta^*) + s_A(A \cond \bC \con \eta^*)  \big\} \Cond  \bC \Big]
\nonumber
\\
&
=
\bigg[
\frac{1}{\EXP \big\{ \mathbb{I}(A = \tilde{a}^2{'})/\gamma_{1A} (Y_1,\tilde{a}^2{'} \cond \bC) \cond \bC \big\}}
\bigg]
\Bigg[
\EXP 
\bigg[
\frac{ \mathbb{I}(A = \tilde{a}^2{'}) Q_2 (Y_2 \cond \bC)  \left\{ 
\begin{array}{l}
    s_{1A} (Y_1 , \tilde{a}^2{'} \cond \bC \con \eta^*) \\
    + s_1 (Y_1 \cond \bC \con \eta^*) \\
    + s_A(\tilde{a}^2{'} \cond \bC \con \eta^*)
\end{array}
\right\}  }{\gamma_{12} (Y_1,Y_2 \cond \bC) \gamma_{1A} (Y_1 , \tilde{a}^2{'} \cond \bC) } 
\COND \bC
\bigg]
\Bigg]
\nonumber
\\
&
\stackrel{\eqref{eq-alpha}}{=}
\frac{1}{\theta_D (\bC)}
\int Q_2 (y_2 \cond \bC) 
\left\{ 
\begin{array}{l}
s_{1A}(y_1 , \tilde{a}^2{'} \cond \bC \con \eta^*) 
\\
+ s_1 (y_1 \cond \bC \con \eta^*) 
\\
+ s_A(\tilde{a}^2{'} \cond \bC \con \eta^*) 
\end{array}
\right\}
f_1 (y_1 \cond \bC) 
f_2 (y_2 \cond \bC)  
\, d(y_1,y_2)
\nonumber
\\
&
=
\frac{1}{\theta_D (\bC)}
\bigg\{
\int Q_2 (y_2 \cond \bC) 	
f_2 (y_2 \cond  \bC) \,  d y_2
\bigg\}
\nonumber
\\
&
\quad \times
\underbrace{
\bigg[
\int
\big\{ s_{1A}(y_1 , \tilde{a}^2{'} \cond \bC \con \eta^*) + s_1 (y_1 \cond \bC \con \eta^*) + s_A(\tilde{a}^2{'} \cond \bC \con \eta^*) \big\}
f_1 (y_1 \cond \bC)  
\, d y_1
\bigg]}_{=(*)}
\nonumber
\\
&
=
\frac{\int \big\{ h(y_1,y_2) - \theta (\bC) \big\} \gamma_{12} (y_1,y_2 \cond \bC) f_1 (y_1 \cond  \bC)
f_2 (y_2 \cond  \bC) \,  d (y_1, y_2)}{\theta_D (\bC)}
\times
(*)
\nonumber
\\
&
=
\frac{\theta_N (\bC) - \theta_D (\bC) \theta (\bC) }{\theta_D (\bC)}
\times
(*)
\nonumber
\\
&
=
0\ .
\label{eq-IF2-term2}
\end{align} 

\item $\IF_2 \times (s_{2A}+s_2)$
\begin{align}
&
\EXP  \Big[ \IF_2 (\bO) \big\{ s_{2A} (Y_2 , A \cond \bC \con \eta^*) + s_2 (Y_2 \cond \bC \con \eta^*) \big\} \Cond  \bC \Big]
\nonumber
\\
&
\stackrel{\eqref{eq-boundary}}{=}
\bigg[
\frac{1}{\EXP \big\{ \mathbb{I}(A = \tilde{a}^2{'})/\gamma_{1A}(Y_1,\tilde{a}^2{'} \cond \bC) \cond \bC \big\}}
\bigg]
\Bigg[
\EXP 
\bigg[
\frac{ \mathbb{I}(A = \tilde{a}^2{'}) Q_2 (Y_2 \cond \bC)  s_2 (Y_2 \cond \bC \con \eta^*)  
}{\gamma_{12}(Y_1,Y_2 \cond \bC) \gamma_{1A}(Y_1 , \tilde{a}^2{'} \cond \bC) } 
\COND \bC
\bigg]
\Bigg]
\nonumber
\\
&
\stackrel{\eqref{eq-alpha}}{=}
\frac{1}{\theta_D (\bC)}
\int Q_2 (y_2 \cond \bC) 
s_2 (y_2 \cond \bC \con \eta^*) 
f_1 (y_1 \cond  \bC) 
f_2 (y_2 \cond  \bC)  
\, d(y_1,y_2)
\nonumber
\\
&
=
\frac{1}{\theta_D (\bC)} 
\int Q_2 (y_2 \cond \bC) 
s_2 (y_2 \cond \bC \con \eta^*) 
f_2(y_2 \cond \bC) 
\,  d y_2 
\nonumber
\\
&
=
\frac{1}{\theta_D(\bC)} 
\int \Big\{ h(y_1,y_2) - \theta(\bC) \Big\}
s_2 (y_2 \cond \bC \con \eta^*) 
\gamma_{12}(y_1, y_2 \cond \bC)
f_1 (y_1 \cond \bC)
f_2 (y_2 \cond \bC) 
\,  d (y_1, y_2) 
\nonumber 
\\
&
=
\frac{ \EXP \big[ \mathbb{I}(A = \tilde{a}^2{'}) \big\{ h(Y_1,Y_2) - \theta (\bC) \big\} s_2(Y_2 \cond \bC) / \gamma_{1A}(Y_1 ,\tilde{a}^2{'} \cond \bC) \cond \bC \big] }{ \EXP \big\{ \mathbb{I}(A = \tilde{a}^2{'}) /\gamma_{1A}(Y_1,\tilde{a}^2{'}  \cond \bC) \cond \bC \big\} }\ .
\label{eq-IF2-term3} 
\end{align} 
\end{itemize}

Combine \eqref{eq-IF2-term1}, \eqref{eq-IF2-term2}, \eqref{eq-IF2-term3}, we get
\begin{align}	\label{eq-IF2} 
&
\EXP \Big[
\IF_2 (\bO) s(\bO \con \eta^*)
\Big]
\\
\nonumber
&
=
\EXP \Bigg[
\eqref{eq-IF2-term1} 		
+ \frac{ \EXP \big[ \mathbb{I}(A = \tilde{a}^2{'}) \big\{ h(Y_1,Y_2) - \theta (\bC) \big\} s_2(Y_2 \cond \bC) /\gamma_{1A}(Y_1 ,\tilde{a}^2{'} \cond \bC) \cond \bC \big] }{ \EXP \big\{ \mathbb{I}(A = \tilde{a}^2{'})/\gamma_{1A}(Y_1,\tilde{a}^2{'} \cond \bC) \cond \bC \big\} }
\Bigg]\ .
\end{align}

Third, we focus on $\EXP \big\{ \IF_3 \cdot s \big\}$. From \eqref{eq-Q-center}, we establish that 
\begin{align*}
& \EXP \big\{ \IF_3 (\bO) \cond  \bC \big\}
=
- \frac{1}{\EXP \big\{ \mathbb{I}(A = \tilde{a}^1)  / \gamma_{2A}(Y_2,\tilde{a}^1 \cond \bC) \cond \bC \big\}}
\EXP \bigg\{
\frac{\mathbb{I}(A = \tilde{a}^1)  Q_2(Y_2 \cond \bC)   }{ \gamma_{12}(Y_1,Y_2 \cond \bC)  \gamma_{2A}(Y_2,\tilde{a}^1 \cond \bC)  }
\COND \bC
\bigg\}
= 0  \ .
\end{align*}
Thus, we find 
\begin{align*}
&
\EXP \Big[
\IF_3 (\bO) s(\bO \con \eta^*)
\Big]
\\
&
=
\EXP \Bigg[
\IF_3 (\bO) 
\bigg\{ 
\begin{array}{l}		
s_{12}(Y_1,Y_2 \cond \bC \con \eta^*)
+
s_{1A}(Y_1,A \cond \bC \con \eta^*)
+
s_{2A}(Y_2,A \cond \bC \con \eta^*)
\\
+  
s_1(Y_1 \cond \bC\con \eta^*)
+ 
s_2(Y_2 \cond \bC\con \eta^*)
+ 
s_A(A \cond \bC \con \eta^*)
\end{array}
\bigg\}
\Bigg]
\\
& \hspace*{2cm}
+
\EXP \Big[
\IF_2 (\bO) 
\cdot  s_C(\bC\con \eta^*) 
\Big]
\\
&
=
\EXP \bigg[ 
\EXP \Big[ 
\IF_3 (\bO) 
\bigg\{ 
\begin{array}{l}		
s_{12}(Y_1,Y_2 \cond \bC \con \eta^*)
+
s_{1A}(Y_1,A \cond \bC \con \eta^*)
+
s_{2A}(Y_2,A \cond \bC \con \eta^*)
\\
+  
s_1(Y_1 \cond \bC\con \eta^*)
+ 
s_2(Y_2 \cond \bC\con \eta^*)
+ 
s_A(A \cond \bC \con \eta^*)
\end{array}
\bigg\}
\Cond  \bC
\Big]
\bigg] \ .
\end{align*} 
Each term is represented as follows:
\begin{itemize}[leftmargin=0cm]

\item $\IF_3 \times s_{12}$
\begin{align}
&
\EXP \Big\{ \IF_3 (\bO) s_{12} (Y_1,Y_2 \cond \bC \con \eta^*) \Cond  \bC \Big\}
\nonumber
\\
&
=
-
\bigg[
\frac{1}{\EXP \big\{ \mathbb{I}(A = \tilde{a}^1)  / \gamma_{2A}(Y_2,\tilde{a}^1 \cond \bC) \cond \bC \big\}}
\bigg]
\Bigg[
\EXP 
\bigg\{
\frac{ \mathbb{I}(A = \tilde{a}^1) Q_2 (Y_2 \cond \bC) s_{12} (Y_1,Y_2 \cond \bC \con \eta^*)   }{\gamma_{12}(Y_1,Y_2 \cond \bC) \gamma_{2A}(Y_2,\tilde{a}^1 \cond \bC) } 
\COND \bC
\bigg\}
\Bigg]
\nonumber
\\
&
\stackrel{\eqref{eq-alpha}}{=}
-
\frac{1}{\theta_D (\bC) f_A (\tilde{a}^1 \cond \bC)}
\int Q_2 (y_2 \cond \bC) 
s_{12} (y_1,y_2 \cond \bC \con \eta^*) 
f_1 (y_1 \cond \bC) 
f_2 (y_2 \cond \bC) 
f_A (\tilde{a}^1 \cond \bC)
\, d(y_1,y_2)
\nonumber
\\
&
=
-
\frac{1}{\theta_D (\bC) }
\int Q_2 (y_2 \cond \bC) 
s_{12} (y_1,y_2 \cond \bC \con \eta^*) 
f_1 (y_1 \cond \bC) 
f_2 (y_2 \cond \bC)  
\, d(y_1,y_2)
\label{eq-IF3-term1}
\\
&
=
-1 \times 
\eqref{eq-IF2-term1}
\ .
\nonumber
\end{align} 

\item $\IF_3 \times (s_{1A} + s_{1A} + s_A) $
\begin{align}
&
\EXP  \Big[ \IF_3 (\bO) \big\{ s_{1A} (Y_1 , \tilde{a}^1 \cond \bC \con \eta^*) + s_{1} (Y_1 \cond \bC \con \eta^*) + s_A(\tilde{a}^1 \cond \bC \con \eta^*)  \big\} \Cond  \bC \Big]
\nonumber
\\
&
=
-
\bigg[
\frac{1}{\EXP \big\{ \mathbb{I}(A = \tilde{a}^1)  / \gamma_{2A}(Y_2, \tilde{a}^1 \cond \bC) \cond \bC \big\}}
\bigg]
\Bigg[
\EXP 
\bigg[
\frac{ \mathbb{I}(A = \tilde{a}^1) Q_2 (Y_2 \cond \bC)  
\left\{
\begin{array}{l}
s_{1A} (Y_1 , \tilde{a}^1 \cond \bC \con \eta^*) \\
+ s_1 (Y_1 \cond \bC \con \eta^*) \\
+ s_A(\tilde{a}^1 \cond \bC \con \eta^*)
\end{array}
\right\}
}{\gamma_{12}(Y_1,Y_2 \cond \bC) \gamma_{2A}(Y_2,\tilde{a}^1 \cond \bC) } 
\COND \bC
\bigg]
\Bigg]
\nonumber
\\
&
\stackrel{\eqref{eq-alpha}}{=}
-
\frac{1}{\theta_D(\bC)}
\int 
Q_2(y_2 \cond \bC)
\left\{
\begin{array}{l}
s_{1A} (y_1 , \tilde{a}^1 \cond \bC \con \eta^*) \\
+ s_1 (y_1 \cond \bC \con \eta^*) \\
+ s_A(\tilde{a}^1 \cond \bC \con \eta^*)
\end{array}
\right\}
f_1(y_1 \cond \bC) 
f_2(y_2 \cond \bC)  
\, d(y_1,y_2)
\nonumber
\\
&
=
-
\frac{1}{\theta_D(\bC)}
\bigg\{
\int 
Q_2(y_2 \cond \bC)
f_2(y_2 \cond  \bC) \,  d y_2
\bigg\}
\nonumber
\\
&
\quad \times  
\underbrace{
\bigg[
\int
\big\{ s_{1A}(y_1 , \tilde{a}^1 \cond \bC \con \eta^*) 
+ s_1 (y_1 \cond \bC \con \eta^*) 
+ s_A( \tilde{a}^1 \cond \bC \con \eta^*) \big\}
f_1(y_1 \cond \bC)  
\, d y_1
\bigg]}_{=(*)}
\nonumber
\\
&
=
-
\frac{\int \big\{ h(y_1,y_2) - \theta (\bC) \big\} \gamma_{12} (y_1,y_2 \cond \bC) f_1 (y_1 \cond  \bC)
f_2 (y_2 \cond  \bC) \,  d (y_1, y_2)}{\theta_D (\bC)}
\times
(*)
\nonumber
\\
&
=
-
\frac{\theta_N (\bC) - \theta_D (\bC) \theta (\bC) }{\theta_D (\bC)}
\times
(*)
\nonumber
\\
&
=
0 
\ .
\label{eq-IF3-term2}
\end{align} 

\item $\IF_3 \times (s_{2A}+s_2)$
\begin{align}
&
\EXP \Big[ \IF_3(\bO) \big\{ s_{2A} (Y_2 , \tilde{a}^1 \cond \bC \con \eta^*) 
+ s_{2} (Y_2 \cond \bC \con \eta^*) \big\} \Cond  \bC \Big]
\nonumber
\\
&
=
-
\bigg[
\frac{1}{\EXP \big\{ \mathbb{I}(A = \tilde{a}^1)  / \gamma_{2A}(Y_2, \tilde{a}^1 \cond \bC) \cond \bC \big\}}
\bigg]
\Bigg[
\EXP 
\bigg[
\frac{ \mathbb{I}(A = \tilde{a}^1) Q_2(Y_2 \cond \bC) 
\left\{
\begin{array}{l}
s_{2A} (Y_2 , \tilde{a}^1 \cond \bC \con \eta^*) 
\\
+ s_2 (Y_2 \cond \bC \con \eta^*)
\end{array}
\right\} 
}{\gamma_{12}(Y_1,Y_2 \cond \bC) \gamma_{2A}(Y_2,\tilde{a}^1 \cond \bC) } 
\COND \bC
\bigg]
\Bigg]
\nonumber
\\
&
\stackrel{\eqref{eq-alpha}}{=}
-
\frac{1}{\theta_D(\bC)}
\int 
Q_2(y_2 \cond \bC)
\big\{ s_{2A} (y_2 , \tilde{a}^1 \cond \bC \con \eta^*) 
+ s_2 (y_2 \cond \bC \con \eta^*)  \big\} 
f_1(y_1 \cond  \bC) 
f_2(y_2 \cond  \bC)  
\, d(y_1,y_2)
\nonumber   
\\
&
=
-
\frac{1}{\theta_D(\bC)}
\bigg[
\int 
Q_2(y_2 \cond \bC)
\big\{ s_{2A} (y_2 , \tilde{a}^1 \cond \bC \con \eta^*) 
+ s_2 (y_2 \cond \bC \con \eta^*)  \big\}
f_2(y_2 \cond    \bC) 
\,  d y_2
\bigg]
\nonumber
\\
&
=
-
\frac{1}{\theta_D(\bC)} 
\int \Big\{ h(y_1,y_2) - \theta^*(\bC) \Big\}
\bigg\{
\begin{array}{l}
 s_{2A} (y_2 , \tilde{a}^1 \cond \bC \con \eta^*) 
 \\
+ s_2 (y_2 \cond \bC \con \eta^*) 
\end{array}
\bigg\}
\nonumber
\\
&
\quad
\times 	
\gamma_{12}(y_1, y_2 \cond \bC)
f_1(y_1 \cond   \bC)	
\gamma_{2A}(y_2 ,\tilde{a}^2{'} \cond \bC)
f_2(y_2 \cond  \bC) 
\,  d (y_1, y_2)  
\nonumber  
\\
&
=
-
\frac{ \EXP \bigg[ \mathbb{I}(A = \tilde{a}^2{'}) \big\{ h(Y_1,Y_2) - \theta (\bC) \big\} 
\left\{
\begin{array}{l}
s_{2A}(Y_2, \tilde{a}^1 \cond \bC \con \eta^*) \\
+s_2(Y_2 \cond \bC) 
\end{array}
\right\}
/ \gamma_{1A}(Y_1 ,a_2 \cond \bC) \COND \bC \bigg] }{ \EXP \big\{ \mathbb{I}(A = \tilde{a}^2{'})  / \gamma_{1A}(Y_1,\tilde{a}^2{'} \cond \bC) \cond \bC \big\} }\ .
\label{eq-IF3-term3} 
\end{align}	 
\end{itemize}

Combine \eqref{eq-IF3-term1}, \eqref{eq-IF3-term2}, \eqref{eq-IF3-term3}, we get
\begin{align}		\label{eq-IF3} 
&		\EXP \Big[
\IF_3(\bO) s(\bO \con \eta^*)
\Big]
\\
\nonumber
&
=
\EXP \Bigg[
- \eqref{eq-IF2-term1} 		
-
\frac{ \EXP \big[ \mathbb{I}(A = \tilde{a}^2{'}) \big\{ h(Y_1,Y_2) - \theta^*(\bC) \big\} 
\left\{
\begin{array}{l}
s_{2A}(Y_2, \tilde{a}^1 \cond \bC \con \eta^*) \\
+s_2(Y_2 \cond \bC) 
\end{array}
\right\}
/ \gamma_{1A}(Y_1 ,\tilde{a}^2{'} \cond \bC) \cond \bC \big] }{ \EXP \big\{ \mathbb{I}(A = \tilde{a}^2{'})  / \gamma_{1A}(Y_1,\tilde{a}^2{'} \cond \bC) \cond \bC \big\} }
\Bigg]\ .
\end{align}

Therefore, we can obtain the representation of $\EXP \big[ \{ \IF_1 + \IF_2 + \IF_3 \} s \big]$ by combining \eqref{eq-IF1}, \eqref{eq-IF2}, and \eqref{eq-IF3}:
\begin{align} \label{eq-IF123}
&
\EXP \Big[
\big\{ \IF_1 (\bO) + \IF_2 (\bO) + \IF_3 (\bO) \big\} s(\bO \con \eta^*)
\Big]
\\
&
=
\EXP 
\Bigg[
\frac{ 
\EXP  \left[
\frac{   \mathbb{I}(A = \tilde{a}^2{'}) }{ \gamma_{1A} (Y_1, \tilde{a}^2{'} \cond \bC \con \eta) }
\left[ 
\begin{array}{l}
     \left\{ 
     \begin{array}{l}
      h(Y_1,Y_2)   - \theta(\bC \con \eta)
     \end{array}
      \right\}
\left\{	
\begin{array}{l}			
s_{12}(Y_1,Y_2 \cond \bC \con \eta)
\\
+
s_1 (Y_1 \cond \bC \con \eta)
\\
+
s_2 (Y_2 \cond \bC \con \eta)
\end{array}
\right\}   
\end{array} 
\right] 
\COND \bC \right]
}{ 
\EXP \big[
\mathbb{I}(A = \tilde{a}^2{'})  / \gamma_{1A} (Y_1, \tilde{a}^2{'} \cond \bC \con \eta)
\cond
\bC
\big]
}
\Bigg]\ .
\nonumber
\end{align}
From similar algebra, we find
\begin{align} \label{eq-IF123 2}
& 
\EXP \Big[
\big\{ \IF_1 (\bO) + \IF_2 (\bO) + \IF_3 (\bO) \big\} s(\bO \con \eta^*)
\Big]
\\
&
=
\EXP \Big[
\big\{ \IF_1' (\bO) + \IF_2' (\bO) + \IF_3' (\bO) \big\} s(\bO \con \eta^*)
\Big]\ .
\nonumber
\end{align}

Lastly, we find $\EXP \big\{ \IF_4 (\bO) \big\} = 0$ and, consequently, 
\begin{align*}
&
\EXP \Big[ \IF_4(\bC) s(\bO \con \eta^*) \Big]
\\
&
=
\EXP \Bigg[
\IF_4 (\bC) 
\bigg\{ 
\begin{array}{l}		
s_{12}(Y_1,Y_2 \cond \bC \con \eta^*)
+
s_{1A}(Y_1,A \cond \bC \con \eta^*)
+
s_{2A}(Y_2,A \cond \bC \con \eta^*)
\\
+  
s_1(Y_1 \cond \bC\con \eta^*)
+ 
s_2(Y_2 \cond \bC\con \eta^*)
+ 
s_A(A \cond \bC \con \eta^*)
\end{array}
\bigg\}
\Bigg]
\\
& \hspace*{2cm}
+
\EXP \Big[
\IF_4(\bO) 
\cdot  s_C(\bC\con \eta^*) 
\Big] 
\\
&
=
\EXP  \Big[
\IF_4(\bC)  s_C(\bC\con \eta^*)
\Big] 
\\
&
=
\EXP  \Big[  \theta (\bC)  s_C(\bC\con \eta^*) \Big]
\\
&
\stackrel{\eqref{eq-theta}}{=}
\EXP \Bigg[  \frac{\EXP \big\{ \mathbb{I}(A = \tilde{a}^2{'}) h(Y_1,Y_2) / \gamma_{1A} (Y_1,\tilde{a}^2{'} \cond \bC) \cond  \bC \big\} }{\EXP \big\{ \mathbb{I}(A = \tilde{a}^2{'}) / \gamma_{1A} (Y_1,\tilde{a}^2{'} \cond \bC) \cond \bC \big\} }  s_C(\bC\con \eta^*) \Bigg] 
\\
&
=
\EXP  \Bigg[  \frac{\EXP \big\{ s_C(\bC\con \eta^*) \mathbb{I}(A = \tilde{a}^2{'}) h(Y_1,Y_2) / \gamma_{1A} (Y_1,\tilde{a}^2{'} \cond \bC)  \cond  \bC \big\} }{\EXP \big\{ \mathbb{I}(A = \tilde{a}^2{'}) / \gamma_{1A} (Y_1,\tilde{a}^2{'} \cond \bC) \cond \bC \big\} }   \Bigg]\ .
\end{align*}

Combining this result with \eqref{eq-IF123} and \eqref{eq-IF123 2}, we establish \eqref{eq-path der}, i.e., $\IF(\bO)$ is the valid influence function for $\psi$ in the model $\mathcal{M}_{\text{sym}}$.

\subsection{Efficient Influence Function}
\label{subsec:eif}

To characterize the efficient influence function, we first characterize the tangent space of the model $\mathcal{M}_{\text{sym}}$, denoted by $\mathcal{T}_{\text{sym}}$, and its orthocomplement, denoted by $\mathcal{T}_{\text{sym}}^\perp$.

A density in the unrestricted model for $\bO$ can be parametrized as
\begin{align*}
&
f(Y_1,Y_2,A,\bC \con \kappa, \eta)
\\
&
\propto
\delta(Y_1,Y_2,A \cond \bC \con \kappa)
\gamma_{12}(Y_1,Y_2 \cond \bC \con \eta)
\gamma_{1A}(Y_1,A \cond \bC \con \eta)
\gamma_{2A}(Y_2,A \cond \bC \con \eta)
\\
&
\hspace*{1cm}
\times 
f_{1}(Y_1 \cond \bC \con \eta)
f_{2}(Y_2 \cond \bC \con \eta)
f_{A}(A \cond \bC \con \eta)
f_{C}(\bC \con \eta)
\end{align*}
where $\delta(Y_1,Y_2,A \cond \bC \con \theta)$ is the 3-way interaction among $(Y_1,Y_2,A)$.

The tangent space $\mathcal{T}_{\text{sym}}$ can be viewed as the span of the score function of $f \in \mathcal{M}_{\text{sym}}$ related to $\eta$ because $\mathcal{T}_{\text{sym}}$ imposed the restriction on $\delta$ as $\delta(y_1,y_2,a \cond \bc \con \kappa)=1$. Therefore, one can establish that 
\begin{align*}
\mathcal{T}_{\text{sym}}
=
\left\{
S(Y_1,Y_2,A,\bC)
\,
\left|
\,
\begin{array}{l}
S(Y_1,Y_2,A,\bC) 
\\
=
S_{12}(Y_1,Y_2,\bC)
+
S_{1A}(Y_1,A,\bC)
+
S_{2A}(Y_2,A,\bC)
+
S_C(\bC)
\\[0.25cm]
\EXP \big\{ 
S_{12}(Y_1,Y_2,\bC)
+
S_{1A}(Y_1,A,\bC)
+
S_{2A}(Y_2,A,\bC)
\cond 
\bC
\big\} \\
= 
\EXP \big\{ S_C(\bC) \big\} = 0
\end{array} 
\right.
\right\}
\ .
\end{align*} 

Next, to characterize $\mathcal{T}_{\text{sym}}^\perp$, we use the result in \citet{TT2012} in which the orthocomplement of the nuisance tangent space for $\kappa$ in the unrestricted model, denoted by $\Lambda_{\kappa}^{\perp}$, is characterized. Of note, the nuisance tangent space for $\kappa$ in the unrestricted model is equivalent to the span of the score functions of $f \in \mathcal{M}_{\text{sym}}$ related to $\eta$, implying $\Lambda_{\kappa}^\perp = \mathcal{T}_{\text{sym}}^\perp$. Therefore, we can characterize $\mathcal{T}_{\text{sym}}^\perp$ by using the result in \citet{TT2012}, which results in $ \mathcal{T}_{\text{sym}}^\perp = \big\{ F \cond D: \text{unrestricted} \big\} $ where  
\begin{align}	\label{eq-F}
F(Y_1,Y_2,A,\bC)
&
=
\frac{
(2A-1) \big\{ D(Y_1,Y_2,\bC) - D^\dagger(Y_1,Y_2,\bC) \big\}
}{
\gamma_{12} ( Y_1,Y_2 \cond \bC)
\gamma_{1A} ( Y_1, A \cond \bC)
\gamma_{2A} ( Y_2, A \cond \bC) 	
f_{A} (A \cond  \bC)
} 
\ ,
\\
D^\dagger(Y_1,Y_2,\bC)
&
=
\EXP^{\dagger}
\big\{
D(Y_1,Y_2,\bC) \cond Y_1, \bC
\big\}
+
\EXP^{\dagger}
\big\{
D(Y_1,Y_2,\bC) \cond Y_2, \bC
\big\}
-
\EXP^{\dagger}
\big\{
D(Y_1,Y_2,\bC) \cond \bC
\big\}
\ ,
\nonumber
\end{align}
where $\EXP^{\dagger}$ is the expectation obtained under the admissible density $f^\dagger(y_1,y_2 \cond \bC) = f_1(y_1 \cond \bC) \cdot f_2(y_2 \cond \bC)$. Then, we can establish that
\begin{align}		\label{eq-tangent orthocomplement}
\mathcal{T}_{\text{sym}}^\perp  
=
\left\{ S (Y_1,Y_2,A,\bC) \left| \,
\begin{array}{l}
\EXP \big\{ S (Y_1,Y_2,A, \bC) \cond Y_1, Y_2 , \bC \big\} 
\\
=
\EXP \big\{ S (Y_1,Y_2,A, \bC) \cond Y_1, A , \bC \big\} 
\\
=
\EXP \big\{ S (Y_1,Y_2,A, \bC) \cond Y_2, A , \bC \big\} 
=0
\end{array}
\right.
\right\}
 \ .
\end{align}
We prove this result. 
\begin{proof}

We show that the left hand side of \eqref{eq-tangent orthocomplement}  is a subset of the right hand side of \eqref{eq-tangent orthocomplement}. We take arbitrary $D$ and the corresponding $F$ from \eqref{eq-F}. Then, we achieve the desired result because
\begin{align*} 
\EXP \Big[
F(Y_1,Y_2,A,\bC)
\, \Big| \, Y_1,Y_2,\bC
\Big]
&
\propto
\sum_{a=0}^{1} (2a-1) D(Y_1,Y_2,\bC) f_1 (Y_1 \cond \bC) f_2 (Y_2 \cond \bC)
\\
&	
\propto
D(Y_1,Y_2,\bC)  \cdot \sum_{a=0}^{1} (2a-1)
=0 
\ ,
\\
\EXP \Big[
F(Y_1,Y_2,A,\bC)
\, \Big| \, Y_1,A,\bC
\Big]
&
\propto 
\int (2A-1) \big\{ D(Y_1,y_2,\bC) - D^\dagger (Y_1,y_2,\bC) \big\} f_1 (Y_1 \cond \bC) f_2 (y_2 \cond \bC) \, d y_2
\\
&
=
(2A-1)
\Big[
\EXP^\dagger \{ D(Y_1,Y_2,\bC) - D\dagger(Y_1,Y_2,\bC)  \cond Y_1, \bC \big\}
\Big]
\\
&
=
(2A-1)
\Big[
\EXP^\dagger \{ D(Y_1,Y_2,\bC) - D (Y_1,Y_2,\bC)  \cond Y_1, \bC \big\}
\Big] 
=
0
\ ,
\\
\EXP \Big[
F(Y_1,Y_2,A,\bC)
\, \Big| \, Y_2,A,\bC
\Big]
&
\propto 
\int (2A-1) \big\{ D(y_1,Y_2,\bC) - D^\dagger (y_1,Y_2,\bC) \big\} f_1 (y_1 \cond \bC) f_2 (Y_2 \cond \bC) \, d y_1
\\
&
=
(2A-1)
\Big[
\EXP^\dagger \{ D(Y_1,Y_2,\bC) - D\dagger(Y_1,Y_2,\bC)  \cond Y_2, \bC \big\}
\Big]
\\
&
=
(2A-1)
\Big[
\EXP^\dagger \{ D(Y_1,Y_2,\bC) - D (Y_1,Y_2,\bC)  \cond Y_2, \bC \big\}
\Big] 
=
0 
\ .
\end{align*}	

Next, we show that the right hand side of \eqref{eq-tangent orthocomplement} is a subset of the left hand side of \eqref{eq-tangent orthocomplement}. Suppose $S(Y_1,Y_2,A,\bC)$ satisfy the right hand side of \eqref{eq-tangent orthocomplement}, and let $D'(Y_1,Y_2, A,\bC)$ be a function satisfying
\begin{align*}
S(Y_1,Y_2,A,\bC) 
=
\frac{
D' (Y_1,Y_2,A,\bC)}{\gamma_{12} ( Y_1, Y_2 \cond \bC) 
\gamma_{1A} ( Y_1, A \cond \bC) 
\gamma_{2A} ( Y_2, A \cond \bC) 
f_{A}^*(A \cond \bC) } 
\ .
\end{align*}
Additionally, note that
\begin{align*}
\Pr(A=a \cond Y_1,Y_2,\bC)
&
\propto
f (Y_1,Y_2,A=a \cond \bC)
\\
&
=
\frac{
\gamma_{12} (Y_1,Y_2 \cond \bC)
\gamma_{1A} (Y_1, a \cond \bC)
\gamma_{2A} (Y_2, a \cond \bC)
f_1 (Y_1 \cond \bC)
f_2 (Y_2 \cond \bC)
f_A (A=a \cond \bC)
}{
\N (\bC)
} 
 \ .
\end{align*}	
Therefore, we find
\begin{align*}
0
&
=
\EXP \big[ S(Y_1,Y_2,A,\bC) 
\cond Y_1, Y_2, \bC \big]
\\
&
=
\gamma_{12} (Y_1,Y_2 \cond \bC)
f_1 (Y_1 \cond \bC)
f_2 (Y_2 \cond \bC)
\left[
\begin{array}{l}
\gamma_{1A} (Y_1, \tilde{a}^2{'} \cond \bC)
\gamma_{2A} (Y_2, \tilde{a}^2{'} \cond \bC)
f_A^*( \tilde{a}^2{'} \cond \bC)
S(Y_1,Y_2,\tilde{a}^2{'},\bC)
\\
+
\gamma_{1A} (Y_1, \tilde{a}^1 \cond \bC)
\gamma_{2A} (Y_2, \tilde{a}^1 \cond \bC)
f_A ( \tilde{a}^1 \cond \bC)
S(Y_1,Y_2,\tilde{a}^1,\bC)			
\end{array}
\right]
\\
&
=D'(Y_1,Y_2,1,\bC)
+
D'(Y_1,Y_2,0,\bC) 
\ .
\end{align*}
By defining $D(Y_1,Y_2,\bC) = D'(Y_1,Y_2,1,\bC)$, we obtain
\begin{align*}
&
(2A-1) D(Y_1,Y_2,A,\bC)
=
\gamma_{12} ( Y_1, Y_2 \cond \bC) 
\gamma_{1A} ( Y_1, A \cond \bC) 
\gamma_{2A} ( Y_2, A \cond \bC) 
f_{A} (A \cond  \bC)
S(Y_1,Y_2,A,\bC) 
\\
&
\Leftrightarrow
\quad
S(Y_1,Y_2,A,\bC) 
=
\frac{
(2A-1)
D (Y_1,Y_2,\bC)}{\gamma_{12} ( Y_1, Y_2 \cond \bC) 
\gamma_{1A} ( Y_1, A \cond \bC) 
\gamma_{2A} ( Y_2, A \cond \bC) 
f_{A}^*(A \cond  \bC) }
\ .
\end{align*} 
Moreover, $\EXP^\dagger \{ D(Y_1,Y_2,\bC) \cond Y_1, \bC \big\} = \EXP^\dagger \{ D(Y_1,Y_2,\bC) \cond Y_2, \bC \big\} = \EXP^\dagger \{ D(Y_1,Y_2,\bC) \cond \bC \big\} = 0 $ because 
\begin{align*}
&
\EXP^\dagger \big\{ D(Y_1,Y_2,\bC) \cond Y_1, \bC \big\}
\\
&
\propto  
\int
    \gamma_{12} ( Y_1, y_2 \cond \bC) 
\gamma_{1A}  ( Y_1, A \cond \bC) 
\gamma_{2A} ( y_2, A \cond \bC) 
f_{A} (A \cond  \bC) 
S(Y_1,y_2,A,\bC) 
f_2 (y_2 \cond  \bC)
\, dy_2
\\
&
=
\int 
S(Y_1,y_2,A,\bC) 
\times
\left\{
\begin{array}{l}			
\gamma_{12} ( Y_1, y_2 \cond \bC) 
\gamma_{1A} ( Y_1, A \cond \bC) 
\gamma_{2A} ( y_2, A \cond \bC) 
\\
\times
f_{1} (Y_1 \cond  \bC)
f_{2} (y_2 \cond  \bC)
f_{A} (A \cond  \bC)
\end{array}
\right\}
\, dy_2
\\
&
=
\EXP \big\{ S(Y_1,Y_2,A,\bC)  \cond Y_1,A,\bC \big\}
=
0
\ ,
\\
&
\EXP^\dagger \big\{ D(Y_1,Y_2,\bC) \cond Y_2y, \bC \big\}
\\
&
\propto  
\int
    \gamma_{12} ( Y_1, Y_2 \cond \bC) 
\gamma_{1A}  ( y_1, A \cond \bC) 
\gamma_{2A} ( Y_2, A \cond \bC) 
f_{A} (A \cond  \bC) 
S(y_1,Y_2,A,\bC) 
f_1 (y_1 \cond  \bC)
\, dy_1
\\
&
=
\int 
S(y_1,Y_2,A,\bC) 
\times
\left\{
\begin{array}{l}			
\gamma_{12} ( y_1, Y_2 \cond \bC) 
\gamma_{1A} ( y_1, A \cond \bC) 
\gamma_{2A} ( Y_2, A \cond \bC) 
\\
\times
f_{1} (y_1 \cond  \bC)
f_{2} (Y_2 \cond  \bC)
f_{A} (A \cond  \bC)
\end{array}
\right\}
\, dy_1
\\
&
=
\EXP \big\{ S(Y_1,Y_2,A,\bC)  \cond Y_2,A,\bC \big\}
=
0 \ .
\end{align*} 
This concludes that any function in the right hand side of \eqref{eq-tangent orthocomplement} can be represented as a function having a form \eqref{eq-F}, indicating that the right hand side of \eqref{eq-tangent orthocomplement} is a subset of the left hand side of \eqref{eq-tangent orthocomplement}. Combining the two results, we achieve \eqref{eq-tangent orthocomplement}. 

\end{proof}

The efficient influence function (EIF), denoted by $\IF^*(\bO)$, for $\psi$ can be obtained by projecting an influence function for $\psi$ on $\mathcal{T}_{\text{sym}}$. To be more specific, let $\Pi \{ F \cond \mathcal{T} \big\}$ be the projection of a function $F$ onto a space $\mathcal{T}$. Then, the EIF for $\psi$ is represented as
\begin{align*}
\IF^*(\bO) 
=
\Pi \big\{ \IF(\bO) \cond \mathcal{T}_{\text{sym}} \big\}
=
\IF(\bO) - \Pi\big\{ \IF(\bO) \cond \mathcal{T}_{\text{sym}}^\perp \big\}
\ .
\end{align*}
To characterize $\Pi\big\{ \IF(\bO) \cond \mathcal{T}_{\text{sym}}^\perp \big\}$, we use the alternating conditional expectations (ACE) algorithm; see \citet{ACE1985}, and Section A.4 of \citet{BKRW1998} for details. Recall that $\mathcal{T}_{\text{sym}}^\perp $ is equivalent to $\mathcal{T}_{\text{sym}}^\perp 
=
\overline{
\mathcal{H}_{12} +
\mathcal{H}_{1A} +
\mathcal{H}_{2A} 
}$  where $\mathcal{H}_{12}$, $\mathcal{H}_{1A}$, $\mathcal{H}_{2A}$ are the collections of mean-zero square-integrable functions conditioning on $(Y_1,Y_2,\bC)$, $(Y_1,A,\bC)$, and $(Y_2,A,\bC)$, respectively. Therefore, the projections on $\mathcal{H}_{12}$, $\mathcal{H}_{1A}$, and $\mathcal{H}_{2A}$ are defined as follows:
\begin{align*}
&
\Pi_{12} \big\{ F(\bO) \big\}
=
\Pi \big\{ F(\bO) \cond \mathcal{H}_{12} \big\}	
=
F(\bO) - \EXP \big\{ F(\bO) \cond Y_1,Y_2,\bC \big\} 
\ ,
\\
&
\Pi_{1A} \big\{ F(\bO) \big\}
=
\Pi  \big\{ F(\bO) \cond \mathcal{H}_{1A} \big\}
= 
F(\bO) - \EXP \big\{ F(\bO) \cond Y_1,A,\bC \big\} 
\ ,
\\
&
\Pi_{2A} \big\{ F(\bO) \big\}
=
\Pi  \big\{ F(\bO) \cond \mathcal{H}_{2A} \big\}
=
F(\bO) - \EXP \big\{ F(\bO) \cond Y_2,A,\bC \big\}  
\ .
\end{align*} 
Additionally, let the following operators be the ``residual'' of a function after projecting on $\mathcal{H}$:
\begin{align*}
&	
\mathcal{Q}_{12} \big\{ F(\bO) \big\}
=
(id - \Pi_{12}) \big\{ F(\bO) \big\}
= 
\EXP \big\{ F(\bO) \cond Y_1,Y_2,\bC \big\} 
\ ,
\\
&
\mathcal{Q}_{1A} \big\{ F(\bO) \big\}
=
(id - \Pi_{1A}) \big\{ F(\bO) \big\}
= 
\EXP \big\{ F(\bO) \cond Y_1,A,\bC \big\} 
\ ,
\\
&
\mathcal{Q}_{2A} \big\{ F(\bO) \big\}
=
(id - \Pi_{2A}) \big\{ F(\bO) \big\}
= 
\EXP \big\{ F(\bO) \cond Y_2,A,\bC \big\}  
\ .
\end{align*}
Then, the EIF is characterized based on the ACE algorithm as follows:
\begin{align*}
& 
\Big\|
\big( \mathcal{Q}_{2A} \circ \mathcal{Q}_{1A} \circ \mathcal{Q}_{12} \big)^m \big\{ \IF(\bO) \big\} 
-
\Pi \big\{ \IF(\bO) \cond \mathcal{H}_{12}^\perp \oplus \mathcal{H}_{1A}^\perp \oplus \mathcal{H}_{2A}^\perp \big\}
\Big\|
\\
&
=
\Big\|
\big( \mathcal{Q}_{2A} \circ \mathcal{Q}_{1A} \circ \mathcal{Q}_{12} \big)^m \big\{ \IF(\bO) \big\} 
-
\Pi \big\{ \IF(\bO) \cond \mathcal{T}_{\text{sym}} \big\}
\Big\|
\rightarrow 0
\text{ as }
m \rightarrow \infty 
\ .
\end{align*}
Therefore, the EIF is characterized as a limit of the alternating projections:
\begin{align*}
\IF^*(\bO)
&
=
\lim_{m \rightarrow \infty}
\big( \mathcal{Q}_{2A} \circ \mathcal{Q}_{1A} \circ \mathcal{Q}_{12} \big)^m \big\{ \IF(\bO) \big\} 
\\
&
=
\lim_{m \rightarrow \infty}
\Big( 
\EXP^{(Y_2,A,\bC)} \circ
\EXP^{(Y_1,A,\bC)} \circ
\EXP^{(Y_1,Y_2,\bC)} \Big)^m \big\{ \IF(\bO) \big\} 
\ ,
\end{align*}
where $\EXP^{(Y_1,Y_2,\bC)} \big\{ F(\bO) \big\} = \EXP \big\{ F(\bO) \cond Y_1,Y_2,\bC \big\}$, and the other two operators are defined in a similar manner.

\subsection{Closed-form Representation of the Efficient Influence Function Under Binary Outcomes}

We provide a closed-form representation of the EIF under binary outcomes.  
Suppose that $Y_1,Y_2 \in \{0,1\}$. Then, an alternative representation of $\mathcal{T}_{\text{sym}}^\perp$ is given as
\begin{align}					\label{eq-binary tangent ortho}
\mathcal{T}_{\text{sym}}^\perp
=
\Bigg\{
\frac{(2Y_1-1)(2Y_2-1)(2A-1) }{ f(Y_1,Y_2,A \cond \bC) } D(\bC)
\COND
D: \text{ unrestricted}
\Bigg\}
\ .
\end{align}
We prove this result.	
\begin{proof}
We first show that the right hand side of \eqref{eq-binary tangent ortho} is a subset of the left hand side. 
\begin{align*}
&
D (\bC)
\EXP \bigg\{ 
\frac{ (2Y_1-1)(2Y_2-1)(2A-1) }{f (Y_1,Y_2,A \cond \bC) } 
\COND Y_1,Y_2,\bC
\bigg\}
=
D (\bC)
(2Y_1-1)(2Y_2-1)
\sum_{a=0}^{1}
(2a-1)
=
0
\ .
\end{align*}
Similarly, 
\begin{align*}
\EXP \bigg\{ 
\frac{ (2Y_1-1)(2Y_2-1)(2A-1) }{f (Y_1,Y_2,A \cond \bC) } 
\COND Y_1,A,\bC
\bigg\}
=
\EXP \bigg\{ 
\frac{ (2Y_1-1)(2Y_2-1)(2A-1) }{f (Y_1,Y_2,A \cond \bC) } 
\COND Y_2,A,\bC
\bigg\}
=
0
\ .
\end{align*}
Therefore, \eqref{eq-tangent orthocomplement} is satisfied, implying that  the right hand side of \eqref{eq-binary tangent ortho} belongs to $\mathcal{T}_{\text{sym}}^\perp$. 

Next, we show that the left hand side of \eqref{eq-binary tangent ortho} is a subset of the right hand side. We take $S(Y_1,Y_2,A,\bC) \in \mathcal{T}_{\text{sym}}^\perp$, and let $D(\bC)$ be $ 
D(\bC)
=
f (1,1,1 \cond \bC) S(1,1,1,\bC)
\
\Leftrightarrow
\
S(1,1,1,\bC) = \frac{D(\bC)}{f (1,1,1 \cond \bC)}$. From straightforward algebra, we find
\begin{align*}
&
\EXP \big\{ S(1,1,A, \bC) \cond Y_1=1,Y_2=1, \bC \big\}
=
f(1,1,1 \cond \bC) S(1,1,1,\bC)
+
f(1,1,0 \cond \bC) S(1,1,0,\bC)
=
0
\\
&
\Rightarrow
S(1,1,0,\bC) 
=
-
\frac{f(1,1,1 \cond \bC)}{f(1,1,0 \cond \bC)} S(1,1,1,\bC)
=
-
\frac{ D(\bC) }{f(1,1,0 \cond \bC)}
\ ,
\\ 
&
\EXP \big\{ S(1,Y_2,1, \bC) \cond Y_1=1,A=1, \bC \big\}
=
f(1,1,1 \cond \bC) S(1,1,1,\bC)
+
f(1,0,1 \cond \bC) S(1,0,1,\bC)
=
0
\\
&
\Rightarrow
S(1,0,1,\bC) 
=
-
\frac{f(1,1,1 \cond \bC)}{f(1,0,1 \cond \bC)} S(1,1,1,\bC)
=
-
\frac{ D(\bC) }{f(1,0,1 \cond \bC)}
\ ,
\\ 
&
\EXP \big\{ S(Y_1,1,1, \bC) \cond Y_2=1,A=1, \bC \big\}
=
f(1,1,1 \cond \bC) S(1,1,1,\bC)
+
f(0,1,1 \cond \bC) S(0,1,1,\bC)
=
0
\\
&
\Rightarrow
S(0,1,1,\bC) 
=
-
\frac{f(1,1,1 \cond \bC)}{f(0,1,1 \cond \bC)} S(1,1,1,\bC)
=
-
\frac{ D(\bC) }{f(0,1,1 \cond \bC)}
\ ,
\\ 
&
\EXP \big\{ S(1,0,A, \bC) \cond Y_1=1,Y_2=0, \bC \big\}
=
f(1,0,1 \cond \bC) S(1,0,1,\bC)
+
f(1,0,0 \cond \bC) S(1,0,0,\bC)
=
0
\\
&
\Rightarrow
S(1,0,0,\bC) 
=
-
\frac{f(1,0,1 \cond \bC)}{f(1,0,0 \cond \bC)} S(1,0,1,\bC)
=
\frac{ D(\bC) }{f(1,0,0 \cond \bC)}
\ ,
\\ 
&
\EXP \big\{ S(0,1,A, \bC) \cond Y_1=0,Y_2=1, \bC \big\}
=
f(0,1,1 \cond \bC) S(0,1,1,\bC)
+
f(0,1,0 \cond \bC) S(0,1,0,\bC)
=
0
\\
&
\Rightarrow
S(0,1,0,\bC) 
=
-
\frac{f(0,1,1 \cond \bC)}{f(0,1,0 \cond \bC)} S(0,1,1,\bC)
=
\frac{ D(\bC) }{f(0,1,0 \cond \bC)}
\ ,
\\
&
\EXP \big\{ S(0,Y_2,1, \bC) \cond Y_1=0,A=1, \bC \big\}
=
f(0,1,1 \cond \bC) S(0,1,1,\bC)
+
f(0,0,1 \cond \bC) S(0,0,1,\bC)
=
0
\\
&
\Rightarrow
S(0,0,1,\bC) 
=
-
\frac{f(0,1,1 \cond \bC)}{f(0,0,1 \cond \bC)} S(0,1,1,\bC)
=
\frac{ D(\bC) }{f(0,0,1 \cond \bC)}
\ ,
\\ 
&
\EXP \big\{ S(0,Y_2,0, \bC) \cond Y_1=0,A=0, \bC \big\}
=
f(0,1,0 \cond \bC) S(0,1,0,\bC)
+
f(0,0,0 \cond \bC) S(0,0,0,\bC)
=
0
\\
&
\Rightarrow
S(0,0,0,\bC) 
=
-
\frac{f(0,1,0 \cond \bC)}{f(0,0,0 \cond \bC)} S(0,1,0,\bC)
=
-
\frac{ D(\bC) }{f(0,0,0 \cond \bC)}
\ .
\end{align*}
Therefore, any function $S \in \mathcal{T}_{\text{sym}}^\perp$ has the following form
\begin{align*}
S(y_1,y_2,a,\bC)
=
\frac{(2y_1-1)(2y_2-1)(2a-1)}{f(y_1,y_2,a \cond \bC)}
f(1,1,1 \cond \bC) S(1,1,1,\bC)
\ ,
\end{align*}
implying that $S$ belongs to the right hand side of \eqref{eq-binary tangent ortho}. This completes the proof.

\end{proof}

Using the form \eqref{eq-binary tangent ortho}, we can characterize the projection of arbitrary function $F$ on $\mathcal{T}_{\text{sym}}^\perp$, which is given below. 
\begin{align} \label{eq-projection binary}
\Pi \big\{ F(\bO) \cond \mathcal{T}_{\text{sym}}^{\perp} \big\}
& 
=
\underbrace{
\frac{ 
\EXP \big\{ F(\bO) v(\bO) \cond \bC \big\}
}{
\EXP \big\{ v^2(\bO) \cond \bC \big\}
}
}_{= D (\bO \con F)}
v(\bO)
=
D(\bO \con h)
v(\bO)
\end{align}	
where $ v(\bO)
= 
(2Y_1-1)(2Y_2-1)(2A-1)
/ f (Y_1,Y_2,A \cond \bC) $.
\begin{proof}

Note that for all $D(\bO)$,
\begin{align*}
&
\EXP \Big[
\big\{ F(\bO) - D(\bC \con F) v(\bO)
\big\}
\big\{ v(\bO) D(\bC)
\big\}  \Big]
\\
&
=
\EXP \Big[
D(\bC)
\EXP \big\{ F(\bO) v(\bO) \cond \bC \big\}
-
D(\bC) D(\bC \con F)
\EXP \big\{ v^2(\bO) \cond \bC \big\}
\Big]
\\
&
=
\EXP \Big[
D(\bC)
\EXP \big\{ F(\bO) v(\bO) \cond \bC \big\}
-
D(\bC) \EXP \big\{ F(\bO) v(\bO) \cond \bC \big\}
\Big]
\\
&
=
0 \ .
\end{align*}
This implies that $F (\bO) - D(\bC \con F) v(\bO) $ is orthogonal to $\mathcal{T}_{\text{sym}}^{\perp}$. Additionally, it is trivial that $D(\bC \con F) v(\bO)  \in \mathcal{T}_{\text{sym}}^{\perp}$ from \eqref{eq-binary tangent ortho}. Therefore, \eqref{eq-projection binary} is the projection of $h$ on $\mathcal{T}_{\text{sym}}^\perp$, i.e., $
\Pi \big\{ F(\bO) \Cond \mathcal{T}_{\text{sym}}^{\perp} \big\}
=
D(\bC \con F)
v(\bO) $.
\end{proof}

Note that
\begin{align*}
&
\EXP \big\{ h(\bO) v (\bO) \cond \bC \big\}
=
\sum_{y_1,y_2,a} (2y_1-1)(2y_2-1)(2a-1) h(y_1,y_2,a,\bC)
\ , 
\\
&
\EXP \big\{ v^2(\bO) \cond \bC \big\}
=
\sum_{y_1,y_2,a} \frac{1}{f(y_1,y_2,a \cond \bC) }
\end{align*}
which results in
\begin{align}
\label{eq-v}
& 
\frac{v(\bO)}{	\EXP \big\{ v^2(\bO) \cond \bC \big\}} 
\\
\nonumber
&
=
\Bigg[ \sum_{y_1,y_2,a} \frac{1}{ \gamma_{12}(y_1,y_2 \cond \bC) 
\gamma_{1A} (y_1,a \cond \bC)
\gamma_{2A} (y_2,a \cond \bC)
f_1 (y_1 \cond  \bC)
f_2 (y_1 \cond  \bC)
f_A (a \cond  \bC)}
\Bigg]^{-1}
\\
\nonumber
&
\hspace*{1cm} \times
\Bigg[ 
\frac{(2Y_1-1)(2Y_2-1)(2A-1)}{ 
\gamma_{12} (Y_1,Y_2 \cond \bC) 
\gamma_{1A} (Y_1,A \cond \bC)
\gamma_{2A} (Y_2,A \cond \bC)
f_1 (Y_1 \cond  \bC)
f_2 (Y_1 \cond  \bC)
f_A (A \cond  \bC)}
\Bigg]
\ .
\end{align}
The closed-form representations of $\theta(\bC)$ in \eqref{eq-theta} and $Q_1$ and $Q_2$ in \eqref{eq-supp-Qfunctions} are 
\begin{align} 
\label{eq-proj-theta}
\theta (\bc) = 
&
\frac{ 
\sum_{y_1,y_2} h(y_1,y_2) \gamma_{12}(y_1,y_2) f_1(y_1 \cond \bc) f_2(y_2 \cond \bc)
}{
\sum_{y_1,y_2} \gamma_{12}(y_1,y_2) f_1(y_1 \cond \bc) f_2(y_2 \cond \bc)
}
\\
\nonumber
&
=
\frac{
\left\{
\begin{array}{ll}
h(1,1) \gamma_{12}(1,1) f_1(1 \cond \bc) f_2 (1 \cond \bc) 
&
+ h(1,0) f_1(1 \cond \bc) f_2 (0 \cond \bc) 
\\
+
h(0,1) f_1(0 \cond \bc) f_2 (1 \cond \bc)
&
+
h(0,0)
f_1(0 \cond \bc) f_2 (0 \cond \bc)
\end{array} 
\right\}
}{ 
\left\{
\begin{array}{ll}
\gamma_{12}(1,1) f_1(1 \cond \bc) f_2 (1 \cond \bc) 
&
+ 
f_1(1 \cond \bc) f_2 (0 \cond \bc) 
\\
+
f_1(0 \cond \bc) f_2 (1 \cond \bc)
&
+
f_1(0 \cond \bc) f_2 (0 \cond \bc)
\end{array}
\right\}
}
\ ,
\\
\label{eq-proj-Q1} 
Q_1(y_1 \cond \bc)
&
=
\left\{ 
\begin{array}{ll}
    \big\{ h(1,1) - \theta(\bc) \big\}
    \gamma_{12}(1,1 \cond \bc) f_2(1 \cond \bc)
    \\
    \quad
    +
    \big\{ h(1,0) - \theta(\bc) \big\} f_2(0 \cond \bc)
    & \text{if }y_1=1
    \\[0.2cm]
    \big\{ h(0,1) - \theta(\bc) \big\} f_2(1 \cond \bc)
    \\
    \quad
    +
    \big\{ h(0,0) - \theta(\bc) \big\} f_2(0 \cond \bc)
    & \text{if }y_1=0
\end{array}
\right.
\ ,
\\
\label{eq-proj-Q2} 
Q_2(y_2 \cond \bc)
&
=
\left\{ 
\begin{array}{ll}
    \big\{ h(1,1) - \theta (\bc) \big\} 
\gamma_{12} (1,1 \cond \bc) f_1 (1 \cond \bc)
    \\
    \quad
+
\big\{ h(0,1) - \theta (\bc) \big\}  f_1 (0 \cond \bc) 
    & \text{if }y_2=1
    \\[0.2cm]
    \big\{ h(1,0) - \theta (\bc) \big\} f_1 (1 \cond \bc)
    \\
    \quad
+
\big\{ h(0,0) - \theta (\bc) \big\}  f_1 (0 \cond \bc) 
    & \text{if }y_2=0
\end{array}
\ .
\right.
\end{align}	
Using the results \eqref{eq-projection binary}, \eqref{eq-v}, \eqref{eq-proj-theta}, \eqref{eq-proj-Q1}, \eqref{eq-proj-Q2},  we can obtain the closed-form representation of $\EXP \big\{ \IF_1 \cdot v \cond \bC \big\}$, $\EXP \big\{ \IF_2 \cdot v \cond \bC \big\}$, $\EXP \big\{ \IF_2 \cdot v \cond \bC \big\}$, $\EXP \big\{ \IF_1' \cdot v \cond \bC \big\}$, $\EXP \big\{ \IF_2' \cdot v \cond \bC \big\}$, $\EXP \big\{ \IF_3' \cdot v \cond \bC \big\}$; see below for details. For simplicity, we consider $\tilde{a}^1=0$ and $\tilde{a}^2{'}=1$:

\begin{itemize}[leftmargin=0cm]
\item $\EXP \big\{ \IF_1 \cdot v \cond \bC \big\}$
\begin{align*}
&
\EXP \big\{ \texttt{IF}_1(\bO) v(\bO) \cond \bC \big\}
\\
&
=
\frac{1}{\EXP \big\{ \mathbb{I}(A = \tilde{a}^1)  / \gamma_{2A} (Y_2,\tilde{a}^1 \cond \bC) \cond  \bC \big\}}
\sum_{y_1,y_2,a}
\frac{
(2y_1-1)(2y_2-1)(2a-1) \mathbb{I}(a = \tilde{a}^1)
\big\{ h(y_1,y_2) - \theta (\bC) \big\}  }{ \gamma_{2A} (y_2,\tilde{a}^1 \cond \bC) } 
\\
&
=
\frac{1}{\EXP \big\{ \mathbb{I}(A = 0)  / \gamma_{2A} (Y_2,0 \cond \bC) \cond  \bC \big\}}
\Bigg[
\frac{ h(0,1) - h(1,1)  }{ \gamma_{2A} (1,0 \cond \bC) } 
+ h(1,0) - h(0,0) 
\Bigg]
\ .
\end{align*}

\item $\EXP \big\{ \IF_2 \cdot v \cond \bC \big\}$
\begin{align*}
&
\EXP \big\{ \texttt{IF}_2(\bO) v(\bO) \cond \bC \big\}
\\
&
=
\frac{1}{ \EXP \big\{ \mathbb{I}(A = \tilde{a}^2{'}) / \gamma_{1A} (Y_1,\tilde{a}^2{'} \cond \bC) \cond  \bC \big\} }
\sum_{y_1,y_2,a}
\frac{ (2y_1-1)(2y_2-1)(2a-1) \mathbb{I}(a = \tilde{a}^2{'}) Q_2 (y_2 \cond \bC) }{\gamma_{12} (y_1,y_2 \cond \bC) \gamma_{1A} (y_1,a_2 \cond \bC)} 
\\
&
=
\frac{1}{ \EXP \big\{ \mathbb{I}(A = 1) / \gamma_{1A} (Y_1,1 \cond \bC) \cond  \bC \big\} }
\Bigg[
\frac{Q_2 (1 \cond \bC)}{\gamma_{12} (1,1 \cond \bC) \gamma_{1A} (1,1 \cond \bC) }
-
\frac{Q_2 (0 \cond \bC)}{ \gamma_{1A} (1,1 \cond \bC) }
-
Q_2 (1 \cond \bC)
+
Q_2 (0 \cond \bC)
\Bigg] 
\ .
\end{align*}

\item $\EXP \big\{ \IF_3 \cdot v \cond \bC \big\}$
\begin{align*}
&
\EXP \big\{ \texttt{IF}_3(\bO) v(\bO) \cond \bC \big\}
\\
&
=
-
\frac{1}{\EXP \big\{ \mathbb{I}(A = \tilde{a}^1)  / \gamma_{2A} (Y_2,\tilde{a}^1 \cond \bC) \cond  \bC \big\}}
\sum_{y_1,y_2,a}
\frac{ (2y_1-1)(2y_2-1)(2a-1) \mathbb{I}(a = \tilde{a}^1) Q_2 (y_2 \cond \bC) }{\gamma_{12} (y_1,y_2 \cond \bC) \gamma_{2A} (y_2,0 \cond \bC)} 
\\
&
= 
\frac{ 
Q_2 (1 \cond \bC) / \gamma_{2A} (1,0 \cond \bC) 
}{\EXP \big\{ \mathbb{I}(A = 0)  / \gamma_{2A} (Y_2,0 \cond \bC) \cond  \bC \big\}}
\Bigg[
\frac{ 1 }{\gamma_{12} (1,1 \cond \bC) }
- 1 
\Bigg]
\ .
\end{align*}

\item $\EXP \big\{ \IF_1' \cdot v \cond \bC \big\}$
\begin{align*}
&
\EXP \big\{ \texttt{IF}_1'(\bO) v(\bO) \cond \bC \big\}
\\
&
=
\frac{1}{\EXP \big\{ \mathbb{I}(A = \tilde{a}^2{'})  / \gamma_{1A} (Y_1,\tilde{a}^2{'} \cond \bC) \cond  \bC \big\}}
\sum_{y_1,y_2,a}
\frac{
(2y_1-1)(2y_2-1)(2a-1) \mathbb{I}(a = \tilde{a}^2{'})
\big\{ h(y_1,y_2) - \theta (\bC) \big\}  }{ \gamma_{1A} (y_1,\tilde{a}^2{'} \cond \bC) } 
\\
&
=
\frac{1}{\EXP \big\{ \mathbb{I}(A = 1)  / \gamma_{1A} (Y_1,1 \cond \bC) \cond  \bC \big\}}
\Bigg[
\frac{ h(1,1) - h(1,0)  }{ \gamma_{1A} (1,1 \cond \bC) } 
+  h(0,0) - h(0,1) 
\Bigg]
\ .
\end{align*}

\item $\EXP \big\{ \IF_2' \cdot v \cond \bC \big\}$
\begin{align*}
&
\EXP \big\{ \texttt{IF}_2'(\bO) v(\bO) \cond \bC \big\}
\\
&
=
-
\frac{1}{ \EXP \big\{ \mathbb{I}(A = \tilde{a}^2{'}) / \gamma_{1A} (Y_1,\tilde{a}^2{'} \cond \bC) \cond  \bC \big\} }
\sum_{y_1,y_2,a}
\frac{ (2y_1-1)(2y_2-1)(2a-1)\mathbb{I}(a = \tilde{a}^2{'})a Q_1 (y_1 \cond \bC) }{\gamma_{12} (y_1,y_2 \cond \bC) \gamma_{1A} (y_1,\tilde{a}^2{'} \cond \bC)}
\\
&
= 
\frac{ Q_1 (1 \cond \bC) / \gamma_{1A} (1,1 \cond \bC)  }{ \EXP \big\{ \mathbb{I}(A = 1) / \gamma_{1A} (Y_1,1 \cond \bC) \cond  \bC \big\} }
\Bigg[
1
-
\frac{1}{\gamma_{12} (1,1 \cond \bC)  } 
\Bigg] 
\ .
\end{align*}

\item $\EXP \big\{ \IF_3' \cdot v \cond \bC \big\}$
\begin{align*}
& 
\EXP \big\{ \texttt{IF}_3'(\bO) v(\bO) \cond \bC \big\}
\\
&
=
\frac{1}{\EXP \big\{ \mathbb{I}(A = \tilde{a}^1)  / \gamma_{2A} (Y_2,\tilde{a}^1 \cond \bC) \cond  \bC \big\}}
\sum_{y_1,y_2,a}
\frac{ (2y_1-1)(2y_2-1)(2a-1) \mathbb{I}(a = \tilde{a}^1) Q_1 (y_1 \cond \bC) }{\gamma_{12} (y_1,y_2 \cond \bC) \gamma_{2A} (y_2,\tilde{a}^1 \cond \bC)} 
\\
&
= 
\frac{1}{\EXP \big\{ \mathbb{I}(A = 0)  / \gamma_{2A} (Y_2,0 \cond \bC) \cond  \bC \big\}}
\Bigg[
-
\frac{Q_1 (1 \cond \bC) }{\gamma_{12} (1,1 \cond \bC) \gamma_{2A} (1,0 \cond \bC) }		
+
\frac{Q_1 (0 \cond \bC) }{ \gamma_{2A} (1,0 \cond \bC) }
+
Q_1 (1 \cond \bC)
-
Q_1 (0 \cond \bC) 
\Bigg]
\ .
\end{align*}

\end{itemize}

\noindent Therefore, we find the projection $\Pi \{ \IF_1 + \IF_2 + \IF_3 \cond \mathcal{T}_{\text{sym}}^\perp \}$ and $\Pi \{ \IF_1' + \IF_2' + \IF_3' \cond \mathcal{T}_{\text{sym}}^\perp \}$ are 
\begin{align*}
& 
\Pi \big\{ \texttt{IF}_1 (\bO) + \texttt{IF}_2 (\bO) + \texttt{IF}_3 (\bO) \cond \mathcal{T}_{\text{sym}}^{\perp} \big\}
\\
& 
=
\left[
\begin{array}{l}
\frac{1}{\EXP  \{ \mathbb{I}(A = 0)  / \gamma_{2A} (Y_2,0 \cond \bC) \cond  \bC  \}}
\Big[
\frac{ h(0,1) - h(1,1)  }{ \gamma_{2A} (1,0 \cond \bC) } 
+ h(1,0) - h(0,0) 
\Big]
\\[0.4cm]
+
\frac{1}{ \EXP  \{ \mathbb{I}(A = 1) / \gamma_{1A} (Y_1,1 \cond \bC) \cond  \bC  \} }
\left[	
\begin{array}{l}
\frac{Q_2 (1 \cond \bC)}{\gamma_{12} (1,1 \cond \bC) \gamma_{1A} (1,1 \cond \bC) }
-
\frac{Q_2 (0 \cond \bC)}{ \gamma_{1A} (1,1 \cond \bC) }
\\
-
Q_2 (1 \cond \bC)
+
Q_2 (0 \cond \bC)
\end{array}				 
\right]
\\[0.8cm]
+
\frac{ 
Q_2 (1 \cond \bC) / \gamma_{2A} (1,0 \cond \bC) 
}{\EXP  \{ \mathbb{I}(A = 0)  / \gamma_{2A} (Y_2,0 \cond \bC) \cond  \bC  \}}
\Big[
\frac{ 1 }{\gamma_{12} (1,1 \cond \bC) }
- 1 
\Big]
\end{array}
\right] 
\frac{v(\bO)}{\EXP \big\{ v^2(\bO) \cond \bC \big\}} 
\ ,
\end{align*}	
and
\begin{align*}
& 
\Pi \big\{ \texttt{IF}_1' (\bO) + \texttt{IF}_2' (\bO) + \texttt{IF}_3' (\bO) \cond \mathcal{T}_{\text{sym}}^{\perp} \big\}
\\
& 
=
\left[
\begin{array}{l}
\frac{1}{\EXP  \{ \mathbb{I}(A = 1)  / \gamma_{1A} (Y_1,0 \cond \bC) \cond  \bC  \}}
\Big[
\frac{ h(1,1) - h(1,0)  }{ \gamma_{1A} (1,1 \cond \bC) } 
+ h(0,0) - h(0,1) 
\Big]
\\[0.8cm]
+
\frac{ 
Q_1 (1 \cond \bC) / \gamma_{1A} (1,1 \cond \bC) 
}{\EXP  \{ \mathbb{I}(A = 1)  / \gamma_{1A} (Y_2,1 \cond \bC) \cond  \bC  \}}
\Big[
1
-
\frac{ 1 }{\gamma_{12} (1,1 \cond \bC) }
\Big]
\\[0.4cm]
+
\frac{1}{ \EXP  \{ \mathbb{I}(A = 0) / \gamma_{2A} (Y_2,0 \cond \bC) \cond  \bC  \} }
\left[
\begin{array}{l}
Q_1 (1 \cond \bC)
-
Q_1 (0 \cond \bC)
\\
-
\frac{Q_1 (1 \cond \bC)}{\gamma_{12} (1,1 \cond \bC) \gamma_{2A} (1,0 \cond \bC) }
+
\frac{Q_1 (0 \cond \bC)}{ \gamma_{2A} (1,0 \cond \bC) }
\end{array}
\right]  
\end{array}
\right] 
\frac{v(\bO)}{\EXP \big\{ v^2(\bO) \cond \bC \big\}} 
\ .
\end{align*}
Again, the projections remain the symmetric structure. From some complex algebra, one can verify that 
\begin{align*}
&
\big\{
\texttt{IF}_1 (\bO)
+
\texttt{IF}_2 (\bO)
+
\texttt{IF}_3 (\bO)
\big\}
-
\Pi \big\{ \texttt{IF}_1 (\bO) + \texttt{IF}_2 (\bO) + \texttt{IF}_3 (\bO) \cond \mathcal{T}_{\text{sym}}^{\perp} \big\}
\\
&
=
\big\{
\texttt{IF}_1' (\bO)
+
\texttt{IF}_2' (\bO)
+
\texttt{IF}_3' (\bO)
\big\}
-
\Pi \big\{ \texttt{IF}_1' (\bO) + \texttt{IF}_2' (\bO) + \texttt{IF}_3' (\bO) \cond \mathcal{T}_{\text{sym}}^{\perp} \big\}
\ .
\end{align*}
Therefore, we find that the EIF does not depend on the choice of $w(\bC)$, and is represented as $\texttt{IF}^*(\bO) = \IF_{YA}^*(\bO) + \IF_4(\bC)$ where
\begin{align*}
& \IF_{YA}^*(\bO)
\\
& 
=		
w(\bC)
\left[
\begin{array}{l}			
\big\{
\texttt{IF}_1 (\bO)
+
\texttt{IF}_2 (\bO)
+
\texttt{IF}_3 (\bO)
\big\}
\\
-
\Pi \big\{  \texttt{IF}_1 (\bO) + \texttt{IF}_2 (\bO) + \texttt{IF}_3 (\bO)  \cond \mathcal{T}_{\text{sym}}^{\perp} \big\}
\end{array}
\right] 
\\
&
\hspace*{0.5cm} 
+		
\big\{ 1- w(\bC) \big\}
\left[
\begin{array}{l}
\big\{
\texttt{IF}_1' (\bO)
+
\texttt{IF}_2' (\bO)
+
\texttt{IF}_3' (\bO)
\big\}
\\
-
\Pi \big\{  \texttt{IF}_1' (\bO) + \texttt{IF}_2' (\bO) + \texttt{IF}_3' (\bO)  \cond \mathcal{T}_{\text{sym}}^{\perp} \big\}
\end{array}
\right] 
\\
& 
=
\big\{
\texttt{IF}_1 (\bO)
+
\texttt{IF}_2 (\bO)
+
\texttt{IF}_3 (\bO)
\big\}
-
\Pi \big\{ \texttt{IF}_1 (\bO) + \texttt{IF}_2 (\bO) + \texttt{IF}_3 (\bO) \cond \mathcal{T}_{\text{sym}}^{\perp} \big\} 
\\
&
=
\Bigg[
\frac{ 1 }{  \EXP \big\{ (1-A)  / \gamma_{2A} (Y_2,0 \cond \bC) \cond  \bC \big\} }  
\frac{  \big\{ h(Y_1,Y_2) - \theta (\bC) \big\} (1-A) }{ \gamma_{2A} (Y_2,0 \cond \bC) }
\Bigg] 
\\
&
\hspace*{1cm}
+ 
\Bigg[
\frac{1}{\EXP \big\{ A/\gamma_{1A} (Y_1,1 \cond \bC) \cond \bC \big\}}
\frac{A Q_2 (Y_2 \cond \bC) }{\gamma_{12} (Y_1,Y_2 \cond \bC) \gamma_{1A} (Y_1,1 \cond \bC)  } 
\Bigg] 
\\
&
\hspace*{1cm} -
\Bigg[ 
\frac{1}{\EXP \big\{ (1-A)  / \gamma_{2A} (Y_2,0 \cond \bC) \cond \bC \big\} }
\frac{ (1-A) Q_2 (Y_2 \cond \bC)  }{\gamma_{12} (Y_1,Y_2 \cond \bC) \gamma_{2A} (Y_2,0 \cond \bC)  } 
\Bigg]
\\
&
\hspace*{1cm} 
-
\left[
\begin{array}{l}
\frac{1}{\EXP  \{ \mathbb{I}(A = 0)  / \gamma_{2A} (Y_2,0 \cond \bC) \cond  \bC  \}}
\Big[
\frac{ h(0,1) - h(1,1)  }{ \gamma_{2A} (1,0 \cond \bC) } 
+ h(1,0) - h(0,0) 
\Big]
\\[0.4cm]
+
\frac{1}{ \EXP  \{ \mathbb{I}(A = 1) / \gamma_{1A} (Y_1,1 \cond \bC) \cond  \bC  \} }
\left[	
\begin{array}{l}
\frac{Q_2 (1 \cond \bC)}{\gamma_{12} (1,1 \cond \bC) \gamma_{1A} (1,1 \cond \bC) }
-
\frac{Q_2 (0 \cond \bC)}{ \gamma_{1A} (1,1 \cond \bC) }
\\
-
Q_2 (1 \cond \bC)
+
Q_2 (0 \cond \bC)
\end{array}				 
\right]
\\[0.8cm]
+
\frac{ 
Q_2 (1 \cond \bC) / \gamma_{2A} (1,0 \cond \bC) 
}{\EXP  \{ \mathbb{I}(A = 0)  / \gamma_{2A} (Y_2,0 \cond \bC) \cond  \bC  \}}
\Big[
\frac{ 1 }{\gamma_{12} (1,1 \cond \bC) }
- 1 
\Big]
\end{array}
\right] 
\frac{v(\bO)}{\EXP \big\{ v^2(\bO) \cond \bC \big\}} 
\ .
\end{align*} 
Also, after some algebra, we find $\text{IF}_{YA}^*(\bO)$ does not have a 3-way interaction, i.e., 
\begin{align*}
&
\frac{ 
\texttt{IF}_{YA}^*(Y_1=1,Y_2=1,A=1 \cond \bC)
\texttt{IF}_{YA}^*(Y_1=0,Y_2=0,A=1 \cond \bC)
}{ 
\texttt{IF}_{YA}^*(Y_1=0,Y_2=1,A=1 \cond \bC)
\texttt{IF}_{YA}^*(Y_1=1,Y_2=0,A=1 \cond \bC)
}
\\
&
=
\frac{ 
\texttt{IF}_{YA}^*(Y_1=1,Y_2=1,A=0 \cond \bC)
\texttt{IF}_{YA}^*(Y_1=0,Y_2=0,A=0 \cond \bC)
}{ 
\texttt{IF}_{YA}^*(Y_1=0,Y_2=1,A=0 \cond \bC)
\texttt{IF}_{YA}^*(Y_1=1,Y_2=0,A=0 \cond \bC)
}
\ ,
\end{align*}
verifying that $\texttt{IF}^*(\bO) \in \mathcal{T}_{\text{sym}}$.

\subsection{Robustness Properties of the Estimator}

Recall that the estimating functions are defined as
\begin{align*}
&
U_{1}^{b} ( \kappa_1 , \omega_1 )
=
g_1(\vec{c}^{\, b} )
\bigg\{
\frac{ \mathbb{I}(a^{b} = \tilde{a}^2{'}) }{  \gamma_{1A} \big( y_1^{b} , \tilde{a}^2{'} \cond \vec{c}^{\, b} \con \omega_1 \big)   }
-
\delta_{1}(\vec{c}^{\, b} \con \kappa_1 )
\bigg\}
\\
&
U_{2}^{b} ( \kappa_2, \omega_2 )
=
g_2(\vec{c}^{\, b} )
\bigg\{
\frac{
\mathbb{I}(a^{b} = \tilde{a}^1) }{  \gamma_{2A} \big( y_2^{b} , \tilde{a}^1 \cond \vec{c}^{\, b} \con \omega_2 \big)    } 
-
\delta_{2} (\vec{c}^{\, b} \con \kappa_2 )
\bigg\}
\ , \\
& 
U_{\psi}^b 
( \psi, 
\kappa_{1},
\kappa_{2},	
\omega_{1}, \omega_{2}, \nu )
\\
&
=
\left[
\begin{array}{l}
w(\vec{c}^{\, b}) 
\frac{1}{ \delta_{2} (\vec{c}^{\, b} \con \kappa_2) }
\frac{  \mathbb{I}(a^{b}=\tilde{a}^1) }{   \gamma_{2A} (y_2^{b} , \tilde{a}^1 \cond \vec{c}^{\, b} \con \omega_2 )  }
\{ h(y_1^{b},y_2^{b}) - \theta(\vec{c}^{\, b} \con \omega_1,\omega_2,\nu ) \}
\\
+
w(\vec{c}^{\, b}) 
\Big[
\frac{1}{ \delta_{1} (\vec{c}^{\, b} \con \kappa_1) }
\frac{  \mathbb{I}(a^{b}=\tilde{a}^2{'})   }{ \gamma_{1A} (y_1^{b} , \tilde{a}^2{'} \cond \vec{c}^{\, b}  \con \omega_1 ) }
-
\frac{1}{ \delta_{2} (\vec{c}^{\, b} \con \kappa_2) }
\frac{  \mathbb{I}(a^{b}=\tilde{a}^1) }{ \gamma_{2A} (y_2^{b} , \tilde{a}^1 \cond \vec{c}^{\, b}  \con \omega_2 )  }
\Big]
\frac{Q_2(y_2^{b} \cond \vec{c}^{\, b} \con \omega_1, \omega_2, \nu) }{\gamma_{12}(y_1^{b},y_2^{b} \cond \vec{c}^{\, b} \con \nu)}
\\ 
+
\big\{ 1 - w(\vec{c}^{\, b})  \big\}
\frac{1}{ \delta_{1} (\vec{c}^{\, b} \con \kappa_1) }
\frac{  \mathbb{I}(a^{b}=\tilde{a}^2{'}) }{  \gamma_{1A} (y_1^{b} , \tilde{a}^2{'} \cond \vec{c}^{\, b}  \con \omega_1 )  }
\{ h(y_1^{b},y_2^{b}) - \theta(\vec{c}^{\, b} \con \omega_1,\omega_2,\nu ) \}
\\
+
\big\{ 1 - w(\vec{c}^{\, b})  \big\}
\Big[
\frac{1}{ \delta_{2} (\vec{c}^{\, b} \con \kappa_2) }
\frac{  \mathbb{I}(a^{b}=\tilde{a}^1)   }{ \gamma_{2A} (y_2^{b} , \tilde{a}^1 \cond \vec{c} \con \omega_2 )  }
-
\frac{1}{ \delta_{1} (\vec{c}^{\, b} \con \kappa_1) }
\frac{  \mathbb{I}(a^{b}=\tilde{a}^2{'})  }{ \gamma_{1A} (y_1^{b} , \tilde{a}^2{'} \cond \vec{c}^{\, b}  \con \omega_1 )  }
\Big]
\frac{Q_1(y_1^{b} \cond \vec{c}^{\, b} \con \omega_1, \omega_2,  \nu) }{\gamma_{12}(y_1^{b},y_2^{b} \cond \vec{c}^{\, b} \con \nu)}
\\ 
+
 \theta(\vec{c}^{\, b} \con \omega_1,\omega_2,\nu )
-
\psi
\end{array}
\right] 
\ ,
\end{align*}
where $\theta(\bc \con \omega_1, \omega_2, \nu)$ is represented as
\begin{align*}
\theta( \bc \con \nu, \omega_1,\omega_2)
=
\frac{
\int h(y_1,y_2) \gamma_{12} (y_1,y_2 \cond \bc \con \nu) 
f_1(y_1 \cond \tilde{a}^1,Y_2=0,\bc \con \omega_1)
f_2(y_2 \cond \tilde{a}^2{'},Y_2=0,\bc \con \omega_2) 
d (y_1,y_2)
}{
\int \gamma_{12} (y_1,y_2 \cond \bc \con \nu) 
f_1(y_1 \cond \tilde{a}^1,Y_2=0,\bc \con \omega_1)
f_2(y_2 \cond \tilde{a}^2{'},Y_2=0,\bc \con \omega_2) 
d (y_1,y_2)
}
\ .
\end{align*}
We use the dagger superscript ($^\dagger$) to denote the solution to the mean-zero moment conditions above, i.e.,
\begin{align*}
    &
    \EXP \big\{ U_{1} ( \kappa_1^\dagger , \omega_1^\dagger ) \big\} = 0
     \ ,
     &&
    \EXP \big\{ U_{2} ( \kappa_2^\dagger , \omega_2^\dagger ) \big\} = 0
     \ ,
     &&
     \EXP \big\{ U_{\psi}
( \psi^\dagger, 
\kappa_{1}^\dagger,
\kappa_{2}^\dagger,	
\omega_{1}^\dagger, 
\omega_{2}^\dagger, 
\nu^\dagger ) \big\} = 0
\ .
\end{align*}

Recall that model $\mathcal{M}^*$ is defined as $\mathcal{M}^* = \mathcal{M}_\gamma \cap \big\{ 
\{ \mathcal{M}_{f_1} \cap \mathcal{M}_{\delta_1}\} \cup \{ \mathcal{M}_{f_2} \cap \mathcal{M}_{\delta_2}\} \big\}$. Therefore, model $\mathcal{M}^*$ is satisfied if and only if either $\mathcal{M}_1^* := \mathcal{M}_\gamma \cap \{ \mathcal{M}_{f_1} \cap \mathcal{M}_{\delta_1}\}$ or $\mathcal{M}_2^* := \mathcal{M}_\gamma \cap \{ \mathcal{M}_{f_2} \cap \mathcal{M}_{\delta_2}\}$, but not necessarily both, is satisfied. Under models $\mathcal{M}_1^*$ and $\mathcal{M}_2^*$, we find
\begin{align*}
    &
    \text{Under $\mathcal{M}_1^*$, it follows that }
    \left\{
    \begin{array}{l}         
    \gamma_{12}(y_1,y_2 \cond \bC \con \nu^\dagger)
    = 
    \gamma_{12}(y_1,y_2 \cond \bC)
    \\
    f_{1}(y_1 \cond \bC \con \omega_1^\dagger)
    =
    f_{1}(y_1 \cond \bC)
    \\
    \delta_1(\bC \con \kappa_1^\dagger) 
=
\theta_D (\bC) f_A(\tilde{a}^2{'} \cond \bC) / \N (\bC) 
    \end{array}
    \ ,
    \right.
    \\
    &
    \text{Under $\mathcal{M}_2^*$, it follows that }
    \left\{
    \begin{array}{l}         
    \gamma_{12}(y_1,y_2 \cond \bC \con \nu^\dagger)
    = 
    \gamma_{12}(y_1,y_2 \cond \bC)
    \\
    f_{2}(y_2 \cond \bC \con \omega_2^\dagger)
    =
    f_{2}(y_2 \cond \bC)
        \\
    \delta_2(\bC \con \kappa_2^\dagger) 
=
\theta_D (\bC) f_A(\tilde{a}^1 \cond \bC) / \N (\bC) 
    \end{array}
    \ .
    \right.
\end{align*}
The results regarding $\delta_1$ and $\delta_2$ are established as follows. The moment restrictions $\EXP \big\{ U_1 (\kappa_1 , \omega_1) \big\} = 0$ and $\EXP \big\{ U_2 (\kappa_2 , \omega_2) \big\} = 0$, which imply
\begin{align}		
&
\delta_1(\bC \con \kappa_1)  
= 
f_A(\tilde{a}^2{'} \cond \bC)
\frac{
\int 
\gamma_{12} (y_1,y_2 \cond \bC)
f_1 (y_1 \cond  \bC)
f_2 (y_2 \cond  \bC)
\frac{ \gamma_{1A}(y_1,\tilde{a}^2{'} \cond \bC) }{\gamma_{1A}(y_1,\tilde{a}^2{'} \cond \bC \con \omega_1)}
\, d (y_1,y_2)
}{
\N (\bC)
} 
\ ,
\label{eq-delta1}
\\
&
\delta_2(\bC \con \kappa_2)  
=
f_A(\tilde{a}^1 \cond \bC)
\frac{
\int 
\gamma_{12} (y_1,y_2 \cond \bC)
f_1 (y_1 \cond  \bC)
f_2 (y_2 \cond  \bC)
\frac{ \gamma_{2A}(y_2,\tilde{a}^1 \cond \bC) }{\gamma_{2A}(y_2,\tilde{a}^1 \cond \bC \con \omega_2)}
\, d (y_1,y_2)
}{
\N (\bC)
} 
\ .
\label{eq-delta2}
\end{align}
Under model $\mathcal{M}_{1}^*$, we have  
\begin{align}	\label{eq-true delta1}
&
\delta_1(\bC \con \kappa_1^\dagger) 
=
\frac{\theta_D (\bC) f_A(\tilde{a}^2{'} \cond \bC)}{ \N (\bC) } 
\ .
\end{align}
Likewise, under model $\mathcal{M}_2^*$, we have
\begin{align}	\label{eq-true delta2}
&
\delta_2(\bC \con \kappa_2^\dagger)
=
\frac{\theta_D (\bC) f_A(\tilde{a}^1 \cond \bC)}{ \N (\bC) }
\ .
\end{align}	
These results agree with \eqref{eq-alpha}. 

Next, $Q_1(y_1 \cond \bC \con \omega_1,\omega_2,\nu)$ and $Q_2(y_2 \cond \bC \con \omega_1,\omega_2,\nu)$ are represented as
\begin{align*}
&
Q_1(y_1 \cond \bC \con \omega_1, \omega_2, \nu)
=
\int  \big\{ 
h(y_1,y_2) - \theta(\bC \con \omega_1, \omega_2, \nu)
\big\}	 	
\gamma_{12}(y_1,y_2 \cond \bC \con \nu)
f_{2} (y_2 \cond  \bC \con \omega_2) \, dy_2
\ ,
\\
&
Q_2(y_2 \cond \bC \con \omega_1, \omega_2,  \nu)
=
\int  \big\{ 
h(y_1,y_2) - \theta(\bC \con \omega_1, \omega_2, \nu)
\big\}	 	
\gamma_{12}(y_1,y_2 \cond \bC \con \nu)
f_{1} (y_1 \cond  \bC \con \omega_1) \, dy_1 
\ .
\end{align*}
Under model $\mathcal{M}_1^*$, we have
\begin{align}	\label{eq-Q1 zero}
&
\int Q_1(y_1 \cond \bC \con \omega_1^\dagger,\omega_2^\dagger, \nu^\dagger)
f_1 (y_1 \cond  \bC) \, dy_1
\nonumber
\\
&
=
\int h(y_1,y_2)  
\gamma_{12} (y_1,y_2 \con \bC)
f_1 ( y_1 \cond  \bC)
f_2(y_2 \cond  \bC \con \omega_2^\dagger) \, d(y_1,y_2)
\nonumber
\\
&
\quad 
-
\theta(\bC \con \omega_1^\dagger,\omega_2^\dagger, \nu^\dagger) 
\int  
\gamma_{12} (y_1,y_2 \con \bC)
f_1 ( y_1 \cond  \bC)
f_2(y_2 \cond  \bC \con \omega_2^\dagger) \, d(y_1,y_2)
\nonumber
\\
&
=
\int h(y_1,y_2) \gamma_{12} (y_1,y_2 \con \bC)
f_1  ( y_1 \cond  \bC)
f_2 (y_2 \cond \bC \con \omega_2^\dagger) \, d(y_1,y_2)
\nonumber
\\
&
\quad
-
\int h(y_1,y_2) \gamma_{12} (y_1,y_2 \con \bC)
f_1  ( y_1 \cond  \bC)
f_2 (y_2 \cond \bC \con \omega_2^\dagger) \, d(y_1,y_2)
\nonumber
\\
&
=
0 \ ,		
\end{align}
and
\begin{align}	\label{eq-Q1 nonzero}
&
\int Q_2(y_2 \cond \bC \con \omega_1^\dagger,\omega_2^\dagger, \nu^\dagger)
f_2 (y_2 \cond \bC) \, dy_2
\nonumber
\\
&
=
\int h(y_1,y_2)  
\gamma_{12} (y_1,y_2 \con \bC)
f_1 ( y_1 \cond  \bC)
f_2(y_2 \cond  \bC) \, d(y_1,y_2)
\nonumber
\\
&
\quad 
-
\theta(\bC \con \omega_1^\dagger,\omega_2^\dagger, \nu^\dagger) 
\int  
\gamma_{12} (y_1,y_2 \con \bC)
f_1 ( y_1 \cond  \bC)
f_2(y_2 \cond  \bC) \, d(y_1,y_2)
\nonumber 
\\
&
=
\theta_D (\bC)
\big\{
\theta(\bC)  
- 
\theta(\bC \con \omega_1^\dagger,\omega_2^\dagger, \nu^\dagger) 
\big\}
\ .
\end{align} 
Likewise, under model $\mathcal{M}_2^*$, we have
\begin{align}	
&
\int Q_2(y_2 \cond \bC \con \omega_1^\dagger,\omega_2^\dagger, \nu^\dagger) 
f_1(y_1 \cond \bC) \, dy_1 
=
\theta_D (\bC)
\big\{
\theta(\bC)  
- 
\theta(\bC \con \omega_1^\dagger,\omega_2^\dagger, \nu^\dagger) 
\big\}
\ ,
\label{eq-Q2 nonzero}
\\
&
\int Q_2(y_2 \cond \bC \con \omega_1^\dagger,\omega_2^\dagger, \nu^\dagger) 
f_2(y_2 \cond \bC) \, dy_2
=
0		
\ .
\label{eq-Q2 zero}
\end{align}  

Next, we study the expectation of components in $U_{\psi}  ( \psi, \kappa_{1}, \kappa_2,\omega_1,\omega_2,\nu )$, which are referred to as $T_1,\ldots,T_5$ below:
\begin{itemize}[leftmargin=0cm]
\item Let $T_1$ be
\begin{align*}
T_1(\bc \con \kappa_2, \omega_1,\omega_2,\nu)
=
\EXP \bigg[
\frac{1}{\delta_{2} (\bc \con \kappa_2) }
\frac{  \mathbb{I}(A=\tilde{a}^1) }{   \gamma_{2A} (Y_2 , \tilde{a}^1 \cond \vec{c} \con \omega_2 )  }
\{ h(Y_1,Y_2) - \theta (\bc \con \omega_1,\omega_2,\nu) \}
\COND
\bc
\bigg]
\ .
\end{align*}
Under model $\mathcal{M}_1^*$, we have
\begin{align*}
T_1(\bc \con \kappa_2^\dagger, \omega_1^\dagger,\omega_2^\dagger,\nu^\dagger)
=  
\frac{1}{\delta_2(\bc \con \kappa_2^\dagger) }
\EXP
\bigg[
\frac{h(Y_1,Y_2) \mathbb{I}(A = \tilde{a}^1) }{\gamma_{2A}(Y_2 , \tilde{a}^1 \cond \bc \con \omega_2^\dagger) }
\COND \bc
\bigg]
-
\theta (\bc \con \omega_1^\dagger,\omega_2^\dagger,\nu^\dagger)
\ .
\end{align*}
Under model $\mathcal{M}_2^*$, we have
\begin{align*}
T_1(\bc \con \kappa_2^\dagger, \omega_1^\dagger,\omega_2^\dagger,\nu^\dagger)
& =  
\frac{1}{\delta_2(\bc \con \kappa_2^\dagger) }
\EXP
\bigg[
\frac{h(Y_1,Y_2) \mathbb{I}(A = \tilde{a}^1) }{\gamma_{2A}(Y_2 , \tilde{a}^1 \cond \bc \con \omega_2^\dagger) }
\COND \bc
\bigg]
-
\theta (\bc \con \omega_1^\dagger,\omega_2^\dagger,\nu^\dagger)
\\
&
=
\theta(\bc) - 	\theta (\bc \con \omega_1^\dagger,\omega_2^\dagger,\nu^\dagger)
\ .
\end{align*}
The second identity is from \eqref{eq-IPW}. 

\item Let $T_2$ be
\begin{align*}
& 
T_2(\bc \con \kappa_1, \kappa_2, \omega_1,\omega_2,\nu)
\\
&
=
\EXP \Bigg[
\bigg[
\frac{1}{\delta_{1} (\bc \con \kappa_1) }
\frac{  \mathbb{I}(A=\tilde{a}^2{'})   }{ \gamma_{1A} (Y_1  , \tilde{a}^2{'} \cond \bc  \con \omega_1 ) }
-
\frac{1}{\delta_{2}(\bc \con \kappa_2) }
\frac{  \mathbb{I}(A=\tilde{a}^1) }{ \gamma_{2A} (Y_2  , \tilde{a}^1 \cond \bc  \con \omega_2 )  }
\bigg]
\frac{Q_2(Y_2  \cond \bc \con \omega_1, \omega_2, \nu) }{\gamma_{12} (Y_1 ,Y_2  \cond \bc \con \nu)}
\COND \bc \Bigg]
\ .
\end{align*}
Under model $\mathcal{M}_1^*$, we have
\begin{align*}
&
T_2(\bc \con \kappa_1^\dagger, \kappa_2^\dagger, \omega_1^\dagger, \omega_2^\dagger, \nu^\dagger)
\\
&
= 
\frac{ f_A(\tilde{a}^2{'} \cond \bc) }{\N(\bc) } 
\frac{1}{ \delta_{1}(\bc \con \kappa_1^\dagger) }
\int Q_2(y_2 \cond \bc \con \omega_1^\dagger,\omega_2^\dagger,\nu^\dagger) 
f_2(y_2 \cond \bc)
\, dy_2
\\
&
\hspace*{0.5cm}
- 
\frac{1}{\delta_{2}(\bc \con \kappa_2^\dagger) }
\int 	
\frac{ 
Q_2(y_2 \cond \bc \con \omega_1^\dagger,\omega_2^\dagger,\nu^\dagger) 
}{ 
\gamma_{2A}(y_2,\tilde{a}^1 \cond \bc \con \omega_2^\dagger) 
}
\frac{ 
\gamma_{2A}(y_2,\tilde{a}^1 \cond \bc)  
f_2(y_2 \cond \bc)
f_A(\tilde{a}^1 \cond \bc)
}{\N(\bc)}
\, d y_2
\\
&
=
\frac{ 
\theta_D (\bc)
\big\{
\theta(\bc)  
- 
\theta(\bc \con \omega_1^\dagger,\omega_2^\dagger, \nu^\dagger) 
\big\}
}{ \theta_D(\bc) }
\\
&
\hspace*{0.5cm}
- 
\frac{1}{\delta_{2}(\bc \con \kappa_2^\dagger) }
\int 	
\frac{ 
\big\{ h(y_1,y_2) - \theta(\bc \con \omega_1^\dagger,\omega_2^\dagger, \nu^\dagger)  \big\} 
}{ 
\gamma_{2A}(y_2,\tilde{a}^1 \cond \bc \con \omega_2^\dagger) 
}
\frac{
\gamma_{12}(y_1,y_2 \cond \bc) 
\gamma_{2A}(y_2,\tilde{a}^1 \cond \bc) 
f_1(y_1 \cond \bc)
f_2(y_2 \cond \bc)
f_A(\tilde{a}^1 \cond \bc)
}{\N(\bc)}
\, d(y_1,y_2)
\\
&
=
\theta(\bc)   
-
\frac{1}{\delta_2(\bc \con \kappa_2^\dagger) }
\EXP
\bigg[
\frac{h(Y_1,Y_2) \mathbb{I}(A = \tilde{a}^1) }{\gamma_{2A}(Y_2 , \tilde{a}^1 \cond \bc \con \omega_2^\dagger) }
\COND \bc
\bigg] 
\ .
\end{align*} 
The identities are obtained from \eqref{eq-delta1}-\eqref{eq-Q2 zero}.

Similarly, under model $\mathcal{M}_2^*$, we have
\begin{align*}
&
T_2(\bc \con \kappa_1^\dagger, \kappa_2^\dagger, \omega_1^\dagger, \omega_2^\dagger, \nu^\dagger)
\\
&
= 
\frac{1}{\delta_{1}(\bc \con \kappa_1^\dagger) }
\int 	
\frac{ 
Q_2(y_2 \cond \bc \con \omega_1^\dagger,\omega_2^\dagger,\nu^\dagger) 
}{ 
\gamma_{1A}(y_1,\tilde{a}^2{'} \cond \bc \con \omega_1^\dagger) 
}
\frac{ 
\gamma_{1A}(y_1,\tilde{a}^2{'} \cond \bc)  
f_1(y_1 \cond \bc)
f_2(y_2 \cond \bc)
f_A(\tilde{a}^2{'} \cond \bc)
}{\N(\bc)}
\, d (y_1,y_2)
\\
&
\hspace*{0.5cm}
- 
\frac{ f_A(\tilde{a}^1 \cond \bc) }{\N(\bc) } 
\frac{1}{ \delta_{2}(\bc \con \kappa_2^\dagger) }
\int Q_2(y_2 \cond \bc \con \omega_1^\dagger,\omega_2^\dagger,\nu^\dagger) 
f_2(y_2 \cond \bc)
\, dy_2
\\
&
=
\frac{f_A(\tilde{a}^2{'} \cond \bc) }{\delta_{1}(\bc \con \kappa_1^\dagger)  \N (\bc)}
\Bigg[
\int \frac{\gamma_{1A}(y_1,\tilde{a}^2{'} \cond \bc)}{\gamma_{1A} (y_1,\tilde{a}^2{'} \cond \bc \con \omega_1^\dagger) }
f_1(y_1 \cond \bc) \, dy_1
\Bigg]
\Bigg[
\int Q_2(y_2 \cond \bc \con \omega_1^\dagger,\omega_2^\dagger,\nu^\dagger) 
f_2(y_2 \cond \bc)
\, dy_2
\Bigg]
\\
&
\hspace*{0.5cm}
- 
\frac{ f_A(\tilde{a}^1 \cond \bc) }{\N(\bc) } 
\frac{1}{ \delta_{2}(\bc \con \kappa_2^\dagger) }
\int Q_2(y_2 \cond \bc \con \omega_1^\dagger,\omega_2^\dagger,\nu^\dagger) 
f_2(y_2 \cond \bc)
\, dy_2
\\
&
=
0
\ .
\end{align*} 
The last equality is satisfied from \eqref{eq-Q2 zero}. 

\item Let $T_3$ be
\begin{align*}
T_3(\bc \con \kappa_1, \omega_1,\omega_2,\nu)
=
\EXP \bigg[
\frac{1}{\delta_{1} (\bc \con \kappa_1) }
\frac{  \mathbb{I}(A=\tilde{a}^2{'}) }{   \gamma_{1A} (Y_1 , \tilde{a}^2{'} \cond \bc \con \omega_2 )  }
\{ h(Y_1,Y_2) - \theta_k (\bc \con \omega_1,\omega_2,\nu) \}
\COND
\bc
\bigg]
\ .
\end{align*}
Under model $\mathcal{M}_1^*$, we have
\begin{align*}
T_3(\bc \con \kappa_1^\dagger, \omega_1^\dagger,\omega_2^\dagger,\nu^\dagger)
& 
=  
\frac{1}{\delta_1(\bc \con \kappa_1^\dagger) }
\EXP
\bigg[
\frac{h(Y_1,Y_2) \mathbb{I}(A = \tilde{a}^2{'}) }{\gamma_{1A}(Y_1 , \tilde{a}^2{'} \cond \bc \con \omega_1^\dagger) }
\COND \bc
\bigg]
-
\theta (\bc \con \omega_1^\dagger,\omega_2^\dagger,\nu^\dagger)
\\
&
=
\theta(\bc) - 	\theta (\bc \con \omega_1^\dagger,\omega_2^\dagger,\nu^\dagger)
\ .
\end{align*}
The second identity is from \eqref{eq-IPW}.

Likewise, under model $\mathcal{M}_2^*$, we have
\begin{align*}
T_3(\bc \con \kappa_1^\dagger, \omega_1^\dagger,\omega_2^\dagger,\nu^\dagger)
& =  
\frac{1}{\delta_1(\bc \con \kappa_1^\dagger) }
\EXP
\bigg[
\frac{h(Y_1,Y_2) \mathbb{I}(A = \tilde{a}^2{'}) }{\gamma_{1A}(Y_1 , \tilde{a}^2{'} \cond \bc \con \omega_1^\dagger) }
\COND \bc
\bigg]
-
\theta (\bc \con \omega_1^\dagger,\omega_2^\dagger,\nu^\dagger)
\ .
\end{align*}

\item Let $T_4$ be
\begin{align*}
& 
T_4(\bc \con \kappa_1, \kappa_2, \omega_1,\omega_2,\nu)
\\
&
=
\EXP \Bigg[
\bigg[
\frac{1}{\delta_{2}(\bc \con \kappa_2) }
\frac{  \mathbb{I}(A=\tilde{a}^1) }{ \gamma_{2A} (Y_2  , \tilde{a}^1 \cond \bc  \con \omega_2 )  }
-
\frac{1}{\delta_{1} (\bc \con \kappa_1) }
\frac{  \mathbb{I}(A=\tilde{a}^2{'})   }{ \gamma_{1A} (Y_1  , \tilde{a}^2{'} \cond \bc  \con \omega_1 ) }
\bigg]
\frac{Q_1(Y_1  \cond \bc \con \omega_1, \omega_2, \nu) }{\gamma_{12} (Y_1 ,Y_2  \cond \bc \con \nu)}
\COND \bc \Bigg]
\ .
\end{align*}
Under model $\mathcal{M}_1^*$, we have
\begin{align*}
&
T_4(\bc \con \kappa_1^\dagger, \kappa_2^\dagger, \omega_1^\dagger, \omega_2^\dagger, \nu^\dagger)
\\
&
= 
\frac{1}{\delta_{2}(\bc \con \kappa_2^\dagger) }
\int 	
\frac{ 
Q_1(y_1 \cond \bc \con \omega_1^\dagger,\omega_2^\dagger,\nu^\dagger) 
}{ 
\gamma_{2A}(y_2,\tilde{a}^1 \cond \bc \con \omega_2^\dagger) 
}
\frac{ 
\gamma_{2A}(y_2,\tilde{a}^1 \cond \bc)  
f_1(y_1 \cond \bc)
f_2(y_2 \cond \bc)
f_A(\tilde{a}^1 \cond \bc)
}{\N(\bc)}
\, d (y_1,y_2)
\\
&
\hspace*{0.5cm}
- 
\frac{ f_A(\tilde{a}^2{'} \cond \bc) }{\N(\bc) } 
\frac{1}{ \delta_{1}(\bc \con \kappa_1^\dagger) }
\int Q_1(y_1 \cond \bc \con \omega_1^\dagger,\omega_2^\dagger,\nu^\dagger) 
f_1(y_1 \cond \bc)
\, dy_1
\\
&
=
\frac{f_A(\tilde{a}^1 \cond \bc) }{\delta_{2}(\bc \con \kappa_2^\dagger)  \N (\bc)}
\Bigg[
\int \frac{\gamma_{2A}(y_2,\tilde{a}^1 \cond \bc)}{\gamma_{2A} (y_2,\tilde{a}^1 \cond \bc \con \omega_2^\dagger) }
f_2(y_2 \cond \bc) \, dy_2
\Bigg]
\Bigg[
\int Q_1(y_1 \cond \bc \con \omega_1^\dagger,\omega_2^\dagger,\nu^\dagger) 
f_1(y_1 \cond \bc)
\, dy_1
\Bigg]
\\
&
\hspace*{0.5cm}
- 
\frac{ f_A(\tilde{a}^2{'} \cond \bc) }{\N(\bc) } 
\frac{1}{ \delta_{1}(\bc \con \kappa_1^\dagger) }
\int Q_1(y_1 \cond \bc \con \omega_1^\dagger,\omega_2^\dagger,\nu^\dagger) 
f_1(y_1 \cond \bc)
\, dy_1
\\
&
=
0 \ .
\end{align*} 
The last equality is satisfied from \eqref{eq-Q1 zero}.

Similarly, under model $\mathcal{M}_2^*$, we have
\begin{align*}
&
T_4(\bc \con \kappa_1^\dagger, \kappa_2^\dagger, \omega_1^\dagger, \omega_2^\dagger, \nu^\dagger)
\\
&
= 
\frac{ f_A(\tilde{a}^1 \cond \bc) }{\N(\bc) } 
\frac{1}{ \delta_{2}(\bc \con \kappa_2^\dagger) }
\int Q_1(y_1 \cond \bc \con \omega_1^\dagger,\omega_2^\dagger,\nu^\dagger) 
f_1(y_1 \cond \bc)
\, dy_1
\\
&
\hspace*{0.5cm}
- 
\frac{1}{\delta_{1}(\bc \con \kappa_1^\dagger) }
\int 	
\frac{ 
Q_1(y_1 \cond \bc \con \omega_1^\dagger,\omega_2^\dagger,\nu^\dagger) 
}{ 
\gamma_{1A}(y_1,\tilde{a}^2{'} \cond \bc \con \omega_1^\dagger) 
}
\frac{ 
\gamma_{1A}(y_1,\tilde{a}^2{'} \cond \bc)  
f_1(y_1 \cond \bc)
f_A(\tilde{a}^2{'} \cond \bc)
}{\N(\bc)}
\, d y_1
\\
&
=
\frac{ 
\theta_D (\bc)
\big\{
\theta(\bc)  
- 
\theta(\bc \con \omega_1^\dagger,\omega_2^\dagger, \nu^\dagger) 
\big\}
}{ \theta_D(\bc) }
\\
&
\hspace*{0.5cm}
- 
\frac{1}{\delta_{1}(\bc \con \kappa_1^\dagger) }
\int 	
\frac{ 
\big\{ h(y_1,y_2) - \theta(\bc \con \omega_1^\dagger,\omega_2^\dagger, \nu^\dagger)  \big\} 
}{ 
\gamma_{1A}(y_1,\tilde{a}^2{'} \cond \bc \con \omega_1^\dagger) 
}
\frac{
\gamma_{12}(y_1,y_2 \cond \bc) 
\gamma_{1A}(y_1,\tilde{a}^2{'} \cond \bc) 
f_1(y_1 \cond \bc)
f_2(y_2 \cond \bc)
f_A(\tilde{a}^2{'} \cond \bc)
}{\N(\bc)}
\, d(y_1,y_2)
\\
&
=
\theta(\bc)   
-
\frac{1}{\delta_1(\bc \con \kappa_1^\dagger) }
\EXP
\bigg[
\frac{h(Y_1,Y_2) \mathbb{I}(A = \tilde{a}^2{'}) }{\gamma_{1A}(Y_1 , \tilde{a}^2{'} \cond \bc \con \omega_1^\dagger) }
\COND \bc
\bigg]  \ .
\end{align*} 
The identities are obtained from \eqref{eq-delta1}-\eqref{eq-Q2 zero}.

\item Let $T_5$ be $ T_5(\bc \con \omega_1,\omega_2,\nu)
=
\theta(\bc \con \omega_1,\omega_2,\nu) - \psi(\omega_1,\omega_2,\nu)$. 
\end{itemize}

Combining the results extablished in $T_1,\ldots,T_5$,  we find the expectation of $\EXP \big\{ U_\psi(\psi,\kappa_1,\kappa_2,\omega_1,\omega_2,\nu) \big\}$ is
\begin{align*}
&
\EXP \big\{ U_\psi(\psi,\kappa_1,\kappa_2,\omega_1,\omega_2,\nu) \big\}
\\
&
=
\EXP \big[
\EXP \big\{ U_\psi(\psi,\kappa_1,\kappa_2,\omega_1,\omega_2,\nu) 
\cond \bC \big\}
\big]
\\
&
=
\EXP
\left[ 
\begin{array}{l}
w(\bC) \big\{ T_1(\bC \con \kappa_2, \omega_1,\omega_2,\nu)
+
T_2(\bC \con \kappa_1,\kappa_2, \omega_1,\omega_2,\nu) 
\big\}
\\	
+ \big\{ 1- w(\bC) \big\} \big\{ T_3(\bC \con \kappa_1, \omega_1,\omega_2,\nu)
+
T_4(\bC \con \kappa_1, \kappa_2, \omega_1,\omega_2,\nu)
\big\}
\\
+ T_5(\bC \con \omega_1,\omega_2,\nu)
\end{array}
\right]  \ .
\end{align*}
Under model $\mathcal{M}_1^*$, we find 
\begin{align*}
&
T_1(\bC \con \kappa_2^\dagger, \omega_1^\dagger,\omega_2^\dagger,\nu^\dagger)
+
T_2(\bC \con \kappa_1^\dagger, \kappa_2^\dagger,  \omega_1^\dagger,\omega_2^\dagger,\nu^\dagger)
=
\theta(\bC) - \theta(\bC \con \omega_1^\dagger,\omega_2^\dagger,\nu^\dagger)
\ ,
\\
&
T_3(\bC \con \kappa_1^\dagger, \omega_1^\dagger,\omega_2^\dagger,\nu^\dagger)
+
T_4(\bC \con \kappa_1^\dagger, \kappa_2^\dagger, \omega_1^\dagger,\omega_2^\dagger,\nu^\dagger)
=
\theta(\bC) - \theta(\bC \con \omega_1^\dagger,\omega_2^\dagger,\nu^\dagger)
\ .
\end{align*}
Likewise, under model $\mathcal{M}_2^*$, we find 
\begin{align*}
&
T_1(\bC \con \kappa_2^\dagger, \omega_1^\dagger,\omega_2^\dagger,\nu^\dagger)
+
T_2(\bC \con \kappa_1^\dagger, \kappa_2^\dagger, \omega_1^\dagger,\omega_2^\dagger,\nu^\dagger)
=
\theta(\bC) - \theta(\bC \con \omega_1^\dagger,\omega_2^\dagger,\nu^\dagger)
\ ,
\\
&
T_3(\bC \con \kappa_1^\dagger, \omega_1^\dagger,\omega_2^\dagger,\nu^\dagger)
+
T_4(\bC \con \kappa_1^\dagger, \kappa_2^\dagger, \omega_1^\dagger,\omega_2^\dagger,\nu^\dagger)
=
\theta(\bC) - \theta(\bC \con \omega_1^\dagger,\omega_2^\dagger,\nu^\dagger)
\ .
\end{align*}
Therefore, under $\mathcal{M}^*=\mathcal{M}_1^* \cup \mathcal{M}_2^*$, we find
\begin{align*}
&
\EXP \big\{ U_\psi(\psi,\kappa_1^\dagger,\kappa_2^\dagger,\omega_1^\dagger,\omega_2^\dagger,\nu^\dagger) \big\}
\\
&
=
\EXP 
\big\{ \theta(\bC) - \theta(\bC \con \omega_1^\dagger,\omega_2^\dagger,\nu^\dagger)
+
\theta(\bC \con \omega_1^\dagger,\omega_2^\dagger,\nu^\dagger) - \psi \big\}
\\
&
=
\EXP 
\big\{ \theta(\bC) 
\big\} - \psi
\ .
\end{align*}
Therefore, the solution to the estimating equation $\EXP \big\{ U_\psi(\psi,\kappa_1^{\,\dagger},\kappa_2^{\,\dagger}, \omega_1^\dagger,\omega_2^\dagger,\nu^\dagger) \big\} = 0$ recovers the true effect $\psi = \psi(\tilde{a}^1,\tilde{a}^2{'}) = \EXP \big\{ \theta(\tilde{a}^1,\tilde{a}^2{'},\bC) \big\}$ under model $\mathcal{M}^*$. This establishes the robustness property of $\widehat{\psi}_{dr}$. 

\subsection{Estimator Under Joint Gaussian Outcomes}

We end the section by presenting representations of the nuisance functions under a simple working model. Suppose that an investigator posits a working model for the outcomes $(Y_1,Y_2)$ as the following joint normal distribution:
\begin{align*}
\begin{pmatrix}
Y_1 \\ Y_2
\end{pmatrix}
\cond (A=a, \bC)
\sim 
N
\left(
\begin{pmatrix}
\mu_{1a}(\bC) \\ 
\mu_{2a}(\bC)
\end{pmatrix}
,
\begin{pmatrix}		
\sigma^2 & 
\rho \sigma^2 \\ 
\rho \sigma^2 & 
\sigma^2
\end{pmatrix}
\right)
\ .
\end{align*}
We suppress covariates $\bC$ for notational brevity hereafter, i.e., $\mu_{1a} := \mu_{1a}(\bC)$ and $\mu_{2a} := \mu_{2a}(\bC)$. Consider the reparametrizations $\nu = \sigma^2 (1-\rho^2) / \rho$, $\xi_{1a} = \mu_{1a} -\rho \mu_{2a}$, and $\xi_{2a} = \mu_{2a} -\rho \mu_{1a}$ for $a=0,1$. In addition, suppose $\xi_{1a}$ and $\xi_{2a}$ are further parametrized by finite-dimensional parameters $\zeta_{1a}$ and $\zeta_{2a}$, respectively. For instance, one may consider the following linear models:
\begin{align*}
    &
    \xi_{1a}(\bC \con \zeta_{1a})
    =
    \xi_{1a}(\bC \con \zeta_{1a})
    =
    \zeta_{1a0} + \zeta_{1aC}^T \bC
    \ ,
    \\
    &
    \xi_{2a}(\bC \con \zeta_{2a})
    =
    \xi_{2a}(\bC \con \zeta_{1a})
    =
    \zeta_{2a0} + \zeta_{2aC}^T \bC
\ .
\end{align*}
Note that the original distribution is written as
\begin{align*}
\begin{pmatrix}
Y_1 \\ Y_2
\end{pmatrix}
\cond (A=a, \bC)
\sim 
N
\left(
\frac{1}{1-\rho^2}
\begin{pmatrix}
\xi_{1a}(\bC) + \rho \xi_{2a}(\bC) \\ 
\xi_{2a}(\bC) + \rho \xi_{1a}(\bC)
\end{pmatrix}
,
\frac{\rho \nu}{1-\rho^2}
\begin{pmatrix}		
1 & 
\rho  \\ 
\rho  & 
1
\end{pmatrix}
\right)\ .
\end{align*}

Then, the odds ratio function and the two baseline densities are represented as 
\begin{align*}
&
\gamma_{12}(y_1,y_2 \cond \bC \con \nu)
=
\exp \bigg( \frac{ y_1 y_2 }{ \nu } \bigg)
\ ,
\\
& 
f_1(y_1 \cond a,Y_2=0,\bC \con \rho, \nu, \zeta_{1a})
=
\frac{1}{\sqrt{2\pi} \sqrt{ \rho \nu }}
\exp \bigg[ 
- \frac{ \{ y_1 - \xi_{1a} (\bC \con \zeta_{1a}) \}^2 }{
2 \rho \nu 
} 
\bigg]
\sim N \big( \xi_{1a}(\bC \con \zeta_{1a}) , \rho \nu \big)
\ ,
\\
& f_2(y_2 \cond a,Y_1=0,\bC  \con \rho, \nu, \zeta_{2a})
=
\frac{1}{\sqrt{2\pi} \sqrt{ \rho \nu } }
\exp \bigg[ 
- \frac{ \{ y_2 - \xi_{2a}(\bC \con \zeta_{2a}) \}^2 }{
 2  \rho \nu 
} 
\bigg]
\sim N \big( \xi_{2a}(\bC \con \zeta_{2a}) , \rho \nu \big)
\ .
\end{align*}
Additionally, $\gamma_{1A}$, $\gamma_{2A}$, $\delta_1$, and $\delta_2$ are represented as 
\begin{align*}
\gamma_{1A}(y_1,a \cond \bC \con \rho, \nu, \zeta_{1\tilde{a}^1}, \zeta_{1\tilde{a}^2{'}})
& = \exp \bigg\{ \frac{ y_1 \mathbb{I}(a=a{_2'}) \{ \xi_{1 \tilde{a}^2{'}}(\bC \con \zeta_{1\tilde{a}_{2}'}) - \xi_{1 \tilde{a}^1} (\bC \con \zeta_{1\tilde{a}_{1}}) \} }{ \rho \nu } \bigg\}
\ ,
\\
\gamma_{2A}(y_2,a \cond \bC \con \rho, \nu, \zeta_{2\tilde{a}^1}, \zeta_{2\tilde{a}^2{'}}) 
&= \exp \bigg\{ \frac{y_2 \mathbb{I}(a=\tilde{a}^1) \{ \xi_{2 \tilde{a}^1}(\bC \con \zeta_{2\tilde{a}^1}) - \xi_{2 \tilde{a}{_2'}}(\bC \con \zeta_{2\tilde{a}_{2}'}) \} }{ \rho \nu} \bigg\}
\ , 
\end{align*}
and
\begin{align}
& \delta_1 (\bC \con \rho, \nu, \zeta_{1\tilde{a}^1}, \zeta_{1\tilde{a}^2{'}}, \zeta_{2\tilde{a}^2{'}} )
\label{eq-delta1-normal}
\\
\nonumber
&
=
\EXP
    \Bigg[ 
    \frac{ \mathbb{I}(A=a_{2}') }{ {\gamma}_{1A}(Y_1, \tilde{a}^2{'} \cond \bC \con \rho, \nu, \zeta_{1\tilde{a}^1}, \zeta_{1\tilde{a}^2{'}})} 
    \, \Bigg| \, \bC
    \Bigg]
    \\
\nonumber
&
    =
    \Pr(A=\tilde{a}^2{'} \cond \bC) 
\EXP
    \Bigg[ 
    \exp \bigg\{ - \frac{ 
Y_1 \{ \xi_{1 \tilde{a}^2{'}}(\bC \con \zeta_{1\tilde{a}^2{'}}) - \xi_{1 \tilde{a}^1}(\bC \con \zeta_{1\tilde{a}^1}) \}
}{ \rho \nu }
\bigg\}
    \, \Bigg| \, A=\tilde{a}^2{'}, \bC
    \Bigg]
\\
\nonumber
&
=
\Pr(A=\tilde{a}^2{'} \cond \bC) 
\exp 
\left[
\begin{array}{l}
-
\bigg\{ 
\frac{ \xi_{1\tilde{a}^2{'}}(\bC \con \zeta_{1\tilde{a}^2{'}}) +\rho \xi_{2\tilde{a}^2{'}}(\bC \con \zeta_{2\tilde{a}^2{'}}) }{1-\rho^2}
\bigg\}
\bigg\{ 
\frac{ 
\xi_{1 \tilde{a}^2{'}}(\bC \con \zeta_{1\tilde{a}^2{'}}) - \xi_{1 \tilde{a}^1}(\bC \con \zeta_{1\tilde{a}^1})
}{ \rho \nu }
\bigg\}
\\
+ \frac{1}{2}
\frac{\rho \nu}{1-\rho^2}
\bigg\{ 
\frac{ 
\xi_{1 \tilde{a}^2{'}}(\bC \con \zeta_{1\tilde{a}^2{'}}) - \xi_{1 \tilde{a}^1}(\bC \con \zeta_{1\tilde{a}^1})
}{ \rho \nu }
\bigg\}^2
\end{array}
\right] 
\\
\nonumber
&
=
\Pr(A=\tilde{a}^2{'} \cond \bC) 
\exp 
\Bigg[
\frac{1}{\rho \nu (1-\rho^2)}
\Bigg[
\begin{array}{l}
0.5 \big\{ 
\xi_{1\tilde{a}^1}^2(\bC \con \zeta_{1\tilde{a}^1})
-
\xi_{1\tilde{a}^2{'}}^2(\bC \con \zeta_{1\tilde{a}^2{'}})
\big\}
\\
+
\rho
\xi_{2\tilde{a}^2{'}}(\bC \con \zeta_{2\tilde{a}^2{'}})
\big\{ 
\xi_{1\tilde{a}^1}(\bC \con \zeta_{1\tilde{a}^1})
-
\xi_{1\tilde{a}^2{'}}(\bC \con \zeta_{1\tilde{a}^2{'}})
\big\}
\end{array}
\Bigg]
\Bigg]
\end{align}
and
\begin{align}
&
\delta_2 (\bC \con \rho, \nu, \zeta_{2\tilde{a}^1}, \zeta_{2\tilde{a}^2{'}}, \zeta_{1\tilde{a}^1} )
\label{eq-delta2-normal}
\\
\nonumber
&
=
\Pr(A=\tilde{a}^1 \cond \bC) 
\exp 
\Bigg[
\frac{1}{\rho \nu (1-\rho^2)}
\Bigg[
\begin{array}{l}
0.5 \big\{ 
\xi_{2\tilde{a}^2{'}}^2(\bC \con \zeta_{2\tilde{a}^2{'}})
-
\xi_{2\tilde{a}^1}^2(\bC \con \zeta_{2\tilde{a}^1})
\big\}
\\
+
\rho
\xi_{1\tilde{a}^1}(\bC \con \zeta_{1\tilde{a}^1})
\big\{ 
\xi_{2\tilde{a}^2{'}}(\bC \con \zeta_{2\tilde{a}^2{'}})
-
\xi_{2\tilde{a}^1}(\bC \con \zeta_{2\tilde{a}^1})
\big\}
\end{array}
\Bigg]
\Bigg]
\end{align}
Note that $\delta_1$ and $\delta_2$ depend on $\zeta_{2\tilde{a}^2{'}}$ and $\zeta_{1\tilde{a}^1}$, respectively. 
One may posit working models for $\delta_1$ and $\delta_2$ and drop the dependencies on $\zeta_{2\tilde{a}^2{'}}$ and $\zeta_{1\tilde{a}^1}$, respectively, i.e., 
\begin{align*}
&
     \delta_1 (\bC \con \rho, \nu, \zeta_{1\tilde{a}^1}, \zeta_{1\tilde{a}^2{'}}, \kappa_1 )
     \\
     &
     =
\Pr(A=\tilde{a}^2{'} \cond \bC) 
\exp 
\Bigg[
\frac{1}{\rho \nu (1-\rho^2)}
\Bigg[
\begin{array}{l}
0.5 \big\{ 
\xi_{1\tilde{a}^1}^2(\bC \con \zeta_{1\tilde{a}^1})
-
\xi_{1\tilde{a}^2{'}}^2(\bC \con \zeta_{1\tilde{a}^2{'}})
\big\}
\\
+
\rho
\eta_1(\bC \con \kappa_1)
\big\{ 
\xi_{1\tilde{a}^1}(\bC \con \zeta_{1\tilde{a}^1})
-
\xi_{1\tilde{a}^2{'}}(\bC \con \zeta_{1\tilde{a}^2{'}})
\big\}
\end{array}
\Bigg]
\Bigg]
\\
&
\delta_2 (\bC \con \rho, \nu, \zeta_{2\tilde{a}^1}, \zeta_{2\tilde{a}^2{'}}, \kappa_2 )
\\
&
=
\Pr(A=\tilde{a}^1 \cond \bC) 
\exp 
\Bigg[
\frac{1}{\rho \nu (1-\rho^2)}
\Bigg[
\begin{array}{l}
0.5 \big\{ 
\xi_{2\tilde{a}^2{'}}^2(\bC \con \zeta_{2\tilde{a}^2{'}})
-
\xi_{2\tilde{a}^1}^2(\bC \con \zeta_{2\tilde{a}^1})
\big\}
\\
+
\rho
\eta_2(\bC \con \kappa_2)
\big\{ 
\xi_{2\tilde{a}^2{'}}(\bC \con \zeta_{2\tilde{a}^2{'}})
-
\xi_{2\tilde{a}^1}(\bC \con \zeta_{2\tilde{a}^1})
\big\}
\end{array}
\Bigg] \Bigg]
\end{align*}
where $\eta_1$ and $\eta_2$ are user-specified functions. Then, from the definitions of $\delta_1$ and $\delta_2$ in \eqref{eq-alpha}, we find the following results hold for any $g$:
\begin{align*}
&
    \EXP \Bigg[ 
     g(\bC) \Bigg[
     \Pr(A=\tilde{a}^2{'} \cond \bC) 
\exp 
\Bigg[
\frac{-\Bigg[
\begin{array}{l}
0.5 \big\{ 
\xi_{1\tilde{a}^1}^2(\bC \con \zeta_{1\tilde{a}^1})
-
\xi_{1\tilde{a}^2{'}}^2(\bC \con \zeta_{1\tilde{a}^2{'}})
\big\}
\\
+
\rho
\eta_1(\bC \con \kappa_1)
\big\{ 
\xi_{1\tilde{a}^1}(\bC \con \zeta_{1\tilde{a}^1})
-
\xi_{1\tilde{a}^2{'}}(\bC \con \zeta_{1\tilde{a}^2{'}})
\big\}
\end{array}
\Bigg]}{\rho \nu (1-\rho^2)}
\Bigg]
     - 1
     \Bigg]
    \Bigg] 
    =
    0
    \\
    &
    \EXP \Bigg[ 
     g(\bC) 
     \Bigg[
     \Pr(A=\tilde{a}^1 \cond \bC) 
\exp 
\Bigg[
\frac{-\Bigg[
\begin{array}{l}
0.5 \big\{ 
\xi_{2\tilde{a}^2{'}}^2(\bC \con \zeta_{2\tilde{a}^2{'}})
-
\xi_{2\tilde{a}^1}^2(\bC \con \zeta_{2\tilde{a}^1})
\big\}
\\
+
\rho
\eta_2(\bC \con \kappa_2)
\big\{ 
\xi_{2\tilde{a}^2{'}}(\bC \con \zeta_{2\tilde{a}^2{'}})
-
\xi_{2\tilde{a}^1}(\bC \con \zeta_{2\tilde{a}^1})
\big\}
\end{array}
\Bigg]}{\rho \nu (1-\rho^2)}
 \Bigg]
 -1
     \Bigg]
    \Bigg] 
    =
    0
\end{align*}

To characterize $\theta$ and $Q$ functions, suppose that $h(y_1,y_2)$ is chosen as $h(y_1,y_2) = y_1$. Under this choice, $\theta$ and $Q$ functions are represented as
\begin{align*}
&
\theta(\tilde{a}^1,\tilde{a}^2{'} , \bC)
=
\frac{ 
{\xi}_{1 \tilde{a}^1}(\bC \con \zeta_{1\tilde{a}_{1}}) + 
{\rho}  {\xi}_{2 \tilde{a}^2{'} }(\bC \con \zeta_{2\tilde{a}_{2}'})
}{
1-\rho^2
}
\\
&
Q_1 (y_1 \cond \bC )
=
\exp
\bigg[
\frac{ \rho y_1
 \big\{
\rho y_1
+ 
2
\xi_{2 \tilde{a}^2{'}} (\bC \con \zeta_{2\tilde{a}_{2}'})
\big\}
 }{2 \rho \nu }
\bigg]
\big\{ y_1 - 
\theta(\tilde{a}^1,\tilde{a}^2{'} , \bC  )
\big\}
\\
&
Q_2 (y_2 \cond \bC)
=
\exp
\bigg[
\frac{ \rho y_2 \big\{
\rho y_2
+ 
2
\xi_{1 \tilde{a}^1} (\bC \con \zeta_{1\tilde{a}_{1}})
\big\}
 }{2 \rho \nu }
\bigg]
\big\{ 
\xi_{1 \tilde{a}^1}(\bC \con \zeta_{1\tilde{a}_{1}}) + \rho y_2
-
\theta(\tilde{a}^1,\tilde{a}^2{'} , \bC )
\big\}
\end{align*}
We then use these estimating functions to estimate $\kappa_1$ and $\kappa_2$.

The estimator for $\psi$ can be obtained by substituting the nuisance functions in the estimating equation with these representations. Let $\widehat{\nu}$, $\widehat{\rho}$, $\widehat{\zeta}_{1a}$, $\widehat{\zeta}_{2a}$, $\widehat{\kappa}_1$, $\widehat{\kappa}_2$ be the estimators of $\nu$, $\rho$, $\zeta_{1a}$, $\zeta_{2a}$, $\kappa_1$, $\kappa_2$ respectively. Then, we obtain
\begin{align*}
    &
\widehat{\xi}_{1a} = \xi_{1a}(\bC \con \widehat{\zeta}_{1a})
\quad , \quad 
\widehat{\xi}_{2a} = \xi_{2a}(\bC \con \widehat{\zeta}_{2a})
\\
&
\widehat{\gamma}_{1A}(y_1,a) = \exp \bigg\{ \frac{ y_1 \mathbb{I}(a=\tilde{a}^2{'}) ( \widehat{\xi}_{1 \tilde{a}^2{'}} - \widehat{\xi}_{1 \tilde{a}^1} ) }{ \widehat{\rho} \widehat{\nu} } \bigg\}
\\
&
\widehat{\gamma}_{2A}(y_2,a) = \exp \bigg\{ \frac{y_2 \mathbb{I}(a=\tilde{a}^1) ( \widehat{\xi}_{2 \tilde{a}^1} - \widehat{\xi}_{2 \tilde{a}^2{'}} ) }{ \widehat{\rho} \widehat{\nu}} \bigg\}
\end{align*}
In addition, from \eqref{eq-delta1-normal} and \eqref{eq-delta2-normal}, we have
\begin{align*}
    \widehat{\delta}_{1} 
    & 
    =
\widehat{\Pr}(A=\tilde{a}^2{'} \cond \bC) 
\exp 
\Bigg[
\frac{0.5 \big\{ 
\widehat{\xi}_{1\tilde{a}^1}^2
-
\widehat{\xi}_{1\tilde{a}^2{'}}^2
\big\}
+
\widehat{\rho}
\eta_1(\bC \con \widehat{\kappa}_1)
\big\{ 
\widehat{\xi}_{1\tilde{a}^1}
-
\widehat{\xi}_{1\tilde{a}^2{'}}
\big\}}{\widehat{\rho} \widehat{\nu} (1-\widehat{\rho}^2)}
\Bigg]
\\
    \widehat{\delta}_{2} 
    & 
    =
\widehat{\Pr}(A=\tilde{a}^1 \cond \bC) 
\exp 
\Bigg[
\frac{0.5 \big\{ 
\widehat{\xi}_{2\tilde{a}^2{'}}^2
-
\widehat{\xi}_{2\tilde{a}^1}^2
\big\}
+
\widehat{\rho}
\eta_2(\bC \con \widehat{\kappa}_2)
\big\{ 
\widehat{\xi}_{2\tilde{a}^2{'}}
-
\widehat{\xi}_{2\tilde{a}^1}
\big\}}{ \widehat{\rho} \widehat{\nu} (1-\widehat{\rho}^2)}
\Bigg]
\end{align*}
Lastly, we define
\begin{align*}
    &
\widehat{\theta}
=
\frac{ 
\widehat{\xi}_{1 \tilde{a}^1} + 
\widehat{\rho}  \widehat{\xi}_{2 \tilde{a}^2{'} }
}{
1-\widehat{\rho}^2
}
\\
&
\widehat{Q}_1 (y_1)
=
\exp
\bigg\{
\frac{ \widehat{\rho} y_1
 \big(
\widehat{\rho} y_1
+ 
2
\widehat{\xi}_{2 \tilde{a}^2{'}}
\big)
 }{2 \widehat{\rho} \widehat{\nu} }
\bigg\}
\big\{ y_1 - 
\widehat{\theta}
\big\}
\\
&
\widehat{Q}_2 (y_2)
=
\exp
\bigg\{
\frac{ \widehat{\rho} y_2 \big(
\widehat{\rho} y_2
+ 
2
\widehat{\xi}_{1 \tilde{a}^1}
\big)
 }{2 \widehat{\rho} \widehat{\nu} }
\bigg\}
\big\{ 
\widehat{\xi}_{1 \tilde{a}^1} + \widehat{\rho} y_2
-
\widehat{\theta}
\big\}
\\
&
\widehat{D}^{b}
=
\frac{  \mathbb{I}(a^{b}=\tilde{a}^2{'})   }{ \widehat{\delta}_{1} \cdot \widehat{\gamma}_{1A} (y_1^{b} , \tilde{a}^2{'} ) }
-
\frac{  \mathbb{I}(a^{b}=\tilde{a}^1) }{ \widehat{\delta}_{2}  \cdot \widehat{\gamma}_{2A} (y_2^{b} , \tilde{a}^1 )  } \ .
\end{align*}
Then, the estimator $\widehat{\psi}_{dr}$ has the following form:
\begin{align*}
\widehat{\psi}_{dr}
=
\frac{1}{B}
\sum_{b=1}^{B}
\left[
    \begin{array}{l}
    w \cdot
    \mathbb{I}(a^{b}=\tilde{a}^1)
\big\{ y_1^{b} - \widehat{\theta} \big\}
/ \big\{ \widehat{\delta}_{2} \cdot \widehat{\gamma}_{2A} (y_2^{b} , \tilde{a}^1 )  \big\}
\\
+
w \cdot \widehat{D}^{b} \cdot
\widehat{Q}_2(y_2^{b}) / \widehat{\gamma}_{12}(y_1^{b},y_2^{b})
\\
+
(1-w) \cdot
\mathbb{I}(a^{b}=\tilde{a}^2{'}) \cdot
\big\{ y_1^{b} - \widehat{\theta} \big\}
/ \big\{ \widehat{\delta}_{1} \cdot \widehat{\gamma}_{1A} (y_1^{b} , \tilde{a}^2{'} )  \big\}
\\ 
-
(1-w) \cdot \widehat{D}^{b} \cdot
\widehat{Q}_1(y_1^{b}) / \widehat{\gamma}_{12}(y_1^{b},y_2^{b})
\\ 
+
\widehat{\theta}
\end{array}
\right] 
\end{align*}

\newpage

\section{Details of the Simulation Study}
\label{sec:sim-details}

\subsection{Data Generating Mechanism}

For each dyad, we first generate two covariates from a bivariate normal distribution as follows:
\begin{align*}
	\bC 
 =
 \begin{pmatrix}
     C_1 \\ C_2
 \end{pmatrix}
	\sim 
	N \left(
		\begin{pmatrix}
			0 \\ 0
		\end{pmatrix}
		,  
		\begin{pmatrix}
			1 & 0.1 \\ 0.1 & 1
		\end{pmatrix}
	\right)  
\end{align*}
We then generate treatment from the following Bernoulli distribution:
\begin{align*}
    A \sim \text{Ber}
\Big(
	\text{expit} \big( 0.1 + 0.3 C_1 - 0.2 C_2 \big)
\Big)
\end{align*}
Lastly, the outcome is generated from the following bivariate normal distribution:
\begin{align*}
&
\begin{pmatrix}
	Y_1 \\ Y_2
	\end{pmatrix}
	\cond (A=a,\bC)
	\sim 
	N \left(
		\begin{pmatrix}
			-0.09375 + 0.3750 a +  C_1 + 0.7250 C_2
   \\
   0.09375 + 0.5625a + 1.1875 C_1 + 0.8375 C_2 
   \end{pmatrix}
		,  
		\begin{pmatrix}
			0.5 & 0.125 \\ 0.125 & 0.5
		\end{pmatrix}
	\right)  
\end{align*}
Under this specification, the estimands have values of
$\EXP \{ Y_{2}(\tilde{a}^{1}=0,\tilde{a}^{2}{'}=0) \} = 0.09375$, $\EXP \{Y_{2}(\tilde{a}^{1}=0,\tilde{a}^{2}{'}=1)\} =  0.59375$, $\EXP \{Y_{2}(\tilde{a}^{1}=1,\tilde{a}^{2}{'}=0)\} =  0.15625$, and $ \EXP \{Y_{2}(\tilde{a}^{1}=1,\tilde{a}^{2}{'}=1)\} =  0.65625$.

\subsection{Summary of the Simulation Study}

\begin{figure}[!htb]
 \centering
 \hspace*{-1in}
 \includegraphics[width=1.25\textwidth]{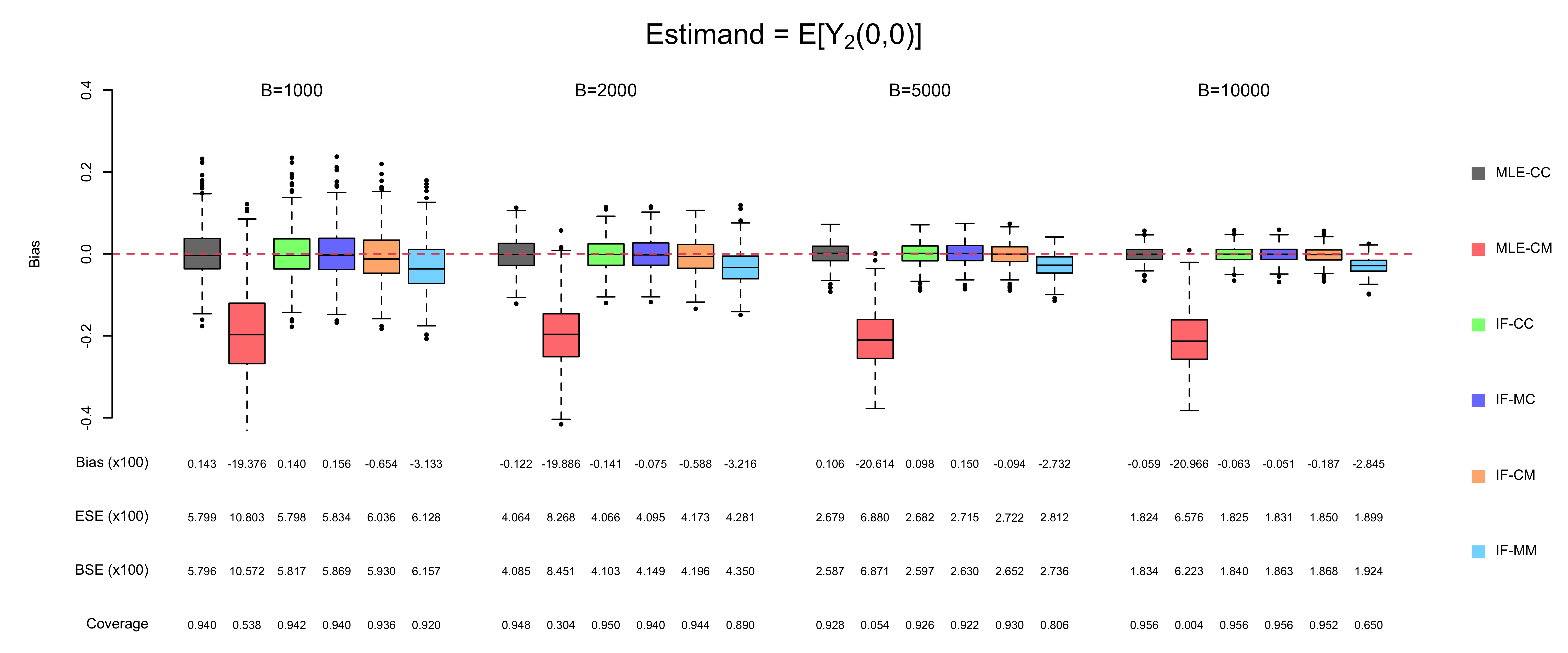}
 \caption{Estimation result for $\EXP \{ Y_{2}(\tilde{a}^{1}=0,\tilde{a}^{2}{'}=0) \}$}
 \label{fig:Simulation1}
 \end{figure}

\begin{figure}[!htb]
 \centering
 \hspace*{-1in}
 \includegraphics[width=1.25\textwidth]{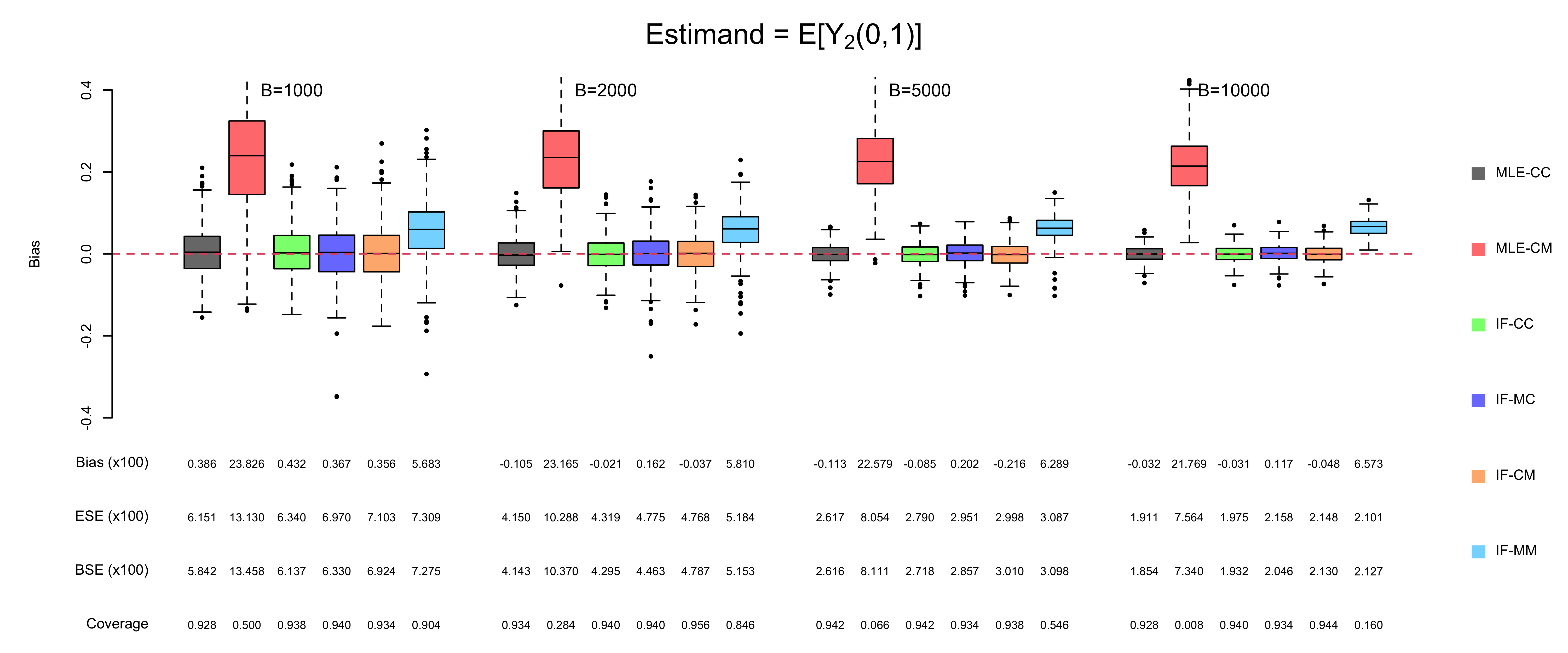}
 \caption{Estimation result for $\EXP \{ Y_{2}(\tilde{a}^{1}=0,\tilde{a}^{2}{'}=1) \}$}
 \label{fig:Simulation2}
 \end{figure}

\begin{figure}[!htb]
 \centering
 \hspace*{-1in}
 \includegraphics[width=1.25\textwidth]{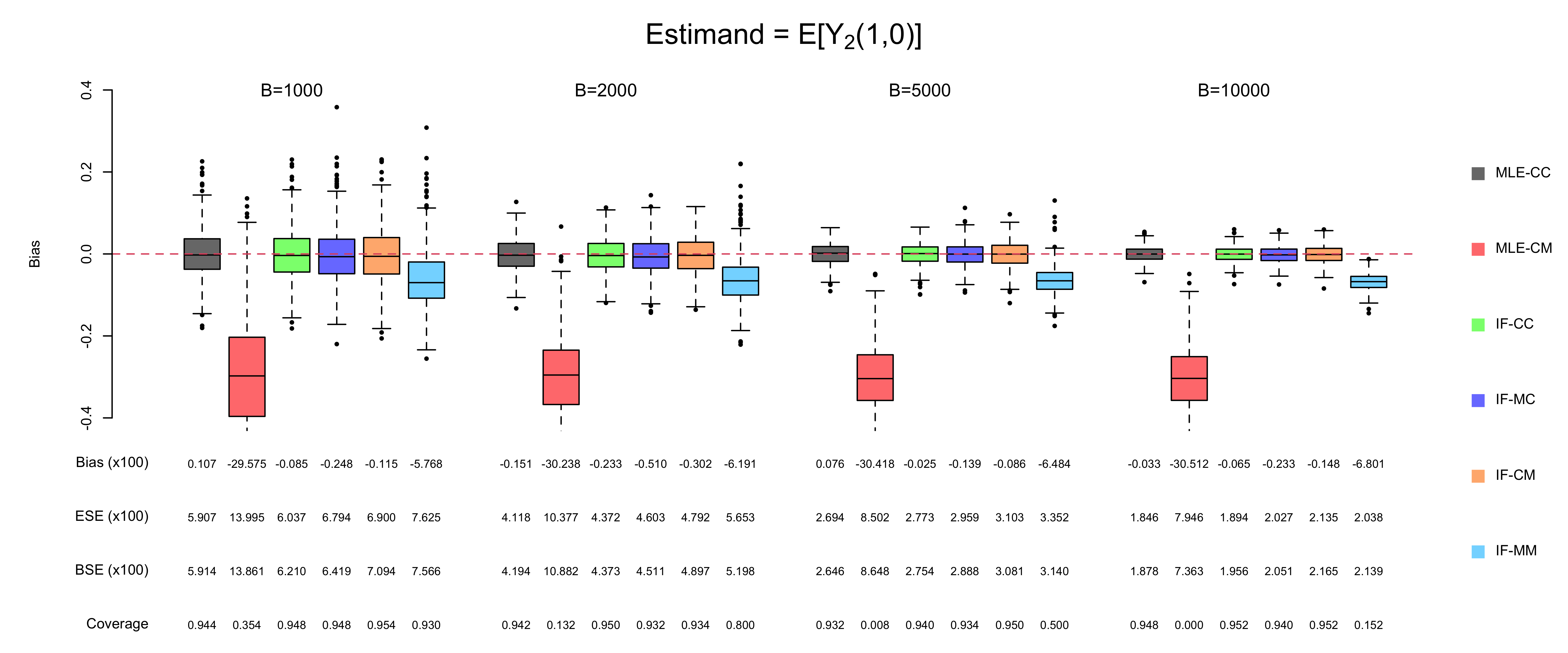}
 \caption{Estimation result for $\EXP \{ Y_{2}(\tilde{a}^{1}=1,\tilde{a}^{2}{'}=0) \}$}
 \label{fig:Simulation3}
 \end{figure}

\begin{figure}[!htb]
 \centering
 \hspace*{-1in}
 \includegraphics[width=1.25\textwidth]{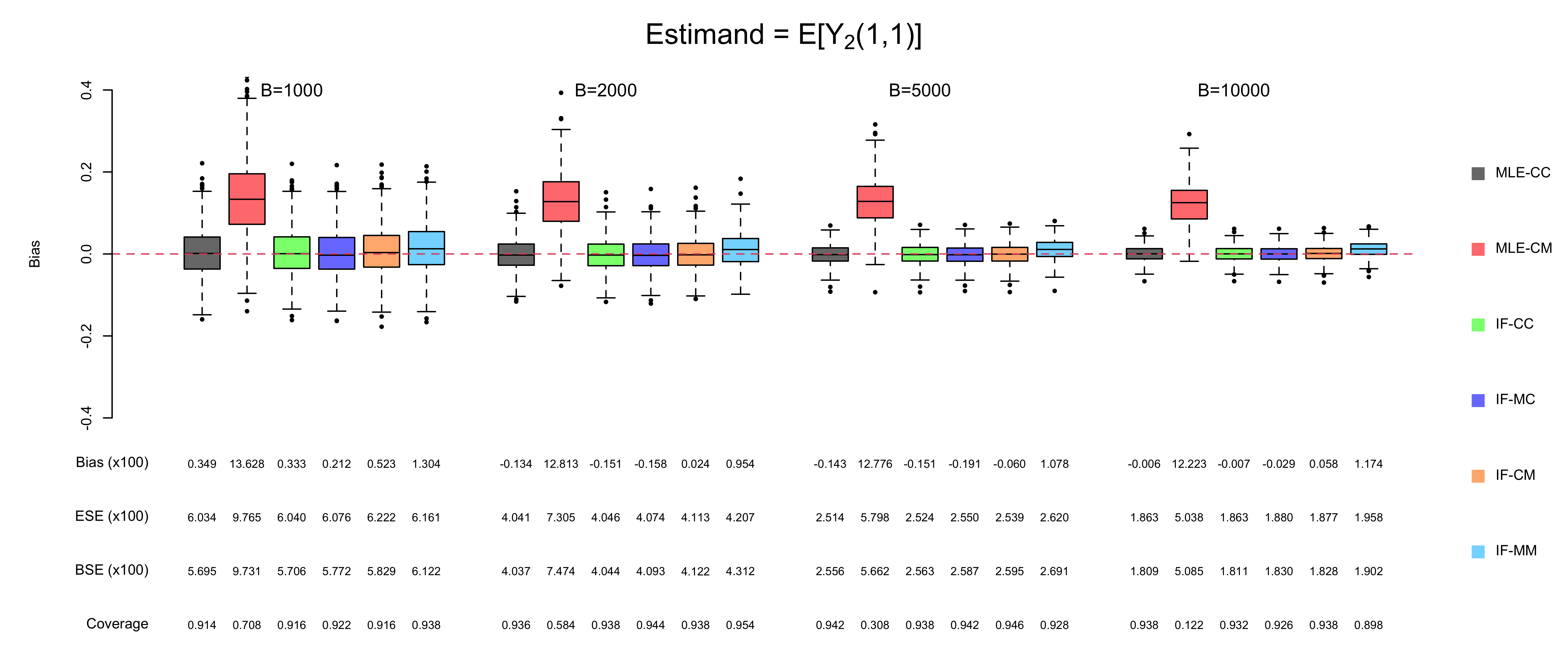}
 \caption{Estimation result for $\EXP \{ Y_{2}(\tilde{a}^{1}=1,\tilde{a}^{2}{'}=1) \}$}
 \label{fig:Simulation4}
 \end{figure}

We present visual summaries of the simulation study in Figures \ref{fig:Simulation1}-\ref{fig:Simulation4}. The boxplots are based on 500 estimates of the MLE and influence function-based estimators. Of note, for the MLE and influence function-based estimator, we considered the following model specification scenarios for $\gamma_{12}(y_1,y_2 \cond \vec{c} ; \nu)$, $f_1(y_1 \cond a, \vec{c} ; \omega_1) := f(y_1 \cond a, Y_2=0,\vec{c} ; \omega_1)$, $f_2(y_2 \cond a, \vec{c} ; \omega_2) := f(y_2 \cond a, Y_1=0,\vec{c} ; \omega_2)$, $\delta_1(\vec{c} ; \kappa_1)$, $\delta_2(\vec{c} ; \kappa_2)$:
\begin{itemize}
    \item For the MLE:
    \begin{enumerate}
    \item[] (MLE-CC) $\gamma_{12}$, $f_{1}$,  $f_{2}$  are correctly specified;
    \item[] (MLE-CM) $\gamma_{12}$, $f_{1}$ are correctly specified, and $f_{2}$ is mis-specified;
\end{enumerate} 
    \item For the influence-function based estimator:
    \begin{enumerate}
    \item[] (IF-CC) $\gamma_{12}$, $f_{1}$, $\delta_1$, $f_{2}$, $\delta_2$ are correctly specified;
    \item[] (IF-CM) $\gamma_{12}$, $f_{1}$, $\delta_1$ are correctly specified, and $f_{2}$, $\delta_2$ are mis-specified;
    \item[] (IF-MC) $\gamma_{12}$, $f_{2}$, $\delta_2$ are correctly specified, and $f_{1}$, $\delta_1$ are mis-specified;
    \item[] (IF-MM) $\gamma_{12}$ is correctly specified, and $f_{1}$, $\delta_1$, $f_{2}$, $\delta_2$ are mis-specified;
\end{enumerate} 
\end{itemize}

The numbers in bias, ESE, BSE, and Coverage rows represent empirical biases, empirical standard errors, the median of bootstrap standard errors, and the empirical coverage rates of 95\% percentile bootstrap confidence intervals.

\clearpage

\section{Replication Code for Data Analysis}

\subsection{SAS code}

The following SAS code assumes that the Wisconsin Longitudinal Study (WLS) data has been downloaded from \url{https://www.ssc.wisc.edu/wlsresearch/data/}.

\noindent In particular, we make use of the "marriage" dataset that contains information on each participant and his/her spouse, as well as the "long" version of the main WLS data. Text in brackets should be replaced to map to your own file library locations.

\lstset{language=SAS, 
breaklines=true,  
basicstyle=\scriptsize,
columns=fixed,
showstringspaces=false,
keepspaces=true,
frame=shadowbox} 

\begin{lstlisting}

OPTIONS nofmterr;

data marriage noformats;
set [marriage].public_grad_marriage_roster_long;
keep=idpub cmdeath cmspdth cmstart mar_num;
run;

data wls_long;
set [long].wls_bl_13_07 (where=(rtype="g") 
keep=idpub rtype z_dglev z_brdxdy z_sexrsp
z_hcb10sp z_hc012sp z_hc013sp z_hb005rec
z_gu002re z_gu003re z_gu004re z_au002re 
z_au003re z_au004re gwiiq_bm z_aa003re 
edsphh ocssphu z_edhhyr z_ocsh57u2 z_aa001re;
run;



proc sql;
create table wls_select AS SELECT
* FROM
wls_long INNER JOIN
marriage on wls_long.idpub = marriage.idpub;
quit;

PROC EXPORT DATA=wls_select
FILE="[output.dta]"
DBMS=STATA REPLACE;
RUN;

\end{lstlisting}


\clearpage

\subsection{R code}

\lstset{language=R, 
breaklines=true,  
basicstyle=\scriptsize,
columns=fixed,
showstringspaces=false,
showspaces=false,
frame=shadowbox,
identifierstyle=\color{blue}\ttfamily,
stringstyle=\color{purple}\ttfamily,
commentstyle=\color{green}\ttfamily,} 

\begin{lstlisting}

##################################################
# Data Cleaning
##################################################

library(haven)
library(dplyr)
library(readr)
library(boot)
library(DescTools)
# Required : output . dta from Appendix 3
wls_data_select <- read_dta("[output.dta]")%>%
  transmute(idpub ,
            rtype ,
            grad_female = ifelse(z_sexrsp ==2 , 1 , 0) ,
            graduate_birth_year = z_brdxdy + 1900 ,
            education_spouse = z_hcb10sp ,
            spouse_age_2011= z_hc012sp ,
            spouse_age_marriage = z_hc013sp ,
            education_graduate = z_dglev ,
            mar_num ,
            graduate_death_year = cmdeath ,
            spouse_death_year = cmspdth ,
            marriage_year = cmstart ,
            z_gu002re ,
            z_gu003re ,
            graduate_age_depressed = z_gu004re ,
            z_au002re ,
            z_au003re ,
            spouse_age_depressed = z_au004re ,
            edsphh ,
            gwiiq_bm ,
            ocssphu ,
            z_edhhyr ,
            z_ocsh57u2 ,
            z_aa001re
  )%>%
  mutate(spouse_age_death = 
           ifelse(spouse_death_year > 0 , 
                  spouse_death_year-(marriage_year - spouse_age_marriage), NA),
         graduate_age_death = 
           ifelse(graduate_death_year > 0 , 
                  graduate_death_year - graduate_birth_year , NA),
         spouse_dead = ifelse(spouse_death_year > 0 , 1 , 0) ,
         graduate_dead = ifelse(graduate_death_year >0 , 1 , 0) ,
         spouse_college = case_when (
           is.na(education_spouse)==1 ~ NA_real_,
           education_spouse <= 0 ~ NA_real_,
           education_spouse >=16 & education_spouse <= 26 ~ 1 ,
           education_spouse == 32 ~ 0 ,
           education_spouse == 35 ~ 0 ,
           education_spouse == 40 ~ 0 ,
           education_spouse >= 45 ~ 1 ,
           TRUE ~ 0
         ),
         graduate_college = case_when (
           is.na(education_graduate)~ NA_real_,
           education_graduate < 2 ~ 0 ,
           education_graduate >= 2 ~ 1
         ),
         graduate_depressed = case_when (
           is.na(z_gu002re)== 1 ~ NA_real_,
           z_gu002re < 0 ~ NA_real_,
           z_gu002re == 1 ~ 1 ,
           z_gu002re == 2 ~ 0 ,
           TRUE ~ 0
         ),
         spouse_depressed = case_when (
           is.na(z_au002re)== 1 ~ NA_real_,
           z_au002re < 0 ~ NA_real_,
           z_au002re == 1 ~ 1 ,
           z_au002re == 2 ~ 0 ,
           TRUE ~ 9999
         ),
         graduate_depressed_reason = case_when (
           is.na(z_gu003re)== 1 ~ NA_real_,
           z_gu003re == 1 | z_gu003re ==2 ~ 1 ,
           z_gu003re == 3 ~ 0 ,
           z_gu003re < 0 ~ 0 ,
           TRUE ~ 9999) ,
         spouse_depressed_reason = case_when (
           is.na(z_au003re)== 1 ~ NA_real_,
           z_au003re == 1 | z_au003re ==2 ~ 1 ,
           z_au003re == 3 ~ 0 ,
           z_au003re < 0 ~ 0 ,
           TRUE ~ 9999) ,
         educ_spouse_hoh = case_when (
           edsphh < 0 ~ NA_real_,
           edsphh >=0 & edsphh <= 11 ~ 0 ,
           edsphh >= 12 & edsphh < 16 ~ 1 ,
           edsphh >=16 ~ 2) ,
         gwiiq_bm ,
         duncan_sei_spouse = case_when (
           ocssphu < 0 ~ NA_real_,
           TRUE ~ ocssphu),
         educ_grad_hoh = case_when (
           z_edhhyr < 0 ~ NA_real_,
           z_edhhyr > 0 & z_edhhyr <=11 ~ 0 ,
           (z_edhhyr >=12 & z_edhhyr < 16) | z_edhhyr == 32 ~ 1 ,
           z_edhhyr >= 16 ~ 2) ,
         duncan_sei_grad = case_when (
           z_ocsh57u2 < 0 ~ NA_real_,
           TRUE ~ z_ocsh57u2),
         spouse_female = case_when (
           z_aa001re == 1 ~ 0 ,
           z_aa001re == 2 ~ 1 ,
           TRUE ~ NA_real_
         ))
analytic <- wls_data_select %>%
  filter(is.na(spouse_college)==0 & is.na(graduate_college)==0) %>%
  filter(is.na(marriage_year)==0) %>%
  filter(is.na(spouse_age_marriage)==0) %>%
  filter(is.na(graduate_depressed)==0 & is.na(spouse_depressed)==0) %>%
  filter(graduate_depressed_reason==0 & spouse_depressed_reason== 0)

# suppressMessages(analytic <- read_csv("WLS_analytic_25August2020.csv"))

analytic2 <- analytic %>% 
  # Down to 5477 dyads with the following filter, 
  # to ensure we have info on the exposure of interest
  filter(is.na(spouse_college)==0 & is.na(graduate_college)==0) %>%
  # Don't lose anyone with the following filter
  # --they all know when they were married!
  filter(is.na(marriage_year)==0) %>%
  # Everyone knows the age of their spouse when they were married too
  filter(is.na(spouse_age_marriage)==0) %>%
  # Down to 2,956 dyads with the following filter,
  # to ensure we have info on the outcome of interest
  filter(is.na(graduate_depressed)==0 & is.na(spouse_depressed)==0) %>%
  # Here ==0 means depressive symptoms not due to external factors 
  # (see #4 above). Down to 2,699 dyads.
  filter(graduate_depressed_reason == 0 & spouse_depressed_reason == 0)

analytic2 <- analytic2 %>%
  dplyr::select(graduate_depressed,
                spouse_depressed,
                graduate_college,
                educ_spouse_hoh,
                duncan_sei_spouse,
                educ_grad_hoh,
                duncan_sei_grad,
                spouse_female,
                grad_female,
                gwiiq_bm) %>%
  filter(complete.cases(.))

analytic2$duncan_sei_spouse <- scale(analytic2$duncan_sei_spouse)
analytic2$duncan_sei_grad   <- scale(analytic2$duncan_sei_grad)
analytic2$gwiiq_bm          <- scale(analytic2$gwiiq_bm)

analytic2$educ_spouse_hoh_M <- as.numeric(analytic2$educ_spouse_hoh==1)
analytic2$educ_spouse_hoh_H <- as.numeric(analytic2$educ_spouse_hoh==2)
analytic2$educ_spouse_hoh_B <- as.numeric(analytic2$educ_spouse_hoh>0)

analytic2$educ_grad_hoh_M <- as.numeric(analytic2$educ_grad_hoh==1)
analytic2$educ_grad_hoh_H <- as.numeric(analytic2$educ_grad_hoh==2)
analytic2$educ_grad_hoh_B <- as.numeric(analytic2$educ_grad_hoh>0)

analytic2 <- analytic2 %>%
  dplyr::select(graduate_depressed,
                spouse_depressed,
                graduate_college,
                educ_spouse_hoh_B,
                duncan_sei_spouse,
                educ_grad_hoh_B,
                duncan_sei_grad,
                spouse_female,
                grad_female,
                gwiiq_bm) %>%
  filter(complete.cases(.))

##################################################
# Define Functions
##################################################

expit      <- function(v){exp(v)/(1+exp(v))}

#####################
# Functions for MLEs
#####################

Likelihood.Threeway <- function(Y1,Y2,
                                A,
                                Cmat.f1,
                                Cmat.f2,
                                Cmat,
                                xi1.coef.cand,
                                xi2.coef.cand,
                                nu1.cand,
                                nu0.cand){
  
  Threeway.OR.under.A1 <- Cmat%*%nu1.cand
  Threeway.OR.under.A0 <- Cmat%*%nu0.cand
  
  Odds1 <- Cmat.f1%*%xi1.coef.cand
  Odds2 <- Cmat.f2%*%xi2.coef.cand
  
  
  denom.vector.under.A1 <- cbind(1,
                                 exp(Odds2),
                                 exp(Odds1),
                                 exp(Threeway.OR.under.A1+Odds1+Odds2))
  denom.under.A1 <- apply(denom.vector.under.A1,1,sum)
  numer.under.A1 <- rep(0,length(Y1))
  
  for(tt in 1:length(Y1)){
    pos <- 1+Y2[tt]+2*Y1[tt]
    numer.under.A1[tt] <- denom.vector.under.A1[tt,pos]
  }
  
  denom.vector.under.A0 <- cbind(1,
                                 exp(Odds2),
                                 exp(Odds1),
                                 exp(Threeway.OR.under.A0+Odds1+Odds2))
  denom.under.A0 <- apply(denom.vector.under.A0,1,sum)
  numer.under.A0 <- rep(0,length(Y1))
  
  for(tt in 1:length(Y1)){
    pos <- 1+Y2[tt]+2*Y1[tt]
    numer.under.A0[tt] <- denom.vector.under.A0[tt,pos]
  }
  
  -sum(log((numer.under.A1/denom.under.A1)*A+
             (numer.under.A0/denom.under.A0)*(1-A)))# -1*log-likelihood
  
} 

Effect.MLE <- function(Cmat.f1,
                       Cmat.f2,
                       xi1.MLE,
                       xi2.MLE,
                       OddsRatio.MLE,
                       a1,a2){
  
  Cmat.f1.view <- Cmat.f1
  Cmat.f2.view <- Cmat.f2
  
  Cmat.f1.view[,2] <- a1
  Cmat.f2.view[,2] <- a2
  
  Odds1.MLE <- Cmat.f1.view%*%xi1.MLE
  Odds2.MLE <- Cmat.f2.view%*%xi2.MLE
  
  denom.vector <- cbind(1,
                        exp(Odds2.MLE),
                        exp(Odds1.MLE),
                        exp(OddsRatio.MLE+Odds1.MLE+Odds2.MLE))
  
  numer.vector <- cbind(0,
                        0,
                        exp(Odds1.MLE),
                        exp(OddsRatio.MLE+Odds1.MLE+Odds2.MLE))
  
  denom <- apply(denom.vector,1,sum)
  numer <- apply(numer.vector,1,sum)
  
  mean(numer/denom)
}

####################################
# Functions for IF-based Estimators
####################################

propensity <- function(Cmat.T){
  if(is.null(dim(Cmat.T)[2])){  # Cmat=vector
    Cmat <- matrix(Cmat.T,length(Cmat.T),1)
  }  
  return(as.numeric(expit(cbind(1,Cmat.T)%*%propensity.coef)))
}

f1   <- function(A,Cmat.T,xi1.cand){ 
  if(is.null(dim(Cmat.T)[2])){  # Cmat=vector
    Cmat.T <- matrix(Cmat.T,length(Cmat.T),1)
  } 
  return(as.numeric(expit(cbind(1,A,Cmat.T)%*%xi1.cand)))
}

f2   <- function(A,Cmat.T,xi2.cand){ 
  if(is.null(dim(Cmat.T)[2])){  # Cmat=vector
    Cmat.T <- matrix(Cmat.T,length(Cmat.T),1)
  } 
  return(as.numeric(expit(cbind(1,A,Cmat.T)%*%xi2.cand)))
}

OR   <- function(Cmat.T,nu.cand){ 
  if(is.null(dim(Cmat.T)[2])){  # Cmat=vector
    Cmat.T <- matrix(Cmat.T,length(Cmat.T),1)
  } 
  return(as.numeric(exp(Cmat.T%*%nu.cand)))
} 

ProbY <- function(A,Cmat,xi1.cand,xi2.cand,nu.cand){
  ProbMat <- 
    cbind((1-f1(A,Cmat,xi1.cand))*(1-f2(A,Cmat,xi2.cand)),
          (1-f1(A,Cmat,xi1.cand))*(f2(A,Cmat,xi2.cand)),
          (f1(A,Cmat,xi1.cand))*(1-f2(A,Cmat,xi2.cand)),
          (f1(A,Cmat,xi1.cand))*(f2(A,Cmat,xi2.cand))*(OR(Cmat,nu.cand)))
  return(ProbMat)
} 

thetaft <- function(Cmat.f1, Cmat.f2, Cmat,
                    xi1.cand, xi2.cand, nu.cand,a1,a2){
  ProbY1.A0 <- f1(a1,Cmat.f1[,-c(1,2)],xi1.cand)
  ProbY2.A1 <- f2(a2,Cmat.f2[,-c(1,2)],xi2.cand)
  Denom <- (1-ProbY1.A0)*(1-ProbY2.A1) + (1-ProbY1.A0)*(ProbY2.A1) + 
    (ProbY1.A0)*(1-ProbY2.A1) + OR(Cmat,nu.cand)*(ProbY1.A0)*(ProbY2.A1)
  Numer <- (ProbY1.A0)*(1-ProbY2.A1) + OR(Cmat,nu.cand)*(ProbY1.A0)*(ProbY2.A1)
  Numer/Denom
}

Effect.DR <- function(Y1,
                      Y2,
                      A,
                      Cmat.f1,
                      Cmat.f2,
                      Cmat.f1.A0,
                      Cmat.f2.A0,
                      Cmat.f1.A1,
                      Cmat.f2.A1,
                      Cmat,
                      Cmat.PS,
                      nu.goodguess=NULL){
  
  ## In fact, we have a (near?) closed-representation:
  xi1.coef.crude <- 
    as.numeric(glm(Y1[Y2==0]~0+Cmat.f1[Y2==0,],family="binomial")$coefficients)
  xi2.coef.crude <- 
    as.numeric(glm(Y2[Y1==0]~0+Cmat.f2[Y1==0,],family="binomial")$coefficients)
  
  Fit.OR0  <- coef(nnet::multinom(as.factor((2*Y1+Y2)[A==0])~0+Cmat[A==0,]))
  Fit.OR1  <- coef(nnet::multinom(as.factor((2*Y1+Y2)[A==1])~0+Cmat[A==1,]))
  nu.crude <- 
    ((Fit.OR1[3,]-Fit.OR1[2,]-Fit.OR1[1,])+
       (Fit.OR0[3,]-Fit.OR0[2,]-Fit.OR0[1,]))/2
  
  nu.cand  <- nu.crude
  xi1.coef.cand <- xi1.coef.crude
  
  profile.likelihood1 <- function(Y1,Y2,Cmat.f1,nu.cand,xi1.coef.cand){
    
    Pr.Y1.0  <- 1-f1(Cmat.f1[,2],
                     Cmat.f1[,-c(1,2)],
                     xi1.coef.cand)
    Pr.Y1.1  <- 1-Pr.Y1.0
    OR.value <- OR(Cmat,nu.cand)
    
    Numer <- (Pr.Y1.0*(1-Y1)+Pr.Y1.1*(Y1))*(OR.value^(Y1*Y2))
    Denom <- Pr.Y1.0 + Pr.Y1.1*(OR.value^(1*Y2))
    
    return(mean(log(Numer/Denom)))
  }
  
  
  profile.likelihood1 <- function(Y1,Y2,Cmat.f1,nu.cand,xi1.coef.cand){
    
    Pr.Y1.0  <- 1-f1(Cmat.f1[,2],
                     Cmat.f1[,-c(1,2)],
                     xi1.coef.cand)
    Pr.Y1.1  <- 1-Pr.Y1.0
    OR.value <- OR(Cmat,nu.cand)
    
    Numer <- (Pr.Y1.0*(1-Y1)+Pr.Y1.1*(Y1))*(OR.value^(Y1*Y2))
    Denom <- Pr.Y1.0 + Pr.Y1.1*(OR.value^(1*Y2))
    
    return(mean(log(Numer/Denom)))
  }
  
  profile.likelihood2 <- function(Y1,Y2,Cmat.f2,nu.cand,xi2.coef.cand){
    
    Pr.Y2.0  <- 1-f2(Cmat.f2[,2],
                     Cmat.f2[,-c(1,2)],
                     xi2.coef.cand)
    Pr.Y2.1  <- 1-Pr.Y2.0
    OR.value <- OR(Cmat,nu.cand)
    
    Numer <- (Pr.Y2.0*(1-Y2)+Pr.Y2.1*(Y2))*(OR.value^(Y1*Y2))
    Denom <- Pr.Y2.0 + Pr.Y2.1*(OR.value^(Y1*1))
    
    return(mean(log(Numer/Denom)))
  }
  
  OddsRatio.Moment <- function(Y1,Y2,Cmat.f1,Cmat.f2,nu.cand){
    
    PMLE1 <- optim(par=c(xi1.coef.crude),
                   function(para){
                     -profile.likelihood1(Y1,Y2,Cmat.f1,
                                          nu.cand,
                                          para)
                   })$par
    
    PMLE2 <- optim(par=c(xi2.coef.crude),
                   function(para){
                     -profile.likelihood2(Y1,Y2,Cmat.f2,
                                          nu.cand,
                                          para)
                   })$par
    
    xi1.coef.PMLE <- PMLE1
    xi2.coef.PMLE <- PMLE2
    OR.value      <- OR(Cmat,
                        nu.cand)
    Residual      <- 
      as.numeric(((Y1 - expit(Cmat.f1%*%xi1.coef.PMLE))*
                    (Y2 - expit(Cmat.f2%*%xi2.coef.PMLE))*(OR.value^(-Y1*Y2))))
    
    return(sum(apply(Cmat*Residual,2,mean)^2))
    
  }
  
  if(!is.null(nu.goodguess)){
    nu.crude <-  nu.goodguess
    
    nu.MLE <- optim(par=nu.crude,
                    fn=function(nu.cand){
                      OddsRatio.Moment(Y1,Y2,Cmat.f1,Cmat.f2,nu.cand)
                    },method="CG")$par
    
  } else {
    
    nu.MLE <- optim(par=nu.crude,
                    fn=function(nu.cand){
                      OddsRatio.Moment(Y1,Y2,Cmat.f1,Cmat.f2,nu.cand)
                    })$par
    
    nu.MLE <- optim(par=nu.MLE,
                    fn=function(nu.cand){
                      OddsRatio.Moment(Y1,Y2,Cmat.f1,Cmat.f2,nu.cand)
                    })$par
    
    nu.MLE <- optim(par=nu.MLE,
                    fn=function(nu.cand){
                      OddsRatio.Moment(Y1,Y2,Cmat.f1,Cmat.f2,nu.cand)
                    },method="CG")$par
    
  }
  
  MLE1 <- optim(par=c(xi1.coef.crude),
                function(para){
                  -profile.likelihood1(Y1,Y2,Cmat.f1,
                                       nu.MLE,
                                       para)
                },method="CG")$par
  
  MLE2 <- optim(par=c(xi2.coef.crude),
                function(para){
                  -profile.likelihood2(Y1,Y2,Cmat.f2,
                                       nu.MLE,
                                       para)
                },method="CG")$par
  
  xi1.coef.MLE <- MLE1
  xi2.coef.MLE <- MLE2
  OR.value     <- OR(Cmat,nu.MLE)
  
  DERIV <- function(p1L,p2L,oL,nuL,Cmat.T){
    Numer <- exp(p1L+nuL+oL) + exp(p1L) + exp(oL) + 1
    Denom <- exp(p2L+nuL+oL) + exp(p2L) + exp(oL) + 1
    DNumer <- Cmat.T*(exp(p1L+nuL+oL)+exp(oL))
    DDenom <- Cmat.T*(exp(p2L+nuL+oL)+exp(oL))
    MAT <- (DNumer*Denom - Numer*DDenom)/(Denom^2)
    MAT/matrix(apply(MAT,2,sd),dim(MAT)[1],dim(MAT)[2],byrow=T)
  }
  
  ################################################ 
  # psi(0,1)
  ################################################ 
  
  xi1.A0.MLE <- expit(Cmat.f1.A0%*%xi1.coef.MLE)
  xi1.A1.MLE <- expit(Cmat.f1.A1%*%xi1.coef.MLE)
  xi2.A0.MLE <- expit(Cmat.f2.A0%*%xi2.coef.MLE)
  xi2.A1.MLE <- expit(Cmat.f2.A1%*%xi2.coef.MLE)
  
  gamma12.MLE <- OR.value^(Y1*Y2)
  f1.A0.MLE   <- xi1.A0.MLE*(Y1) + (1-xi1.A0.MLE)*(1-Y1)
  f2.A1.MLE   <- xi2.A1.MLE*(Y2) + (1-xi2.A1.MLE)*(1-Y2)
  f1.A1.MLE   <- xi1.A1.MLE*(Y1) + (1-xi1.A1.MLE)*(1-Y1)
  f2.A0.MLE   <- xi2.A0.MLE*(Y2) + (1-xi2.A0.MLE)*(1-Y2)
  gamma1A.MLE <- ((f1.A1.MLE/f1.A0.MLE)*((1-xi1.A0.MLE)/(1-xi1.A1.MLE)))^(A)
  gamma2A.MLE <- ((f2.A0.MLE/f2.A1.MLE)*((1-xi2.A1.MLE)/(1-xi2.A0.MLE)))^(1-A)
  
  propensity.fit <- glm(A~0+Cmat.PS,family="binomial")
  propensity.MLE <- expit(Cmat.PS%*%propensity.fit$coefficients)
  
  delta1.Moment <- function(delta1.xi2.A1.coef.cand){
    
    Cmat.1C <- Cmat.f1.A1[,-2]  
    
    xi21.temp <- as.numeric(Cmat.1C%*%delta1.xi2.A1.coef.cand)
    xi10.temp <- as.numeric(Cmat.f1.A0%*%xi1.coef.MLE)
    xi11.temp <- as.numeric(Cmat.f1.A1%*%xi1.coef.MLE)
    nu.temp   <- as.numeric(Cmat%*%nu.MLE)
    delta1    <- propensity.MLE*
      ((exp(xi10.temp+nu.temp+xi21.temp)+exp(xi10.temp)+exp(xi21.temp)+1)/
         (exp(xi11.temp+nu.temp+xi21.temp)+exp(xi11.temp)+exp(xi21.temp)+1))
    
    sum((apply(DERIV(xi10.temp,xi11.temp,xi21.temp,nu.temp,Cmat.1C)*
                 as.numeric(A/gamma1A.MLE - delta1),2,mean))^2)
  }
  
  delta1.xi2.A1.coef.MLE <- 
    optim(c(sum(xi2.coef.crude[1:2]), 
            rep(mean(xi2.coef.crude[-c(1,2)]),dim(Cmat.f1.A1)[2]-2)),
          delta1.Moment)$par
  
  xi21.temp  <- as.numeric(Cmat.f1.A1[,-2]%*%delta1.xi2.A1.coef.MLE)
  xi10.temp  <- as.numeric(Cmat.f1.A0%*%xi1.coef.MLE)
  xi11.temp  <- as.numeric(Cmat.f1.A1%*%xi1.coef.MLE)
  nu.temp    <- as.numeric(Cmat%*%nu.MLE)
  delta1.MLE <- propensity.MLE*
    ((exp(xi10.temp+nu.temp+xi21.temp)+exp(xi10.temp)+exp(xi21.temp)+1)/
       (exp(xi11.temp+nu.temp+xi21.temp)+exp(xi11.temp)+exp(xi21.temp)+1))
  
  delta2.Moment <- function(delta2.xi1.A0.coef.cand){
    
    Cmat.1C <- Cmat.f2.A1[,-2]  
    
    xi10.temp <- as.numeric(Cmat.1C%*%delta2.xi1.A0.coef.cand)
    xi21.temp <- as.numeric(Cmat.f2.A1%*%xi2.coef.MLE)
    xi20.temp <- as.numeric(Cmat.f2.A0%*%xi2.coef.MLE)
    nu.temp   <- as.numeric(Cmat%*%nu.MLE)
    delta2    <- (1-propensity.MLE)*
      ((exp(xi21.temp+nu.temp+xi10.temp)+exp(xi21.temp)+exp(xi10.temp)+1)/
         (exp(xi20.temp+nu.temp+xi10.temp)+exp(xi20.temp)+exp(xi10.temp)+1))
    
    sum((apply(DERIV(xi21.temp,xi20.temp,xi10.temp,nu.temp,Cmat.1C)*
                 as.numeric((1-A)/gamma2A.MLE - delta2),2,mean))^2)
  }
  
  delta2.xi1.A0.coef.MLE <- 
    optim(c(mean(xi1.coef.crude[1:2]), 
            rep(mean(xi1.coef.crude[-c(1,2)]),dim(Cmat.f2.A0)[2]-2)),
          delta2.Moment)$par
  
  xi10.temp  <- as.numeric(Cmat.f2.A1[,-2]%*%delta2.xi1.A0.coef.MLE)
  xi21.temp  <- as.numeric(Cmat.f2.A1%*%xi2.coef.MLE)
  xi20.temp  <- as.numeric(Cmat.f2.A0%*%xi2.coef.MLE)
  nu.temp    <- as.numeric(Cmat%*%nu.MLE)
  delta2.MLE <- (1-propensity.MLE)*
    ((exp(xi21.temp+nu.temp+xi10.temp)+exp(xi21.temp)+exp(xi10.temp)+1)/
       (exp(xi20.temp+nu.temp+xi10.temp)+exp(xi20.temp)+exp(xi10.temp)+1))
  
  theta.MLE   <- thetaft(Cmat.f1, Cmat.f2, Cmat,
                         xi1.coef.MLE, xi2.coef.MLE, nu.MLE,0,1)
  
  Q1.MLE      <- 
    Y1*(1-theta.MLE)*(OR.value*xi2.A1.MLE) + Y1*(1-theta.MLE)*(1-xi2.A1.MLE) + 
    (1-Y1)*(0-theta.MLE)*(xi2.A1.MLE) + (1-Y1)*(0-theta.MLE)*(1-xi2.A1.MLE) 
  Q2.MLE      <- 
    Y2*(1-theta.MLE)*(OR.value*xi1.A0.MLE) + Y2*(-theta.MLE)*(1-xi1.A0.MLE) + 
    (1-Y2)*(1-theta.MLE)*xi1.A0.MLE + (1-Y2)*(-theta.MLE)*(1-xi1.A0.MLE) 
  
  piece1 <- (Y1-theta.MLE)*(1-A)/delta2.MLE/gamma2A.MLE
  piece2 <- (A/(delta1.MLE*gamma1A.MLE)-
               (1-A)/(delta2.MLE*gamma2A.MLE))*(Q2.MLE/gamma12.MLE)
  piece3 <- (Y1-theta.MLE)*A/delta1.MLE/gamma1A.MLE
  piece4 <- ((1-A)/(delta2.MLE*gamma2A.MLE)-
               A/(delta1.MLE*gamma1A.MLE))*(Q1.MLE/gamma12.MLE)
  piece5 <- theta.MLE 
  
  w.opt <- optim(par=rep(0.5,dim(Cmat.PS)[2]),
                 fn=function(para){
                   weight <- Cmat.PS%*%para
                   
                   sd(weight*piece1+weight*piece2+
                        (1-weight)*piece3+(1-weight)*piece4)
                 },method="CG")
  
  weight <- Cmat.PS%*%w.opt$par 
  
  psiC.01 <- mean(weight*piece1+weight*piece2+
                    (1-weight)*piece3+(1-weight)*piece4+
                    piece5) 
  
  w.opt <- optim(par=0.5,
                 fn=function(para){
                   weight <- para
                   
                   sd(weight*piece1+weight*piece2+
                        (1-weight)*piece3+(1-weight)*piece4)
                 },method="CG")
  
  weight <- w.opt$par 
  
  psi.01 <- mean(weight*piece1+weight*piece2+
                   (1-weight)*piece3+(1-weight)*piece4+
                   piece5)
  
  ################################################ 
  # psi(1,0)
  ################################################  
  
  xi1.A0.MLE <- expit(Cmat.f1.A0%*%xi1.coef.MLE)
  xi1.A1.MLE <- expit(Cmat.f1.A1%*%xi1.coef.MLE)
  xi2.A0.MLE <- expit(Cmat.f2.A0%*%xi2.coef.MLE)
  xi2.A1.MLE <- expit(Cmat.f2.A1%*%xi2.coef.MLE)
  
  gamma12.MLE <- OR.value^(Y1*Y2)
  f1.A0.MLE   <- xi1.A0.MLE*(Y1) + (1-xi1.A0.MLE)*(1-Y1)
  f2.A1.MLE   <- xi2.A1.MLE*(Y2) + (1-xi2.A1.MLE)*(1-Y2)
  f1.A1.MLE   <- xi1.A1.MLE*(Y1) + (1-xi1.A1.MLE)*(1-Y1)
  f2.A0.MLE   <- xi2.A0.MLE*(Y2) + (1-xi2.A0.MLE)*(1-Y2)
  gamma1A.MLE <- ((f1.A0.MLE/f1.A1.MLE)*((1-xi1.A1.MLE)/(1-xi1.A0.MLE)))^(1-A)
  gamma2A.MLE <- ((f2.A1.MLE/f2.A0.MLE)*((1-xi2.A0.MLE)/(1-xi2.A1.MLE)))^(A)
  
  propensity.fit <- glm(A~0+Cmat.PS,family="binomial")
  propensity.MLE <- expit(Cmat.PS%*%propensity.fit$coefficients)
  
  delta1.Moment <- function(delta1.xi2.A0.coef.cand){
    
    Cmat.1C <- Cmat.f1.A0[,-2] 
    
    xi20.temp <- as.numeric(Cmat.1C%*%delta1.xi2.A0.coef.cand)
    xi10.temp <- as.numeric(Cmat.f1.A0%*%xi1.coef.MLE)
    xi11.temp <- as.numeric(Cmat.f1.A1%*%xi1.coef.MLE)
    nu.temp   <- as.numeric(Cmat%*%nu.MLE)
    delta1    <- (1-propensity.MLE)*
      ((exp(xi11.temp+nu.temp+xi20.temp)+exp(xi11.temp)+exp(xi20.temp)+1)/
         (exp(xi10.temp+nu.temp+xi20.temp)+exp(xi10.temp)+exp(xi20.temp)+1))
    
    sum((apply(DERIV(xi11.temp,xi10.temp,xi20.temp,nu.temp,Cmat.1C)*
                 as.numeric((1-A)/gamma1A.MLE - delta1),2,mean))^2)
  }
  
  delta1.xi2.A0.coef.MLE <- 
    optim(c(sum(xi2.coef.crude[1:2]), 
            rep(mean(xi2.coef.crude[-c(1,2)]),dim(Cmat.f1.A0)[2]-2)),
          delta1.Moment)$par
  
  xi20.temp  <- as.numeric(Cmat.f1.A0[,-2]%*%delta1.xi2.A0.coef.MLE)
  xi10.temp  <- as.numeric(Cmat.f1.A0%*%xi1.coef.MLE)
  xi11.temp  <- as.numeric(Cmat.f1.A1%*%xi1.coef.MLE)
  nu.temp    <- as.numeric(Cmat%*%nu.MLE)
  delta1.MLE <- (1-propensity.MLE)*
    ((exp(xi11.temp+nu.temp+xi20.temp)+exp(xi11.temp)+exp(xi20.temp)+1)/
       (exp(xi10.temp+nu.temp+xi20.temp)+exp(xi10.temp)+exp(xi20.temp)+1))
  
  delta2.Moment <- function(delta2.xi1.A1.coef.cand){
    
    Cmat.1C <- Cmat.f2.A0[,-2] 
    
    xi11.temp <- as.numeric(Cmat.1C%*%delta2.xi1.A1.coef.cand)
    xi21.temp <- as.numeric(Cmat.f2.A1%*%xi2.coef.MLE)
    xi20.temp <- as.numeric(Cmat.f2.A0%*%xi2.coef.MLE)
    nu.temp   <- as.numeric(Cmat%*%nu.MLE)
    delta2    <- (propensity.MLE)*
      ((exp(xi20.temp+nu.temp+xi11.temp)+exp(xi20.temp)+exp(xi11.temp)+1)/
         (exp(xi21.temp+nu.temp+xi11.temp)+exp(xi21.temp)+exp(xi11.temp)+1))
    
    sum((apply(DERIV(xi20.temp,xi21.temp,xi11.temp,nu.temp,Cmat.1C)*
                 as.numeric((A)/gamma2A.MLE - delta2),2,mean))^2)
  }
  
  delta2.xi1.A1.coef.MLE <- 
    optim(c(mean(xi1.coef.crude[1:2]), 
            rep(mean(xi1.coef.crude[-c(1,2)]),dim(Cmat.f2.A1)[2]-2)),
          delta2.Moment)$par
  
  xi11.temp  <- as.numeric(Cmat.f2.A0[,-2]%*%delta2.xi1.A1.coef.MLE)
  xi21.temp  <- as.numeric(Cmat.f2.A1%*%xi2.coef.MLE)
  xi20.temp  <- as.numeric(Cmat.f2.A0%*%xi2.coef.MLE)
  nu.temp    <- as.numeric(Cmat%*%nu.MLE)
  delta2.MLE <- (propensity.MLE)*
    ((exp(xi20.temp+nu.temp+xi11.temp)+exp(xi20.temp)+exp(xi11.temp)+1)/
       (exp(xi21.temp+nu.temp+xi11.temp)+exp(xi21.temp)+exp(xi11.temp)+1))
  
  theta.MLE   <- thetaft(Cmat.f1, Cmat.f2, Cmat,
                         xi1.coef.MLE, xi2.coef.MLE, nu.MLE,1,0)
  
  Q1.MLE      <- 
    Y1*(1-theta.MLE)*(OR.value*xi2.A0.MLE) + Y1*(1-theta.MLE)*(1-xi2.A0.MLE) + 
    (1-Y1)*(0-theta.MLE)*(xi2.A0.MLE) + (1-Y1)*(0-theta.MLE)*(1-xi2.A0.MLE) 
  Q2.MLE      <- 
    Y2*(1-theta.MLE)*(OR.value*xi1.A1.MLE) + Y2*(-theta.MLE)*(1-xi1.A1.MLE) + 
    (1-Y2)*(1-theta.MLE)*xi1.A1.MLE + (1-Y2)*(-theta.MLE)*(1-xi1.A1.MLE) 
  
  piece1 <- (Y1-theta.MLE)*(A)/delta2.MLE/gamma2A.MLE
  piece2 <- ((1-A)/(delta1.MLE*gamma1A.MLE)-
               (A)/(delta2.MLE*gamma2A.MLE))*(Q2.MLE/gamma12.MLE)
  piece3 <- (Y1-theta.MLE)*(1-A)/delta1.MLE/gamma1A.MLE
  piece4 <- ((A)/(delta2.MLE*gamma2A.MLE)-
               (1-A)/(delta1.MLE*gamma1A.MLE))*(Q1.MLE/gamma12.MLE)
  piece5 <- theta.MLE
  
  w.opt <- optim(par=rep(0.5,dim(Cmat.PS)[2]),
                 fn=function(para){
                   weight <- Cmat.PS%*%para
                   
                   sd(weight*piece1+weight*piece2+
                        (1-weight)*piece3+(1-weight)*piece4)
                 },method="CG")
  
  weight <- Cmat.PS%*%w.opt$par
  
  psiC.10 <- mean(weight*piece1+weight*piece2+
                    (1-weight)*piece3+(1-weight)*piece4+
                    piece5)
  
  w.opt <- optim(par=0.5,
                 fn=function(para){
                   weight <- para
                   
                   sd(weight*piece1+weight*piece2+
                        (1-weight)*piece3+(1-weight)*piece4)
                 },method="CG")
  
  weight <- w.opt$par
  
  psi.10 <- mean(weight*piece1+weight*piece2+
                   (1-weight)*piece3+(1-weight)*piece4+
                   piece5) 
  
  ################################################ 
  # psi(0,0)
  ################################################  
  
  xi1.A0.MLE <- expit(Cmat.f1.A0%*%xi1.coef.MLE)
  xi1.A1.MLE <- expit(Cmat.f1.A1%*%xi1.coef.MLE)
  xi2.A0.MLE <- expit(Cmat.f2.A0%*%xi2.coef.MLE)
  xi2.A1.MLE <- expit(Cmat.f2.A1%*%xi2.coef.MLE)
  
  gamma12.MLE <- OR.value^(Y1*Y2)
  f1.A0.MLE   <- xi1.A0.MLE*(Y1) + (1-xi1.A0.MLE)*(1-Y1)
  f2.A1.MLE   <- xi2.A1.MLE*(Y2) + (1-xi2.A1.MLE)*(1-Y2)
  f1.A1.MLE   <- xi1.A1.MLE*(Y1) + (1-xi1.A1.MLE)*(1-Y1)
  f2.A0.MLE   <- xi2.A0.MLE*(Y2) + (1-xi2.A0.MLE)*(1-Y2)
  gamma1A.MLE <- ((f1.A1.MLE/f1.A0.MLE)*((1-xi1.A0.MLE)/(1-xi1.A1.MLE)))^(A)
  gamma2A.MLE <- ((f2.A1.MLE/f2.A0.MLE)*((1-xi2.A0.MLE)/(1-xi2.A1.MLE)))^(A)
  
  propensity.fit <- glm(A~0+Cmat.PS,family="binomial")
  propensity.MLE <- expit(Cmat.PS%*%propensity.fit$coefficients)
  
  delta1.MLE <- (1-propensity.MLE)
  delta2.MLE <- (1-propensity.MLE)
  
  theta.MLE   <- thetaft(Cmat.f1, Cmat.f2, Cmat,
                         xi1.coef.MLE, xi2.coef.MLE, nu.MLE,0,0)
  
  Q1.MLE      <- 
    Y1*(1-theta.MLE)*(OR.value*xi2.A0.MLE) + Y1*(1-theta.MLE)*(1-xi2.A0.MLE) + 
    (1-Y1)*(0-theta.MLE)*(xi2.A0.MLE) + (1-Y1)*(0-theta.MLE)*(1-xi2.A0.MLE) 
  Q2.MLE      <- 
    Y2*(1-theta.MLE)*(OR.value*xi1.A0.MLE) + Y2*(-theta.MLE)*(1-xi1.A0.MLE) + 
    (1-Y2)*(1-theta.MLE)*xi1.A0.MLE + (1-Y2)*(-theta.MLE)*(1-xi1.A0.MLE) 
  
  piece1 <- (Y1-theta.MLE)*(1-A)/delta2.MLE/gamma2A.MLE
  piece2 <- ((1-A)/(delta1.MLE*gamma1A.MLE)-
               (1-A)/(delta2.MLE*gamma2A.MLE))*(Q2.MLE/gamma12.MLE)
  piece3 <- (Y1-theta.MLE)*(1-A)/delta1.MLE/gamma1A.MLE
  piece4 <- ((1-A)/(delta2.MLE*gamma2A.MLE)-
               (1-A)/(delta1.MLE*gamma1A.MLE))*(Q1.MLE/gamma12.MLE)
  piece5 <- theta.MLE
  
  w.opt <- optim(par=rep(0.5,dim(Cmat.PS)[2]),
                 fn=function(para){
                   weight <- Cmat.PS%*%para
                   
                   sd(weight*piece1+weight*piece2+
                        (1-weight)*piece3+(1-weight)*piece4)
                 },method="CG")
  
  weight <- Cmat.PS%*%w.opt$par
  
  psiC.00 <- mean(weight*piece1+weight*piece2+
                    (1-weight)*piece3+(1-weight)*piece4+
                    piece5) 
  
  w.opt <- optim(par=0.5,
                 fn=function(para){
                   weight <- para
                   
                   sd(weight*piece1+weight*piece2+
                        (1-weight)*piece3+(1-weight)*piece4)
                 },method="CG")
  
  weight <- w.opt$par
  
  psi.00 <- mean(weight*piece1+weight*piece2+
                   (1-weight)*piece3+(1-weight)*piece4+
                   piece5) 
  
  ################################################ 
  # psi(1,1)
  ################################################  
  
  xi1.A0.MLE <- expit(Cmat.f1.A0%*%xi1.coef.MLE)
  xi1.A1.MLE <- expit(Cmat.f1.A1%*%xi1.coef.MLE)
  xi2.A0.MLE <- expit(Cmat.f2.A0%*%xi2.coef.MLE)
  xi2.A1.MLE <- expit(Cmat.f2.A1%*%xi2.coef.MLE)
  
  gamma12.MLE <- OR.value^(Y1*Y2)
  f1.A0.MLE   <- xi1.A0.MLE*(Y1) + (1-xi1.A0.MLE)*(1-Y1)
  f2.A1.MLE   <- xi2.A1.MLE*(Y2) + (1-xi2.A1.MLE)*(1-Y2)
  f1.A1.MLE   <- xi1.A1.MLE*(Y1) + (1-xi1.A1.MLE)*(1-Y1)
  f2.A0.MLE   <- xi2.A0.MLE*(Y2) + (1-xi2.A0.MLE)*(1-Y2)
  gamma1A.MLE <- ((f1.A0.MLE/f1.A1.MLE)*((1-xi1.A1.MLE)/(1-xi1.A0.MLE)))^(1-A)
  gamma2A.MLE <- ((f2.A0.MLE/f2.A1.MLE)*((1-xi2.A1.MLE)/(1-xi2.A0.MLE)))^(1-A)
  
  propensity.fit <- glm(A~0+Cmat.PS,family="binomial")
  propensity.MLE <- expit(Cmat.PS%*%propensity.fit$coefficients)
  
  delta1.MLE <- (propensity.MLE)
  delta2.MLE <- (propensity.MLE)
  
  theta.MLE   <- thetaft(Cmat.f1, Cmat.f2, Cmat,
                         xi1.coef.MLE, xi2.coef.MLE, nu.MLE,1,1)
  
  Q1.MLE      <- 
    Y1*(1-theta.MLE)*(OR.value*xi2.A1.MLE) + Y1*(1-theta.MLE)*(1-xi2.A1.MLE) + 
    (1-Y1)*(0-theta.MLE)*(xi2.A1.MLE) + (1-Y1)*(0-theta.MLE)*(1-xi2.A1.MLE) 
  Q2.MLE      <- 
    Y2*(1-theta.MLE)*(OR.value*xi1.A1.MLE) + Y2*(-theta.MLE)*(1-xi1.A1.MLE) + 
    (1-Y2)*(1-theta.MLE)*xi1.A1.MLE + (1-Y2)*(-theta.MLE)*(1-xi1.A1.MLE) 
  
  piece1 <- (Y1-theta.MLE)*(A)/delta2.MLE/gamma2A.MLE
  piece2 <- ((A)/(delta1.MLE*gamma1A.MLE)-
               (A)/(delta2.MLE*gamma2A.MLE))*(Q2.MLE/gamma12.MLE)
  piece3 <- (Y1-theta.MLE)*(A)/delta1.MLE/gamma1A.MLE
  piece4 <- ((A)/(delta2.MLE*gamma2A.MLE)-
               (A)/(delta1.MLE*gamma1A.MLE))*(Q1.MLE/gamma12.MLE)
  piece5 <- theta.MLE
  
  w.opt <- optim(par=rep(0.5,dim(Cmat.PS)[2]),
                 fn=function(para){
                   weight <- Cmat.PS%*%para
                   
                   sd(weight*piece1+weight*piece2+
                        (1-weight)*piece3+(1-weight)*piece4)
                 },method="CG")
  
  weight <- Cmat.PS%*%w.opt$par
  
  psiC.11 <- mean(weight*piece1+weight*piece2+
                    (1-weight)*piece3+(1-weight)*piece4+
                    piece5)
  
  w.opt <- optim(par=0.5,
                 fn=function(para){
                   weight <- para
                   
                   sd(weight*piece1+weight*piece2+
                        (1-weight)*piece3+(1-weight)*piece4)
                 },method="CG")
  
  weight <- w.opt$par
  
  psi.11 <- mean(weight*piece1+weight*piece2+
                   (1-weight)*piece3+(1-weight)*piece4+
                   piece5) 
  
  c(psiC.00,psiC.01,psiC.10,psiC.11,
    nu.MLE)
  
}

##################################################
# Model Specification
##################################################

y1_model <- 
  as.formula(spouse_depressed~graduate_college+ 
               spouse_female+duncan_sei_spouse+educ_spouse_hoh_B)
y2_model <- 
  as.formula(graduate_depressed~graduate_college+gwiiq_bm+ 
               grad_female+duncan_sei_grad+educ_grad_hoh_B)

Cmat.f1.A0.input <- Cmat.f1.A1.input <- 
  Cmat.f1.input <- model.matrix(y1_model,data=analytic2)
Cmat.f2.A0.input <- Cmat.f2.A1.input <- 
  Cmat.f2.input <- model.matrix(y2_model,data=analytic2)

Cmat.f1.A0.input[,2] <- 0
Cmat.f2.A0.input[,2] <- 0

Cmat.f1.A1.input[,2] <- 1
Cmat.f2.A1.input[,2] <- 1

Cmat.input <- 
  model.matrix(~gwiiq_bm+grad_female+duncan_sei_grad+educ_grad_hoh_B+
                 duncan_sei_spouse+educ_spouse_hoh_B,data=analytic2)
Cmat.PS.NoIQ.input <- 
  model.matrix(~grad_female+duncan_sei_grad+educ_grad_hoh_B,data=analytic2) 
# Ego's IQ is omitted due to positivity violation


A.input  <- analytic2$graduate_college
Y1.input <- analytic2$spouse_depressed   # This is opposite to the paper
Y2.input <- analytic2$graduate_depressed # This is opposite to the paper
N  <- length(Y1.input)

Y1 <- Y1.input
Y2 <- Y2.input
A  <- A.input
Cmat.f1 <- Cmat.f1.input
Cmat.f2 <- Cmat.f2.input
Cmat.f1.A0 <- Cmat.f1.A0.input
Cmat.f2.A0 <- Cmat.f2.A0.input
Cmat.f1.A1 <- Cmat.f1.A1.input
Cmat.f2.A1 <- Cmat.f2.A1.input
Cmat <- Cmat.input
Cmat.PS.NoIQ <- Cmat.PS.NoIQ.input

##################################################
# Falsification Test
################################################## 

xi1.coef.crude <- 
  as.numeric(glm(Y1[Y2==0]~0+Cmat.f1[Y2==0,],family="binomial")$coefficients)
xi2.coef.crude <- 
  as.numeric(glm(Y2[Y1==0]~0+Cmat.f2[Y1==0,],family="binomial")$coefficients)

Fit.OR0  <- (coef(nnet::multinom(as.factor((2*Y1+Y2)[A==0])~0+Cmat[A==0,])))
Fit.OR1  <- (coef(nnet::multinom(as.factor((2*Y1+Y2)[A==1])~0+Cmat[A==1,])))
nu1.crude <- as.numeric(Fit.OR1[3,]-Fit.OR1[2,]-Fit.OR1[1,])
nu0.crude <- as.numeric(Fit.OR0[3,]-Fit.OR0[2,]-Fit.OR0[1,])  

OPTIM.Threeway <- 
  optim(par = c(xi1.coef.crude,
                xi2.coef.crude,
                nu1.crude,
                nu0.crude),
        function(par){
          xi1.coef.cand <- par[1:length(xi1.coef.crude)]
          xi2.coef.cand <- par[length(xi1.coef.crude)+
                                 1:length(xi2.coef.crude)]
          nu1.cand <- par[length(xi1.coef.crude)+
                            length(xi2.coef.crude)+
                            1:length(nu1.crude)]
          nu0.cand <- par[length(xi1.coef.crude)+
                            length(xi2.coef.crude)+
                            length(nu1.crude)+1:length(nu0.crude)]
          
          Likelihood.Threeway(Y1,Y2,A,Cmat.f1,Cmat.f2,Cmat,
                              xi1.coef.cand,
                              xi2.coef.cand,
                              nu1.cand,
                              nu0.cand)
        },
        method="CG")

OPTIM.Twoway <- 
  optim(par = c(xi1.coef.crude,
                xi2.coef.crude,
                (nu1.crude+nu0.crude)/2),
        function(par){
          xi1.coef.cand <- par[1:length(xi1.coef.crude)]
          xi2.coef.cand <- par[length(xi1.coef.crude)+
                                 1:length(xi2.coef.crude)]
          nu1.cand <- par[length(xi1.coef.crude)+
                            length(xi2.coef.crude)+
                            1:length(nu1.crude)]
          
          Likelihood.Threeway(Y1,Y2,A,Cmat.f1,Cmat.f2,Cmat,
                              xi1.coef.cand,
                              xi2.coef.cand,
                              nu1.cand,
                              nu1.cand)
        },
        method="CG") 

LogLik.Threeway <- 
  -Likelihood.Threeway(Y1,Y2,A,Cmat.f1,Cmat.f2,Cmat,
                       OPTIM.Threeway$par[1:length(xi1.coef.crude)],
                       OPTIM.Threeway$par[length(xi1.coef.crude)+
                                            1:length(xi2.coef.crude)],
                       OPTIM.Threeway$par[length(xi1.coef.crude)+
                                            length(xi2.coef.crude)+
                                            1:length(nu1.crude)],
                       OPTIM.Threeway$par[length(xi1.coef.crude)+
                                            length(xi2.coef.crude)+
                                            length(nu1.crude)+
                                            1:length(nu0.crude)])
LogLik.Twoway <- 
  -Likelihood.Threeway(Y1,Y2,A,Cmat.f1,Cmat.f2,Cmat,
                       OPTIM.Twoway$par[1:length(xi1.coef.crude)],
                       OPTIM.Twoway$par[length(xi1.coef.crude)+
                                          1:length(xi2.coef.crude)],
                       OPTIM.Twoway$par[length(xi1.coef.crude)+
                                          length(xi2.coef.crude)+
                                          1:length(nu1.crude)],
                       OPTIM.Twoway$par[length(xi1.coef.crude)+
                                          length(xi2.coef.crude)+
                                          1:length(nu1.crude)])

LL.stat <- -2*(LogLik.Twoway - LogLik.Threeway) # 4.795045
qchisq(0.95,df=length(nu1.crude)) # 14.06714
length(nu1.crude) # 7
pchisq(LL.stat,df=length(nu1.crude)) # 0.3150417

##################################################
# MLE
################################################## 

xi1.MLE <- OPTIM.Twoway$par[1:length(xi1.coef.crude)]
xi2.MLE <- OPTIM.Twoway$par[length(xi1.coef.crude)+
                              1:length(xi2.coef.crude)]
nu.MLE  <- OPTIM.Twoway$par[length(xi1.coef.crude)+
                              length(xi2.coef.crude)+
                              1:length(nu1.crude)]

OddsRatio.MLE <- Cmat%*%nu.MLE

psi.MLE.00 <- Effect.MLE(Cmat.f1, Cmat.f2, xi1.MLE, xi2.MLE, OddsRatio.MLE, 0,0)
psi.MLE.01 <- Effect.MLE(Cmat.f1, Cmat.f2, xi1.MLE, xi2.MLE, OddsRatio.MLE, 0,1)
psi.MLE.10 <- Effect.MLE(Cmat.f1, Cmat.f2, xi1.MLE, xi2.MLE, OddsRatio.MLE, 1,0)
psi.MLE.11 <- Effect.MLE(Cmat.f1, Cmat.f2, xi1.MLE, xi2.MLE, OddsRatio.MLE, 1,1)

EFF.MLE <- c(psi.MLE.00,
             psi.MLE.01,
             psi.MLE.10,
             psi.MLE.11) # 0.1636600 0.1680488 0.1439494 0.1479268

EFF.MLE[3] - EFF.MLE[1] #  -0.01971058
EFF.MLE[4] - EFF.MLE[3] #  0.003977379

##################################################
# IF-based Estimator
################################################## 

EFF.DR <- Effect.DR(Y1.input,
                    Y2.input,
                    A.input,
                    Cmat.f1.input,
                    Cmat.f2.input,
                    Cmat.f1.A0.input,
                    Cmat.f2.A0.input,
                    Cmat.f1.A1.input,
                    Cmat.f2.A1.input,
                    Cmat.input,
                    Cmat.PS.NoIQ.input,
                    nu.goodguess = NULL)

EFF.DR[1:4] # 0.1622819 0.1657524 0.1504987 0.1594098 
EFF.DR[3]-EFF.DR[1] # -0.01178319 
EFF.DR[4]-EFF.DR[3] # 0.008911084 

##################################################
# Bootstrap (change Num.Boot to a larger numer)
################################################## 

Num.Boot <- 2
Boot.Est <- matrix(0,Num.Boot,8)

for(boot in 1:Num.Boot){
  pos <- sample(1:N,N,replace=T)
  
  Boot.OPTIM.Twoway <- 
    optim(par = c(xi1.MLE,
                  xi2.MLE,
                  nu.MLE),
          function(par){
            xi1.coef.cand <- par[1:length(xi1.coef.crude)]
            xi2.coef.cand <- par[length(xi1.coef.crude)+
                                   1:length(xi2.coef.crude)]
            nu1.cand <- par[length(xi1.coef.crude)+
                              length(xi2.coef.crude)+
                              1:length(nu1.crude)]
            
            Likelihood.Threeway(Y1[pos],
                                Y2[pos],
                                A[pos],
                                Cmat.f1[pos,],
                                Cmat.f2[pos,],
                                Cmat[pos,],
                                xi1.coef.cand,
                                xi2.coef.cand,
                                nu1.cand,
                                nu1.cand)
          },
          method="CG")
  
  Boot.xi1.MLE <- Boot.OPTIM.Twoway$par[1:length(xi1.coef.crude)]
  Boot.xi2.MLE <- Boot.OPTIM.Twoway$par[length(xi1.coef.crude)+
                                          1:length(xi2.coef.crude)]
  Boot.nu.MLE  <- Boot.OPTIM.Twoway$par[length(xi1.coef.crude)+
                                          length(xi2.coef.crude)+
                                          1:length(nu1.crude)]
  
  Boot.OR.MLE <- Cmat[pos,]%*%Boot.nu.MLE
  
  ## (0,0)
  Boot.psi.MLE.00 <- 
    Effect.MLE(Cmat.f1[pos,], Cmat.f2[pos,], 
               Boot.xi1.MLE, Boot.xi2.MLE, Boot.OR.MLE, 0,0)
  Boot.psi.MLE.01 <- 
    Effect.MLE(Cmat.f1[pos,], Cmat.f2[pos,], 
               Boot.xi1.MLE, Boot.xi2.MLE, Boot.OR.MLE, 0,1)
  Boot.psi.MLE.10 <- 
    Effect.MLE(Cmat.f1[pos,], Cmat.f2[pos,], 
               Boot.xi1.MLE, Boot.xi2.MLE, Boot.OR.MLE, 1,0)
  Boot.psi.MLE.11 <- 
    Effect.MLE(Cmat.f1[pos,], Cmat.f2[pos,], 
               Boot.xi1.MLE, Boot.xi2.MLE, Boot.OR.MLE, 1,1)
  
  Boot.MLE <- c(Boot.psi.MLE.00,
                Boot.psi.MLE.01,
                Boot.psi.MLE.10,
                Boot.psi.MLE.11)
  
  Boot.IF <- Effect.DR(Y1.input[pos],
                       Y2.input[pos],
                       A.input[pos],
                       Cmat.f1.input[pos,],
                       Cmat.f2.input[pos,],
                       Cmat.f1.A0.input[pos,],
                       Cmat.f2.A0.input[pos,],
                       Cmat.f1.A1.input[pos,],
                       Cmat.f2.A1.input[pos,],
                       Cmat.input[pos,],
                       Cmat.PS.NoIQ.input[pos,],
                       nu.goodguess = NULL)
  
  Boot.Est[boot,] <- c(Boot.MLE,Boot.IF[1:4])
}

##################################################
# Summary
################################################## 

EFF.DE.MLE <- EFF.MLE[3] - EFF.MLE[1]
EFF.DE.DR  <- EFF.DR[3]  - EFF.DR[1]

EFF.IE.MLE <- EFF.MLE[4] - EFF.MLE[3]
EFF.IE.DR  <- EFF.DR[4]  - EFF.DR[3]

Boot.MLE.Mat <- Boot.Est[,1:4]
Boot.DR.Mat  <- Boot.Est[,4+1:4]

Boot.DE.MLE <- Boot.MLE.Mat[,3]-Boot.MLE.Mat[,1]
Boot.IE.MLE <- Boot.MLE.Mat[,4]-Boot.MLE.Mat[,3]

Boot.DE.DR  <- Boot.DR.Mat[,3] -Boot.DR.Mat[,1]
Boot.IE.DR  <- Boot.DR.Mat[,4] -Boot.DR.Mat[,3]

paste( sprintf("%0.4f (%0.4f, %0.4f)", 
               EFF.DE.MLE, 
               quantile(Boot.DE.MLE,0.025), 
               quantile(Boot.DE.MLE,0.975)), " & ",
       sprintf("%0.4f (%0.4f, %0.4f)", 
               EFF.DE.DR, 
               quantile(Boot.DE.DR,0.025), 
               quantile(Boot.DE.DR,0.975)) )

paste( sprintf("%0.4f (%0.4f, %0.4f)", 
               EFF.IE.MLE, 
               quantile(Boot.IE.MLE,0.025), 
               quantile(Boot.IE.MLE,0.975)), " & ",
       sprintf("%0.4f (%0.4f, %0.4f)", 
               EFF.IE.DR, 
               quantile(Boot.IE.DR,0.025), 
               quantile(Boot.IE.DR,0.975)))

\end{lstlisting}

\clearpage

\bibliography{references}

\end{document}